\documentclass[a4paper,10pt,dvipsnames]{article}
\usepackage[utf8]{inputenc}
\usepackage[english]{babel}
\usepackage{amsthm}
\usepackage{amsfonts}
\usepackage{amssymb}	
\usepackage{amsmath}
\allowdisplaybreaks
\usepackage{chemarr}
\usepackage{slashed}
\usepackage{enumerate}
\usepackage{graphicx}
\usepackage{systeme}
\usepackage{dsfont}
\usepackage{relsize}
\usepackage[
  top=1.35cm,
  bottom=1.35cm,
  left=1.35cm,
  right=1.35cm
]{geometry}
\usepackage{float}
\usepackage{tikz}
\usepackage[labelformat=simple]{subcaption}

\usepackage{bmpsize}
\usepackage{epstopdf}
\usepackage{bbm}
\usepackage{etoolbox}
\usepackage{titling}
\newcommand*\samethanks[1][\value{footnote}]{\footnotemark[#1]}
\usepackage{blindtext,graphicx}
\usepackage[absolute]{textpos}
\usepackage[hang,flushmargin]{footmisc}

\usepackage{xfrac}
\usepackage{nicefrac}
\usepackage{esvect}
\usepackage{cases}
\usepackage{empheq}
\usepackage{stmaryrd}
\usepackage{cancel}
\usepackage[linguistics]{forest}
\usepackage{url}
\usepackage[font=small]{caption}
\usepackage{array}
\usepackage{dsfont}
\usepackage{bbold}
\usepackage[most]{tcolorbox}
\usepackage{dashrule}
\usepackage{pifont}
\usepackage{fancybox}
\usepackage{fontawesome}
\usepackage{setspace}
\usepackage{diagbox}
\usepackage{booktabs}
\usepackage{mathabx}
\usepackage{multirow}
\usepackage[locale=US,exponent-product=\cdot,per-mode=fraction]{siunitx}

\renewcommand{\.}{\hspace*{0.07em}}

\definecolor{myred1}{HTML}{c0392b}
\definecolor{mygreen1}{HTML}{27ae60}
\definecolor{mygreen2}{HTML}{16a085}
\definecolor{myblue1}{HTML}{2980b9}
\newlength{\figsizeA}
\newlength{\figsizeB}
\newlength{\figsizeC}
\newlength{\figsizeD}
\newlength{\figsizeE}
\newlength{\figsizeF}
\makeatletter
\newcommand{\thickhline}{%
\noalign {\ifnum 0=`}\fi \hrule height 2pt
\futurelet \reserved@a \@xhline
}
\newcolumntype{"}{@{\hskip\tabcolsep\vrule width 2pt\hskip\tabcolsep}}
\makeatother

\usepackage{rotating}

\usepackage[numbers]{natbib}  

\title{Time domain analysis of microstructured materials through the reduced relaxed micromorphic model}
\author{
Gianluca Rizzi\.\thanks{Faculty of Architecture and Civil Engineering, TU Dortmund, August-Schmidt-Str. 8, 44227 Dortmund, Germany}
\quad and \quad
Angela Madeo\.\samethanks[1]
}
\thanksmarkseries{arabic}
\date{\today}
\begin{document}
\maketitle
\vspace{-10pt}
\begin{abstract}
    Microstructured materials, such as architected metamaterials and phononic crystals, exhibit complex wave propagation phenomena due to their internal structure.
    While full-scale numerical simulations can capture these effects, they are computationally demanding, especially in time-domain analyses.
    To overcome this limitation, effective continuum models have been developed to approximate the macroscopic behavior of these materials while retaining key microscale effects.
    In this work we investigate the time-domain dynamic response of microstructured materials and focus on their effective micromorphic counterparts.
    We compare direct numerical simulations of discrete microstructures with predictions from micromorphic models to assess their accuracy in capturing transient wave phenomena.
    Our findings provide new insights into the applicability and limitations of micromorphic models in time-dependent analyses, contributing to the development of improved predictive tools for metamaterial design and engineering applications.
\end{abstract}

%
%
%
%
%
%
\section{Introduction}

Microstructured materials, characterized by internal architectures at the micro- or meso-scale, exhibit mechanical responses that deviate significantly from those predicted by classical continuum theories.
These materials include composites, lattice structures, metamaterials, and architected solids, whose responses to dynamic loading are governed not only by the properties of the base materials, but also by the geometry and arrangement of their microco-components.
Capturing the response of such complex material requires a modeling framework that surpass conventional elasticity, particularly in the dynamics of wave propagation scenarios \cite{hill1963elastic,chen2001dispersive,boutin2014large,craster2010high,andrianov2008higher,hu2017nonlocal,willis2009exact,willis2011effective,willis2012construction,srivastava2014limit,sridhar2018general,srivastava2017evanescent,schwan2021extended,lakes2023experimental,lakes1987foam,miniaci2016spider,krushynska2018labyrinthine,krushynska2017spider}.
Among the most promising generalized continuum theories is the micromorphic model, which introduces additional degrees of freedom to describe micro-deformations independently of the classical macroscopic displacement.
This enriched kinematic description enables the micromorphic theory to incorporate the effects of micro-inertia, micro-stiffness, and internal length scales, offering a powerful homogenized representation of microstructured media under dynamic loading.
The specific model we use in this work is the so-called Reduced Relaxed Micromorphic Model (RRMM), which has been already proved succesfully in several works \cite{demetriou2024reduced,rizzi2021exploring,rizzi2022boundary,rizzi2022metamaterial,rizzi2022towards,ramirez2023multi,barbagallo2019relaxed,demore2022unfolding}.
The present study investigates the time-domain dynamic behavior of microstructured materials and their effective micromorphic counterparts. Emphasis is placed on the transient response, including wave propagation and dispersive phenomena, analysed through direct numerical simulations. The objective is twofold: to assess the ability of the micromorphic model to replicate the behavior of the underlying microstructure, and to identify the regimes where such effective descriptions remain valid. By establishing quantitative comparisons between the detailed microstructural model and its micromorphic equivalent, this work provides insights into the limits and fidelity of micromorphic homogenization in the context of time-domain dynamics.

\section{An effective micromorphic continuum model}
\label{sec:RRMM}
Here we introduce the equilibrium equations, the associated boundary conditions, and the constitutive relations for a reduced version for dynamic applications \cite{rizzi2021exploring,demore2022unfolding,rizzi2022metamaterial,rizzi2022boundary,rizzi2022towards,voss2023modeling,ramirez2023multi,ramirez2024effective,demetriou2024reduced,demetriou2025effective} of the relaxed micromorphic model (RRMM) \cite{neff2014unifying,voss2023modeling,gourgiotis2024green}.
The kinetic ($K$) and elastic ($W$) energy densities are defined, respectively, as
\begin{align}
    K (\dot{u},\nabla \dot{u}, \dot{P})
    \coloneqq
     & \,
    \dfrac{1}{2}\rho \, \langle \dot{u},\dot{u} \rangle
    + \dfrac{1}{2} \langle \mathbb{J}_{\rm m}  \, \text{sym} \, \dot{P}, \text{sym} \, \dot{P} \rangle
    + \dfrac{1}{2} \langle \mathbb{J}_{\rm c} \, \text{skew} \, \dot{P}, \text{skew} \, \dot{P} \rangle
    \notag
    \\*
     &
    + \dfrac{1}{2} \langle \mathbb{T}_{\rm e} \, \text{sym}\nabla \dot{u}, \text{sym}\nabla \dot{u} \rangle
    + \dfrac{1}{2} \langle \mathbb{T}_{\rm c} \, \text{skew}\nabla \dot{u}, \text{skew}\nabla \dot{u} \rangle
    \, ,
    \label{eq:Energy_RRMM}
    \\
    W (\nabla u, P)
    \coloneqq
     & \,
    \dfrac{1}{2} \langle \mathbb{C}_{\rm e} \, \text{sym}\left(\nabla u -  \, P \right), \text{sym}\left(\nabla u -  \, P \right) \rangle
    + \dfrac{1}{2} \langle \mathbb{C}_{\rm c} \, \text{skew}\left(\nabla u -  \, P \right), \text{skew}\left(\nabla u -  \, P \right) \rangle
    \notag
    \\*
     &
    + \dfrac{1}{2} \langle \mathbb{C}_{\rm m} \, \text{sym}  \, P,\text{sym}  \, P \rangle
    \, ,
    \notag
\end{align}
where $\langle\cdot , \cdot\rangle$ denote the scalar product, the dot represents a time derivative, $u \in \mathbb{R}^{3}$ is the macroscopic displacement field, $P \in \mathbb{R}^{3\times 3}$ is the non-symmetric micro-distortion tensor, $\rho$ is the macroscopic apparent density, $\mathbb{J}_{\rm m}$, $\mathbb{J}_{\rm c}$, $\mathbb{T}_{\rm e}$, $\mathbb{T}_{\rm c}$, are 4th order micro-inertia tensors, and $\mathbb{C}_{\rm e}$, $\mathbb{C}_{\rm m}$, $\mathbb{C}_{\rm c}$ are 4th order elasticity tensors.
In the remainder of this work, we will focus on problems under the plane strain assumption and consider all the micro-inertia and elasticity tensors to possess tetragonal symmetry.
These assumptions reduce the number of independent parameters for each fourth-order tensor to three for those with minor symmetry ($\mathbb{J}_{\rm m}$, $\mathbb{T}_{\rm e}$, $\mathbb{C}_{\rm e}$, $\mathbb{C}_{\rm m}$), and to one for those with minor antisymmetry ($\mathbb{J}_{\rm c}$, $\mathbb{T}_{\rm c}$, $\mathbb{C}_{\rm c}$). For more details we refer to \cite{voss2023modeling,ramirez2024effective,demetriou2024reduced}.
Using a variational approach while considering the independent kinematical fields $u$ and $P$, it is possible to uniquely define both the equilibrium equations (without volume forces) and the boundary conditions in strong form as
\begin{align}
    \text{Div}\,\widetilde{\sigma} -\rho\,\ddot{u} + \text{Div}\,\widehat{\sigma} = 0
    \qquad\qquad
    \text{and}
    \qquad\qquad
    \widetilde{\sigma} - s - \overline{\sigma} = 0
     &  & \text{in} \,\, \Omega
    \, ,
    \label{eq:equi_equa_RRMM}
\end{align}
where
\begin{align}
    \widetilde{\sigma}
     &
    \coloneqq
    \mathbb{C}_{\rm e}\,\text{sym}(\nabla u-P) + \mathbb{C}_{\rm c}\,\text{skew}(\nabla u-P)
    \, ,
     &
    \widehat{\sigma}
     &
    \coloneqq
    \mathbb{T}_{\rm e}\,\text{sym} \, \nabla\ddot{u} + \mathbb{T}_{\rm c}\,\text{skew} \, \nabla\ddot{u}
    \,,
    \label{eq:equiSigAll_RRMM}
    \\*[5pt]
    s
     &
    \coloneqq
    \mathbb{C}_{\rm m}\, \text{sym} \, P
    \, ,
    \qquad\qquad\qquad
     &
    \overline{\sigma}
     &
    \coloneqq
    \mathbb{J}_{\rm m}\,\text{sym} \, \ddot{P} + \mathbb{J}_{\rm c}\,\text{skew} \, \ddot{P}
    \, .
\end{align}
The associated homogeneous Neumann boundary conditions are
\begin{align}
    \widetilde{t} \coloneqq \left(\widetilde{\sigma} + \widehat{\sigma} \right) \, n = 0
    \qquad\qquad\qquad\qquad
    \text{on} \,\, \partial\Omega \, ,
    \label{eq:traction_RRMM}
\end{align}
where $\widetilde{t}$ are the generalized traction and $n$ is the normal to the boundary.

\subsection{A recall to the classical Cauchy model}
\label{sec:cau}

We briefly recall the equilibrium equations, the associated homogeneous Neumann boundary conditions, and the constitutive relations for the classical Cauchy model
\begin{align}
    \text{Div} \, \sigma - \rho\,\ddot{u} = 0  \qquad \text{in} \,\, \Omega
    \,,
     &  &
    t \coloneqq \sigma \, n = 0  \qquad \text{on} \,\, \partial \Omega
    \,,
     &  &
    \sigma \coloneqq \mathbb{C}\, \text{sym}\nabla u
    \,,
    \label{eq:equi_equa_trac_cau}
\end{align}
where $\mathbb{C}$ is the classical 4th order elasticity tensor, $\rho$ is the density, and $t$ is the classical traction associated with the homogeneous Neumann boundary conditions.
For a tetragonal class of symmetry and under the plane strain hypothesis, $\mathbb{C}$ end up depending only on three parameters (see \cite{voss2023modeling,ramirez2024effective,demetriou2024reduced} for more details).
Within the long-wavelength limit, the RRMM reduces to a Cauchy model that we identify with the subscript M to indicate the ``macro'' scale at which it operates ($\mathbb{C}_{\rm M}$).
In this limit we can express the macro elasticity tensor as a function of the elasticity tensors of the RRMM as (see \cite{voss2023modeling,rizzi2021exploring} for more details)
\begin{align}
    \mathbb{C}_{\rm M}
    =
    \mathbb{C}_{\rm m}
    (\mathbb{C}_{\rm e} + \mathbb{C}_{\rm m})^{-1}
    \mathbb{C}_{\rm e}
    \, .
    \label{eq:macro_relation}
\end{align}

A Cauchy continuum with elasticity tensor $\mathbb{C}_{\rm M}$ can thus be seen as the long-wavelength limit of the RRMM.

\section{A \textit{labyrinthine} and a \textit{four-resonator} metamaterials}

In this work we explore metamaterials featuring labyrinthine (unit cell $\mathcal{L}$) \cite{voss2023modeling,hermann2024design} and four-resonator (unit cell $\mathcal{R}$) \cite{demetriou2024reduced,demore2022unfolding,ramirez2024effective,rizzi2021exploring,rizzi2022towards,rizzi2022metamaterial} unit cells, which are designed to manipulate wave propagation through tailored geometries and resonant mechanisms.
The labyrinthine metamaterial (metamaterial $\mathcal{L}$) allows wave control via complex pathing (Bragg scattering), while the four-resonator metamaterial (metamaterial $\mathcal{R}$) exploits multiple resonance effects to achieve tunable dynamic properties (local resonances).
These designs offer insights into advanced metamaterial engineering for applications in vibration attenuation and wave guiding.
In this work we commit to investigate the bulk response of such metamaterials: to isolate the intrinsic metamaterial behavior, outer boundary effects are minimized by limiting the simulation time such that wave reflections from the boundaries are either avoided or kept negligible.

\subsection{A \textit{labyrinthine} unit cell}

\begin{figure}[H]
    \centering
    \begin{minipage}[m]{0.4\textwidth}
        \centering
        {\renewcommand{\arraystretch}{1.1}
            \begin{tabular}{|c|c|c|}
                \hline
                $\kappa$ [GPa]    & $\mu$ [GPa] & $\nu$ [-] \\ \hline
                2.42              & 0.52        & 0.4       \\ \hline
                $\rho$ [kg/m$^3$] & $a$ [mm]    & $b$ [mm]  \\ \hline
                1230              & 50          & 1.25      \\ \hline
            \end{tabular}
        }
    \end{minipage}
    \hspace{0.75cm}
    \centering
    \begin{minipage}[m]{\figsizeA}
        \centering
        \includegraphics[width=0.85\textwidth]{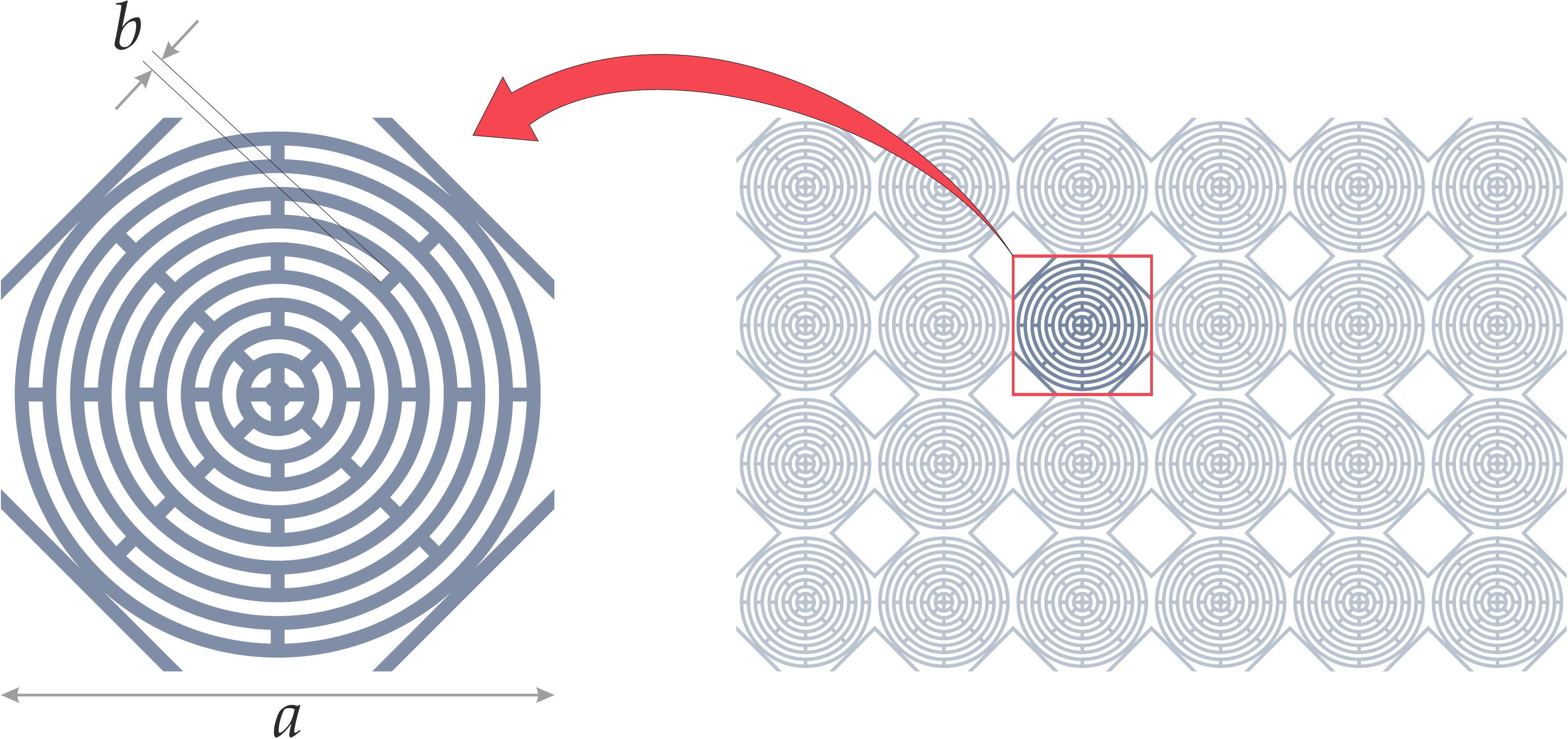}
    \end{minipage}
    \caption{Geometry and material parameters for the unit cell $\mathcal{L}$.}
    \label{fig:maze_unit_cell}
\end{figure}

In Fig.~\ref{fig:maze_unit_cell}, it is possible to find the geometry and the material properties of the unit cell $\mathcal{L}$ we test in Section \ref{sec:maze}.

In Table \ref{tab:coeff_maze}, we list the material parameters that characterize the equivalent RRMM and its longwavelenght limit macro Cauchy.

\begin{table}[H]
    \centering
    {\renewcommand{\arraystretch}{1.1}
        \begin{tabular}{cc|c|c|c|c|cc}
            \hline
            \multicolumn{1}{|c|}{$\rho$ [kg/m$^3$]}              & $\kappa_{\rm m}$ [Pa]   & $\mu_{\rm m}$ [Pa]        & $\mu^{*}_{\rm m}$ [Pa]  & $\kappa_{\rm e}$ [Pa]  & $\mu_{\rm e}$ [Pa]                      & $\mu^{*}_{\rm e}$ [Pa] & \multicolumn{1}{|c|}{$\mu_{\rm c}$ [Pa]}
            \\\hline
            \multicolumn{1}{|c|}{$541$}                          & $5.81 \times 10^6$      & $5.13 \times 10^8$        & $5.13 \times 10^8$      & $2.56 \times 10^6$     & $1.18 \times 10^6$                      & $5.38 \times 10^5$     & \multicolumn{1}{|c|}{$243$}
            \\\hline
            \multicolumn{1}{|c|}{$\overline{\kappa}_{\eta}$ [-]} & $\overline{\eta}_1$ [-] & $\overline{\eta}^{*}$ [-] & $\overline{\eta}_2$ [-] & $ \kappa_{\eta}$ [-]   & $\eta_1$ [-]                            & $\eta^{*}$ [-]         & \multicolumn{1}{|c|}{$\eta_2$ [-]}
            \\\hline
            \multicolumn{1}{|c|}{$0.09$}                         & $0.07$                  & $0.04$                    & $0$                     & $0.07$                 & $4.18$                                  & $4.17$                 & \multicolumn{1}{|c|}{$2.72 \times 10^{-5}$}
            \\\hline
                                                &        & $\kappa_{\rm M}$ [Pa]     & $\mu_{\rm M}$ [Pa]      & $\mu^{*}_{\rm M}$ [Pa] &  $L_{\rm c} = a$ [m] &       &
            \\\cline{3-6}

                                                                 &                         & $1.78 \times 10^{6}$      & $1.18 \times 10^{6}$    & $5.37 \times 10^{5}$   & 0.05                                    &                        &
            \\\cline{3-6}
        \end{tabular}
    }
    \caption{Material parameters for the metamaterial $\mathcal{L}$ effective RRMM and macro Cauchy}
    \label{tab:coeff_maze}
\end{table}

\begin{figure}[H]
    \centering
    \includegraphics[width=\linewidth]{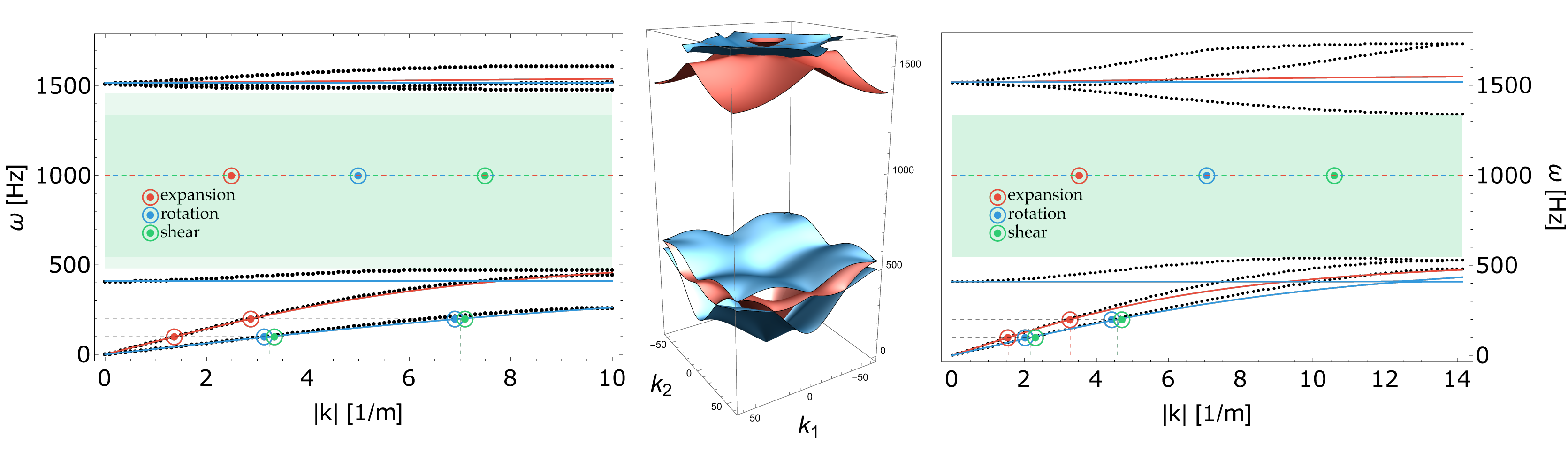}
    \caption{
        (\textit{left}) Dispersion diagram for a direction of propagation along the $x_1$ axis ($k_2=0$);
        (\textit{center}) dispersion diagram for all direction of propagation;
        (\textit{right}) dispersion diagram for a direction of propagation parallel to the $x_1 = x_2$ axis ($k_1=k_2$).
        The dotted lines come from a Bloch-Floquet analysis done on the unit cell $\mathcal{L}$, while the solid lines (blue for shear mdoes and red for pressure modes) from the RRMM with the parameters of Table~\ref{tab:coeff_maze}.
    }
    \label{fig:maze_disp}
\end{figure}

On the left and right panels of Fig.~\ref{fig:maze_disp}, we display the dispersion curves for a direction of propagation along the $x_1$ axis ($k_2=0$) and for a direction of propagation parallel to the $x_1 = x_2$ axis ($k_1=k_2$), respectively, superimposing the ones from the RRMM (solid lines obtined with the parameters form Table~\ref{tab:coeff_maze}) to the one coming from the unit cell $\mathcal{L}$ (solid lines).
In the middle of the same figure, we also report the dispersion diagram for all the direction of propagation coming from the unit cell $\mathcal{L}$.

\subsection{A \textit{four-resonator} unit cell}

\begin{figure}[H]
    \centering
    \begin{minipage}[m]{0.4\textwidth}
        \centering
        {\renewcommand{\arraystretch}{1.1}
            \begin{tabular}{|c|c|c|}
                \hline
                $\kappa$ [GPa]    & $\mu$ [GPa] & $\nu$ [-]        \\ \hline
                2.42              & 0.52        & 0.4              \\ \hline
                $\rho$ [kg/m$^3$] & $a$ [mm]    & $e_g~|~e_p$ [mm] \\ \hline
                1230              & 20          & 0.35~$|$~0.25    \\ \hline
            \end{tabular}
        }
    \end{minipage}%
    \hspace{0.75cm}
    \centering
    \begin{minipage}[m]{\figsizeA}
        \centering
        \includegraphics[width=0.85\textwidth]{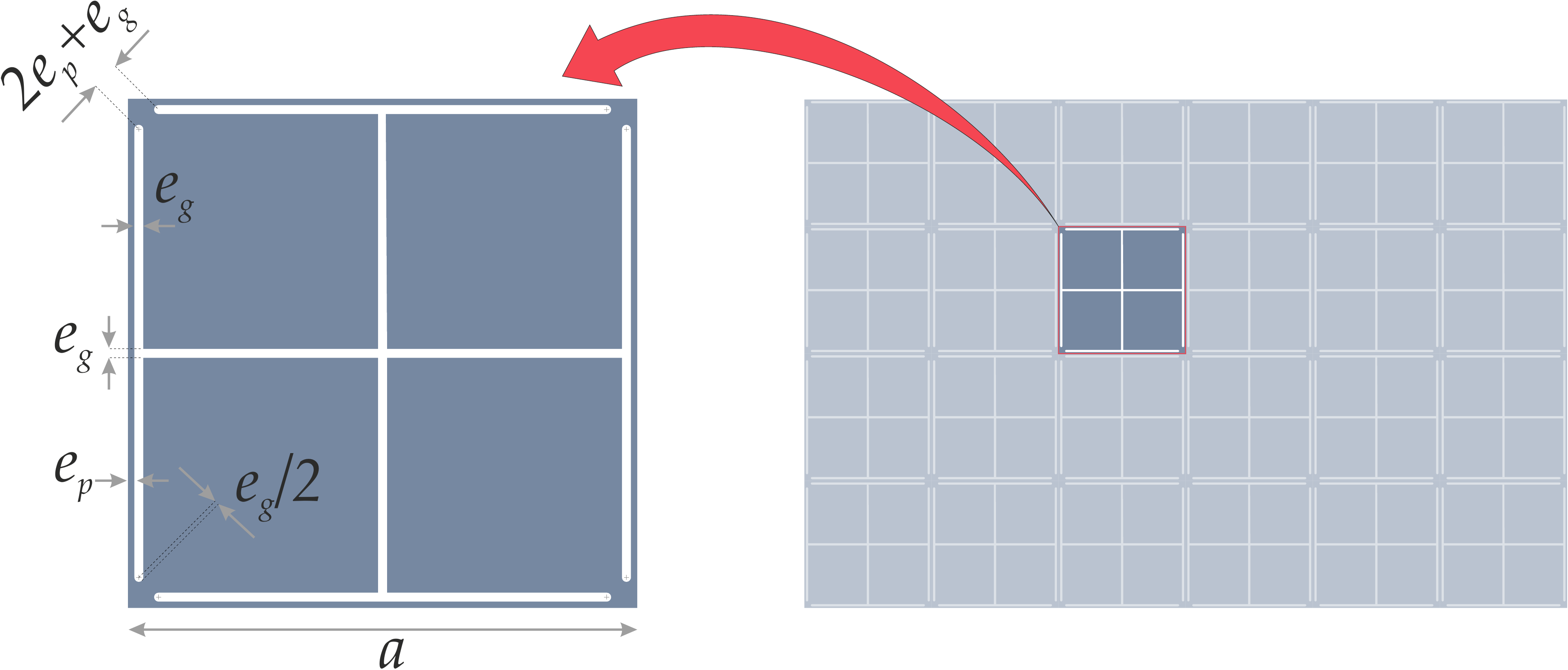}
    \end{minipage}%
    \caption{Geometry and material parameters for the unit cell $\mathcal{R}$.}
    \label{fig:reso_unit_cell}
\end{figure}

In Fig.~\ref{fig:reso_unit_cell}, it is possible to find the geometry and the material properties of the unit cell $\mathcal{R}$ we test in Section \ref{sec:reso}.

In Table \ref{tab:coeff_reso} we list the material parameters that characterize the equivalent RRMM and its longwavelenght limit macro Cauchy.

\begin{table}[H]
    \centering
    {\renewcommand{\arraystretch}{1.1}
        \begin{tabular}{cc|c|c|c|c|cc}
            \hline
            \multicolumn{1}{|c|}{$\rho$ [kg/m$^3$]}              & \multicolumn{1}{c|}{$\kappa_{\rm m}$ [Pa]}   & $\mu_{\rm m}$ [Pa]        & $\mu^{*}_{\rm m}$ [Pa]  & $\kappa_{\rm e}$ [Pa]  & $\mu_{\rm e}$ [Pa]                     & \multicolumn{1}{c|}{$\mu^{*}_{\rm e}$ [Pa]} & \multicolumn{1}{c|}{$\mu_{\rm c}$ [Pa]}
            \\\hline
            \multicolumn{1}{|c|}{$1074$}                         & \multicolumn{1}{c|}{$6.20 \times 10^7$}      & $6.04 \times 10^7$        & $5.70 \times 10^8$      & $3.42 \times 10^7$     & $3.26 \times 10^7$                     & \multicolumn{1}{c|}{$1.63 \times 10^4$}     & \multicolumn{1}{c|}{$152$}
            \\\hline
            \multicolumn{1}{|c|}{$\overline{\kappa}_{\eta}$ [-]} & \multicolumn{1}{c|}{$\overline{\eta}_1$ [-]} & $\overline{\eta}^{*}$ [-] & $\overline{\eta}_2$ [-] & $ \kappa_{\eta}$ [-]   & $\eta_1$ [-]                           & \multicolumn{1}{c|}{$\eta^{*}$ [-]}         & \multicolumn{1}{c|}{$\eta_2$ [-]}
            \\\hline
            \multicolumn{1}{|c|}{$0.14$}                         & \multicolumn{1}{c|}{$0.07$}                  & $0$                       & $0.05$                  & $25.84$                & $24.91$                                & \multicolumn{1}{c|}{$232.77$}               & \multicolumn{1}{c|}{$4.40 \times 10^{-4}$}
            \\\hline
                                                &                             & $\kappa_{\rm M}$ [Pa]     & $\mu_{\rm M}$ [Pa]      & $\mu^{*}_{\rm M}$ [Pa] & $L_{\rm c} = a$ [m] &                            & 
            \\\cline{3-6}

                                                                 &                                              & $2.20 \times 10^{7}$      & $2.12 \times 10^{7}$    & $1.63 \times 10^{4}$   & 0.02                                   &                                             &
            \\\cline{3-6}
        \end{tabular}
    }
    \caption{Material parameters for the metamaterial $\mathcal{R}$ effective RRMM and macro Cauchy.}
    \label{tab:coeff_reso}
\end{table}

On the left and right panels of Fig.~\ref{fig:maze_disp}, we display the dispersion curves for a direction of propagation along the $x_1$ axis ($k_2=0$) and for a direction of propagation parallel to the $x_1 = x_2$ axis ($k_1=k_2$), respectively, superimposing the ones from the RRMM (solid lines obtined with the parameters form Table~\ref{tab:coeff_maze}) to the one coming from the unit cell $\mathcal{R}$ (solid lines).
In the middle of the same figure, we also report the dispersion diagram for all the direction of propagation coming from the unit cell $\mathcal{R}$.

\begin{figure}[H]
    \centering
    \includegraphics[width=\linewidth]{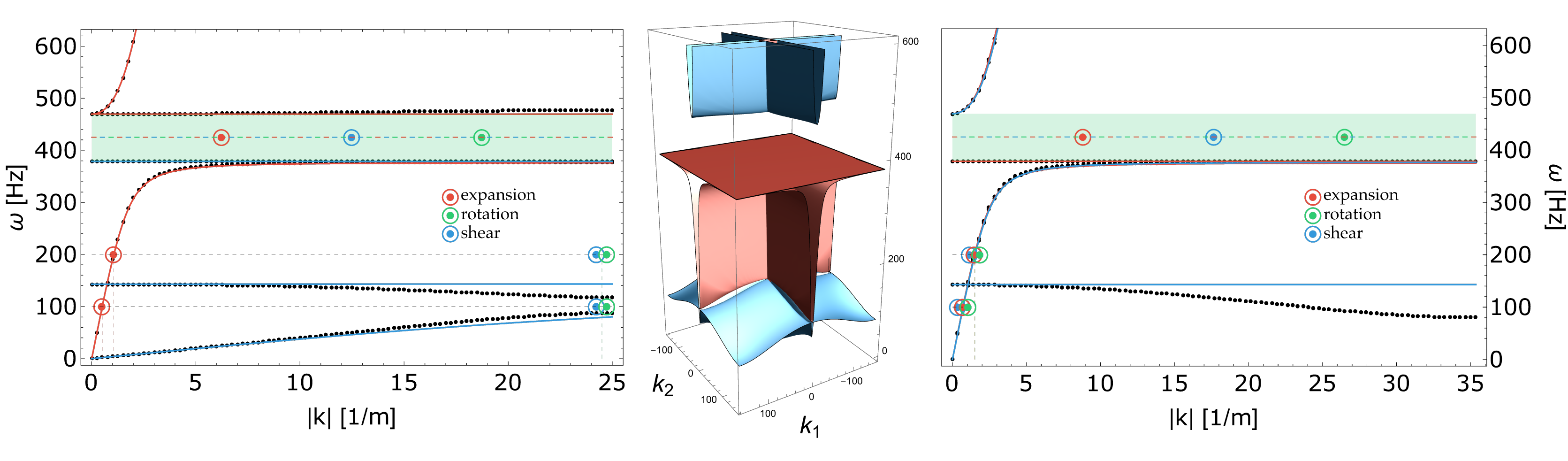}
    \caption{
        (\textit{left}) Dispersion diagram for a direction of propagation along the $x_1$ axis ($k_2=0$);
        (\textit{center}) dispersion diagram for all direction of propagation;
        (\textit{right}) dispersion diagram for a direction of propagation parallel to the $x_1 = x_2$ axis ($k_1=k_2$).
        The dotted lines come from a Bloch-Floquet analysis done on the unit cell $\mathcal{R}$, while the solid lines (blue for shear mdoes and red for pressure modes) from the RRMM with the parameters of Table~\ref{tab:coeff_reso}.
    }
    \label{fig:reso_disp}
\end{figure}

\section{Case studies set-up and time-dependent load conditions}

\begin{figure}[H]
    \begin{table}[H]
        \centering
        \begin{tabular}{cccccccc}
            RRMM                                                     &  &  & Microstructured                                          &  &  & Cauchy                                                   \\
            \includegraphics[width=\figsizeB]{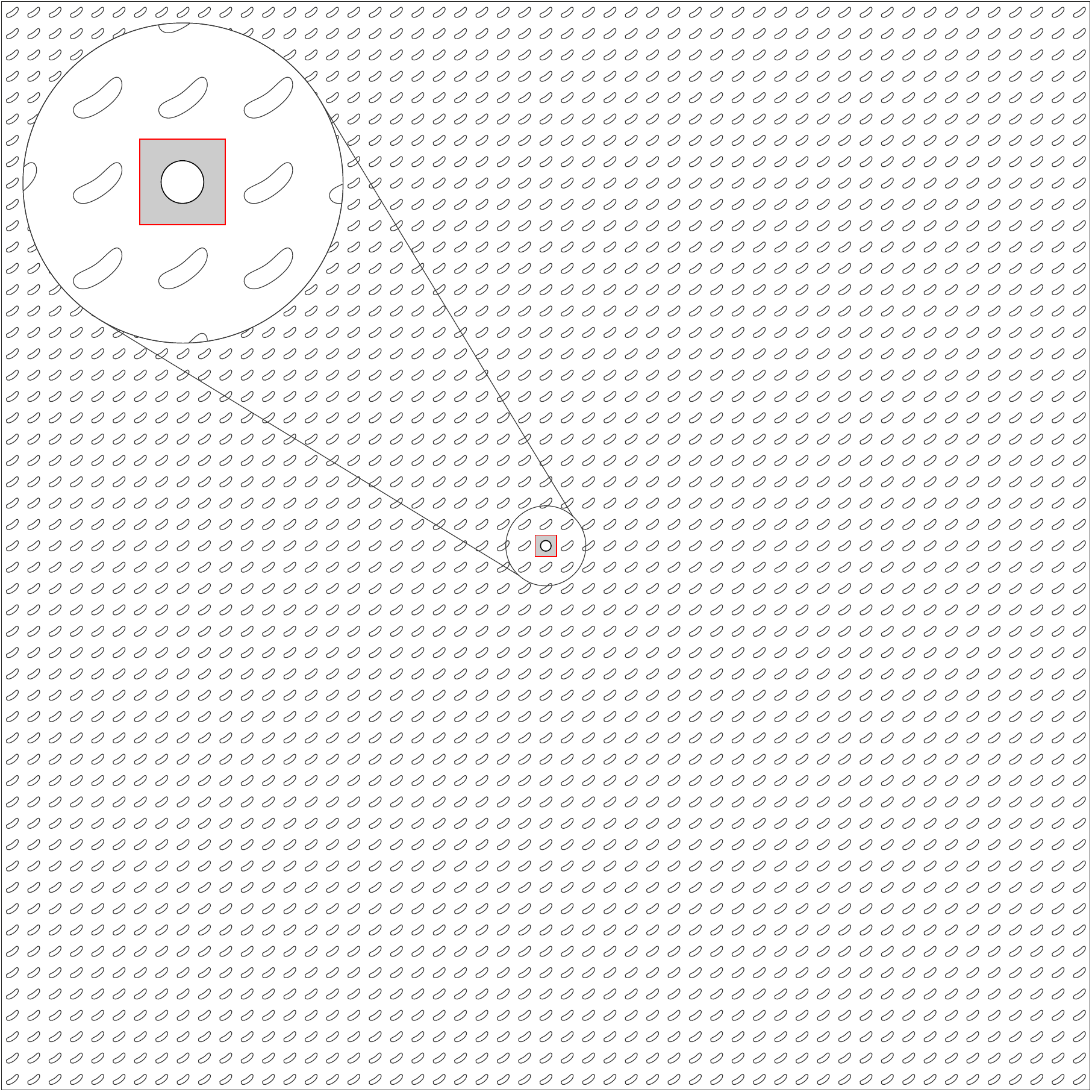} &  &  & \includegraphics[width=\figsizeB]{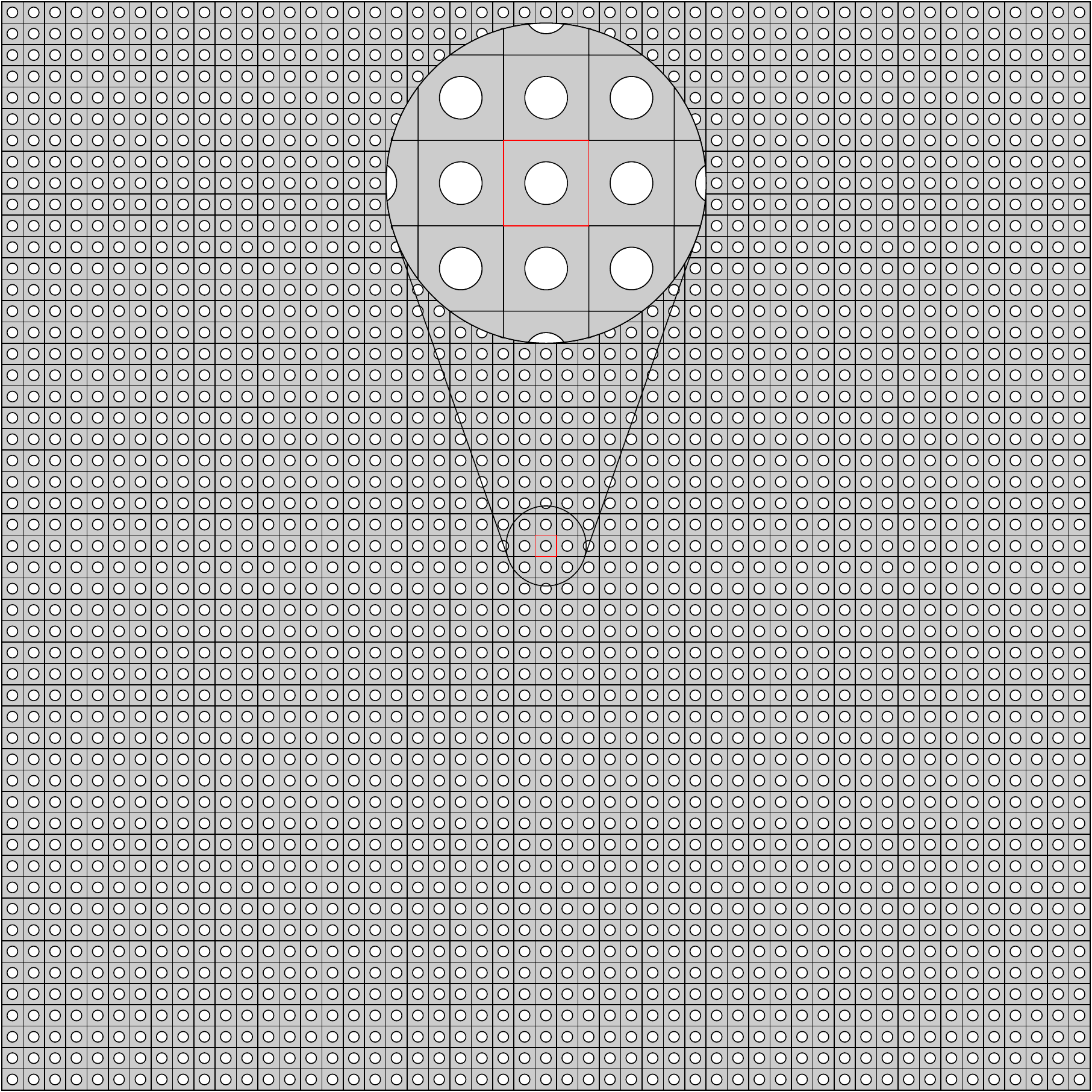} &  &  & \includegraphics[width=\figsizeB]{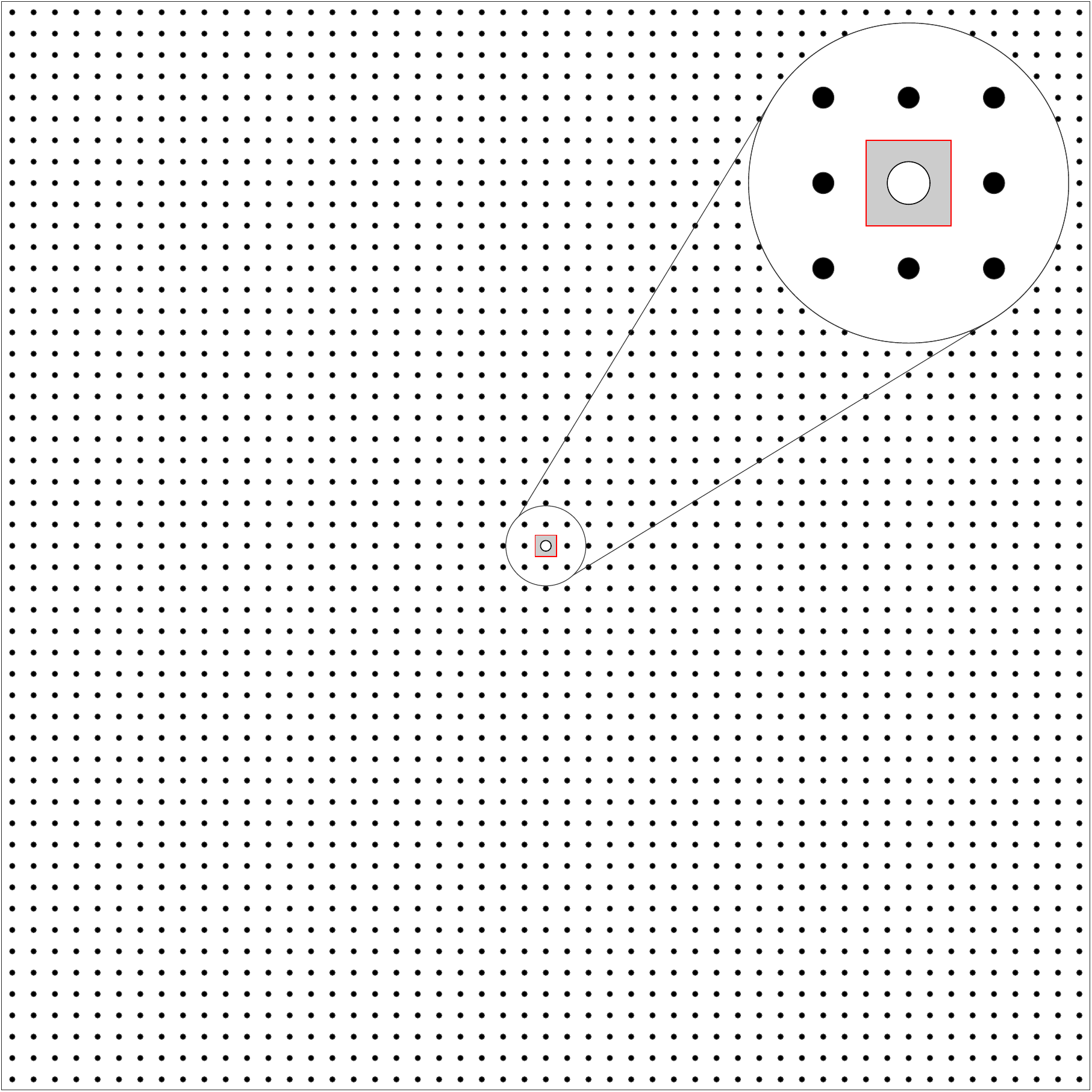}
        \end{tabular}
    \end{table}
    \caption{Scheme of the general set-up for the loading cases:
        (\textit{left}) domain made up of effective RRMM material at which center is placed a unit cell;
        (\textit{center}) domain made up of metamaterials unit cells;
        (\textit{right}) domain made up of effective macro Cauchy material at which center is placed a unit cell.
    }
    \label{fig:gen_set-up}
\end{figure}

We want to provide a set of dynamical tests to assess the capability of the RRMM to represent the behavior of multiple metamaterials, even in the transient regime.
To achieve this, we set up simulations in which a unit cell (or a 3\texttimes 3 cluster of unit cells) is embedded at the center of three domains: a domain made up of a repetition of the same unit cell at the center whose size is 51 \texttimes 51 unit cell big, one made up of the equivalent RRMM, and one made up of the equivalent macro Cauchy model (see Fig.~\ref{fig:gen_set-up}).
A perfect contact condition is enforced between the unit cell (or a 3\texttimes 3 cluster) and the outer domains, ensuring continuity of traction and displacement, and a time-dependent load is applied to some internal boundary of the unit cell to ensure agreement between the load conditions across the three cases.

The specific particularization of the boundary on which the load is applied and of the boundary on which the perfect contact condition is enforces are exemplified in Section \ref{sec:load_boundary}.
The perfect contact condition requires the continuity of displacement ($u^{+} = u^{-}$), which are identically defined in all the models, and the continuity of traction ($t^{+} = t^{-}$), taking in consideration that in the RRMM it is required to use the generalized traction $\widetilde{t}$ ($\widetilde{t}^{+} = t^{-}$).
The load conditions can be summarized as
\begin{gather}
    u(x_1,x_2,\tau) =
    u_0 \, f_{j}(x_1,x_2) \, \text{sin} \left(2\pi \, \omega_0 \, \tau \right)
    \qquad
    \text{with}
    \qquad
    j = \{{\rm r,h,s}\} \, ,
    \notag
    \\*[1.5mm]
    f_{\rm r}(x_1,x_2) = \left(x_2, - x_1\right)^{\rm T} \, ,
    \qquad
    f_{\rm h}(x_1,x_2) = \left(x_1, x_2\right)^{\rm T} \, ,
    \qquad
    f_{\rm s}(x_1,x_2) = \left(x_2, x_1\right)^{\rm T} \, ,
    \label{eq:load_gene}
    \\*[1.5mm]
    \widehat{u}(x_1,x_2,\omega) =
    - i \, \frac{1}{2} u_0 \, \delta (\omega - \omega_0) \, f_{j}(x_1,x_2) \, ,
    \notag
\end{gather}
where $\{{\rm r,h,s}\}$ identify ``rotation'', ``hydrostatic'', and ``shear'', respectively, $u_0$ is a fix amplitude
($u_0 = 0.51$ for our examples),
$\omega_0$ is a fix frequency, $\tau$ is the time, $\widehat{u}(x_1,x_2,\omega)$ is the Fourier transform of $ u(x_1,x_2,\tau)$, and $\delta$ is the Dirac delta function.
Given eq.(\ref{eq:load_gene})$_3$, it is clear that the frequency spectrum of all the different loads consists of a single frequency, i.e. $\omega = \omega_0$.

\subsection{Load and interface conditions for the \textit{labyrinthine} and the \textit{four-resonator} unit cell}
\label{sec:load_boundary}

In Table~\ref{tab:maze_load_def} we show a depiction of the boundary on which the central unit cell $\mathcal{L}$ is loaded, along with its deformed states at different times. The deformed state has been amplified for a better illustration.

\begin{table}[H]
    \centering
    {\centering
        \renewcommand{\arraystretch}{1.1}
        \begin{tabular}{c|ccccc|}
            \cline{2-6}
                                                                                                       &
            Load                                                                                       &
            $t=0.0010$ s                                                                               &
            $t=0.0026$ s                                                                               &
            $t=0.0060$ s                                                                               &
            $t=0.0076$ s
            \\
            \hline
            \multicolumn{1}{|c|}{\rotatebox{90}{\hspace{0.45cm}Rotation}}                              &
            \includegraphics[width=\figsizeC]{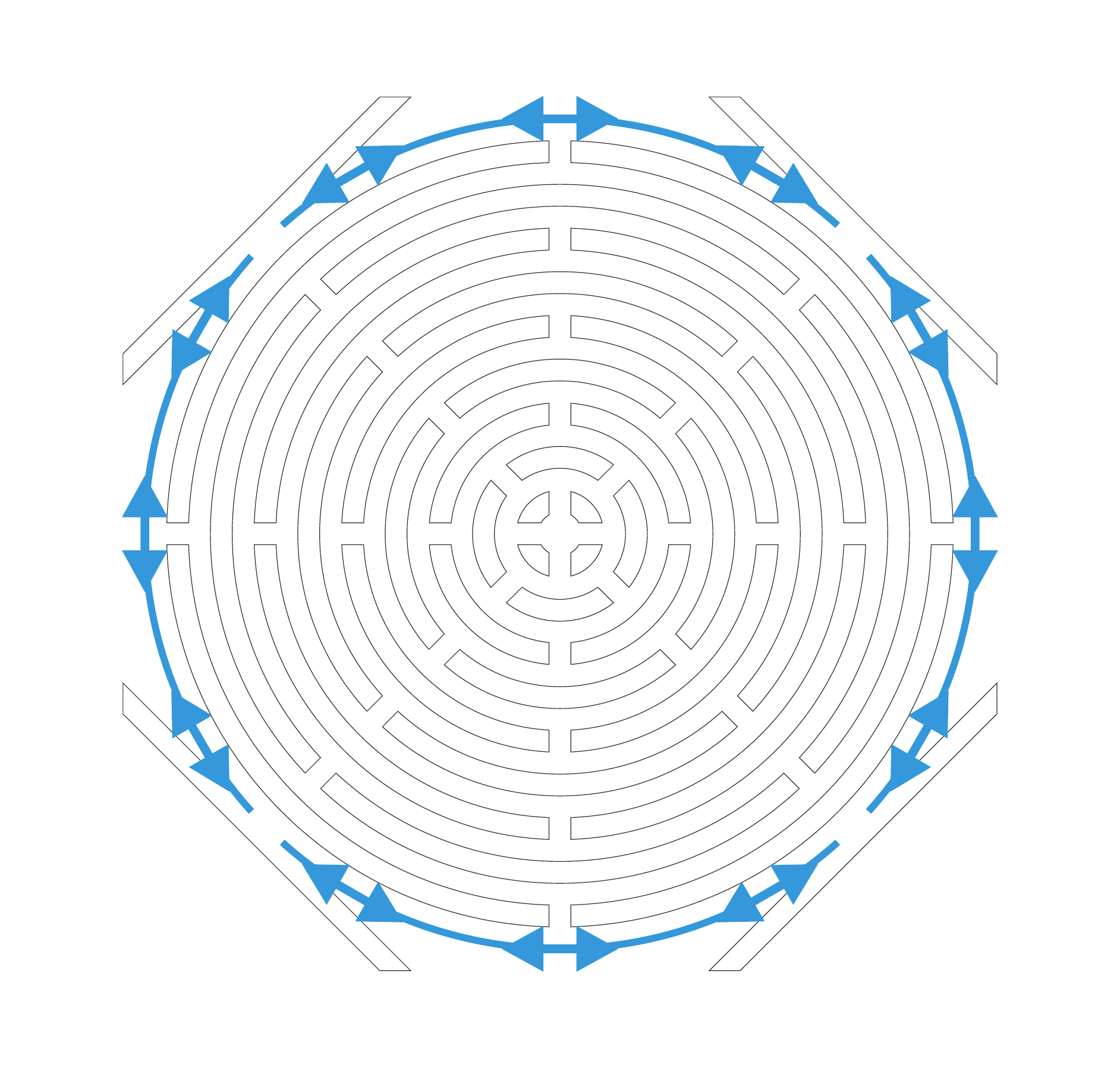}             &
            \includegraphics[width=\figsizeC]{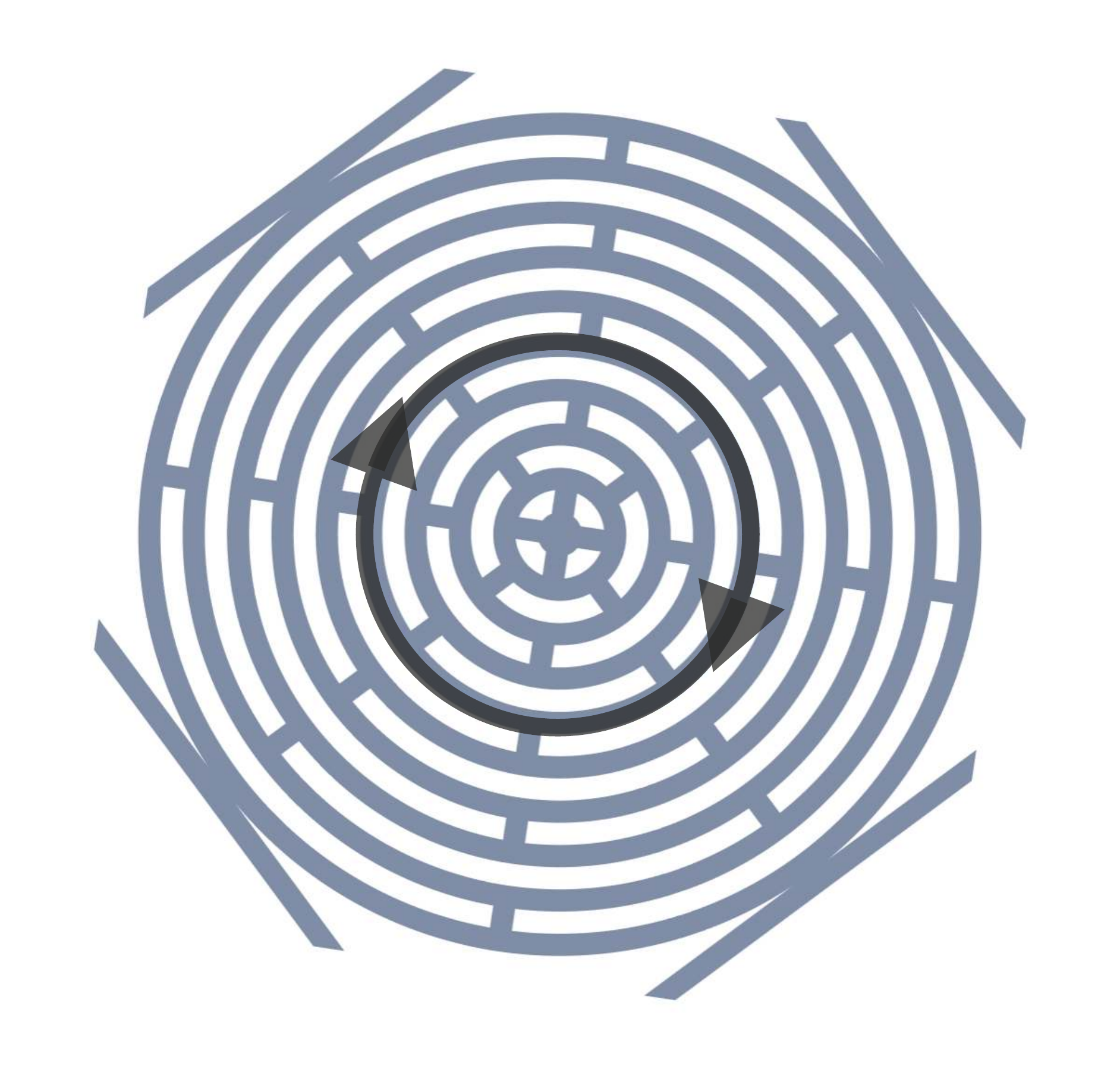} &
            \includegraphics[width=\figsizeC]{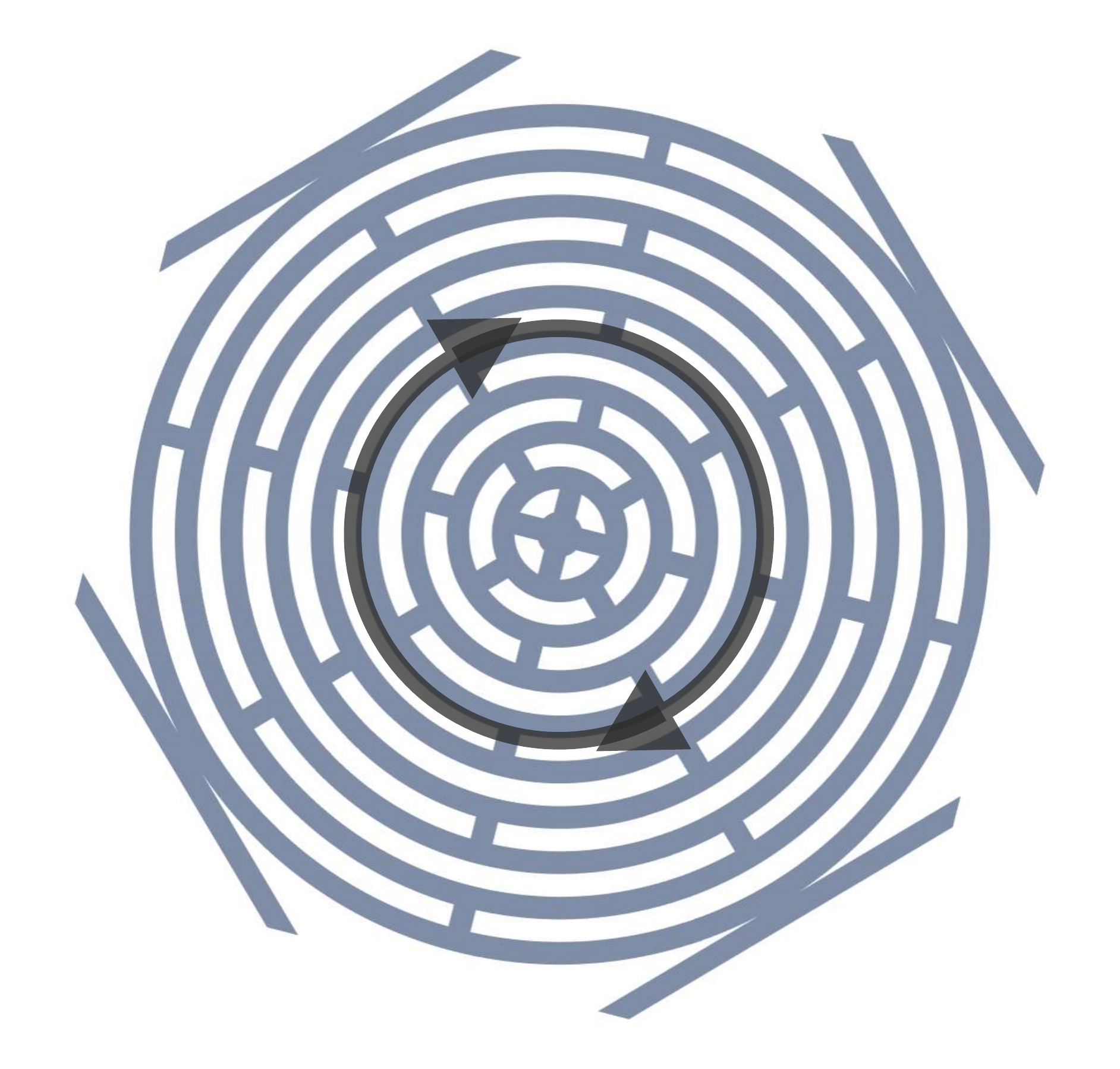} &
            \includegraphics[width=\figsizeC]{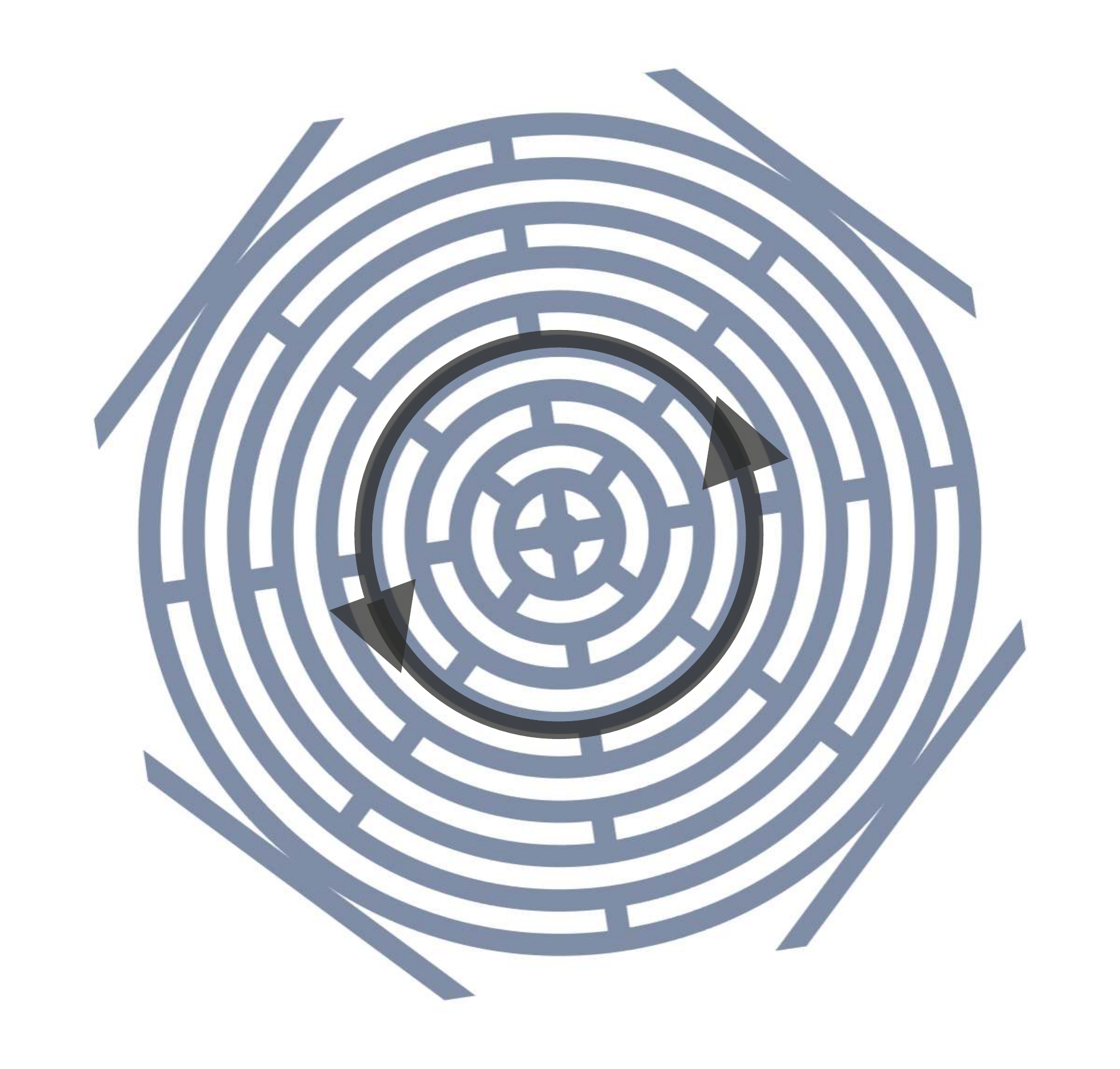} &
            \includegraphics[width=\figsizeC]{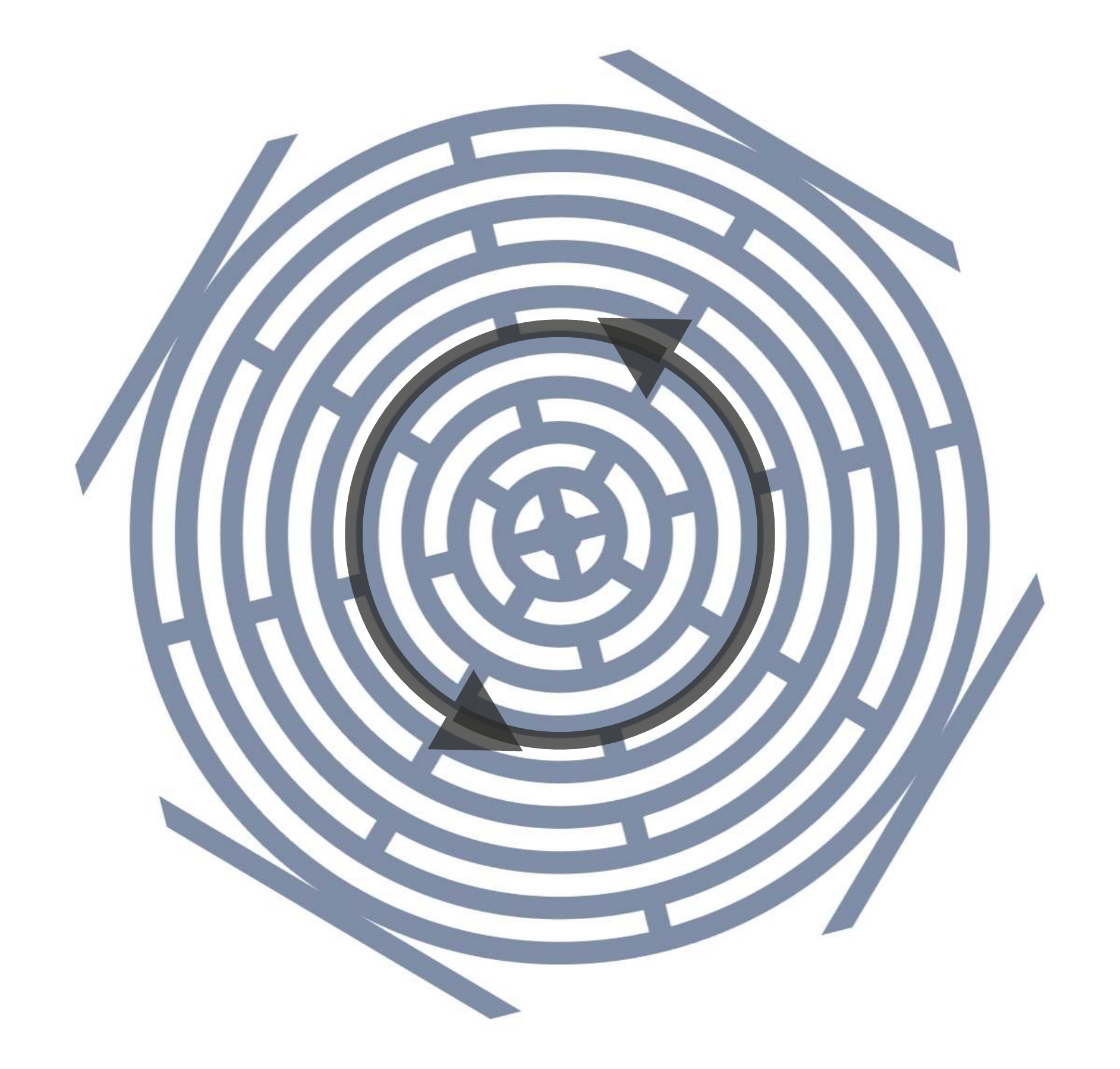}
            \\
            \hline
            \multicolumn{1}{|c|}{\rotatebox{90}{\hspace{0.4cm}Hydrostatic}}                            &
            \includegraphics[width=\figsizeC]{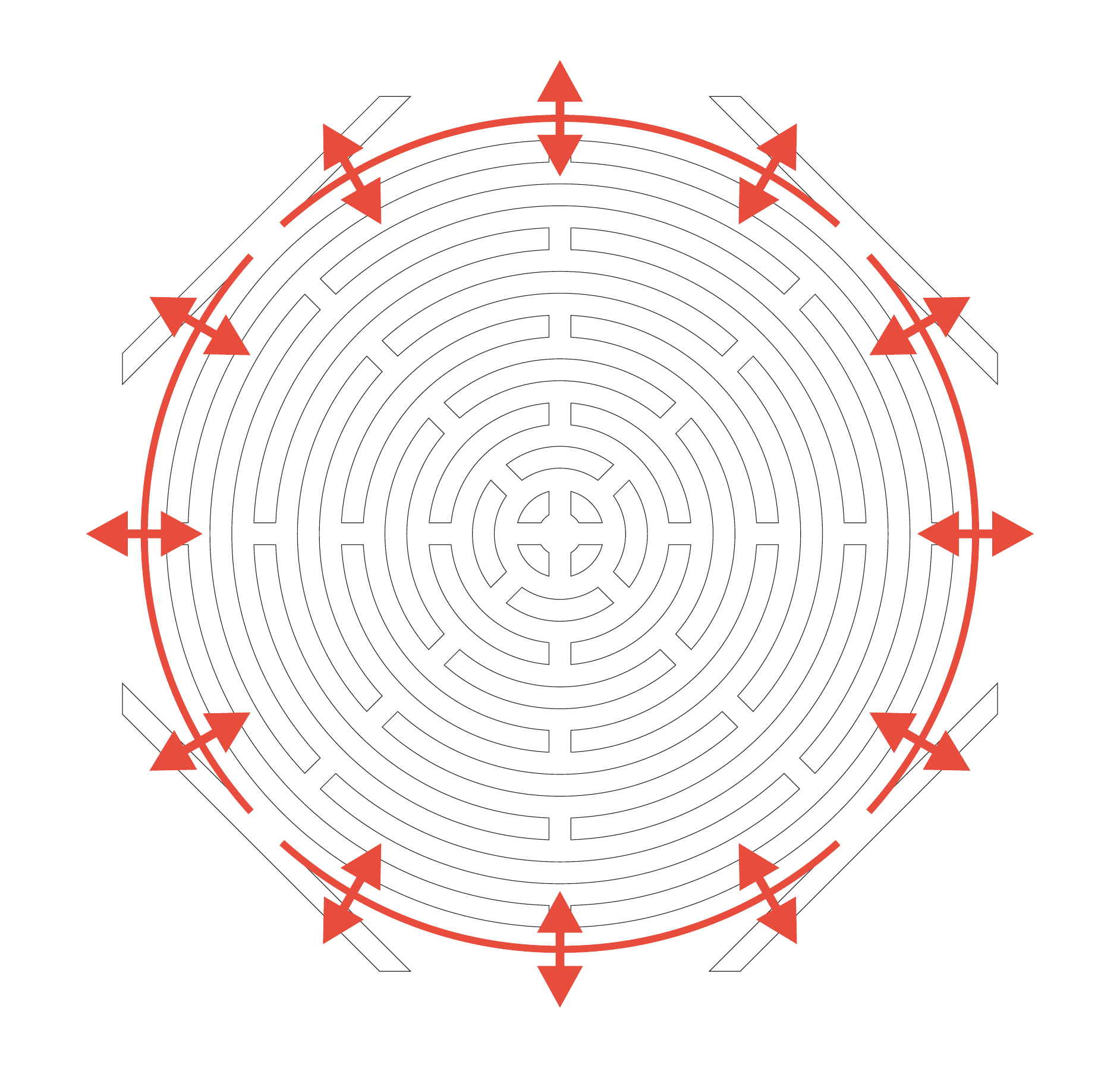}             &
            \includegraphics[width=\figsizeC]{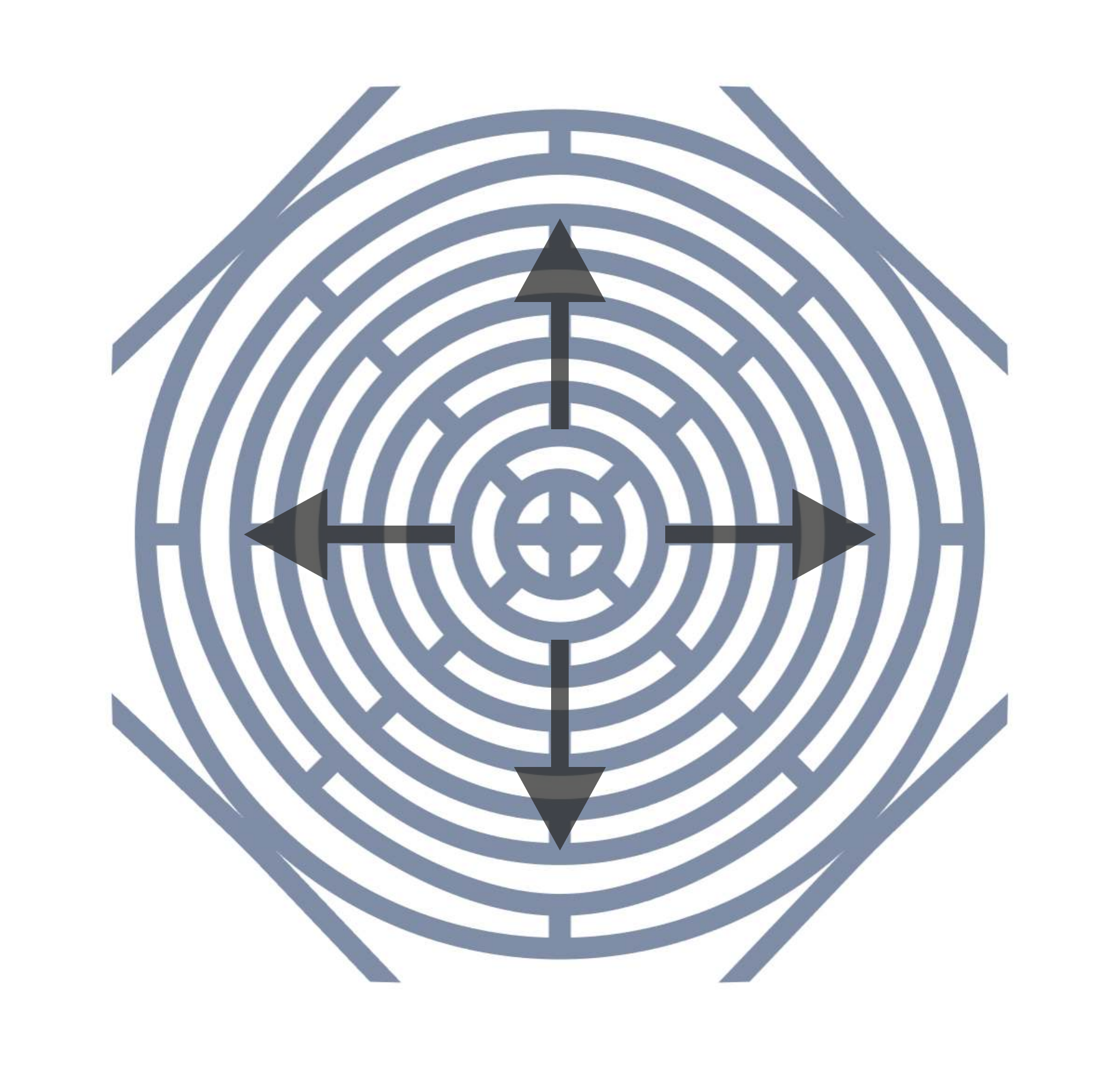} &
            \includegraphics[width=\figsizeC]{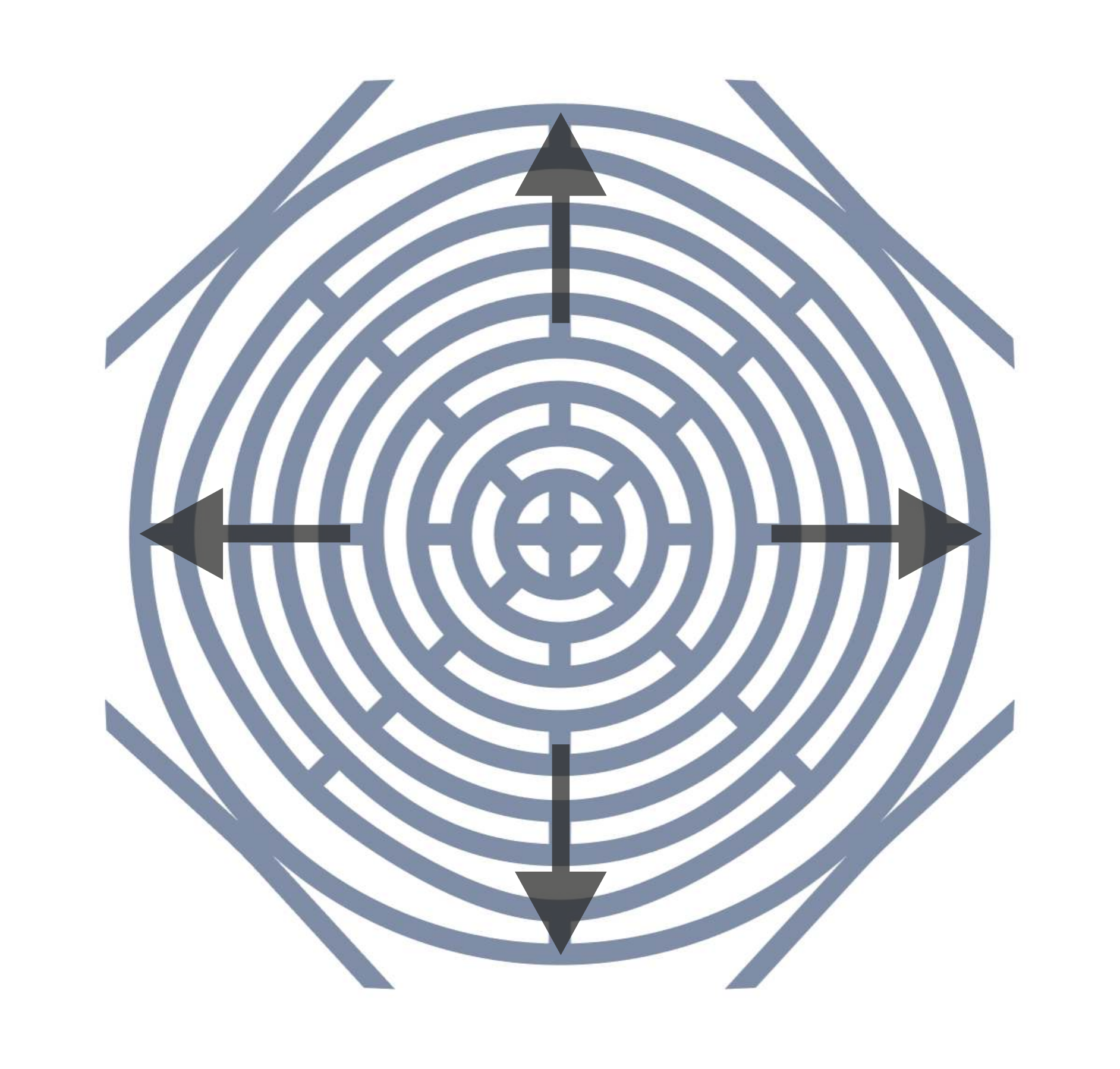} &
            \includegraphics[width=\figsizeC]{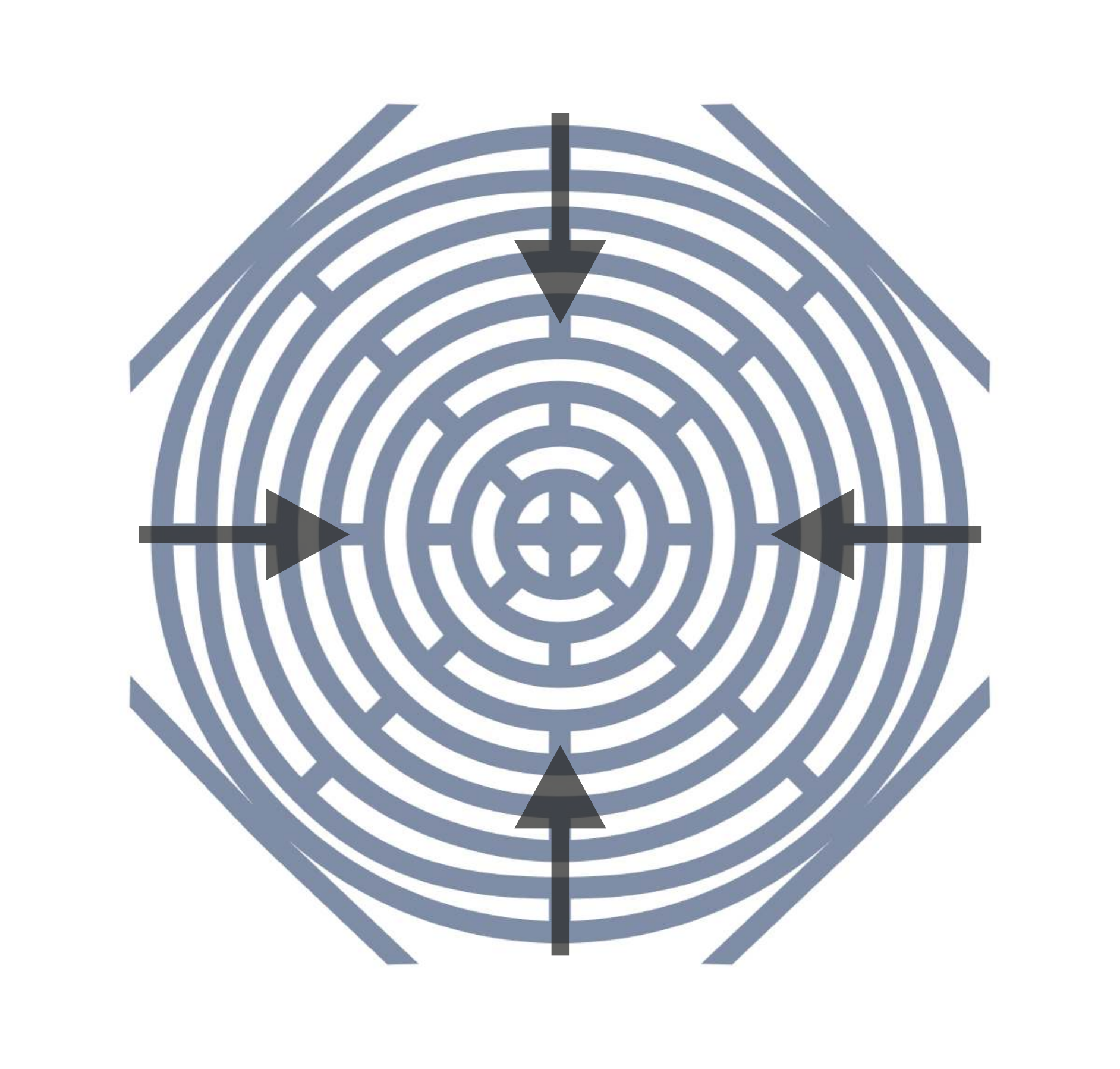} &
            \includegraphics[width=\figsizeC]{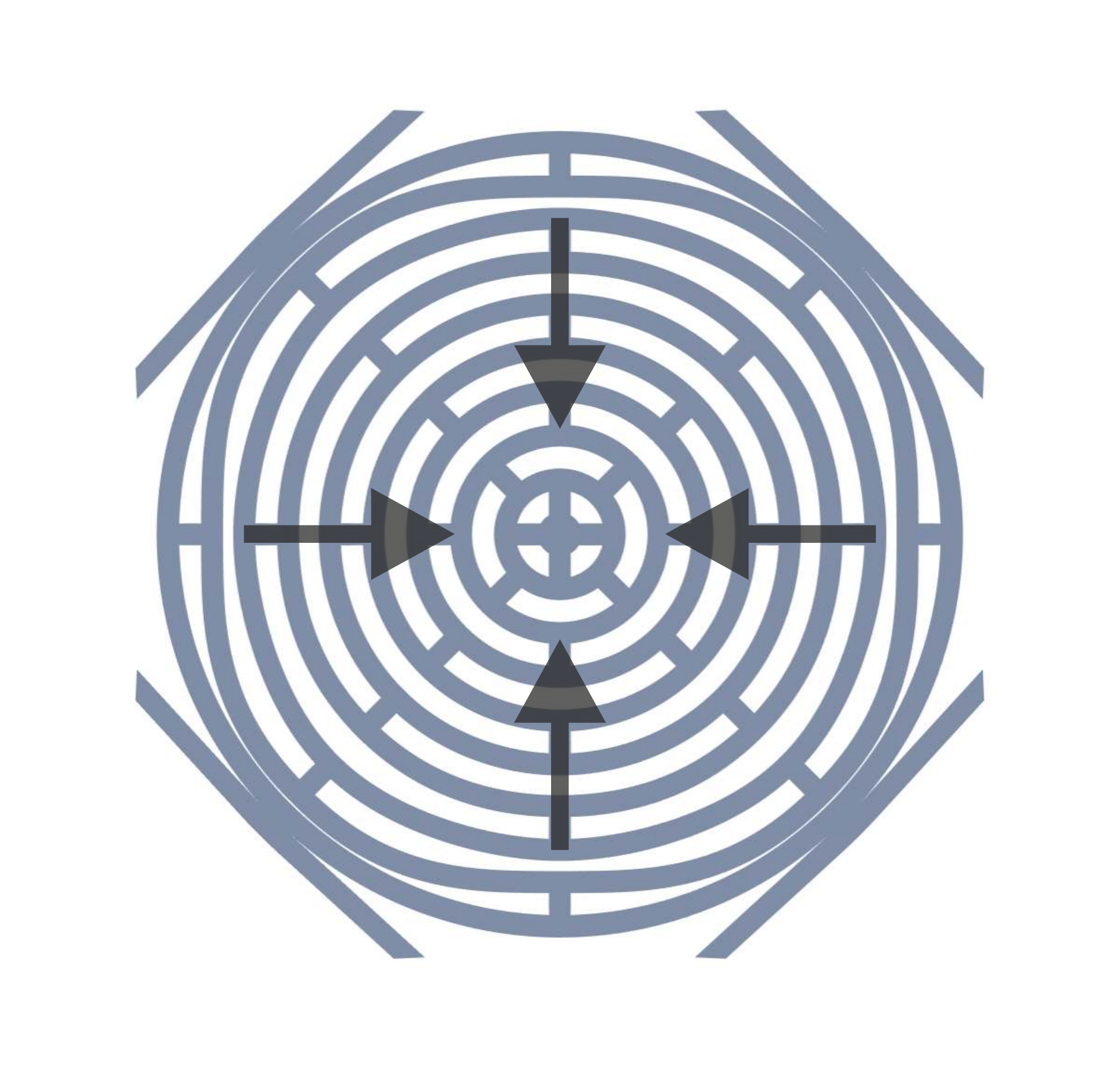}
            \\
            \hline
            \multicolumn{1}{|c|}{\rotatebox{90}{\hspace{0.75cm}Shear}}                                 &
            \includegraphics[width=\figsizeC]{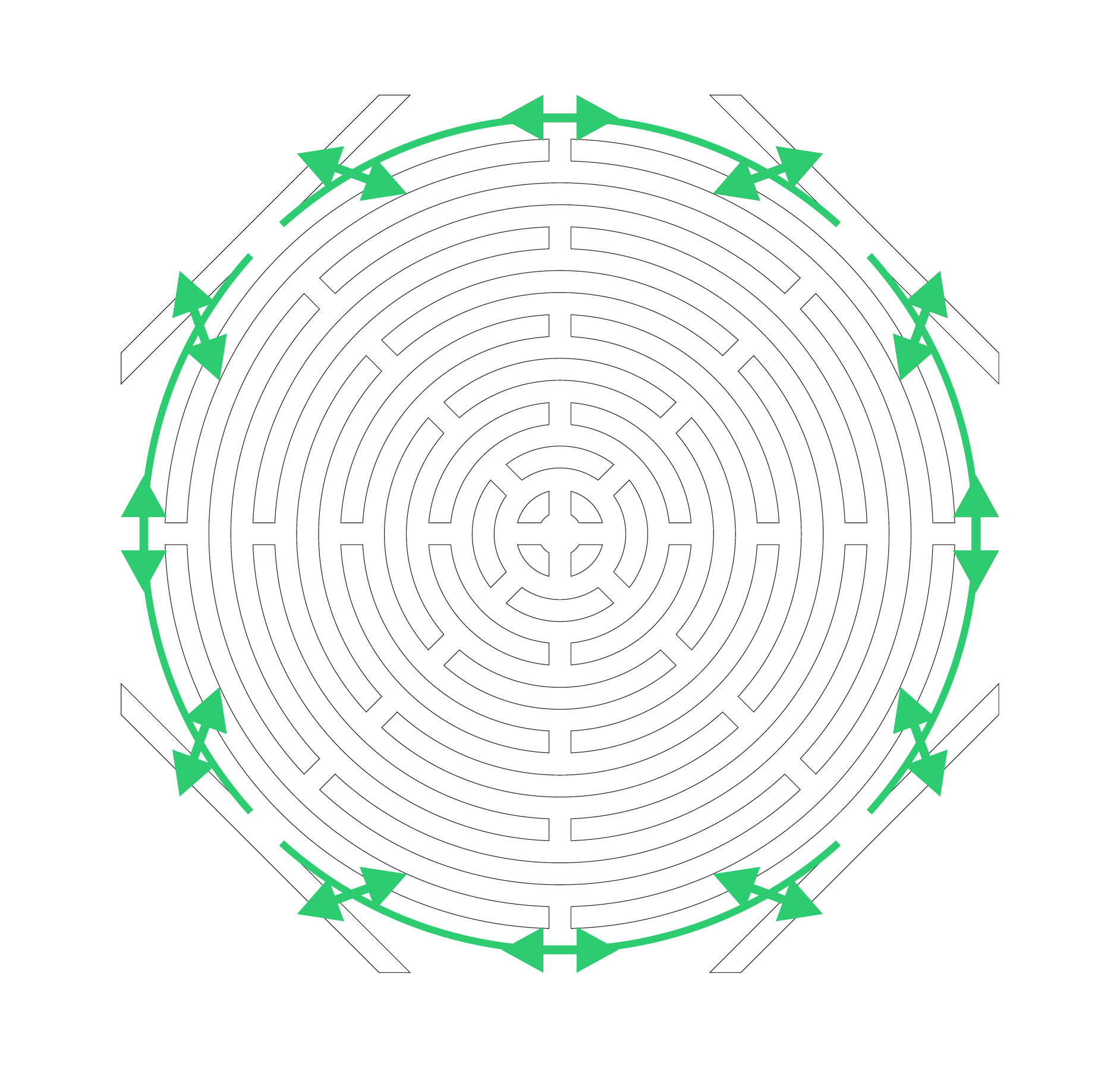}             &
            \includegraphics[width=\figsizeC]{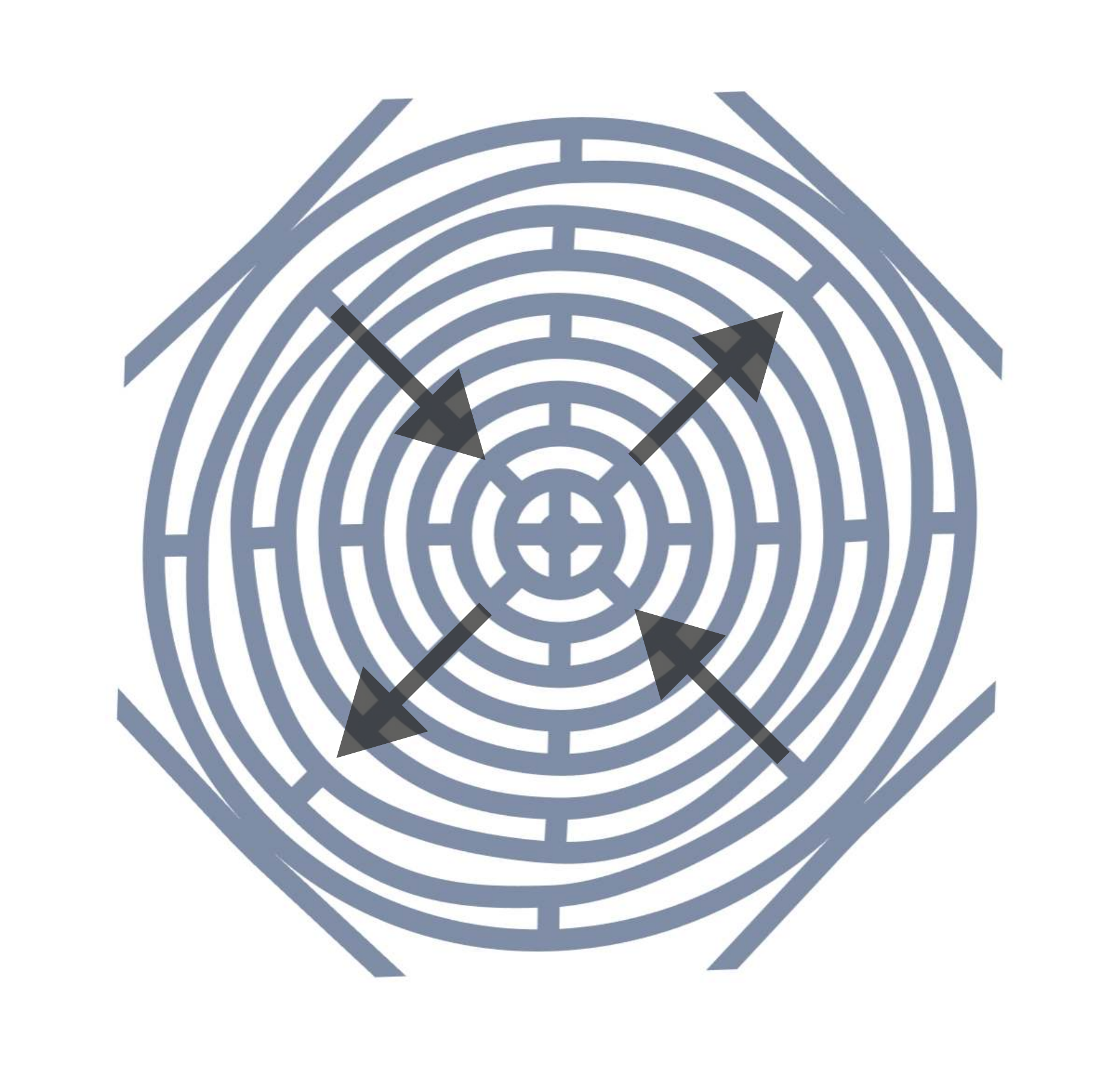} &
            \includegraphics[width=\figsizeC]{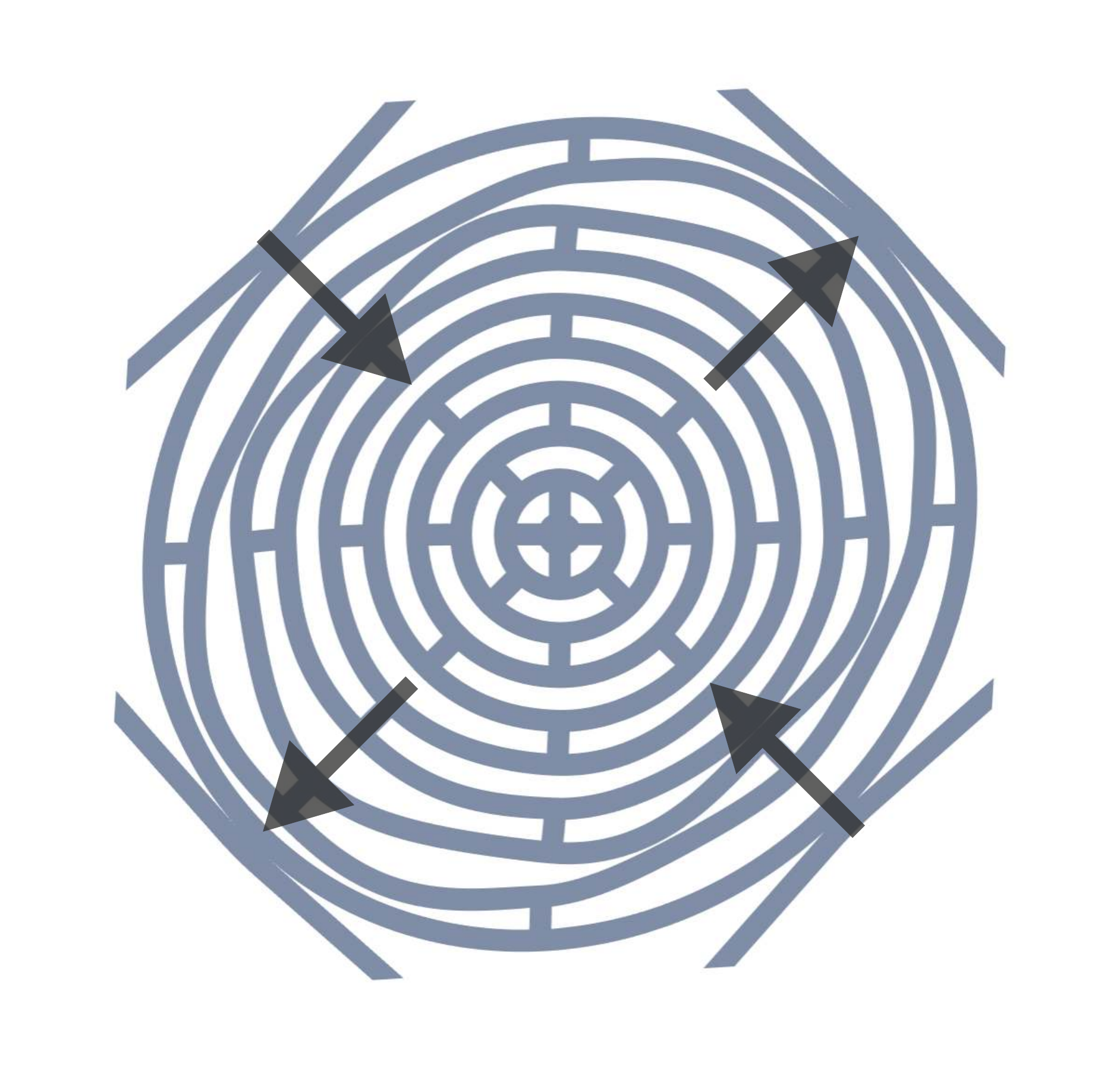} &
            \includegraphics[width=\figsizeC]{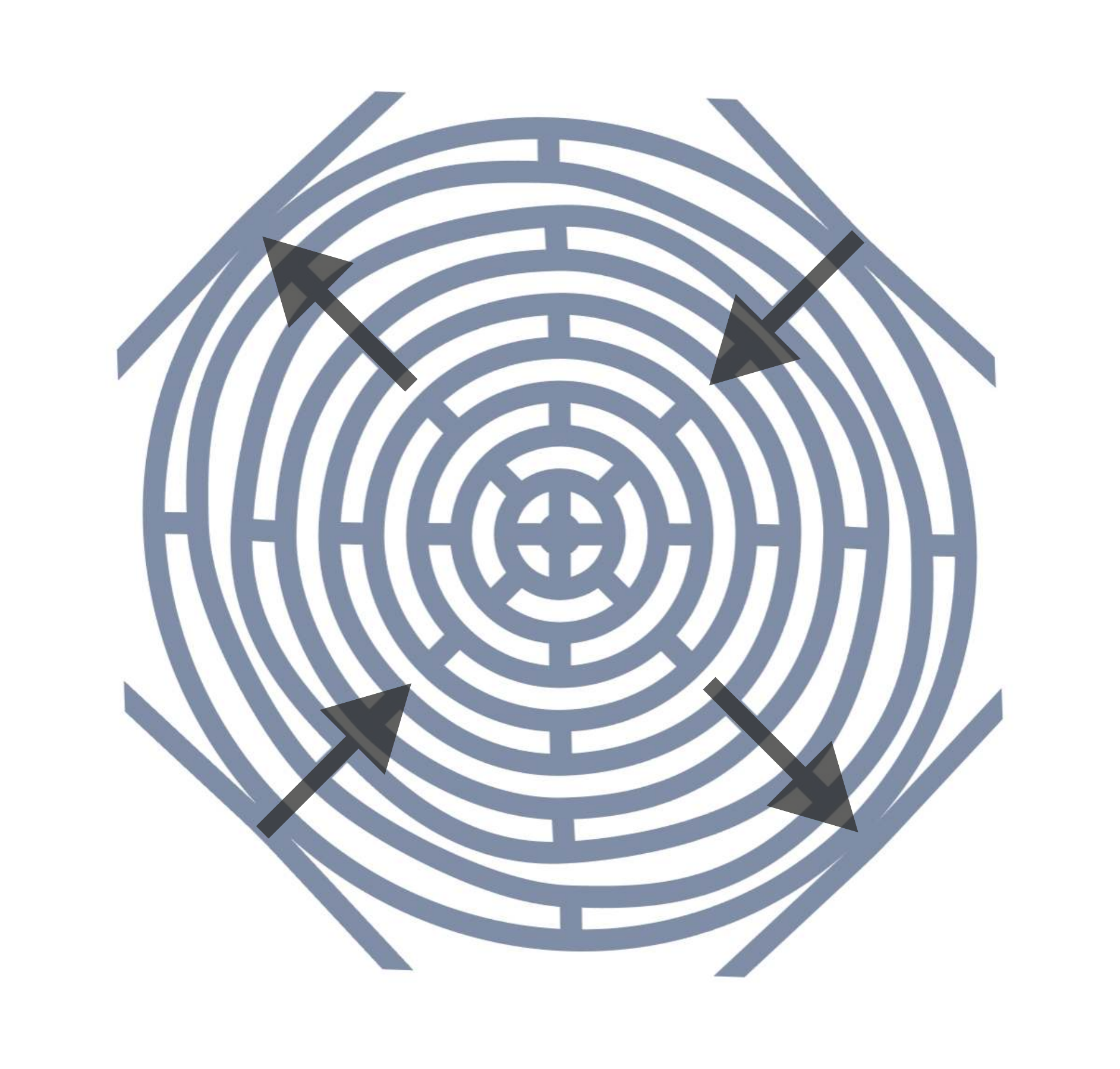} &
            \includegraphics[width=\figsizeC]{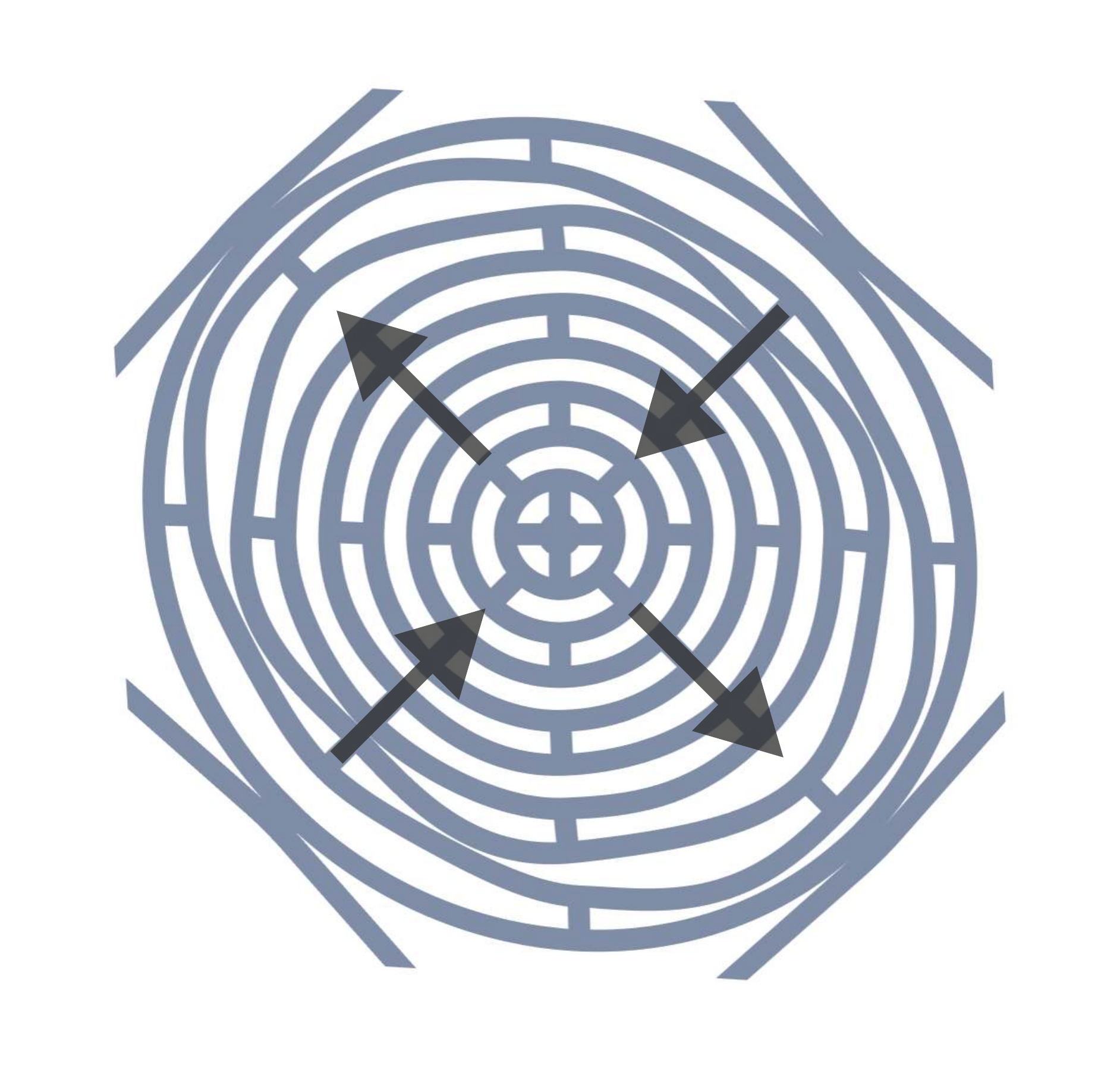}
            \\
            \hline
        \end{tabular}
    }
    \caption{Load conditions for:
        (\textit{first row}) the rotation test and some examples of deformed central unit cell;
        (\textit{second row}) the hydrostatic test and some examples of deformed central unit cell;
        (\textit{third row}) the shear test and some examples of deformed central unit cell.
        The deformation are accentuated to better show the direction of loading.}
    \label{tab:maze_load_def}
\end{table}

In Table~\ref{tab:reso_load_def} we show a depiction of the boundary on which the central unit cell $\mathcal{R}$ is loaded, along with its deformed states at different times. The deformed state has been amplified for a better illustration.

\begin{table}[H]
    \centering
    {\centering
        \renewcommand{\arraystretch}{1.1}
        \begin{tabular}{c|ccccc|}
            \cline{2-6}
                                                                                                       &
            Load                                                                                       &
            $t=0.0010$ s                                                                               &
            $t=0.0026$ s                                                                               &
            $t=0.0060$ s                                                                               &
            $t=0.0076$ s
            \\
            \hline
            \multicolumn{1}{|c|}{\rotatebox{90}{\hspace{0.45cm}Rotation}}                              &
            \includegraphics[width=\figsizeC]{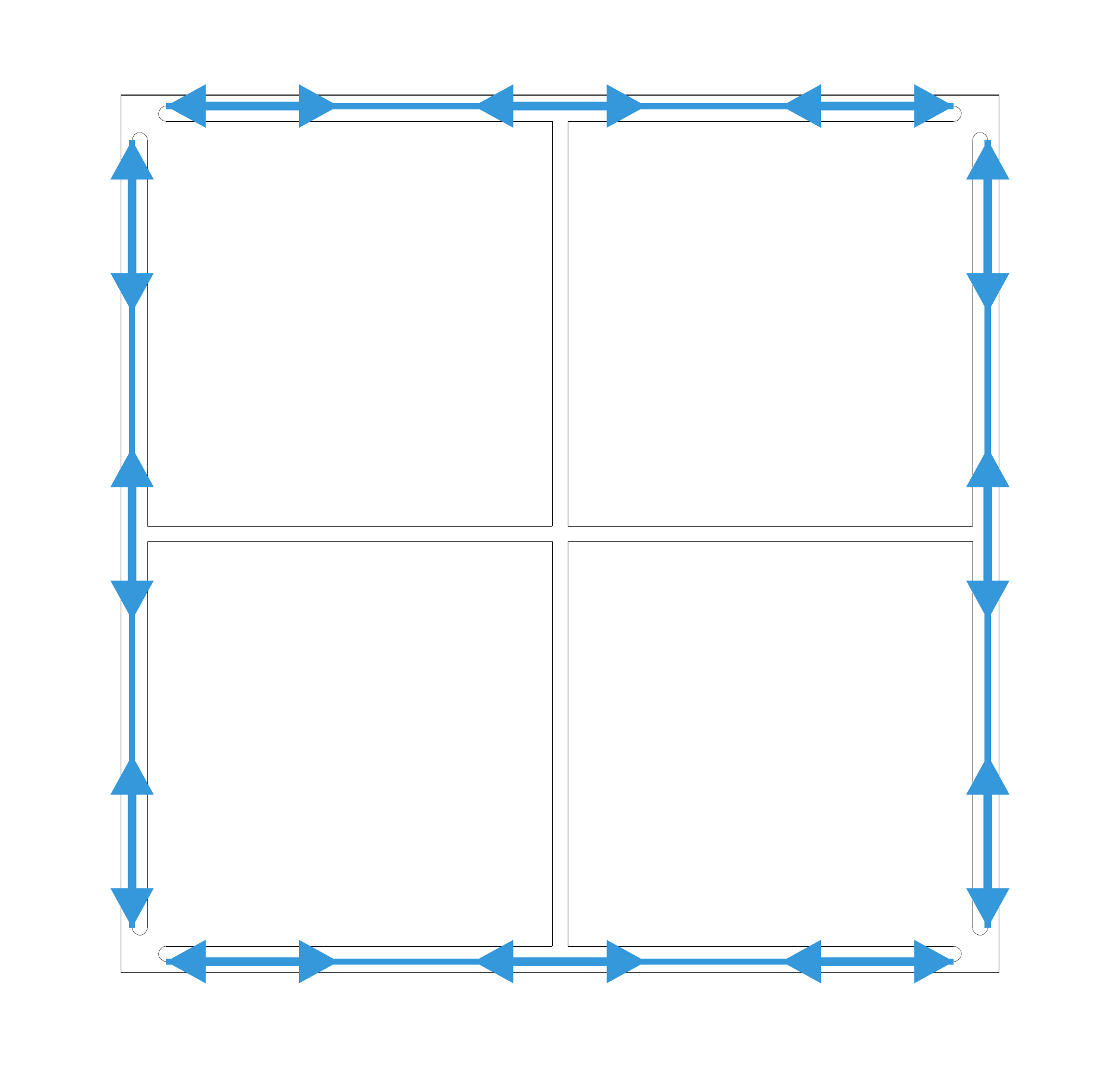}             &
            \includegraphics[width=\figsizeC]{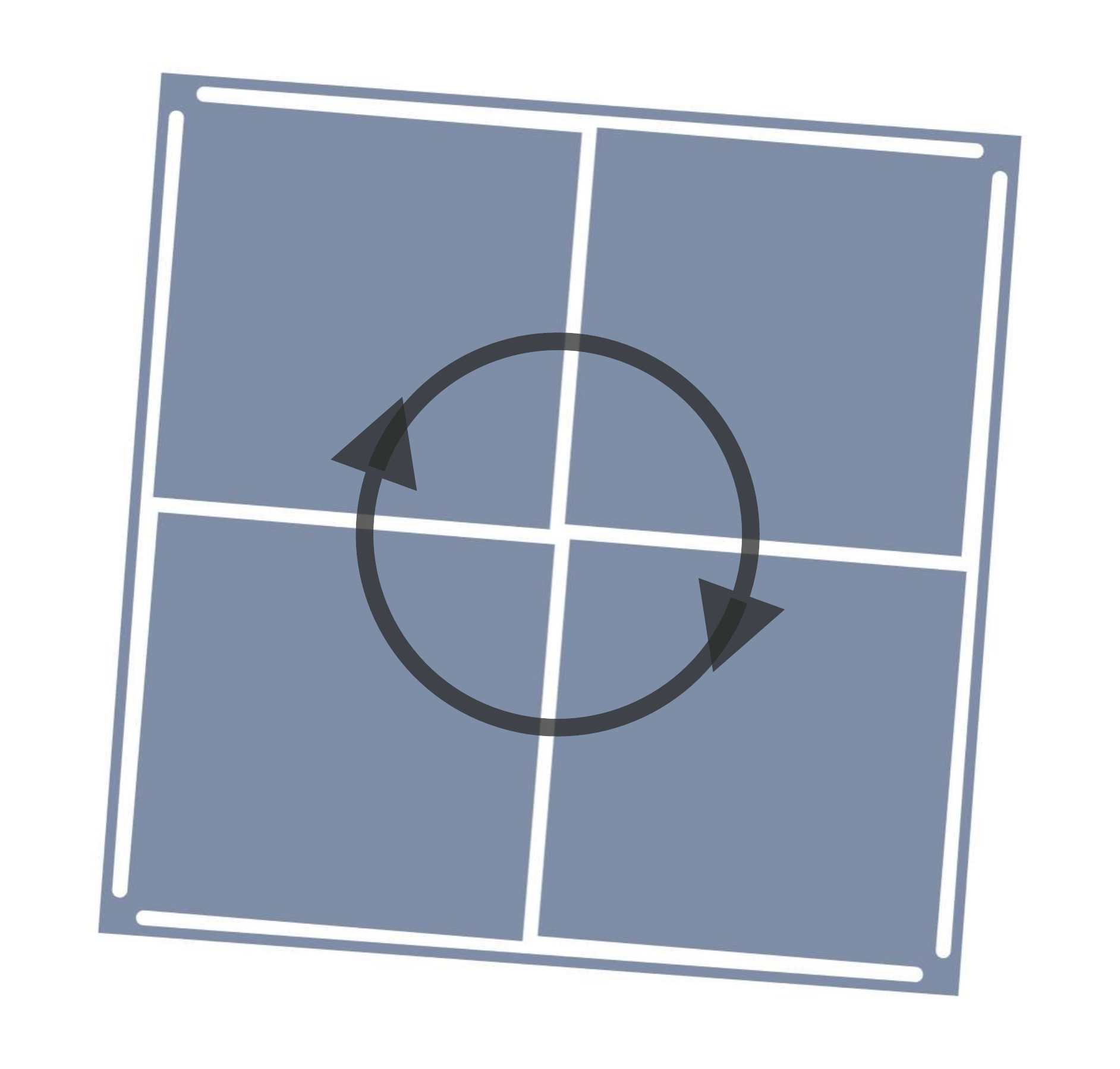} &
            \includegraphics[width=\figsizeC]{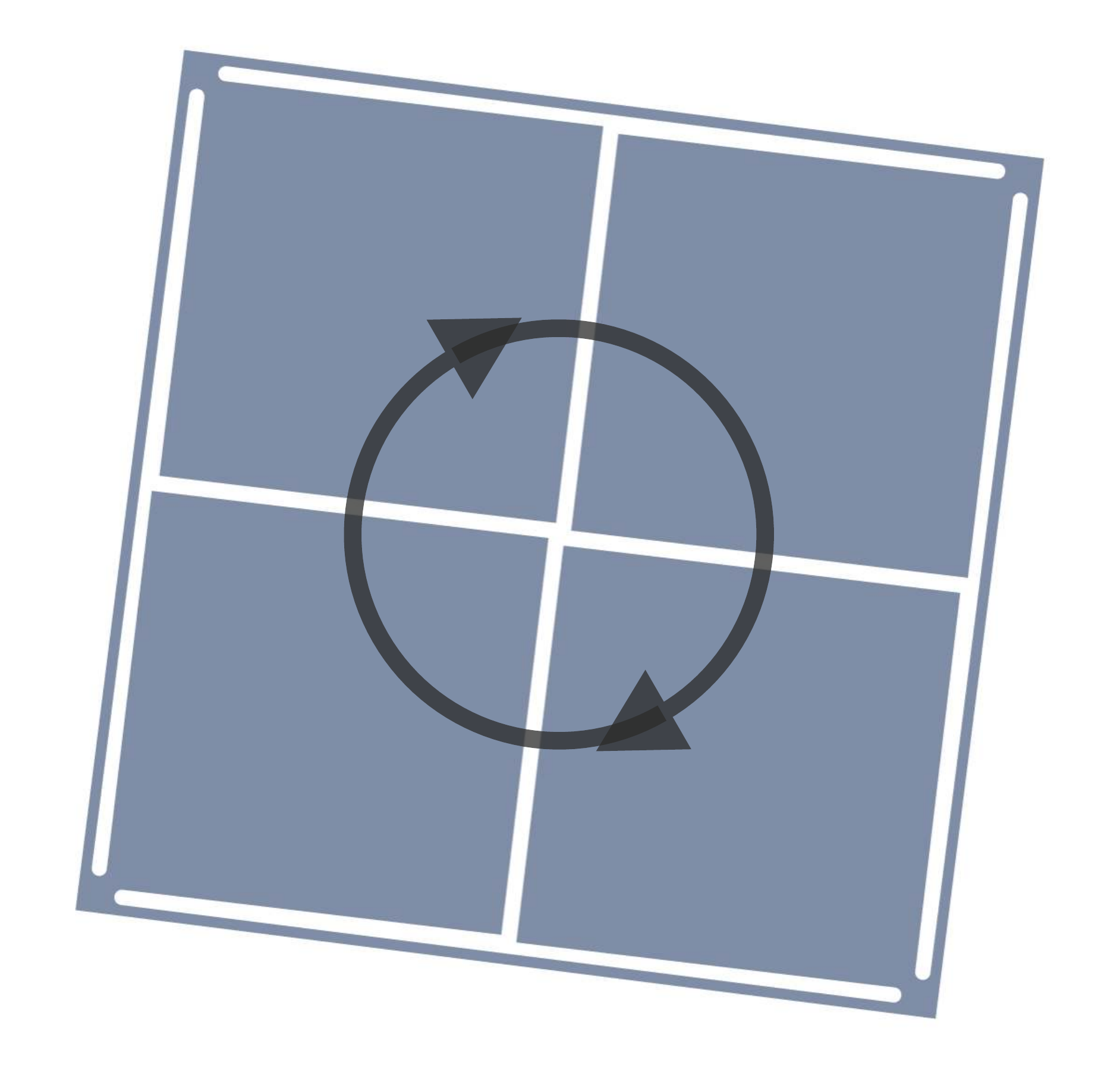} &
            \includegraphics[width=\figsizeC]{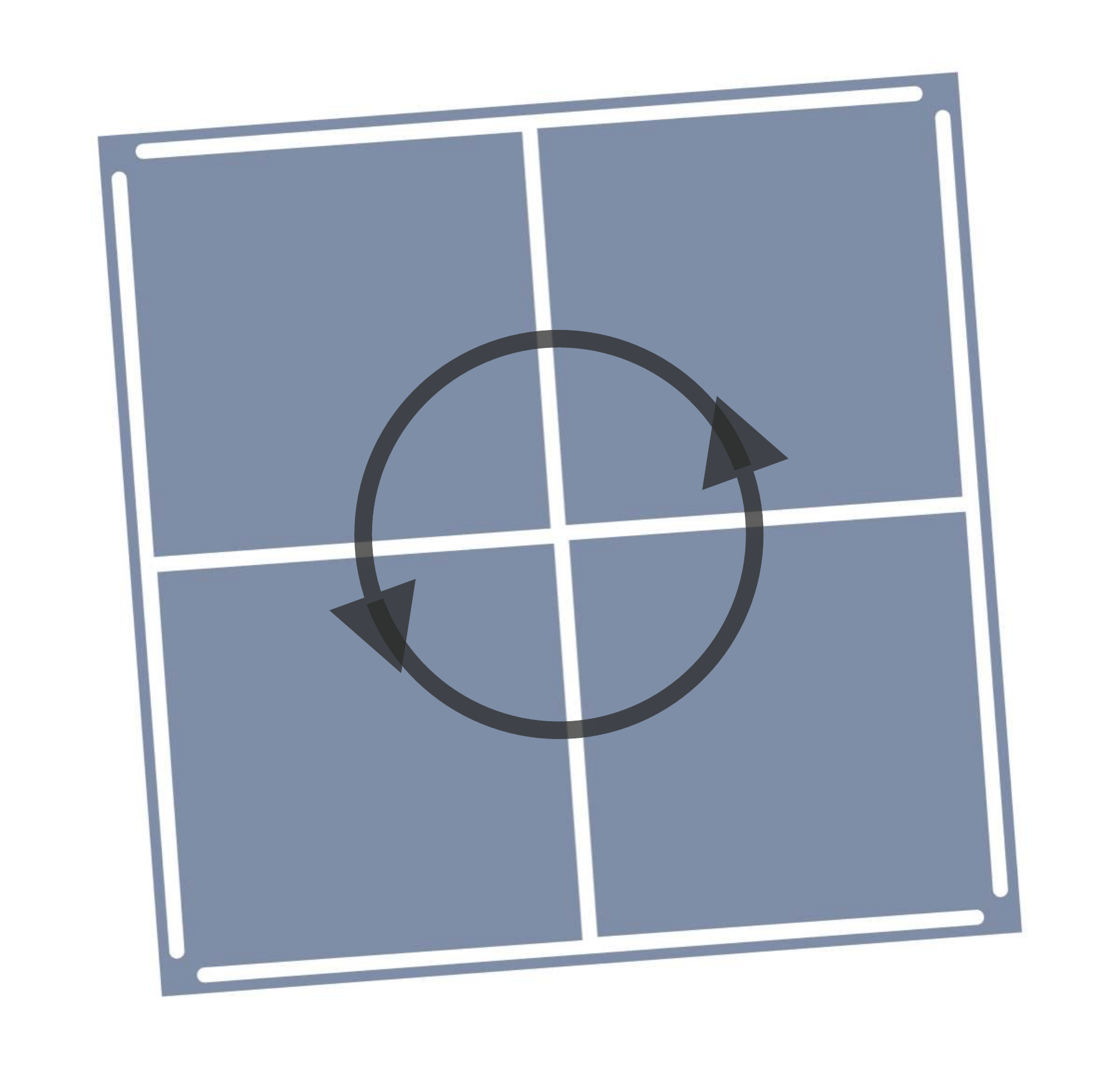} &
            \includegraphics[width=\figsizeC]{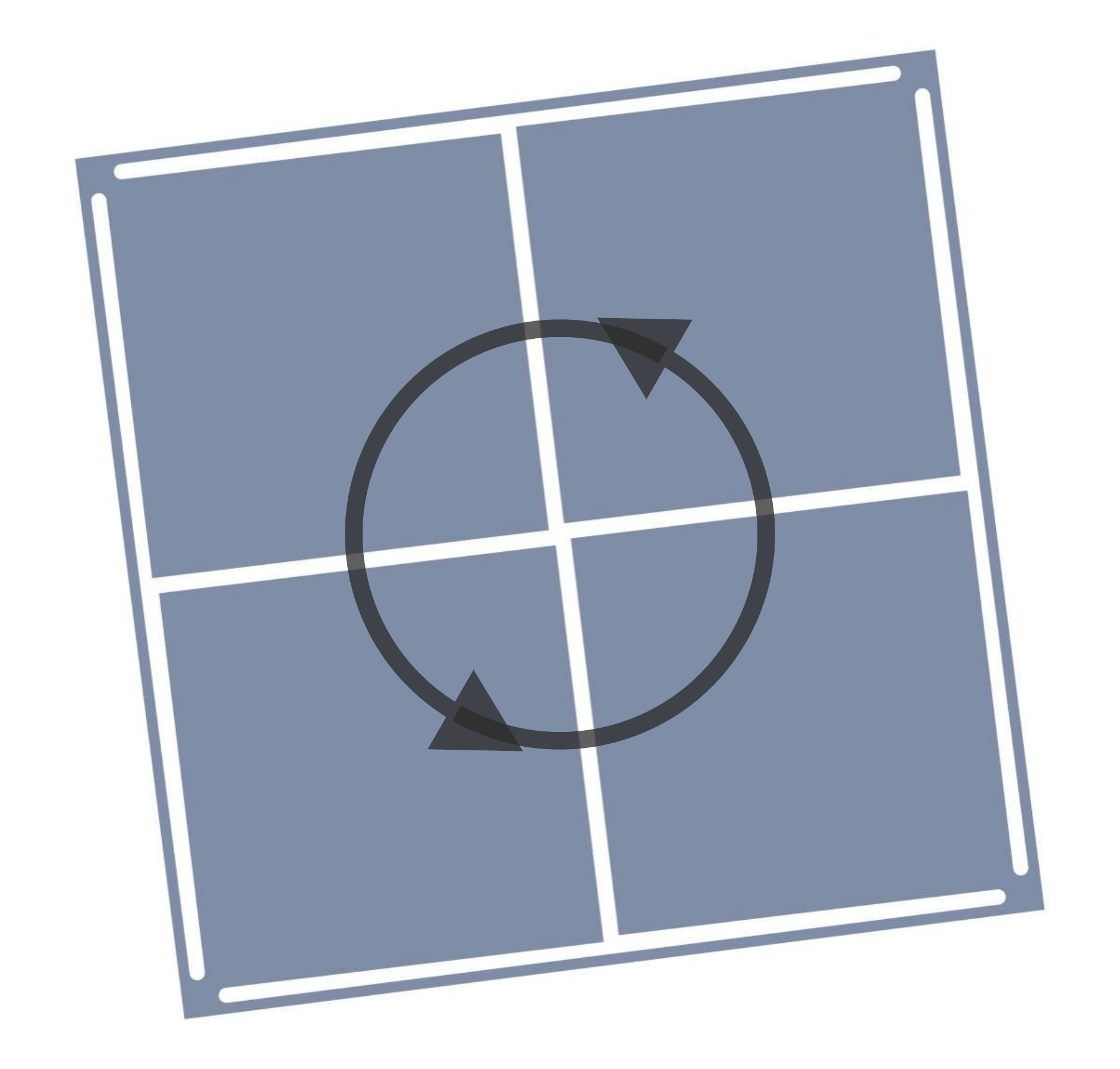}
            \\
            \hline
            \multicolumn{1}{|c|}{\rotatebox{90}{\hspace{0.4cm}Hydrostatic}}                            &
            \includegraphics[width=\figsizeC]{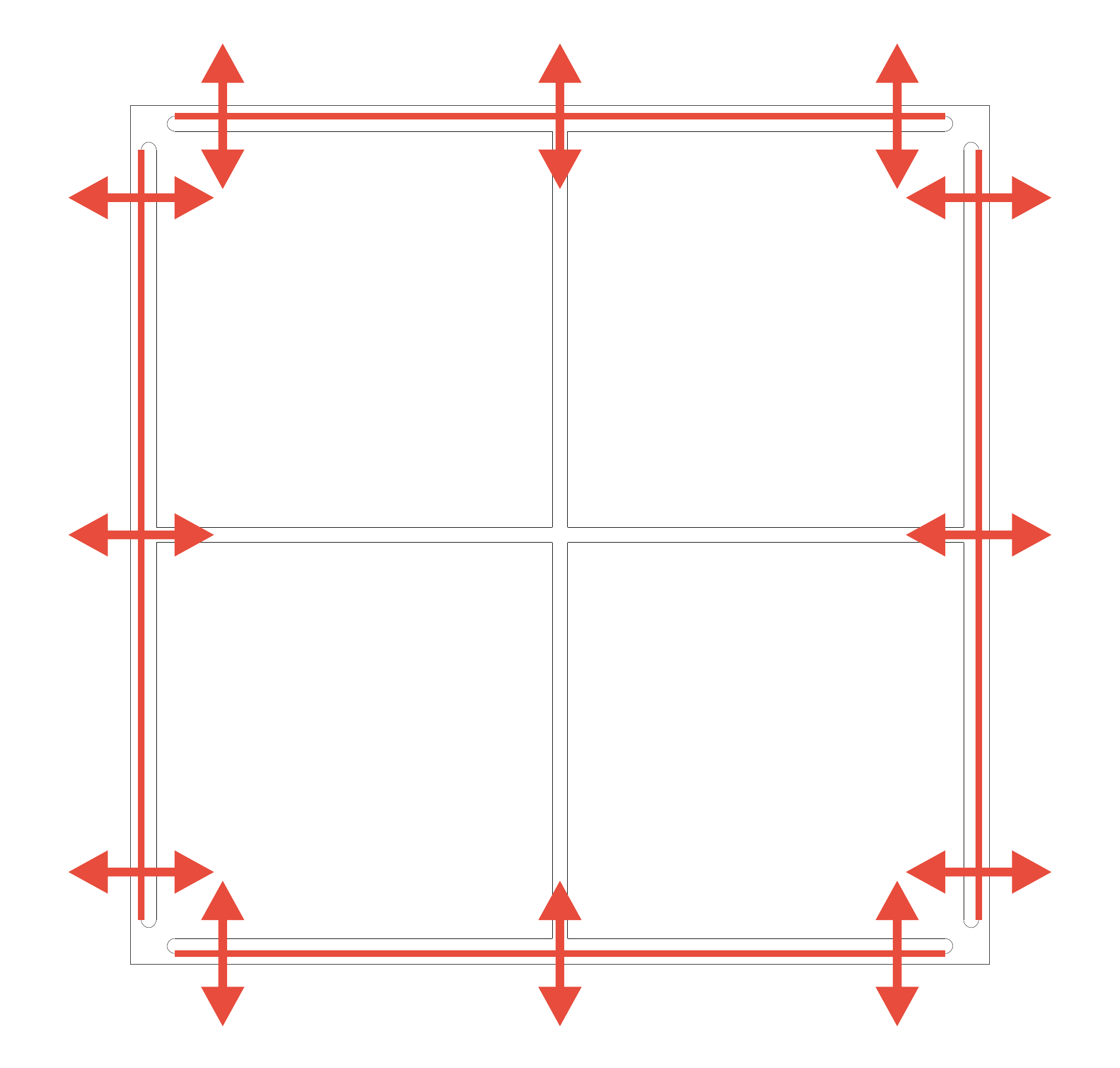}             &
            \includegraphics[width=\figsizeC]{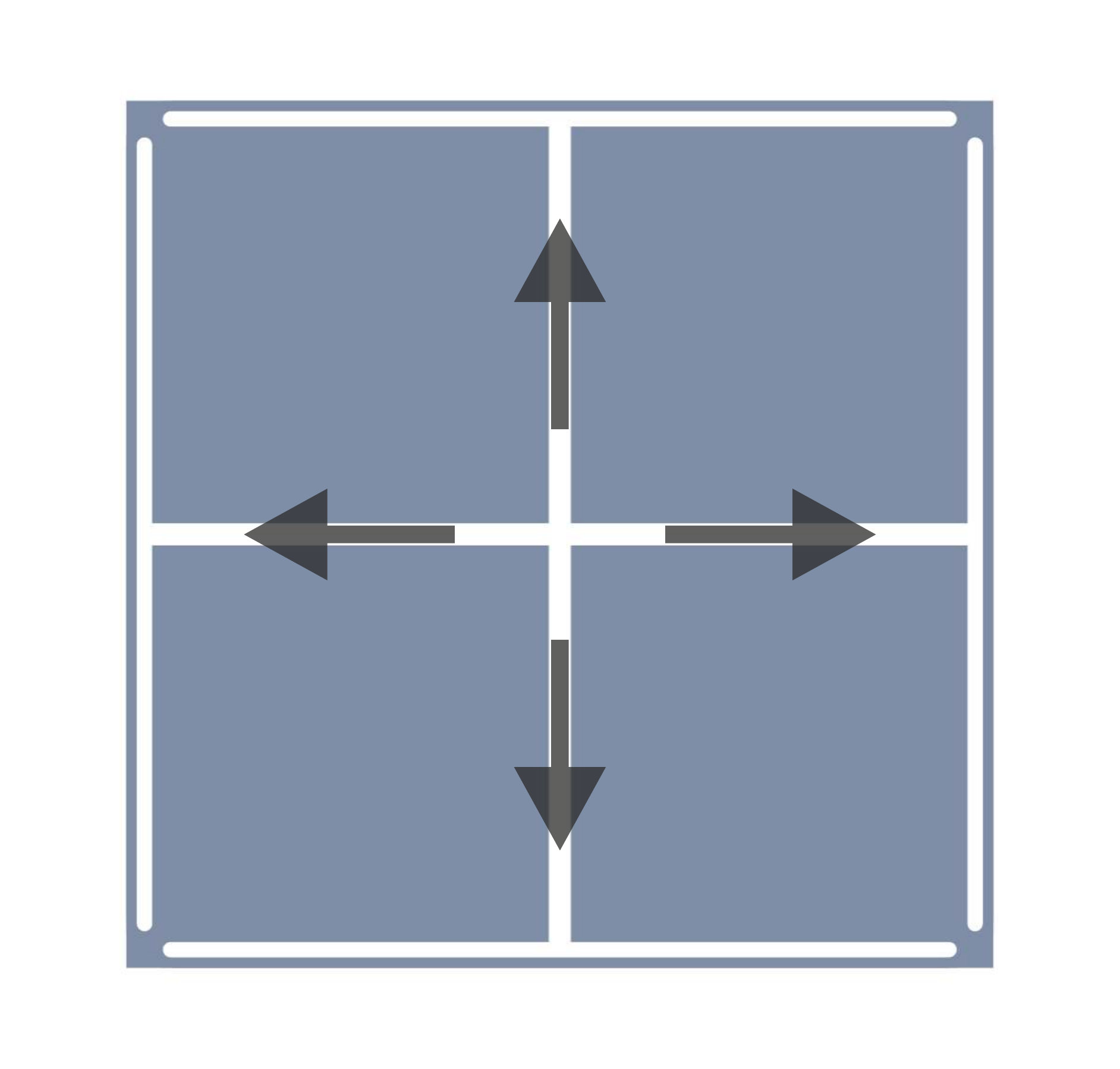} &
            \includegraphics[width=\figsizeC]{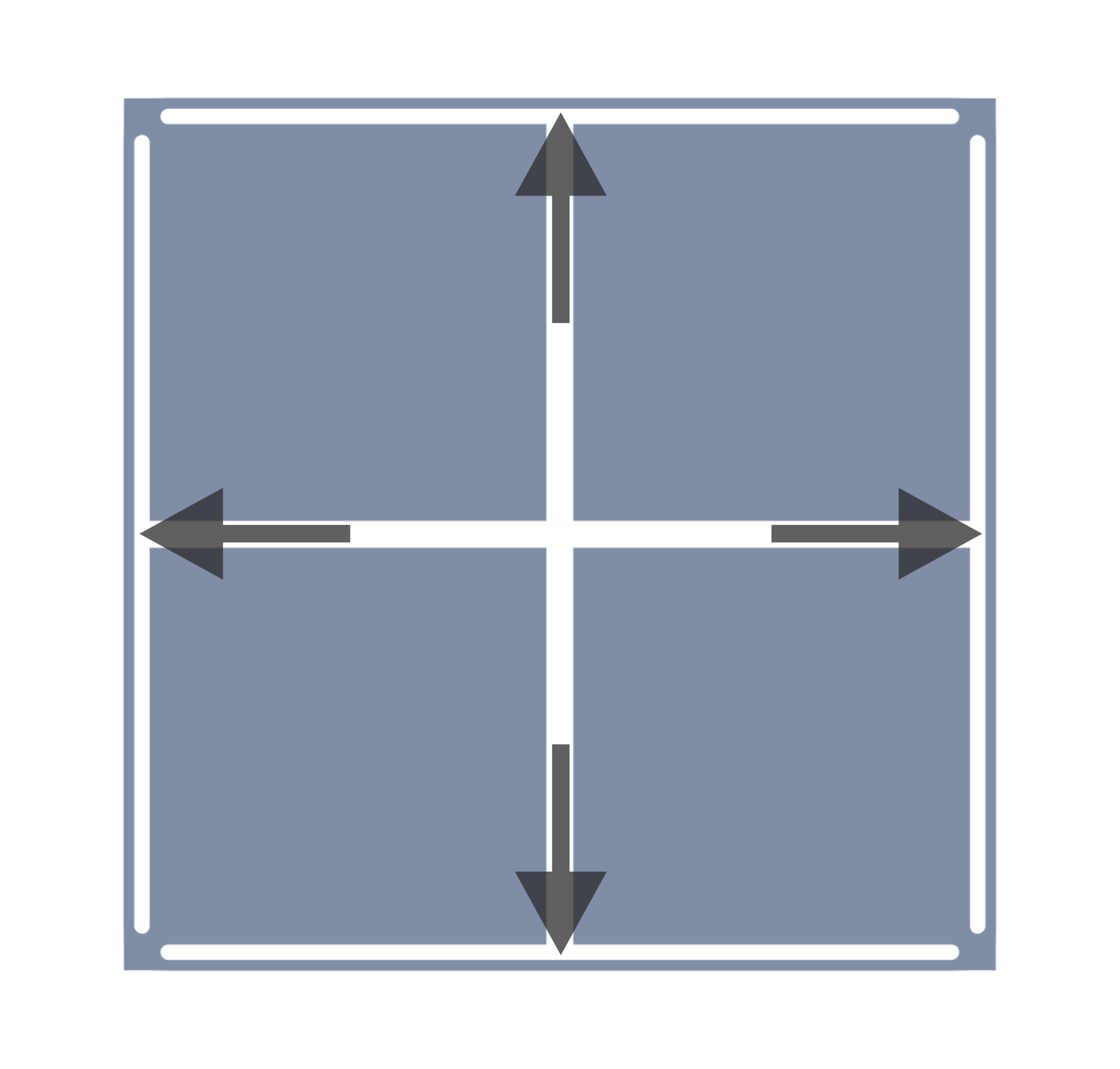} &
            \includegraphics[width=\figsizeC]{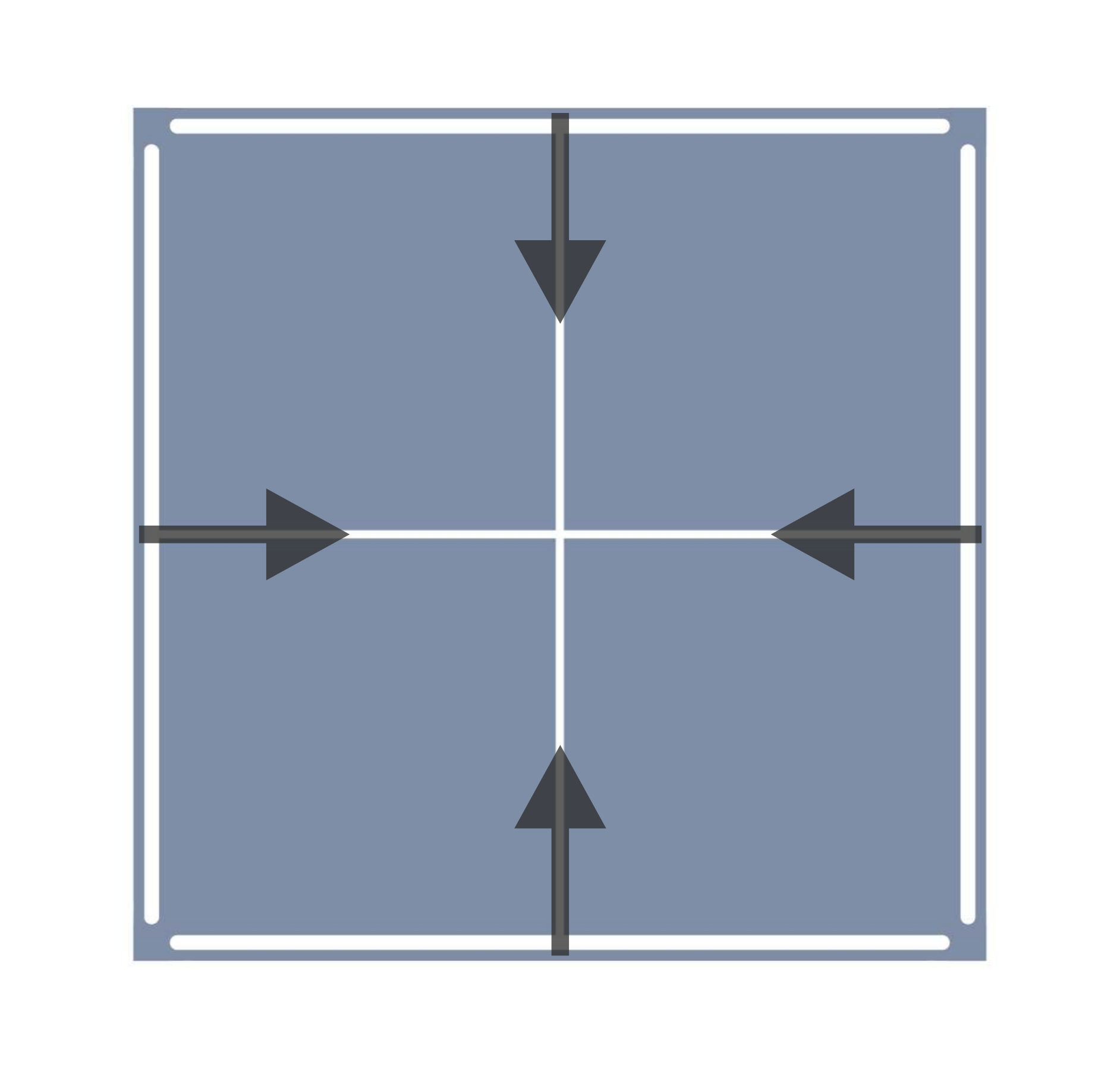} &
            \includegraphics[width=\figsizeC]{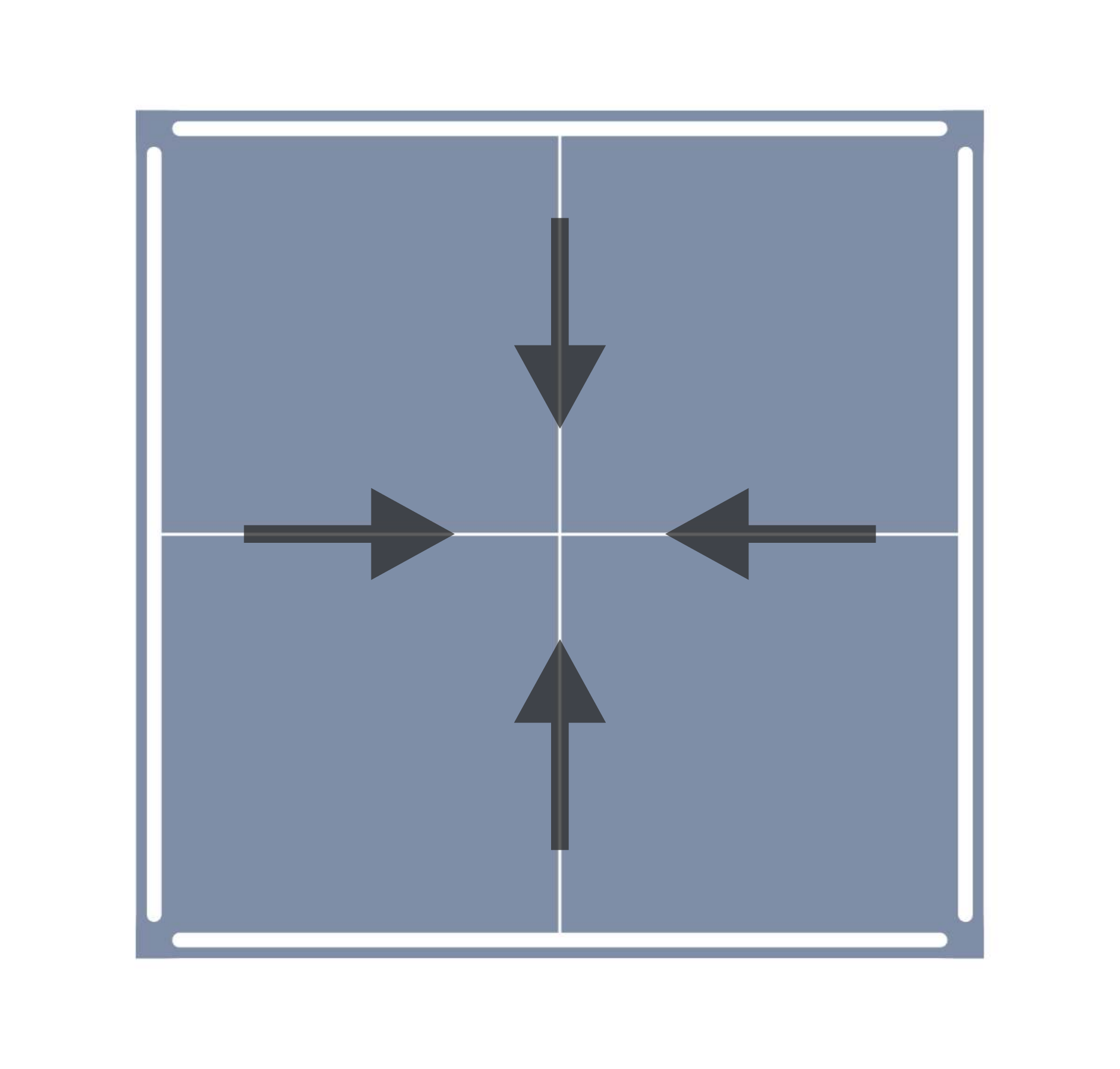}
            \\
            \hline
            \multicolumn{1}{|c|}{\rotatebox{90}{\hspace{0.75cm}Shear}}                                 &
            \includegraphics[width=\figsizeC]{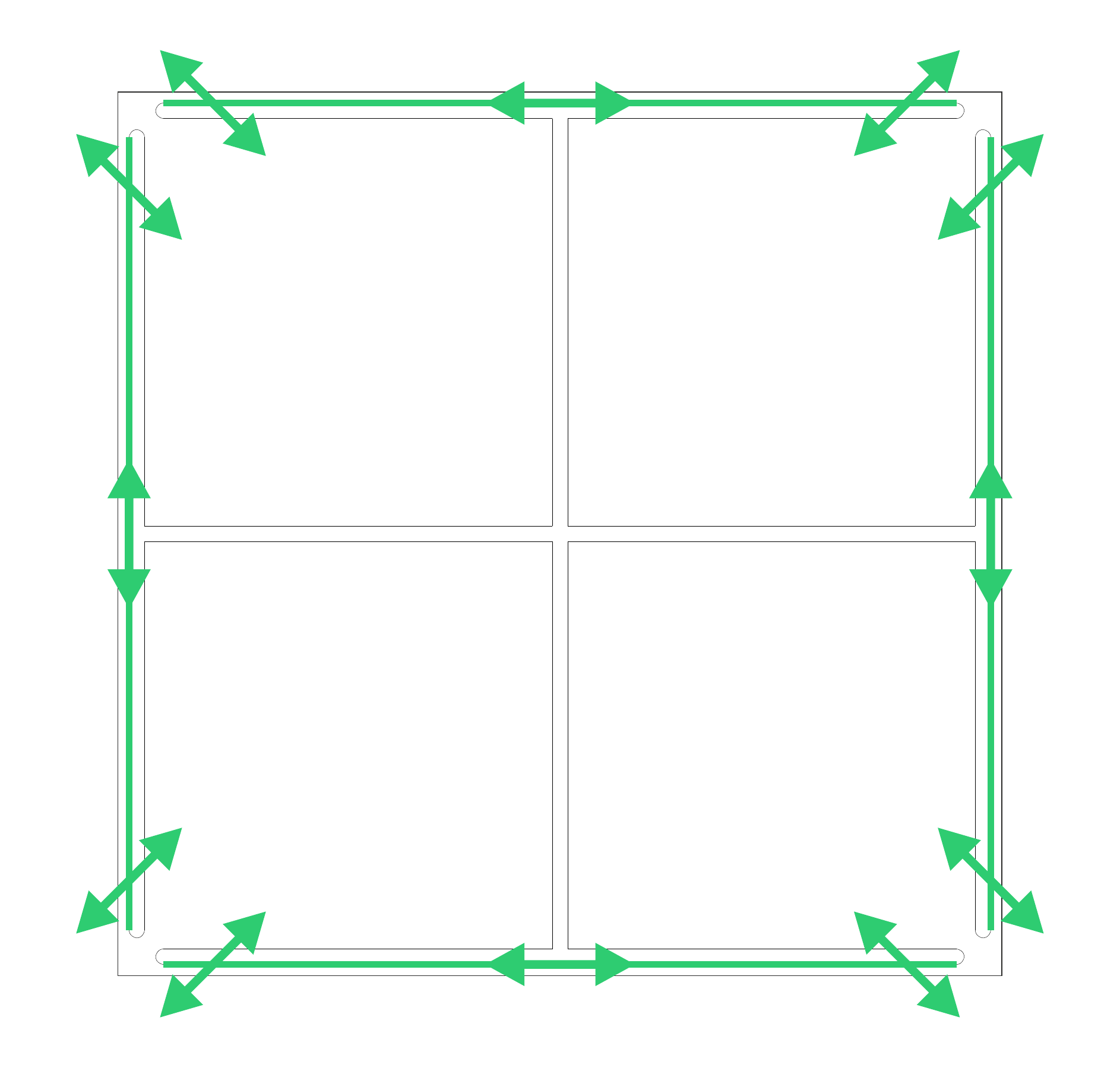}             &
            \includegraphics[width=\figsizeC]{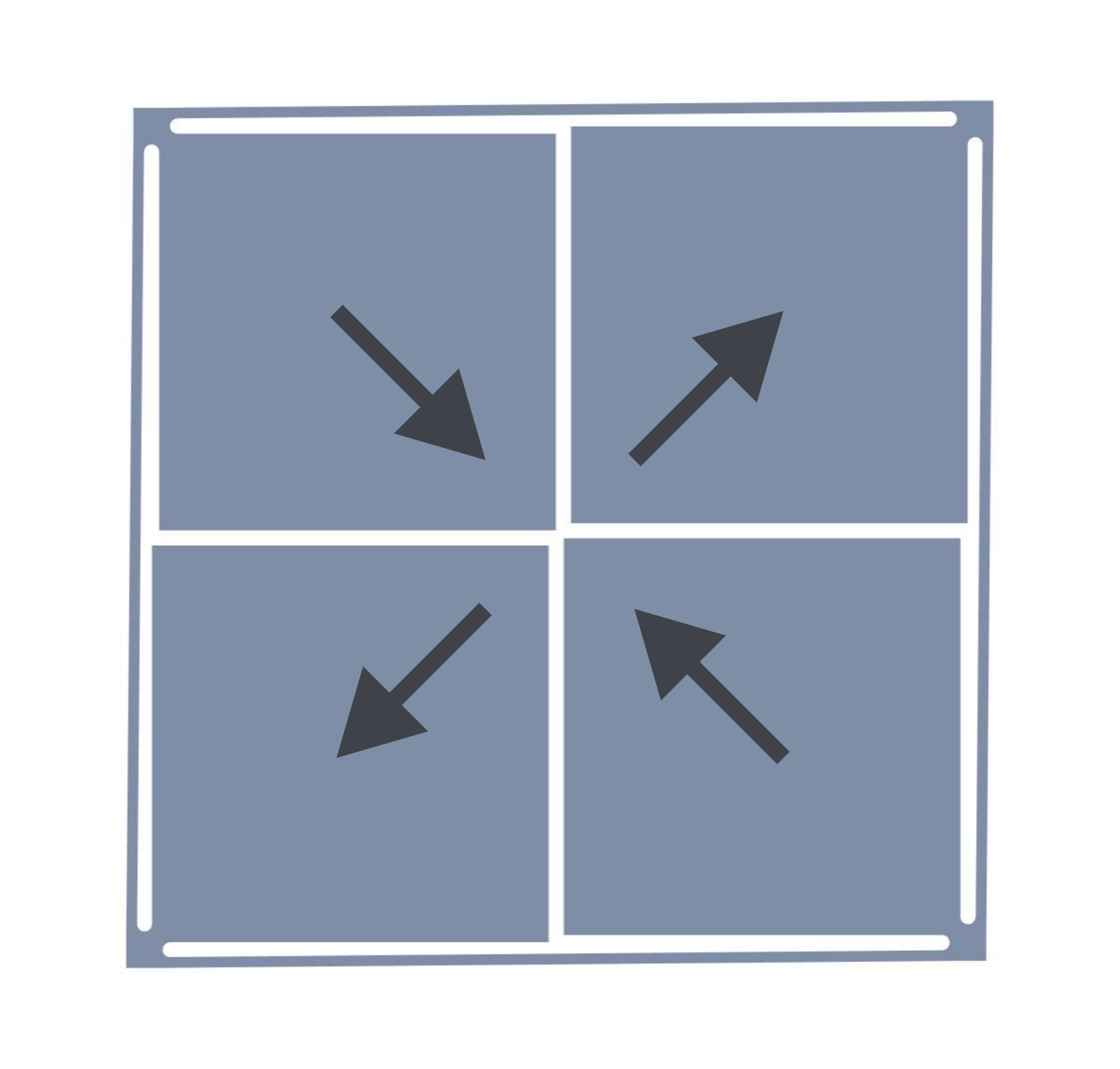} &
            \includegraphics[width=\figsizeC]{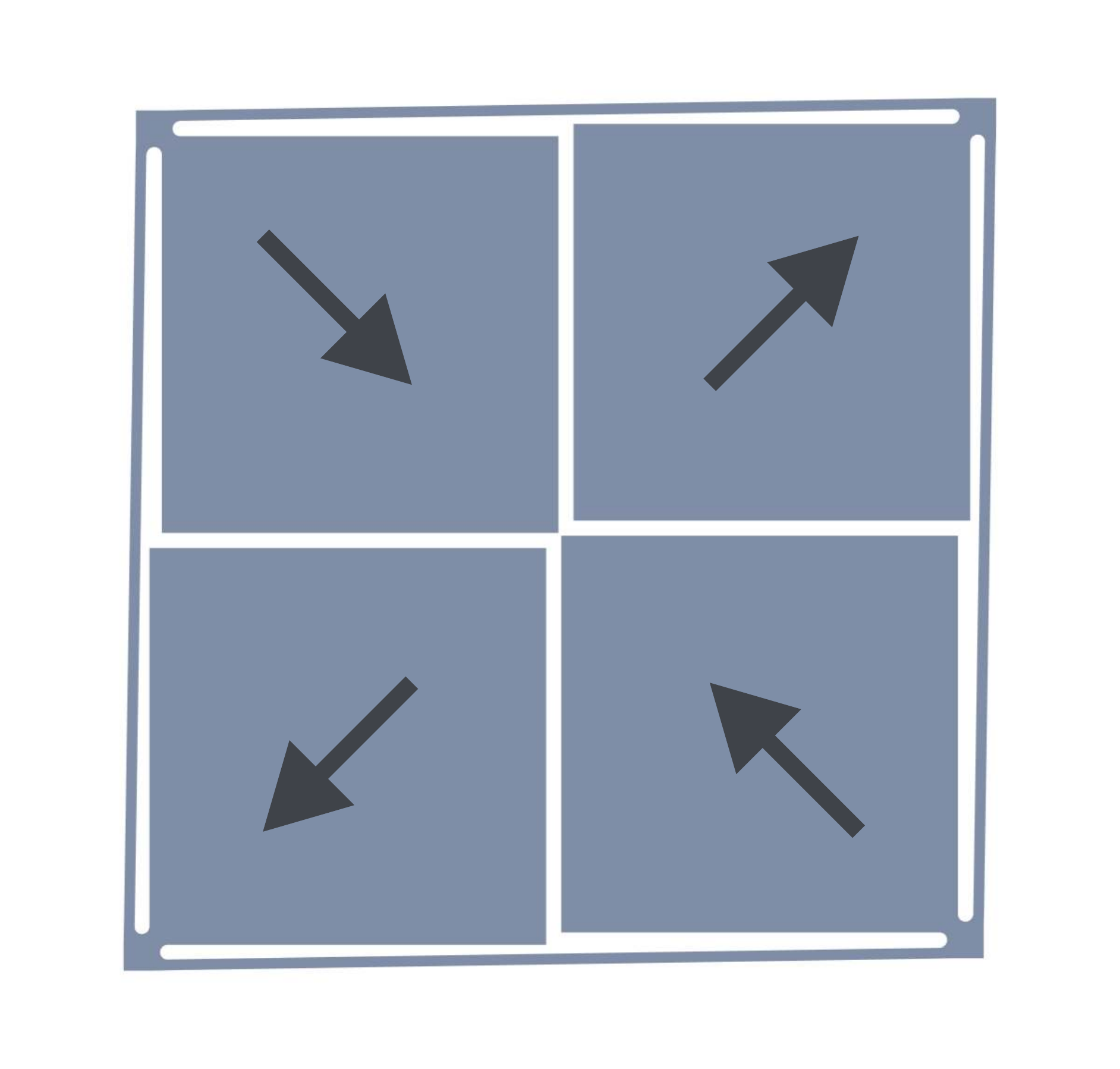} &
            \includegraphics[width=\figsizeC]{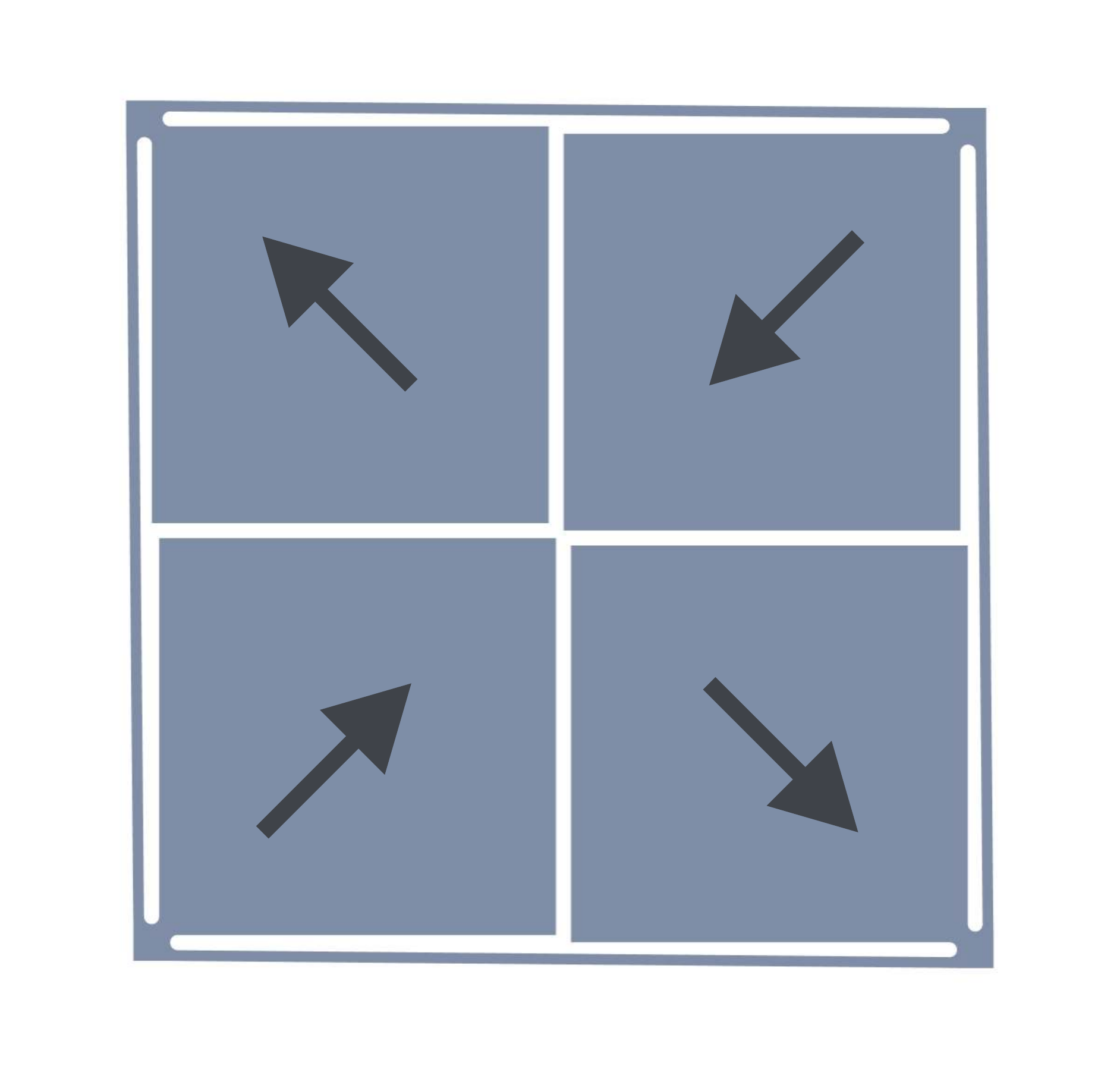} &
            \includegraphics[width=\figsizeC]{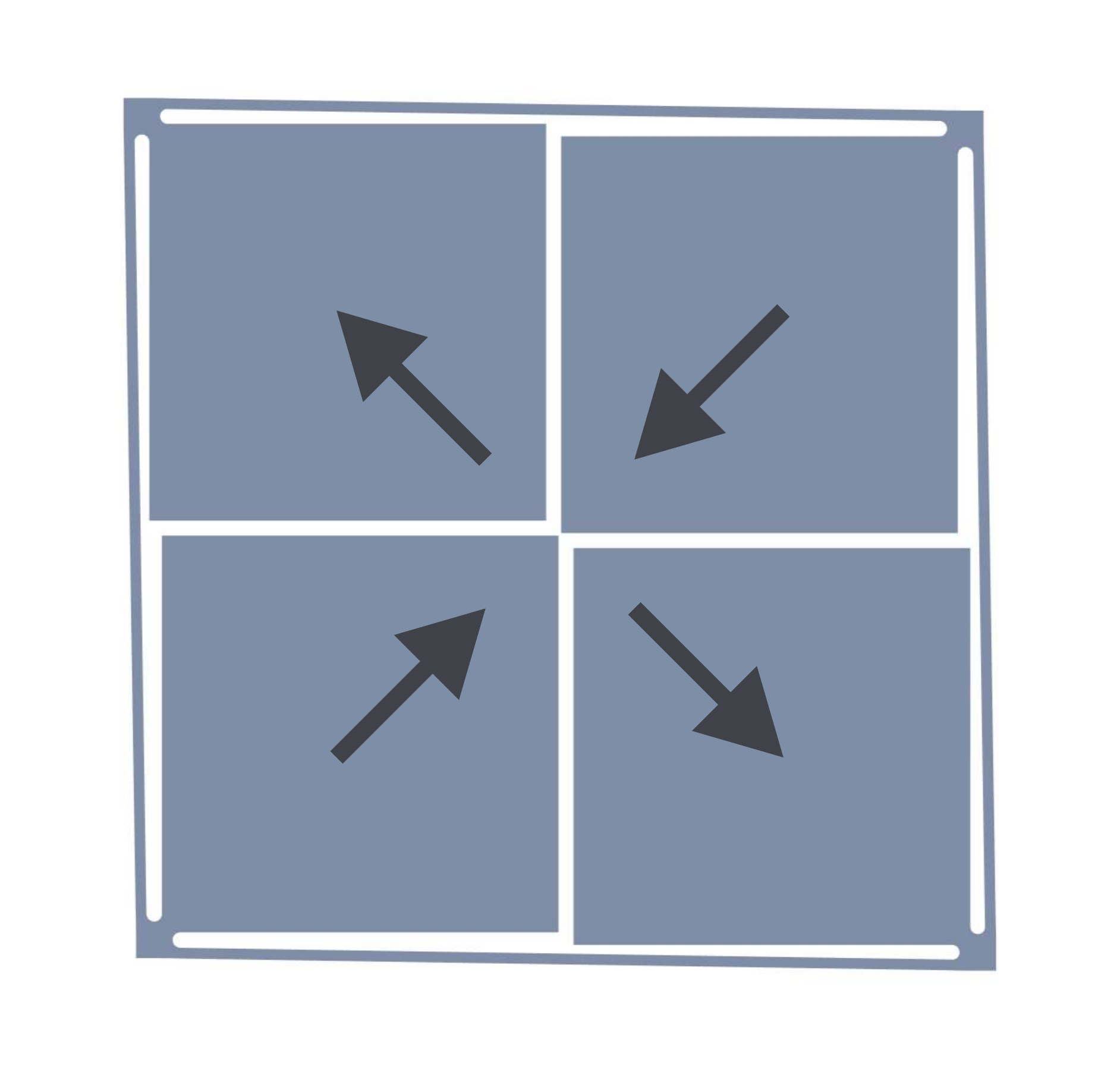}
            \\
            \hline
        \end{tabular}
    }
    \caption{Load conditions for:
        (\textit{first row}) the rotation test and some examples of deformed central unit cell;
        (\textit{second row}) the hydrostatic test and some examples of deformed central unit cell;
        (\textit{third row}) the shear test and some examples of deformed central unit cell.
        The deformation are accentuated to better show the direction of loading.}
    \label{tab:reso_load_def}
\end{table}

\begin{figure}[H]
    \centering
    \includegraphics[width=\figsizeD]{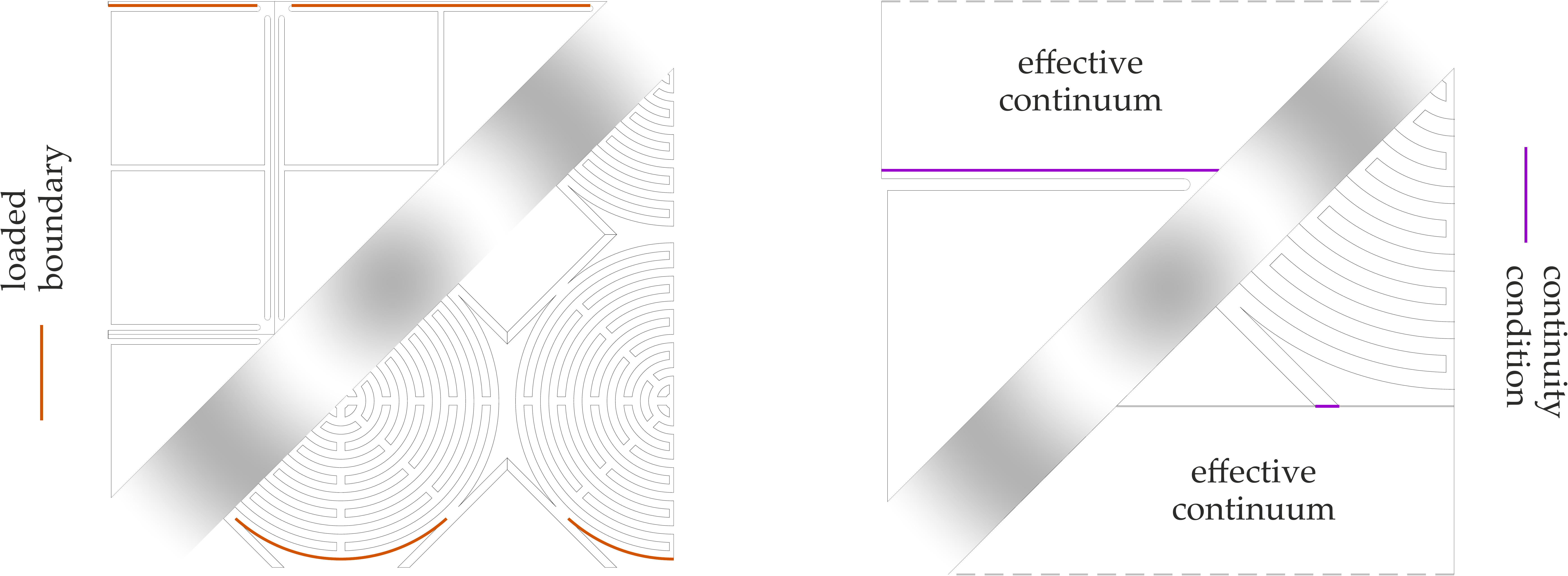}
    \caption{
        (\textit{left}) Boundary load for the study case that have a 3\texttimes 3 cluster of unit cells at the center of the domain for boththe labyrintyn unit cell and the unit cell $\mathcal{R}$;
        (\textit{right}) examplification of the interface between a unit cell (or a cluster) and the outer domain.
        Only the essential domain has been reported here.
    }
    \label{fig:load_3x3}
\end{figure}

Another load case that we will study is the one in which the central unit cell is replaced with a 3\texttimes 3 cluster of unit cells, which allows for a more distributed contact interface.
In the left panel of Fig.~\ref{fig:load_3x3} we report the boundary loaded for both the cells $\mathcal{L}$ cluster and the cells $\mathcal{R}$ cluster, and in the right panle an exemplary case for the interface boundary (just one unit cell and only the essential domains).

The perfect contact condition correspond to the enforcement of the continuity of displacement and traction
\begin{align}
    u_{\rm eff} = u_{\rm mic} &  & \text{and} &  & t_{\rm eff} = t_{\rm mic}\, ,
\end{align}
where $u_{\rm eff}$ and $t_{\rm eff}$ are the displacement and the traction, respectively, either in the RRMM $(u,\widetilde{t}\;\!)$ or the macro Cauchy $(u,t_{\rm M})$, and $u_{\rm mic}$ and $t_{\rm mic}$ are the displacement and the traction of the contact boundary of the central unit cell (or the 3\texttimes 3 cluster), respectively.

\subsection{Boundary conditions on a symmetry and on an antisymmetry plane for a micromorphic model}

All the boundary problems we presented have either a symmetry or an antisymmetry along some planes.
To take advantage of these properties, it is possible to simulate one-eighth of the full domain and apply special boundary conditions on the boundaries identified by the symmetry or antisymmetry planes.

\begin{figure}[H]
    \begin{table}[H]
        \centering
        \begin{tabular}{cccccccc}
            Rotation                                                  &  &  & Hydrostatic &  &  & Shear \\
            \includegraphics[width=\figsizeB]{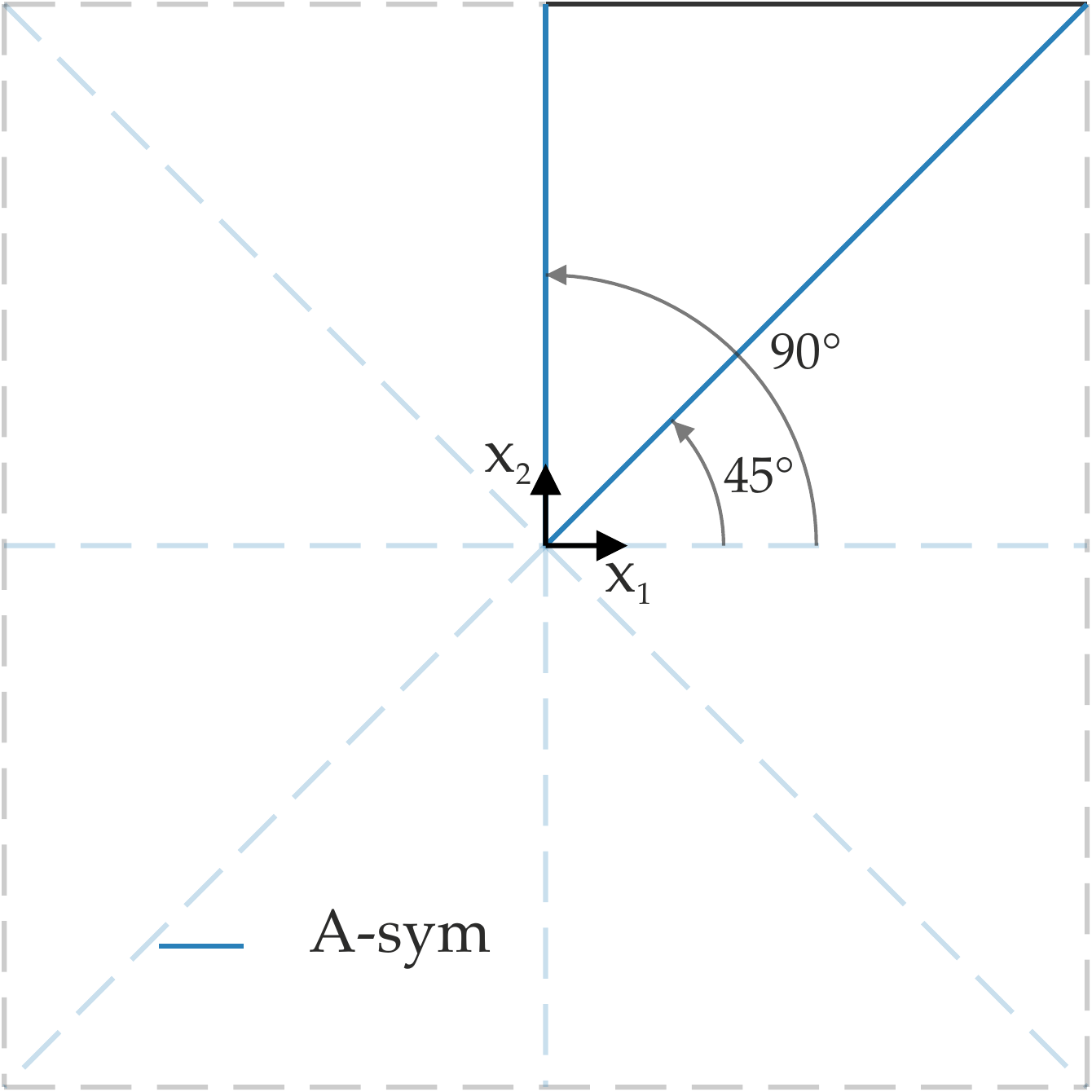} &  &  &
            \includegraphics[width=\figsizeB]{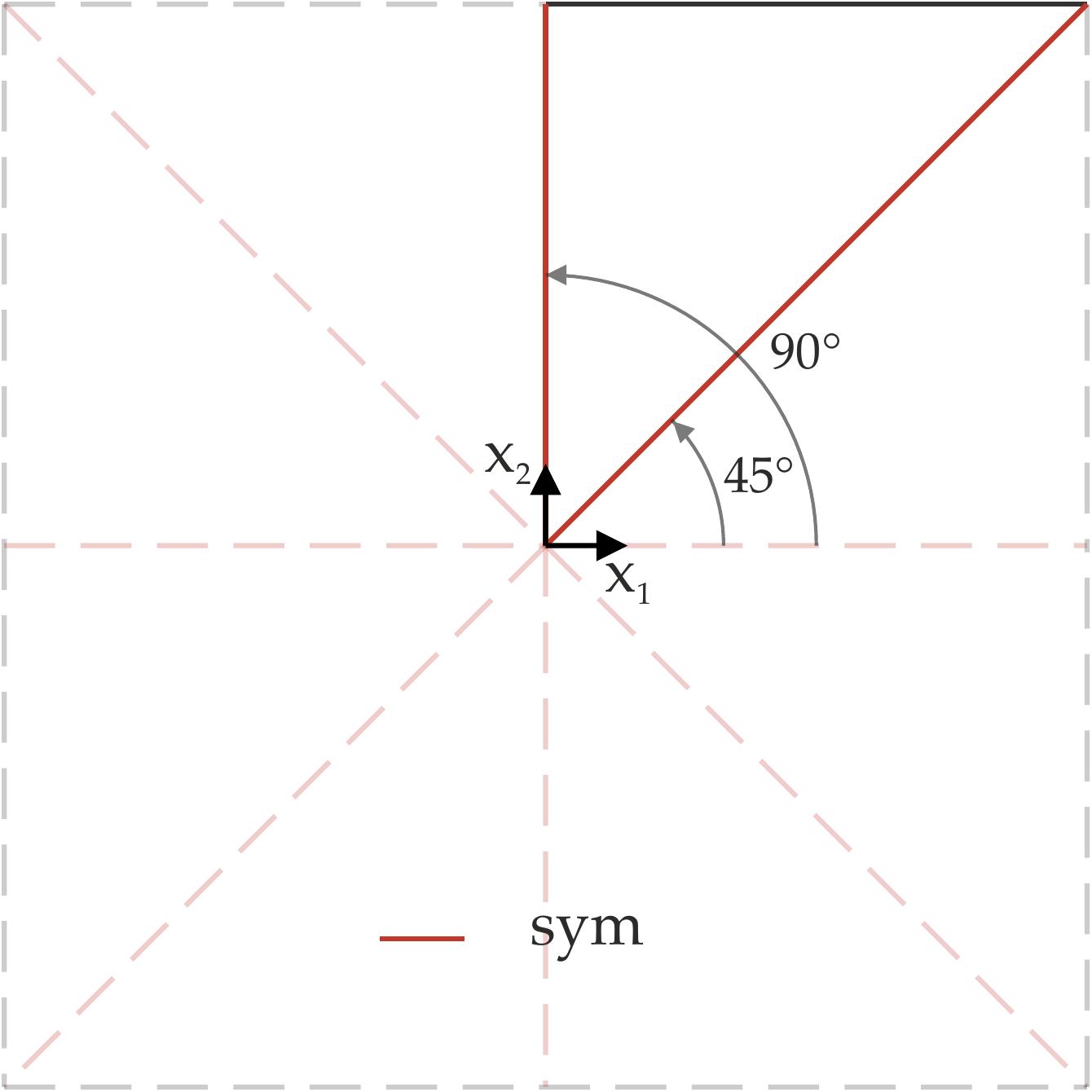}  &  &  &
            \includegraphics[width=\figsizeB]{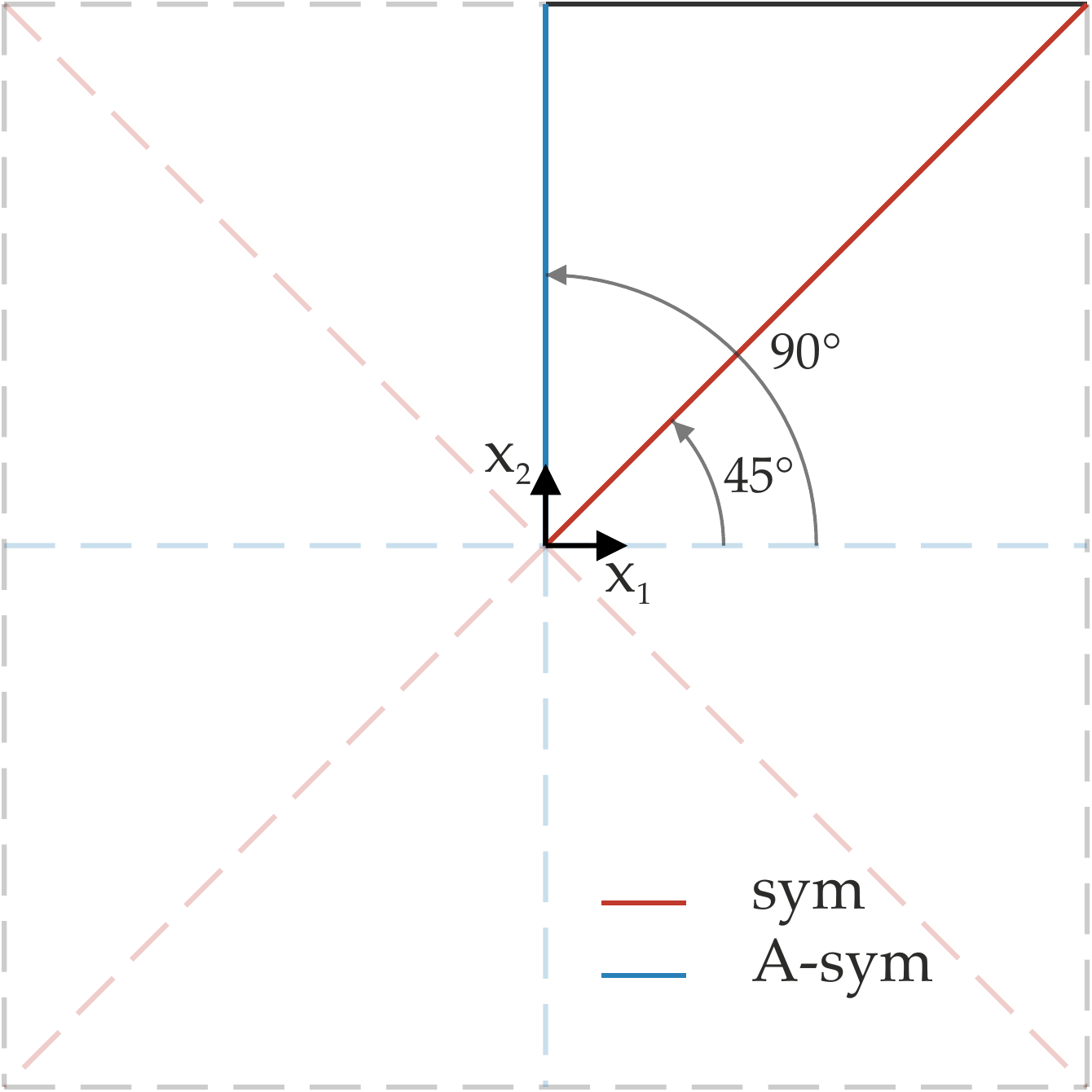}
        \end{tabular}
    \end{table}
    \caption{Scheme of the symmetry conditions:
        (\textit{Rotation}) all the axes of geometrical symmetry require an antisymmetry condition;
        (\textit{Hydrostatic}) all the axes of geometrical symmetry require a symmetry condition;
        (\textit{Shear}) the vertical the axes of geometrical symmetry require an antisymmetry condition, while the oblique a symmetry one.
    }
    \label{fig:symm_cond}
\end{figure}

The symmetry and antisymmetry conditions for the displacement ($u$) and the micro-distortion tensor ($P$) under the plane strain hypothesis can be summarized as (see also \cite{demore2022unfolding} for the symmetry conditions)
\begin{align}
     &
    \text{sym conditions}
     &   &
    \langle u, n \rangle
     &   &
    \text{and}
     &   &
    \langle P, n \otimes t_{\alpha} \rangle
     &   &
    \text{and}
     &   &
    \langle P, t_{\alpha} \otimes n \rangle
    \label{eq:symm_cond_gen}
    \\
     &
    \text{asym conditions}
     &   &
    \langle u, t_{\alpha} \rangle
     &   &
    \text{and}
     &   &
    \langle P, n \otimes n \rangle
     &   &
    \text{and}
     &   &
    \langle P, t_{\alpha} \otimes t_{\alpha} \rangle
    \notag
\end{align}
where $n$ is the normal to the plane, and $t_{\alpha}$ is the unitary tangent vector in the plane that identifies the ``plane strain'' plane.
For the axis parallel to $x_1$ (we will refer to this as 90$^{\circ}$ according to Fig.~\ref{fig:symm_cond}) and for the axis such that $x_1=x_2$ (we will refer to this as 45$^{\circ}$ according to Fig.~\ref{fig:symm_cond}), eq.(\ref{eq:symm_cond_gen}) particularize to
\begin{table}[H]
    \centering
    {\renewcommand{\arraystretch}{1.1}
        \begin{tabular}{cc|c|c|}
            \cline{3-4}
                                                                                &     & Symmetry                             & Antisymmetry                           \\ \cline{3-4}\vspace{0.5mm} \\[-12.25pt]\hline
            \multicolumn{1}{|c|}{\multirow{2}{*}{\rotatebox{90}{90$^{\circ}$}}} & $u$ & $u_1=0$                              & $u_2=0$                                \\ \cline{2-4}
            \multicolumn{1}{|c|}{}                                              & $P$ & $P_{12}=P_{21}=0$                    & $P_{11}=P_{22}=0$                      \\ \hline\hline
            \multicolumn{1}{|c|}{\multirow{2}{*}{\rotatebox{90}{45$^{\circ}$}}} & $u$ & $u_2=u_1$                            & $u_2=-u_2$                             \\ \cline{2-4}
            \multicolumn{1}{|c|}{}                                              & $P$ & $P_{21} = P_{12}$, $P_{22} = P_{11}$ & $P_{21} = -P_{12}$, $P_{22} = -P_{11}$ \\ \hline
        \end{tabular}
    }
    \caption{Summary of the symmetry and antisymmetry conditions for displacement and micro-distortion associated with the axes and the reference systems shown in Fig.~\ref{fig:symm_cond} under the plane strain assumption.}
    \label{tab:summ_sym_cond}
\end{table}

\subsection{Capability measure and its average}
\label{sec:disp_ave}

In order to assess the capability of the effective models to capture the behaviour of the corresponding metamaterial, we usede the respective norm of the displacement field $\lVert u \rVert$ as measure of comaprison.
The norm has been made dimensionless ($\lVert\widetilde{u} \rVert$) by scaling it with respect to the amplitude of the load ($u_0$): the order between taking the norm and the scaling is of course ininflunet as it can be seen in eq.(\ref{eq:avg_operator})$_1$.
In some cases, in order to filter out the contributions from effects below the scale of the size of the unit cell in the microstructures, we have defined an average operator.
Differently from the previous measure, here the order matters, and the average of $\lVert\widetilde{u} \rVert$ (expressed here as $\lVert\overline{u} \rVert$) has been evaluated by first taking the average of the dimensionless displacement component by component, and only then we take the norm as it can be seen in eq.(\ref{eq:avg_operator})$_2$
\begin{align}
    \begin{cases}
        \widetilde{u}_1  \coloneqq \frac{u_1}{u_0} \\[10pt]
        \widetilde{u}_2  \coloneqq \frac{u_2}{u_0}
    \end{cases}
     &                &
    \Longleftrightarrow
     &                &
    \lVert \widetilde{u} \rVert = \frac{\lVert u\rVert}{u_0} \, ,
     & \hspace{1.5cm} &
    \begin{cases}
        \overline{u}_1 \coloneqq \frac{1}{S} \int_{S} \lVert \widetilde{u}_1 \rVert \, \text{d}s \\[10pt]
        \overline{u}_2 \coloneqq \frac{1}{S} \int_{S} \lVert \widetilde{u}_2 \rVert \, \text{d}s
    \end{cases}
     &                &
    \Longrightarrow
     &                &
    \lVert\overline{u} \rVert = \sqrt{\overline{u}_1^2 + \overline{u}_2^2} \, ,
    \label{eq:avg_operator}
\end{align}
where $u_1$ and $u_2$ are the components of the displaement field and $S = a^2$ is the area of the square identifyng the size of the unit cells.

\begin{figure}[H]
    \begin{table}[H]
        \centering
        \begin{tabular}{
            >{\centering\arraybackslash}m{4cm}
            m{0.3cm}
            >{\centering\arraybackslash}m{4cm}
            m{0.3cm}
            >{\centering\arraybackslash}m{4cm}
            }
            $u_1$                                                                                &   & $u_2$ &  & $\lVert u \rVert$ \\[5pt]
            \includegraphics[width=\figsizeE]{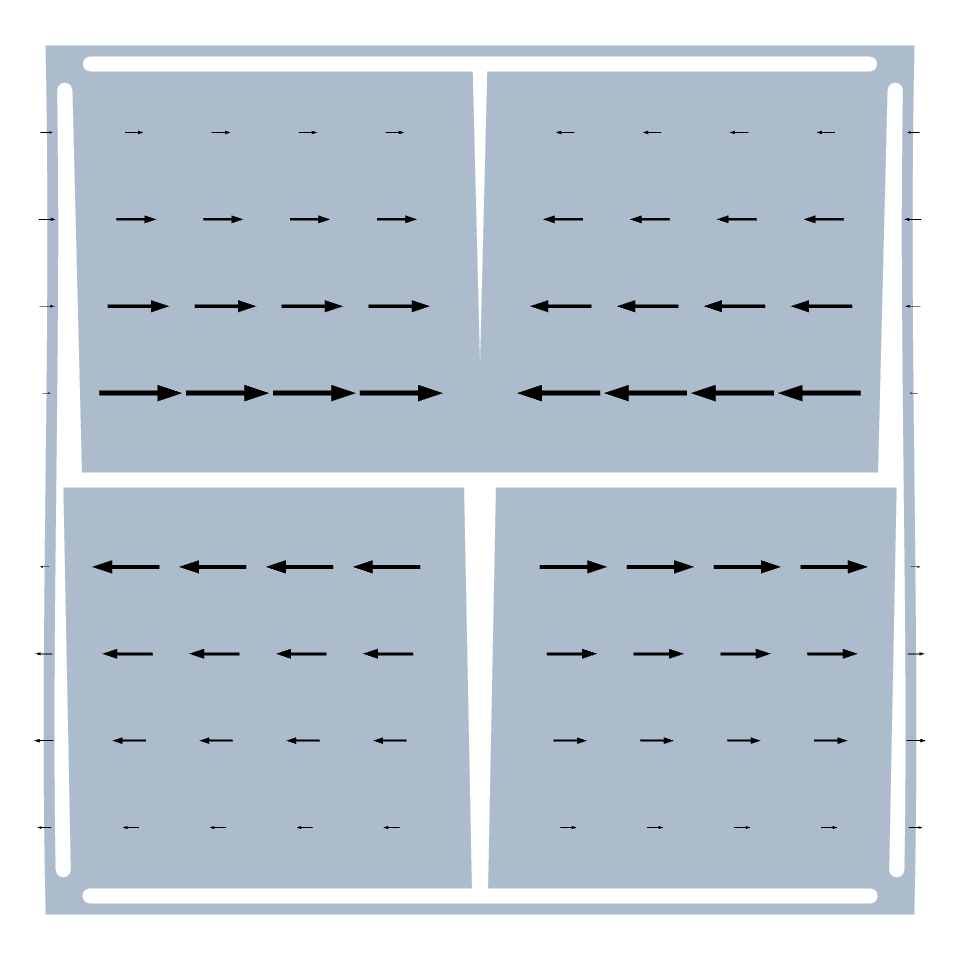} & + &
            \includegraphics[width=\figsizeE]{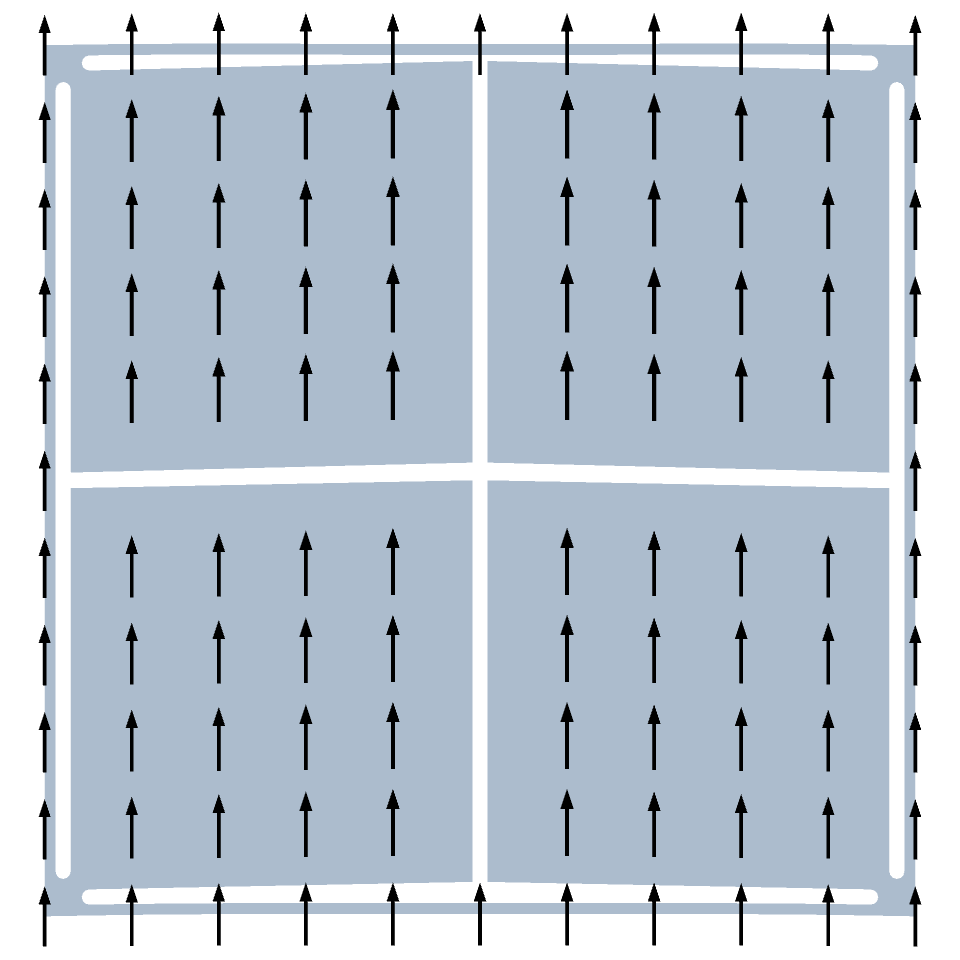} & = &
            \includegraphics[width=\figsizeE]{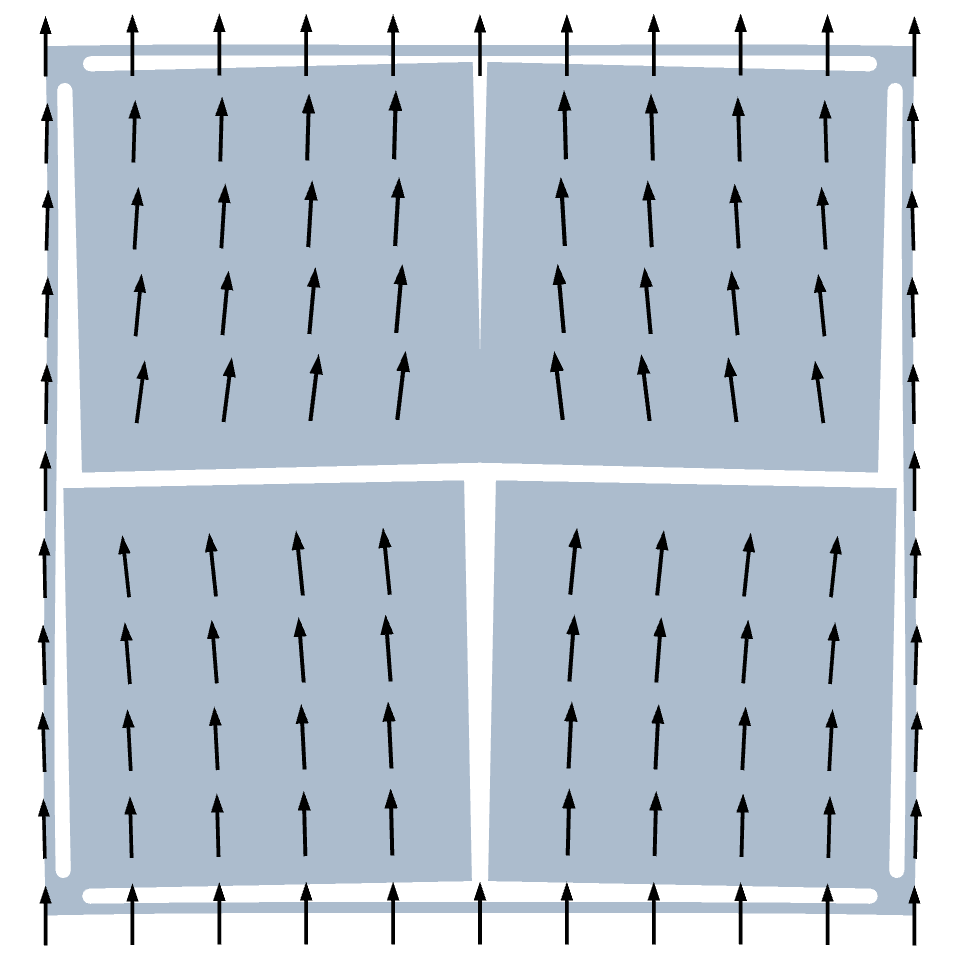}                                    \\[5pt]
            \includegraphics[width=\figsizeE]{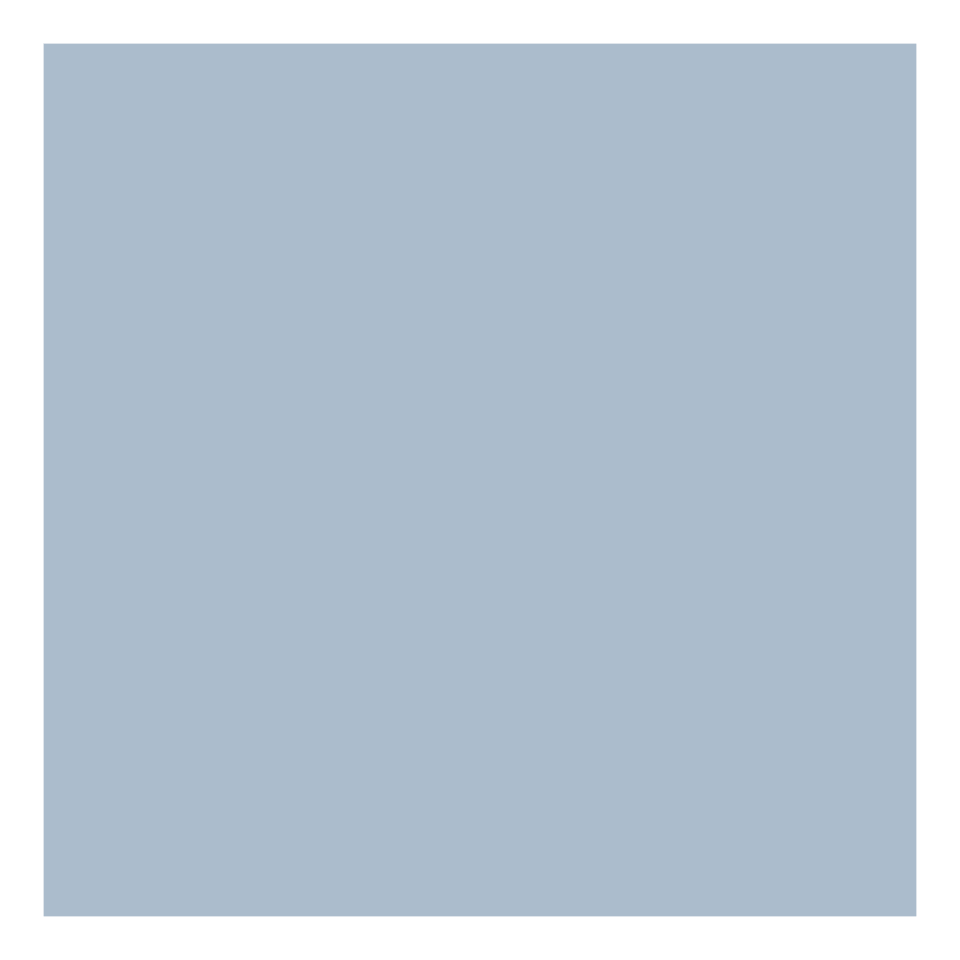} & + &
            \includegraphics[width=\figsizeE]{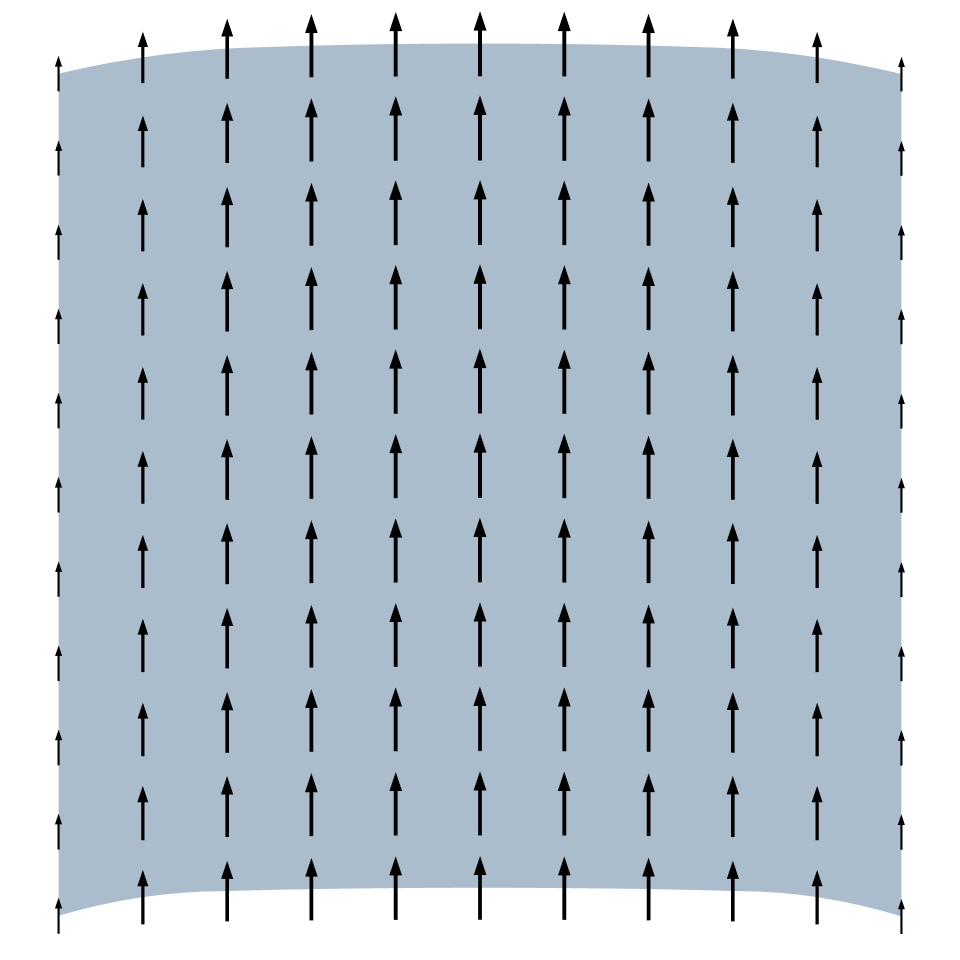} & = &
            \includegraphics[width=\figsizeE]{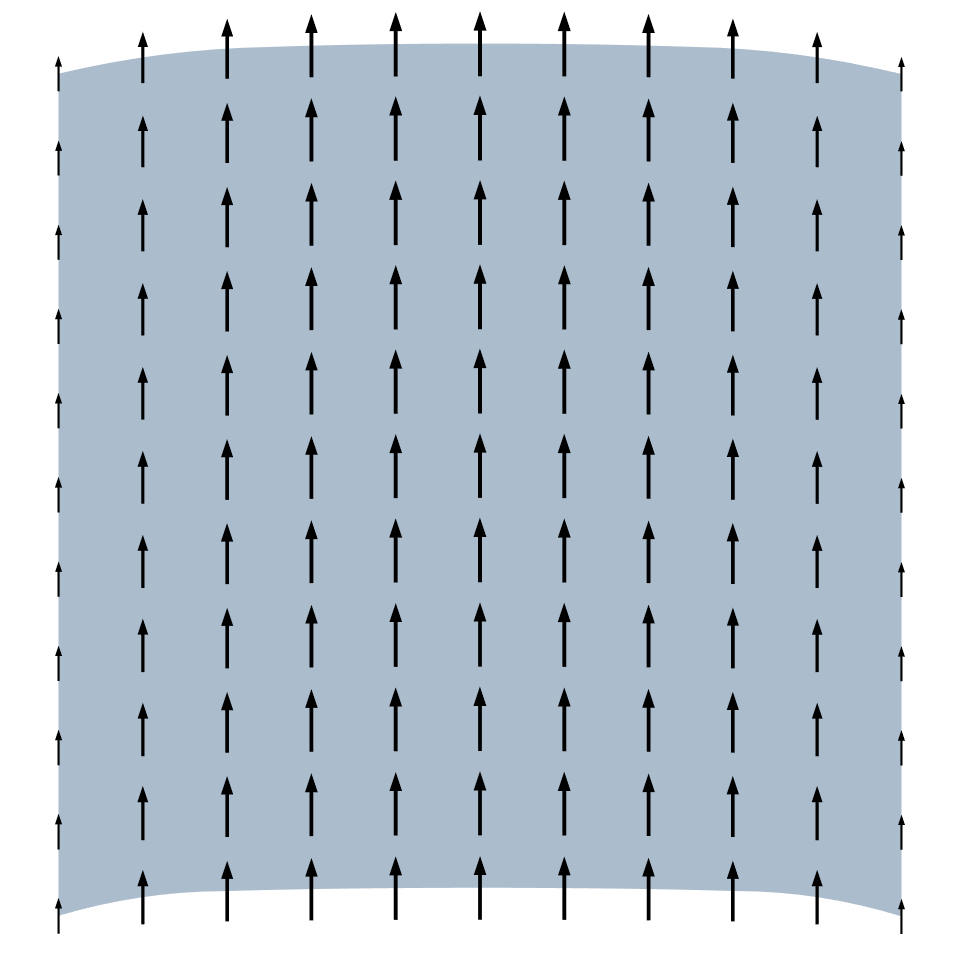}
        \end{tabular}
    \end{table}
    \caption{
        Arrow plot of the displacement field's components $u_1$, $u_2$ and norm $\lVert u \rVert$ for the unit cell $\mathcal{R}$ and its RRMM counterpart.
        It is possible to see that the contribution of the component $u_1$ is localised in the resonator, and it is negligable in the RRMM.
        This effect at sub-cell level cannot be captured by the effective model here proposed, but it is possible to remove it with the average operator $\lVert \overline{u} \rVert$ defined in eq.(\ref{eq:avg_operator})$_2$.
        The specific unit cells reported is the fourth one starting from the central one along the vertical axis of symmetry for the 100 Hz hydrostatic test case (Fig.~\ref{tab:reso_expa_figu_100}), and both the deformation and lenght of the arrows have been scaled for a better visual effect.
    }
    \label{fig:unit_cell_filed}
\end{figure}

In Fig.~\ref{fig:unit_cell_filed} it is possible to see an arrow plot of the displacement field's and norm for the unit cell $\mathcal{R}$ and its RRMM counterpart, for the 100 Hz hydrostatic test case.

It is possible to see how in this case the contribution of the component $u_1$ is localised in the resonator, and it is negligable in the RRMM.
This contribution at sub-cell level cannot be captured by the effective models here proposed, but it is possible to remove it with the average operator $\lVert \overline{u} \rVert$ defined in eq.(\ref{eq:avg_operator})$_2$.

\section{Time-dependent response for a \textit{labyrinthine} unit cell}
\label{sec:maze}

\begin{table}[H]
    \centering
    \begin{tabular}{c|c|c|c|c}
        \cline{2-4}
                                                                                                          & RRMM
                                                                                                          & Microstructure
                                                                                                          & Macro Cauchy
                                                                                                          & \multicolumn{1}{c}{}
        \\ \hline
        \multicolumn{1}{|c|}{\rotatebox{90}{\makebox[\figsizeF][c]{$t= 0.005$ s}}}                        &
        \includegraphics[width=\figsizeF]{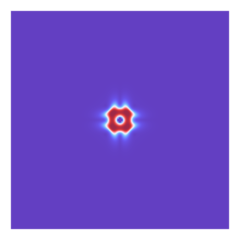} &
        \includegraphics[width=\figsizeF]{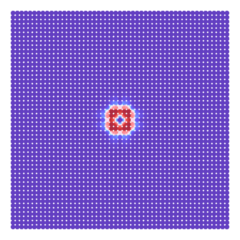} &
        \includegraphics[width=\figsizeF]{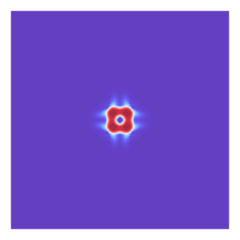} &
        \multicolumn{1}{c|}{}
        \\ \cline{1-4}
        \multicolumn{1}{|c|}{\rotatebox{90}{\makebox[\figsizeF][c]{$t= 0.010$ s}}}                        &
        \includegraphics[width=\figsizeF]{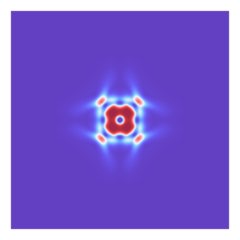} &
        \includegraphics[width=\figsizeF]{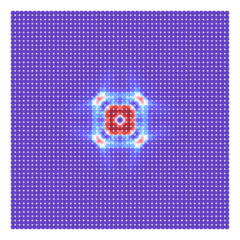} &
        \includegraphics[width=\figsizeF]{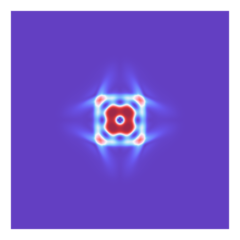} &
        \multicolumn{1}{c|}{\rotatebox{90}{\makebox[\figsizeF][c]{Rotation load - \fbox{1\texttimes 1} - $\omega = 100$ Hz}}}
        \\ \cline{1-4}
        \multicolumn{1}{|c|}{\rotatebox{90}{\makebox[\figsizeF][c]{$t= 0.015$ s}}}                        &
        \includegraphics[width=\figsizeF]{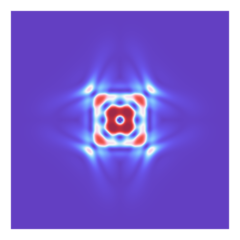} &
        \includegraphics[width=\figsizeF]{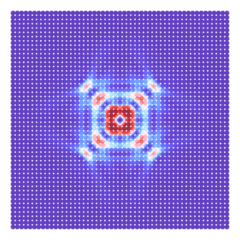} &
        \includegraphics[width=\figsizeF]{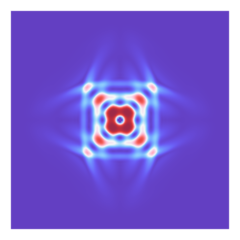} &
        \multicolumn{1}{c|}{}
        \\ \hline
    \end{tabular}
    \caption{
        Dimensionless displacement norm ($\lVert \widetilde{u} \rVert$):
        (\textit{first column}) equivalent RRMM,
        (\textit{second column}) metamaterial $\mathcal{L}$,
        (\textit{third column}) macro Cauchy,
        for a rotation load (Sec.~\ref{sec:load_boundary}) for $\omega = 100$ Hz with a single central unit cell.
    }
    \label{tab:maze_rota_figu_100}
\end{table}

\begin{table}[H]
    \centering
    \begin{tabular}{c|c|c|c|c}
        \cline{2-4}
                                                                                                          & RRMM
                                                                                                          & Microstructure
                                                                                                          & Macro Cauchy
                                                                                                          & \multicolumn{1}{c}{}
        \\ \hline
        \multicolumn{1}{|c|}{\rotatebox{90}{\makebox[\figsizeF][c]{$t= 0.005$ s}}}                        &
        \includegraphics[width=\figsizeF]{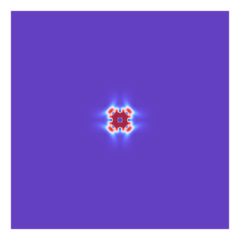} &
        \includegraphics[width=\figsizeF]{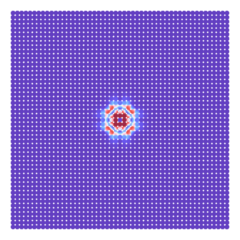} &
        \includegraphics[width=\figsizeF]{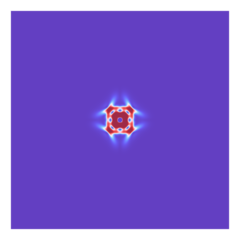} &
        \multicolumn{1}{c|}{}
        \\ \cline{1-4}
        \multicolumn{1}{|c|}{\rotatebox{90}{\makebox[\figsizeF][c]{$t= 0.010$ s}}}                        &
        \includegraphics[width=\figsizeF]{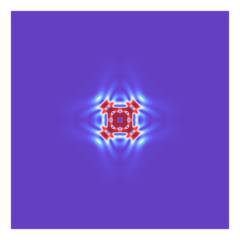} &
        \includegraphics[width=\figsizeF]{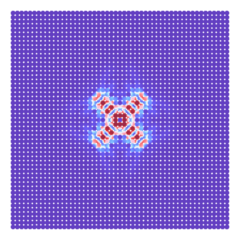} &
        \includegraphics[width=\figsizeF]{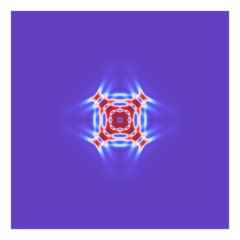} &
        \multicolumn{1}{c|}{\rotatebox{90}{\makebox[\figsizeF][c]{Rotation load - \fbox{1\texttimes 1} - $\omega = 200$ Hz}}}
        \\ \cline{1-4}
        \multicolumn{1}{|c|}{\rotatebox{90}{\makebox[\figsizeF][c]{$t= 0.015$ s}}}                        &
        \includegraphics[width=\figsizeF]{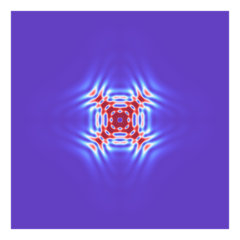} &
        \includegraphics[width=\figsizeF]{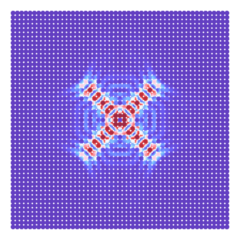} &
        \includegraphics[width=\figsizeF]{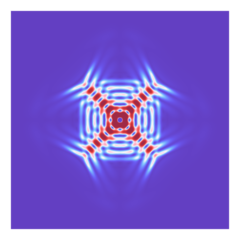} &
        \multicolumn{1}{c|}{}
        \\ \hline
    \end{tabular}
    \caption{
        Dimensionless displacement norm ($\lVert \widetilde{u} \rVert$):
        (\textit{first column}) equivalent RRMM,
        (\textit{second column}) metamaterial $\mathcal{L}$,
        (\textit{third column}) macro Cauchy,
        for a rotation load (Sec.~\ref{sec:load_boundary}) for $\omega = 200$ Hz with a single central unit cell.
    }
    \label{tab:maze_rota_figu_200}
\end{table}

\begin{table}[H]
    \centering
    \begin{tabular}{c|c|c|c|c}
        \cline{2-4}
                                                                                                           & RRMM
                                                                                                           & Microstructure
                                                                                                           & Macro Cauchy
                                                                                                           & \multicolumn{1}{c}{}
        \\ \hline
        \multicolumn{1}{|c|}{\rotatebox{90}{\makebox[\figsizeF][c]{$t= 0.005$ s}}}                         &
        \includegraphics[width=\figsizeF]{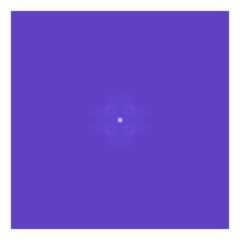} &
        \includegraphics[width=\figsizeF]{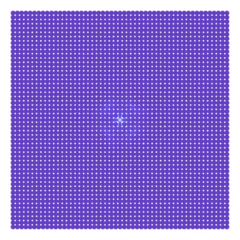} &
        \includegraphics[width=\figsizeF]{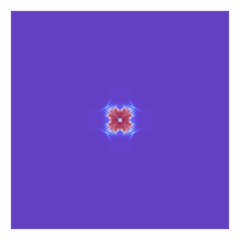} &
        \multicolumn{1}{c|}{}
        \\ \cline{1-4}
        \multicolumn{1}{|c|}{\rotatebox{90}{\makebox[\figsizeF][c]{$t= 0.010$ s}}}                         &
        \includegraphics[width=\figsizeF]{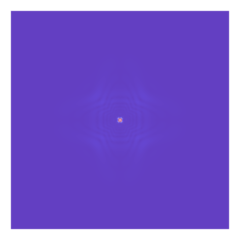} &
        \includegraphics[width=\figsizeF]{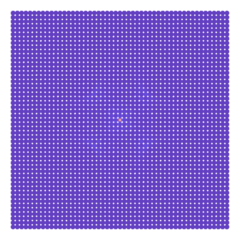} &
        \includegraphics[width=\figsizeF]{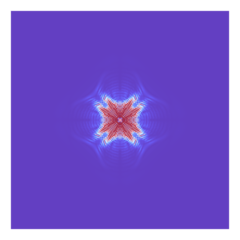} &
        \multicolumn{1}{c|}{\rotatebox{90}{\makebox[\figsizeF][c]{Rotation load - \fbox{1\texttimes 1} - $\omega = 1000$ Hz}}}
        \\ \cline{1-4}
        \multicolumn{1}{|c|}{\rotatebox{90}{\makebox[\figsizeF][c]{$t= 0.015$ s}}}                         &
        \includegraphics[width=\figsizeF]{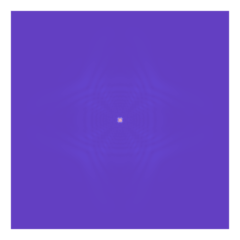} &
        \includegraphics[width=\figsizeF]{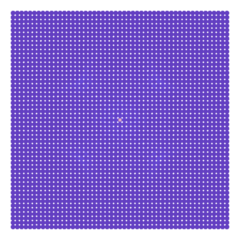} &
        \includegraphics[width=\figsizeF]{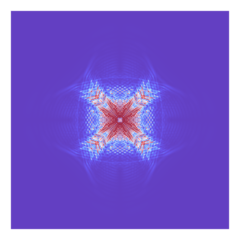} &
        \multicolumn{1}{c|}{}
        \\ \hline
    \end{tabular}
    \caption{
        Dimensionless displacement norm ($\lVert \widetilde{u} \rVert$):
        (\textit{first column}) equivalent RRMM,
        (\textit{second column}) metamaterial $\mathcal{L}$,
        (\textit{third column}) macro Cauchy,
        for a rotation load (Sec.~\ref{sec:load_boundary}) for $\omega = 1000$ Hz with a single central unit cell.
    }
    \label{tab:maze_rota_figu_1000}
\end{table}

In Figs.~\ref{tab:maze_rota_figu_100}--\ref{tab:maze_rota_figu_1000}, we observe that the RRMM shows good agreement at 100 Hz and 1000 Hz, while minor quantitative differences arise at 200 Hz.

\begin{table}[H]
    \centering
    \begin{tabular}{c|c|c|c|c}
        \cline{2-4}
                                                                                                              & RRMM
                                                                                                              & Microstructure
                                                                                                              & Macro Cauchy
                                                                                                              & \multicolumn{1}{c}{}
        \\ \hline
        \multicolumn{1}{|c|}{\rotatebox{90}{\makebox[\figsizeF][c]{$t= 0.005$ s}}}                            &
        \includegraphics[width=\figsizeF]{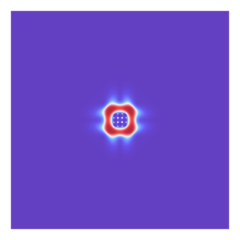} &
        \includegraphics[width=\figsizeF]{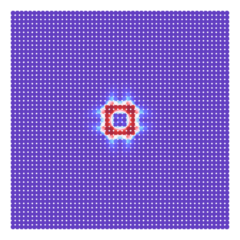} &
        \includegraphics[width=\figsizeF]{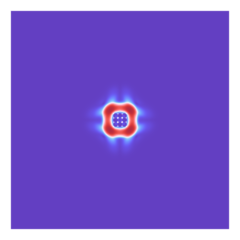} &
        \multicolumn{1}{c|}{}
        \\ \cline{1-4}
        \multicolumn{1}{|c|}{\rotatebox{90}{\makebox[\figsizeF][c]{$t= 0.010$ s}}}                            &
        \includegraphics[width=\figsizeF]{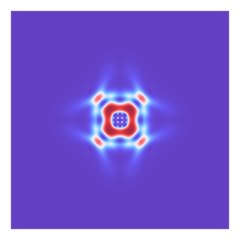} &
        \includegraphics[width=\figsizeF]{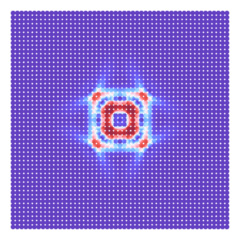} &
        \includegraphics[width=\figsizeF]{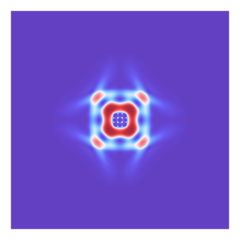} &
        \multicolumn{1}{c|}{\rotatebox{90}{\makebox[\figsizeF][c]{Rotation load - \fbox{3\texttimes 3} - $\omega = 100$ Hz}}}
        \\ \cline{1-4}
        \multicolumn{1}{|c|}{\rotatebox{90}{\makebox[\figsizeF][c]{$t= 0.015$ s}}}                            &
        \includegraphics[width=\figsizeF]{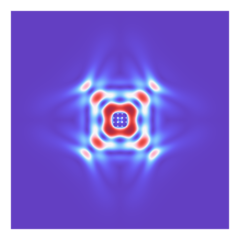} &
        \includegraphics[width=\figsizeF]{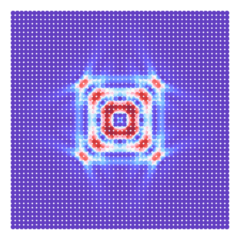} &
        \includegraphics[width=\figsizeF]{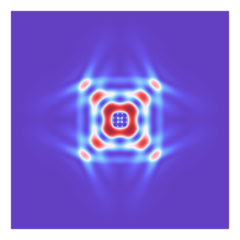} &
        \multicolumn{1}{c|}{}
        \\ \hline
    \end{tabular}
    \caption{
        Dimensionless displacement norm ($\lVert \widetilde{u} \rVert$):
        (\textit{first column}) equivalent RRMM,
        (\textit{second column}) metamaterial $\mathcal{L}$,
        (\textit{third column}) macro Cauchy,
        for a rotation load (Sec.~\ref{sec:load_boundary}) for $\omega = 100$ Hz with a 3\texttimes 3 central cluster.
    }
    \label{tab:maze_rota_figu_100_3x3}
\end{table}

\begin{table}[H]
    \centering
    \begin{tabular}{c|c|c|c|c}
        \cline{2-4}
                                                                                                              & RRMM
                                                                                                              & Microstructure
                                                                                                              & Macro Cauchy
                                                                                                              & \multicolumn{1}{c}{}
        \\ \hline
        \multicolumn{1}{|c|}{\rotatebox{90}{\makebox[\figsizeF][c]{$t= 0.005$ s}}}                            &
        \includegraphics[width=\figsizeF]{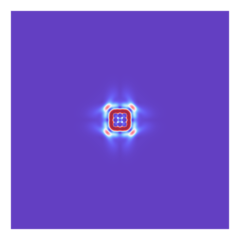} &
        \includegraphics[width=\figsizeF]{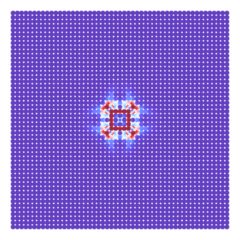} &
        \includegraphics[width=\figsizeF]{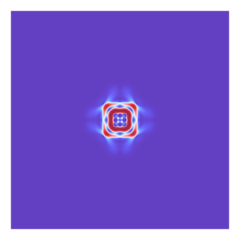} &
        \multicolumn{1}{c|}{}
        \\ \cline{1-4}
        \multicolumn{1}{|c|}{\rotatebox{90}{\makebox[\figsizeF][c]{$t= 0.010$ s}}}                            &
        \includegraphics[width=\figsizeF]{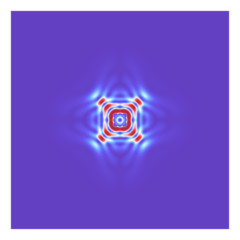} &
        \includegraphics[width=\figsizeF]{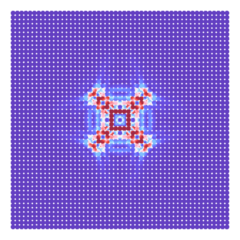} &
        \includegraphics[width=\figsizeF]{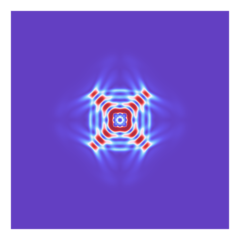} &
        \multicolumn{1}{c|}{\rotatebox{90}{\makebox[\figsizeF][c]{Rotation load - \fbox{3\texttimes 3} - $\omega = 200$ Hz}}}
        \\ \cline{1-4}
        \multicolumn{1}{|c|}{\rotatebox{90}{\makebox[\figsizeF][c]{$t= 0.015$ s}}}                            &
        \includegraphics[width=\figsizeF]{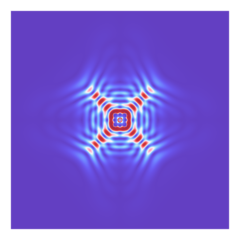} &
        \includegraphics[width=\figsizeF]{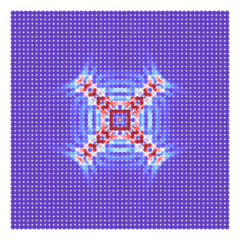} &
        \includegraphics[width=\figsizeF]{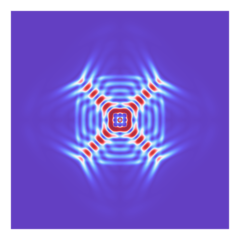} &
        \multicolumn{1}{c|}{}
        \\ \hline
    \end{tabular}
    \caption{
        Dimensionless displacement norm ($\lVert \widetilde{u} \rVert$):
        (\textit{first column}) equivalent RRMM,
        (\textit{second column}) metamaterial $\mathcal{L}$,
        (\textit{third column}) macro Cauchy,
        for a rotation load (Sec.~\ref{sec:load_boundary}) for $\omega = 200$ Hz with a 3\texttimes 3 central cluster.
    }
    \label{tab:maze_rota_figu_200_3x3}
\end{table}

\begin{table}[H]
    \centering
    \begin{tabular}{c|c|c|c|c}
        \cline{2-4}
                                                                                                               & RRMM
                                                                                                               & Microstructure
                                                                                                               & Macro Cauchy
                                                                                                               & \multicolumn{1}{c}{}
        \\ \hline
        \multicolumn{1}{|c|}{\rotatebox{90}{\makebox[\figsizeF][c]{$t= 0.005$ s}}}                             &
        \includegraphics[width=\figsizeF]{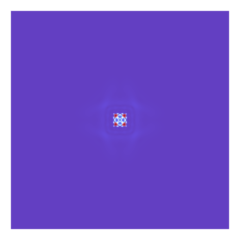} &
        \includegraphics[width=\figsizeF]{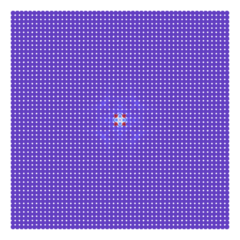} &
        \includegraphics[width=\figsizeF]{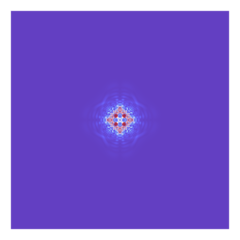} &
        \multicolumn{1}{c|}{}
        \\ \cline{1-4}
        \multicolumn{1}{|c|}{\rotatebox{90}{\makebox[\figsizeF][c]{$t= 0.010$ s}}}                             &
        \includegraphics[width=\figsizeF]{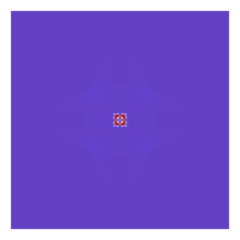} &
        \includegraphics[width=\figsizeF]{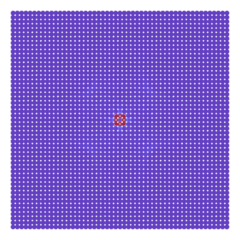} &
        \includegraphics[width=\figsizeF]{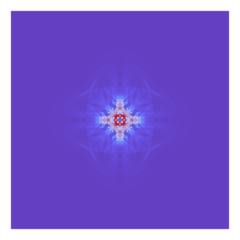} &
        \multicolumn{1}{c|}{\rotatebox{90}{\makebox[\figsizeF][c]{Rotation load - \fbox{3\texttimes 3} - $\omega = 1000$ Hz}}}
        \\ \cline{1-4}
        \multicolumn{1}{|c|}{\rotatebox{90}{\makebox[\figsizeF][c]{$t= 0.015$ s}}}                             &
        \includegraphics[width=\figsizeF]{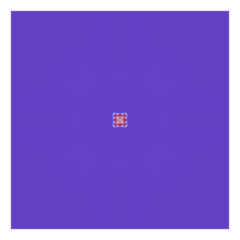} &
        \includegraphics[width=\figsizeF]{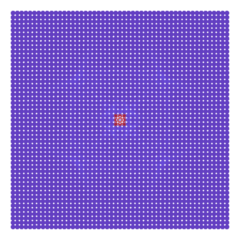} &
        \includegraphics[width=\figsizeF]{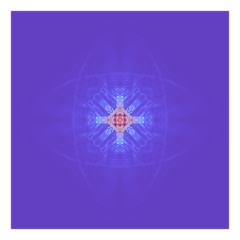} &
        \multicolumn{1}{c|}{}
        \\ \hline
    \end{tabular}
    \caption{
        Dimensionless displacement norm ($\lVert \widetilde{u} \rVert$):
        (\textit{first column}) equivalent RRMM,
        (\textit{second column}) metamaterial $\mathcal{L}$,
        (\textit{third column}) macro Cauchy,
        for a rotation load (Sec.~\ref{sec:load_boundary}) for $\omega = 1000$ Hz with a 3\texttimes 3 central cluster.
    }
    \label{tab:maze_rota_figu_1000_3x3}
\end{table}

In Figs.~\ref{tab:maze_rota_figu_100_3x3}--\ref{tab:maze_rota_figu_1000_3x3}, we still observe that the RRMM shows good agreement at 100 Hz and 1000 Hz, while the quantitative differences at 200 Hz significantly decrease.

\begin{table}[H]
    \centering
    \begin{tabular}{c|c|c|c|c|c}
        \cline{2-5}
                                                                                                              & RRMM
                                                                                                              & Avg. Microstr.
                                                                                                              & Microstructure
                                                                                                              & Macro Cauchy
                                                                                                              & \multicolumn{1}{c}{}
        \\ \hline
        \multicolumn{1}{|c|}{\rotatebox{90}{\makebox[\figsizeF][c]{$t= 0.005$ s}}}                            &
        \includegraphics[width=\figsizeF]{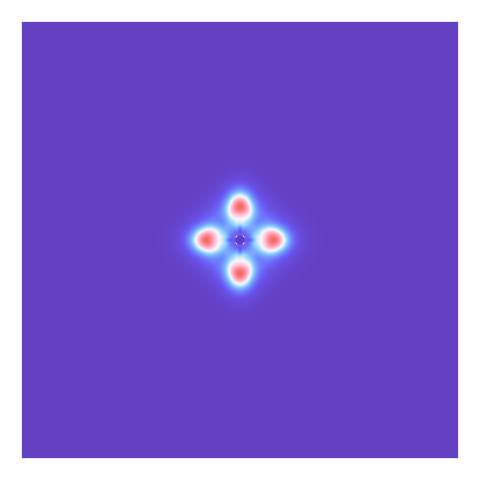}     &
        \includegraphics[width=\figsizeF]{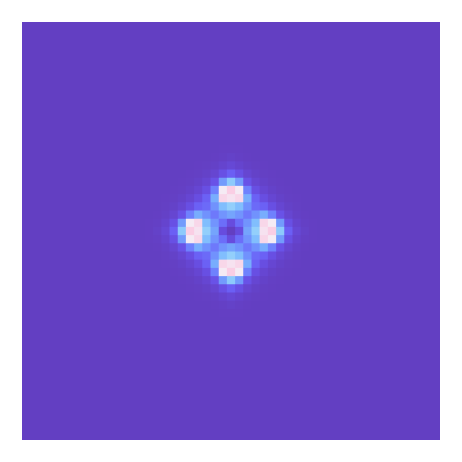} &
        \includegraphics[width=\figsizeF]{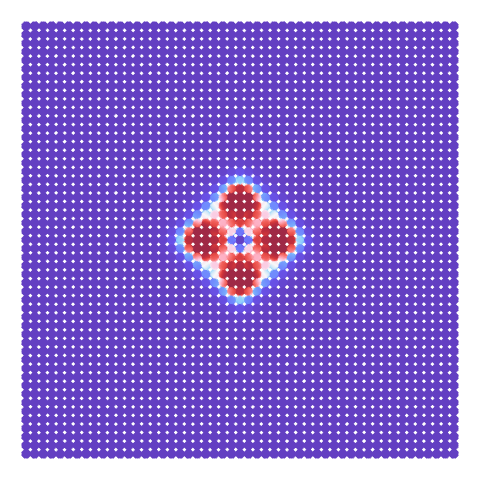}     &
        \includegraphics[width=\figsizeF]{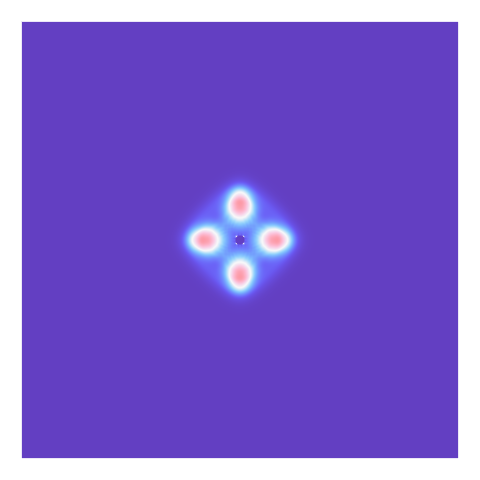}     &
        \multicolumn{1}{c|}{}
        \\ \cline{1-5}
        \multicolumn{1}{|c|}{\rotatebox{90}{\makebox[\figsizeF][c]{$t= 0.010$ s}}}                            &
        \includegraphics[width=\figsizeF]{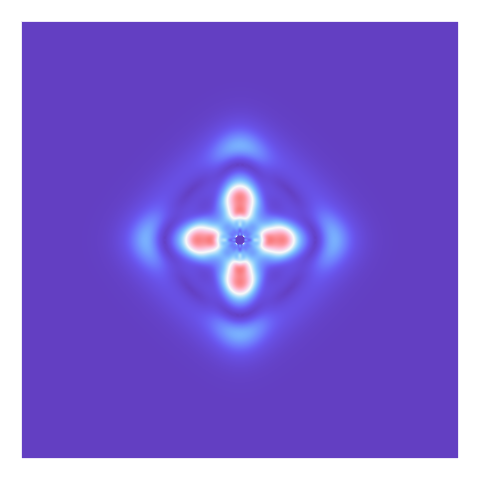}     &
        \includegraphics[width=\figsizeF]{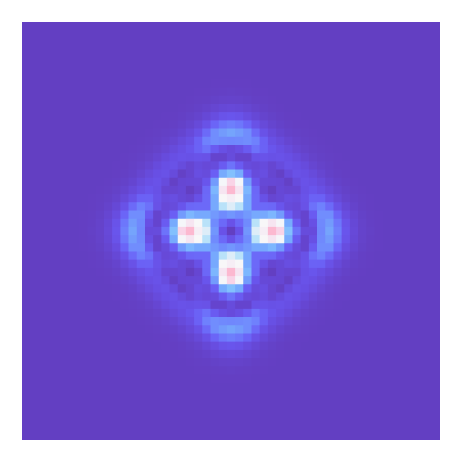} &
        \includegraphics[width=\figsizeF]{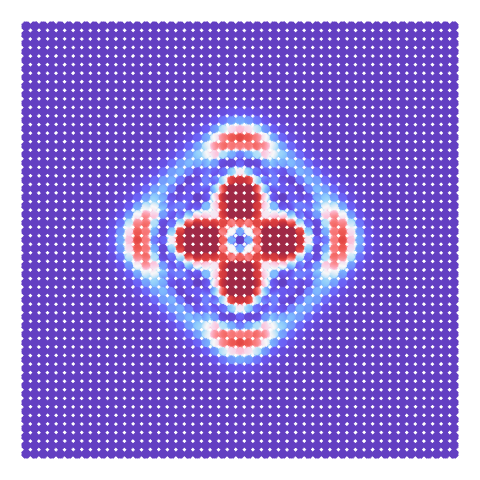}     &
        \includegraphics[width=\figsizeF]{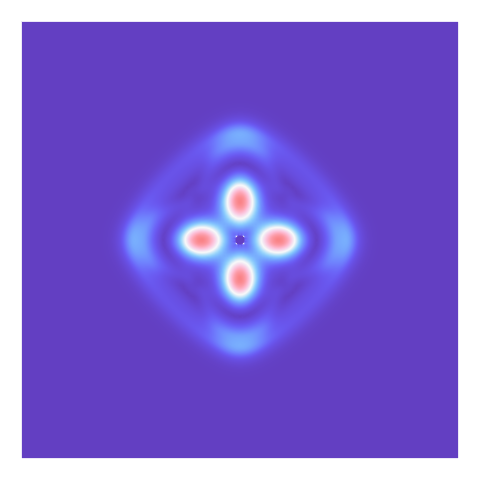}     &
        \multicolumn{1}{c|}{\rotatebox{90}{\makebox[\figsizeF][c]{Hydrostatic load - \fbox{1\texttimes 1} - $\omega = 100$ Hz}}}
        \\ \cline{1-5}
        \multicolumn{1}{|c|}{\rotatebox{90}{\makebox[\figsizeF][c]{$t= 0.015$ s}}}                            &
        \includegraphics[width=\figsizeF]{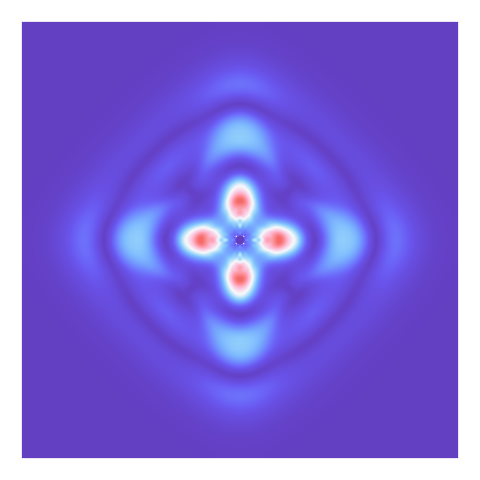}     &
        \includegraphics[width=\figsizeF]{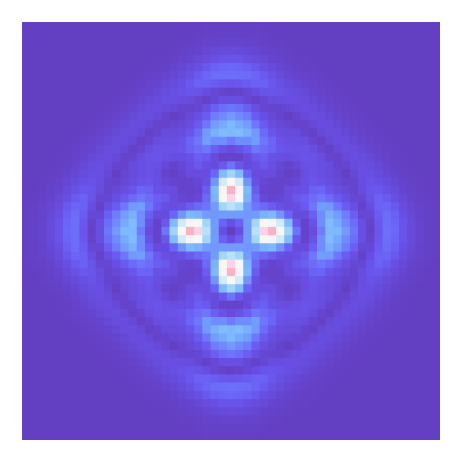} &
        \includegraphics[width=\figsizeF]{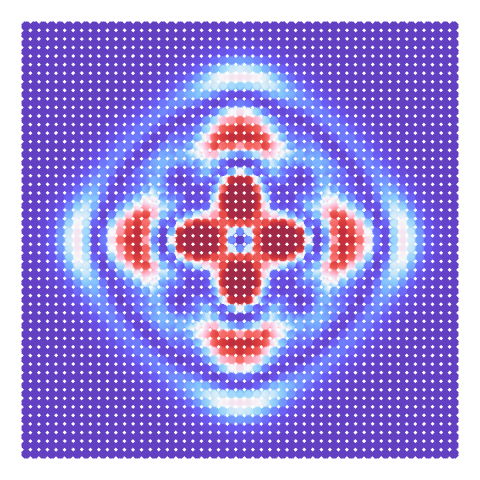}     &
        \includegraphics[width=\figsizeF]{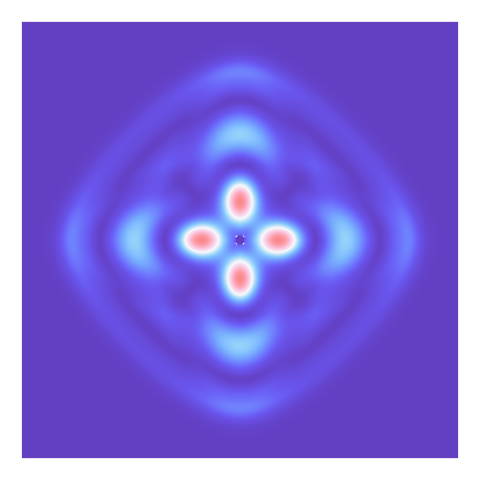}     &
        \multicolumn{1}{c|}{}
        \\ \hline
    \end{tabular}
    \caption{
        Dimensionless displacement norm ($\lVert \widetilde{u} \rVert$):
        (\textit{first column}) equivalent RRMM,
        (\textit{second column}) average metamaterial displacement ($\lVert\overline{u} \rVert$),
        (\textit{third column}) metamaterial $\mathcal{L}$,
        (\textit{fourth column}) macro Cauchy,
        for a hydrostatic load (Sec.~\ref{sec:load_boundary}) for $\omega = 100$ Hz with a single central unit cell.
    }
    \label{tab:maze_expa_figu_100}
\end{table}

\begin{table}[H]
    \centering
    \begin{tabular}{c|c|c|c|c|c}
        \cline{2-5}
                                                                                                              & RRMM
                                                                                                              & Avg. Microstr.
                                                                                                              & Microstructure
                                                                                                              & Macro Cauchy
                                                                                                              & \multicolumn{1}{c}{}
        \\ \hline
        \multicolumn{1}{|c|}{\rotatebox{90}{\makebox[\figsizeF][c]{$t= 0.005$ s}}}                            &
        \includegraphics[width=\figsizeF]{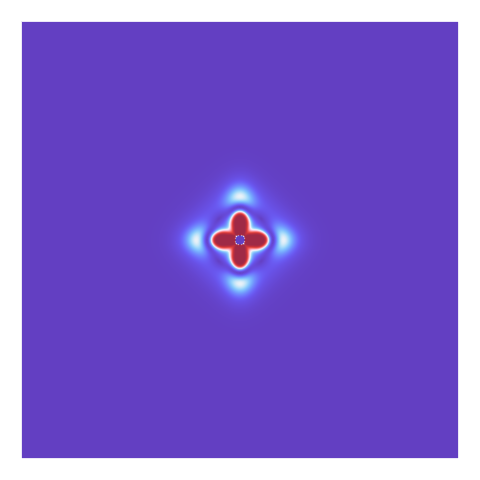}     &
        \includegraphics[width=\figsizeF]{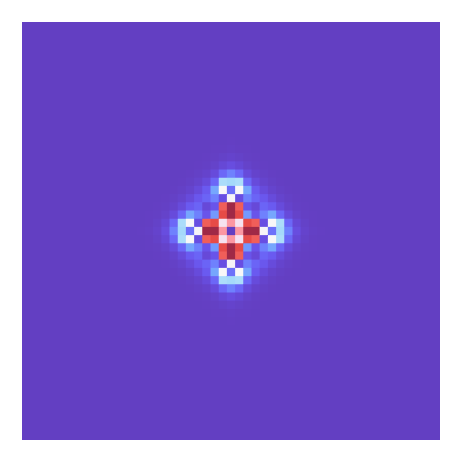} &
        \includegraphics[width=\figsizeF]{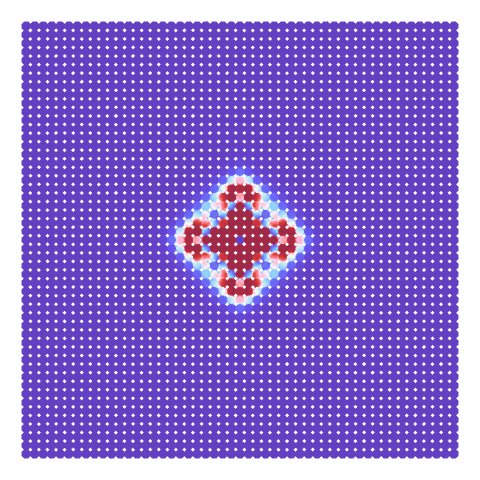}     &
        \includegraphics[width=\figsizeF]{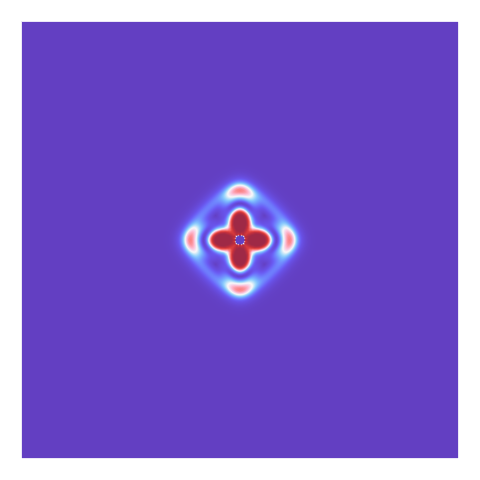}     &
        \multicolumn{1}{c|}{}
        \\ \cline{1-5}
        \multicolumn{1}{|c|}{\rotatebox{90}{\makebox[\figsizeF][c]{$t= 0.010$ s}}}                            &
        \includegraphics[width=\figsizeF]{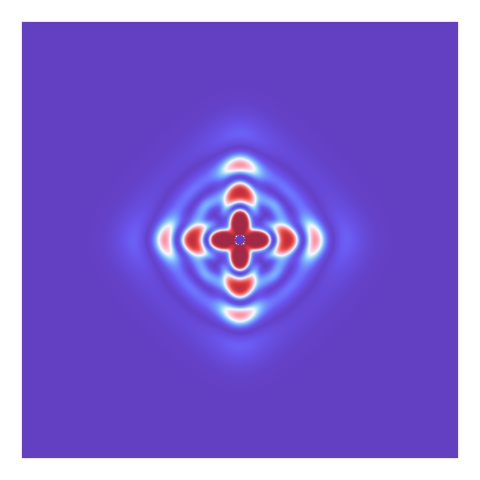}     &
        \includegraphics[width=\figsizeF]{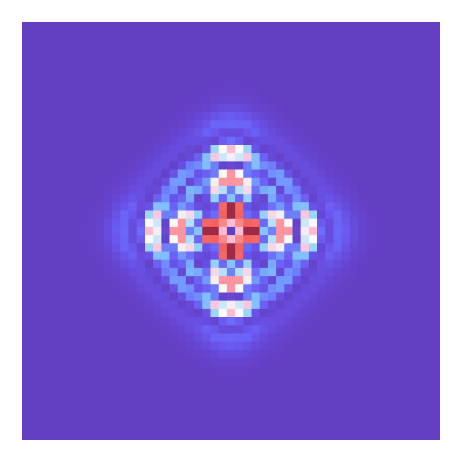} &
        \includegraphics[width=\figsizeF]{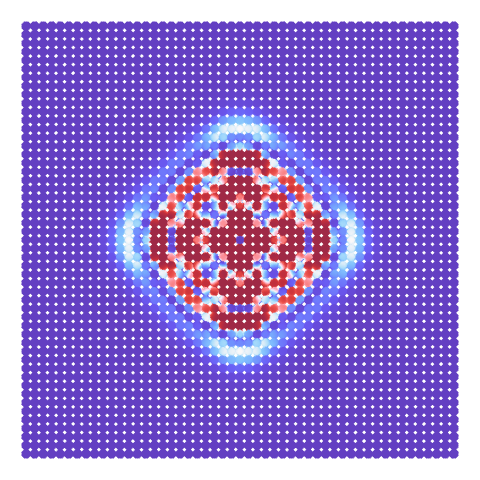}     &
        \includegraphics[width=\figsizeF]{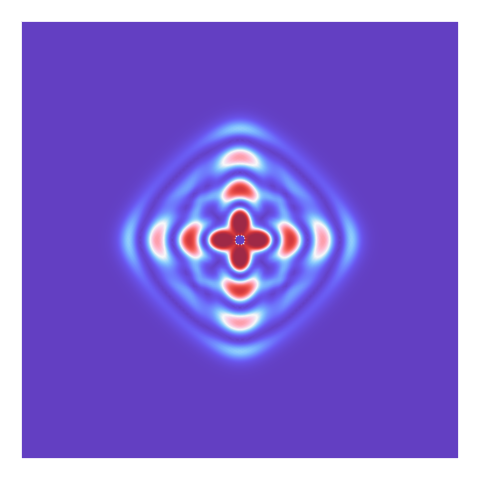}     &
        \multicolumn{1}{c|}{\rotatebox{90}{\makebox[\figsizeF][c]{Hydrostatic load - \fbox{1\texttimes 1} - $\omega = 200$ Hz}}}
        \\ \cline{1-5}
        \multicolumn{1}{|c|}{\rotatebox{90}{\makebox[\figsizeF][c]{$t= 0.015$ s}}}                            &
        \includegraphics[width=\figsizeF]{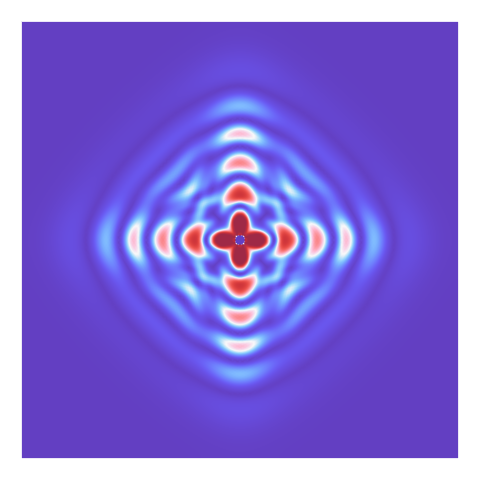}     &
        \includegraphics[width=\figsizeF]{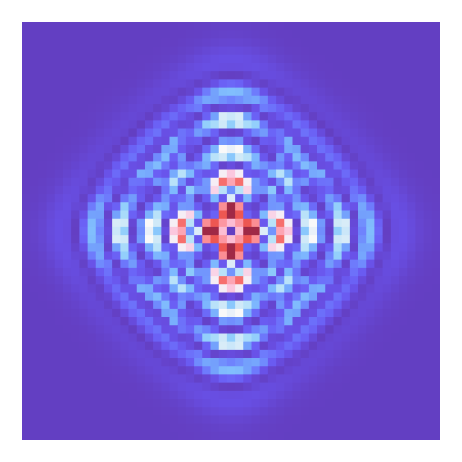} &
        \includegraphics[width=\figsizeF]{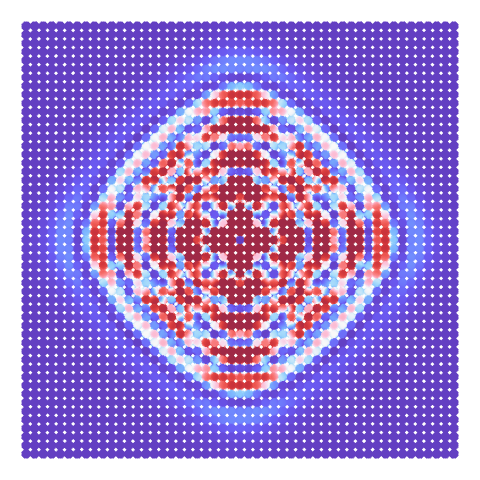}     &
        \includegraphics[width=\figsizeF]{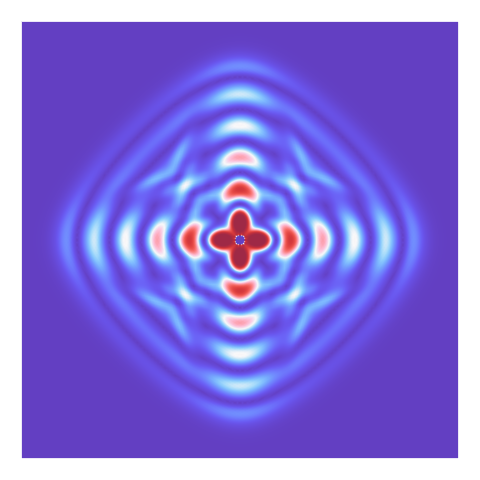}     &
        \multicolumn{1}{c|}{}
        \\ \hline
    \end{tabular}
    \caption{
        Dimensionless displacement norm ($\lVert \widetilde{u} \rVert$):
        (\textit{first column}) equivalent RRMM,
        (\textit{second column}) average metamaterial displacement ($\lVert\overline{u} \rVert$),
        (\textit{third column}) metamaterial $\mathcal{L}$,
        (\textit{fourth column}) macro Cauchy,
        for a hydrostatic load (Sec.~\ref{sec:load_boundary}) for $\omega = 200$ Hz with a single central unit cell.
    }
    \label{tab:maze_expa_figu_200}
\end{table}

\begin{table}[H]
    \centering
    \begin{tabular}{c|c|c|c|c}
        \cline{2-4}
                                                                                                           & RRMM
                                                                                                           & Microstructure
                                                                                                           & Macro Cauchy
                                                                                                           & \multicolumn{1}{c}{}
        \\ \hline
        \multicolumn{1}{|c|}{\rotatebox{90}{\makebox[\figsizeF][c]{$t= 0.005$ s}}}                         &
        \includegraphics[width=\figsizeF]{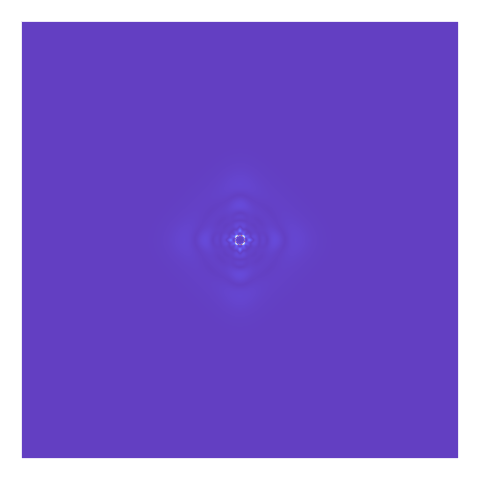} &
        \includegraphics[width=\figsizeF]{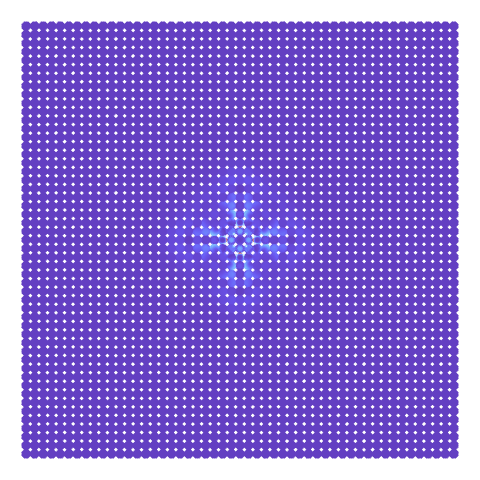} &
        \includegraphics[width=\figsizeF]{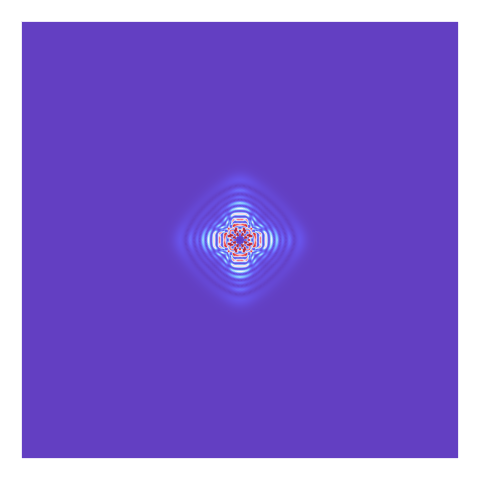} &
        \multicolumn{1}{c|}{}
        \\ \cline{1-4}
        \multicolumn{1}{|c|}{\rotatebox{90}{\makebox[\figsizeF][c]{$t= 0.010$ s}}}                         &
        \includegraphics[width=\figsizeF]{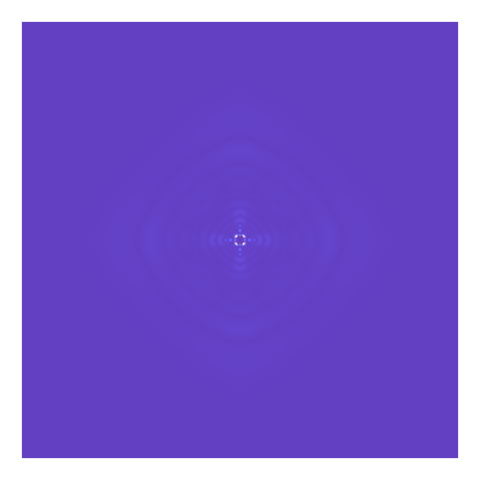} &
        \includegraphics[width=\figsizeF]{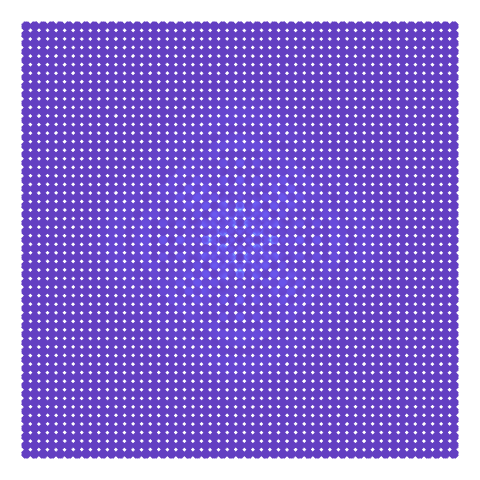} &
        \includegraphics[width=\figsizeF]{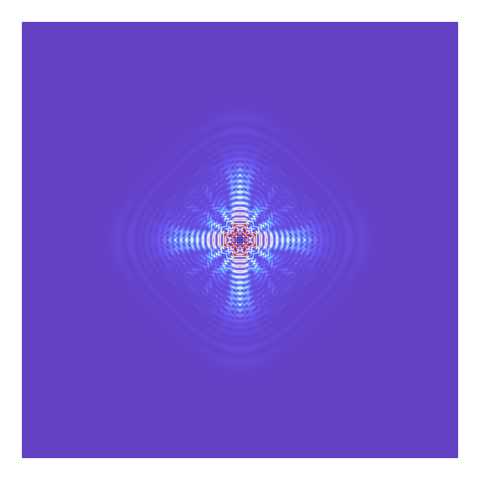} &
        \multicolumn{1}{c|}{\rotatebox{90}{\makebox[\figsizeF][c]{Hydrostatic load - \fbox{1\texttimes 1} - $\omega = 1000$ Hz}}}
        \\ \cline{1-4}
        \multicolumn{1}{|c|}{\rotatebox{90}{\makebox[\figsizeF][c]{$t= 0.015$ s}}}                         &
        \includegraphics[width=\figsizeF]{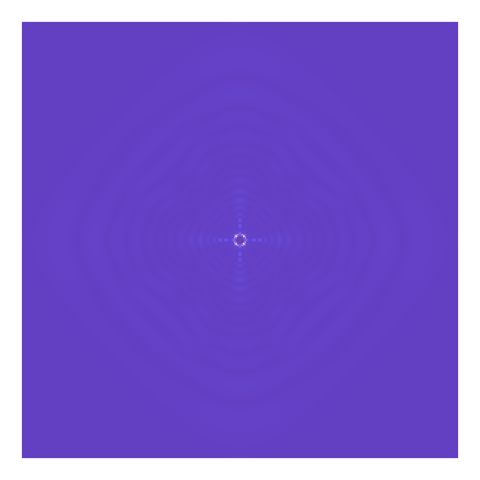} &
        \includegraphics[width=\figsizeF]{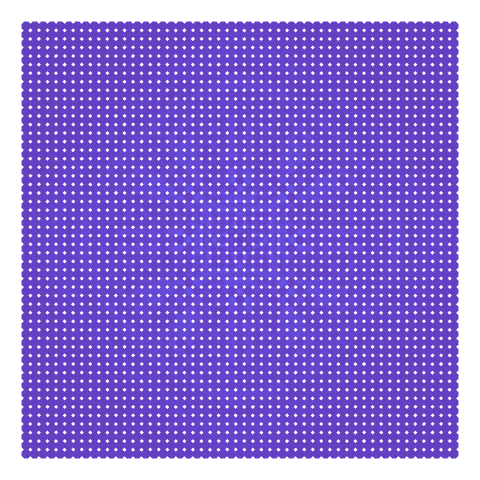} &
        \includegraphics[width=\figsizeF]{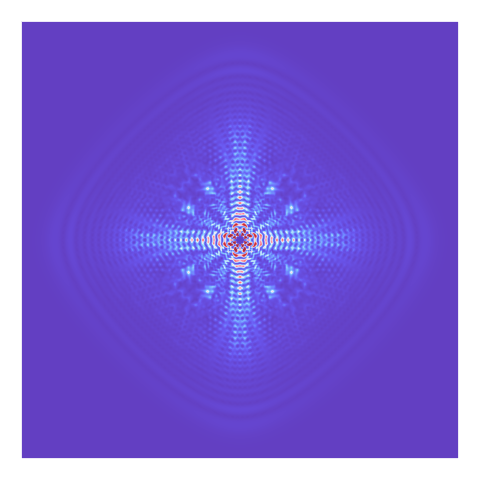} &
        \multicolumn{1}{c|}{}
        \\ \hline
    \end{tabular}
    \caption{
        Dimensionless displacement norm ($\lVert \widetilde{u} \rVert$):
        (\textit{first column}) equivalent RRMM,
        (\textit{second column}) metamaterial $\mathcal{L}$,
        (\textit{third column}) macro Cauchy,
        for a hydrostatic load (Sec.~\ref{sec:load_boundary}) for $\omega = 1000$ Hz with a single central unit cell.
    }
    \label{tab:maze_expa_figu_1000}
\end{table}

In Figs.~\ref{tab:maze_expa_figu_100}--\ref{tab:maze_expa_figu_1000}, we observe that the RRMM shows good qualitative agreement (highlighted by the comparison with the average response of the microstructure) while exhibiting quantitative differences at all frequencies exept for 1000 Hz (bandgap frequency), at which the response is correctly captured.

\begin{table}[H]
    \centering
    \begin{tabular}{c|c|c|c|c}
        \cline{2-4}
                                                                                                              & RRMM
                                                                                                              & Microstructure
                                                                                                              & Macro Cauchy
                                                                                                              & \multicolumn{1}{c}{}
        \\ \hline
        \multicolumn{1}{|c|}{\rotatebox{90}{\makebox[\figsizeF][c]{$t= 0.005$ s}}}                            &
        \includegraphics[width=\figsizeF]{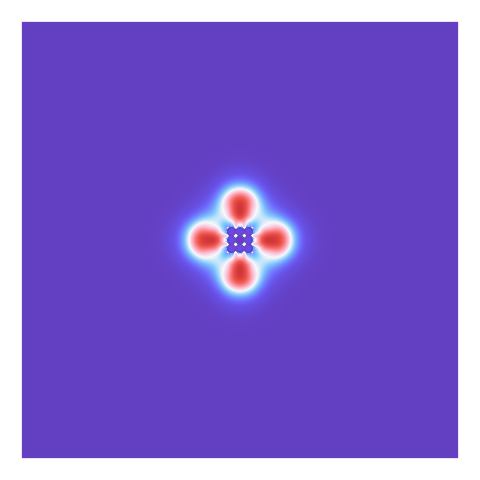} &
        \includegraphics[width=\figsizeF]{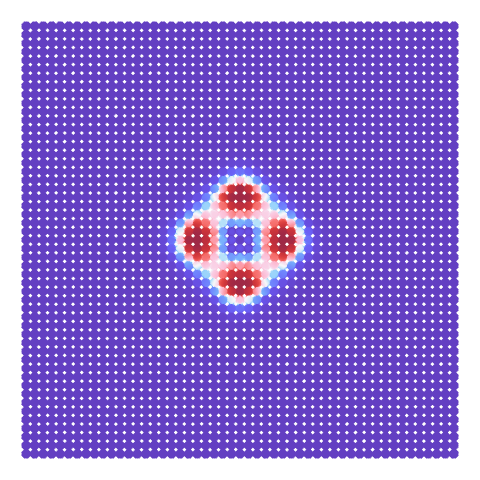} &
        \includegraphics[width=\figsizeF]{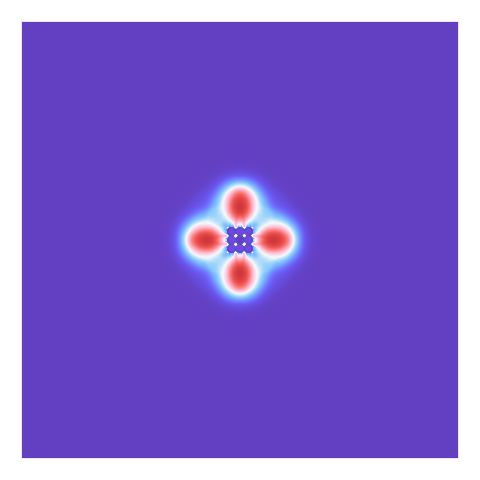} &
        \multicolumn{1}{c|}{}
        \\ \cline{1-4}
        \multicolumn{1}{|c|}{\rotatebox{90}{\makebox[\figsizeF][c]{$t= 0.010$ s}}}                            &
        \includegraphics[width=\figsizeF]{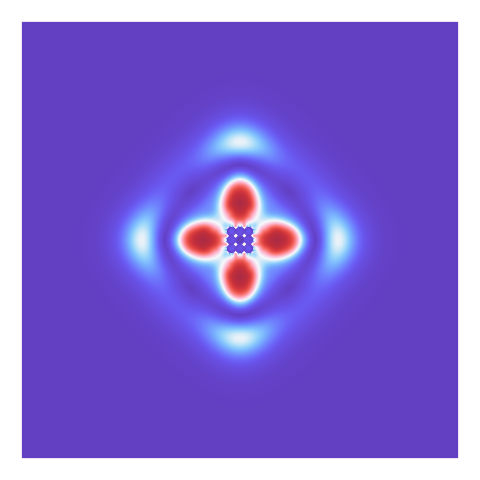} &
        \includegraphics[width=\figsizeF]{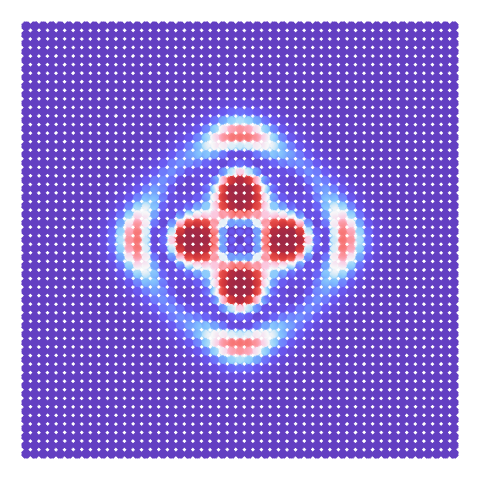} &
        \includegraphics[width=\figsizeF]{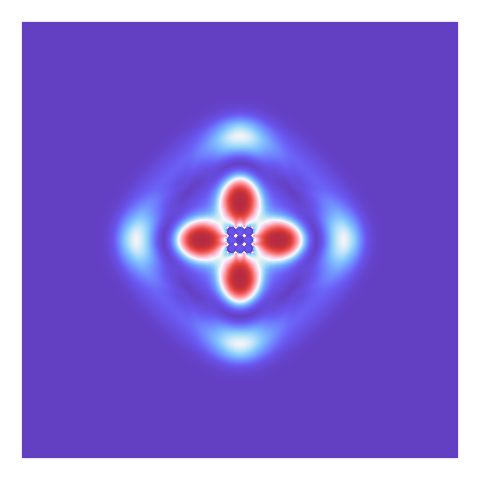} &
        \multicolumn{1}{c|}{\rotatebox{90}{\makebox[\figsizeF][c]{Hydrostatic load - \fbox{3\texttimes 3} - $\omega = 100$ Hz}}}
        \\ \cline{1-4}
        \multicolumn{1}{|c|}{\rotatebox{90}{\makebox[\figsizeF][c]{$t= 0.015$ s}}}                            &
        \includegraphics[width=\figsizeF]{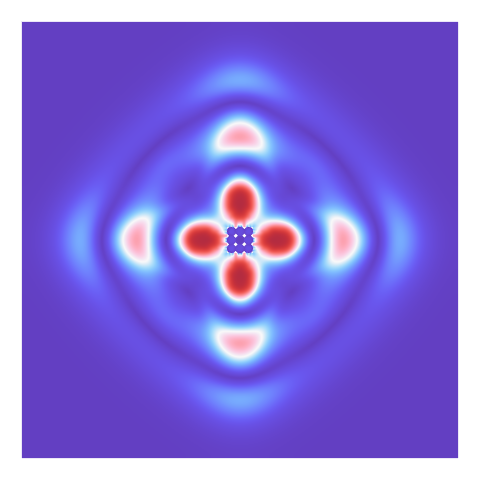} &
        \includegraphics[width=\figsizeF]{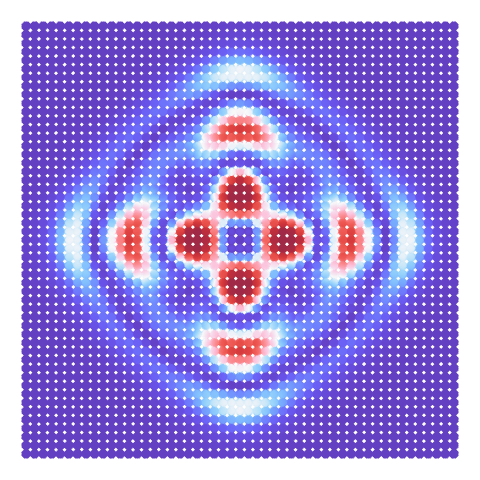} &
        \includegraphics[width=\figsizeF]{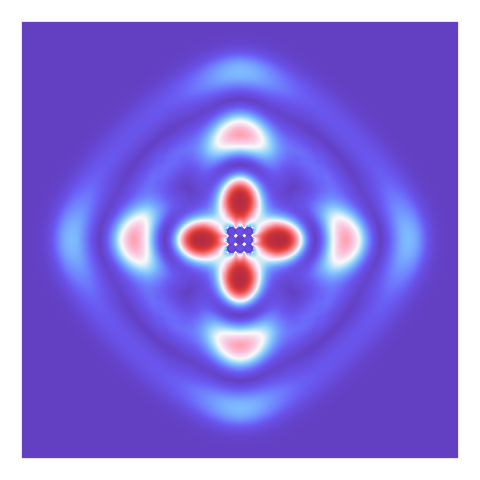} &
        \multicolumn{1}{c|}{}
        \\ \hline
    \end{tabular}
    \caption{
        Dimensionless displacement norm ($\lVert \widetilde{u} \rVert$):
        (\textit{first column}) equivalent RRMM,
        (\textit{second column}) metamaterial $\mathcal{L}$,
        (\textit{third column}) macro Cauchy,
        for a hydrostatic load (Sec.~\ref{sec:load_boundary}) for $\omega = 100$ Hz with a 3\texttimes 3 central cluster.
    }
    \label{tab:maze_expa_figu_100_3x3}
\end{table}

\begin{table}[H]
    \centering
    \begin{tabular}{c|c|c|c|c|c}
        \cline{2-5}
                                                                                                                  & RRMM
                                                                                                                  & Avg. Microstr.
                                                                                                                  & Microstructure
                                                                                                                  & Macro Cauchy
                                                                                                                  & \multicolumn{1}{c}{}
        \\ \hline
        \multicolumn{1}{|c|}{\rotatebox{90}{\makebox[\figsizeF][c]{$t= 0.005$ s}}}                                &
        \includegraphics[width=\figsizeF]{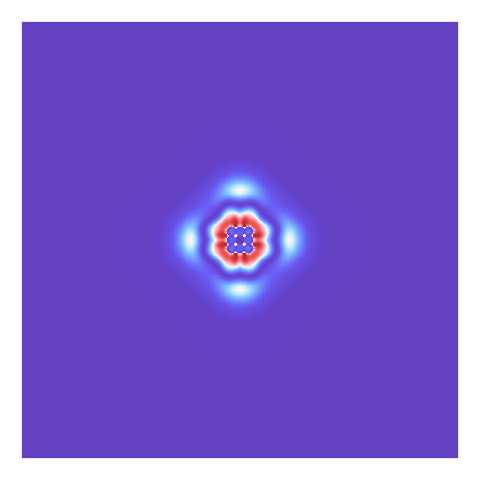}     &
        \includegraphics[width=\figsizeF]{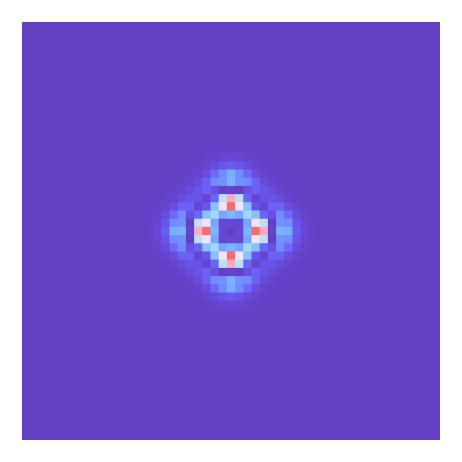} &
        \includegraphics[width=\figsizeF]{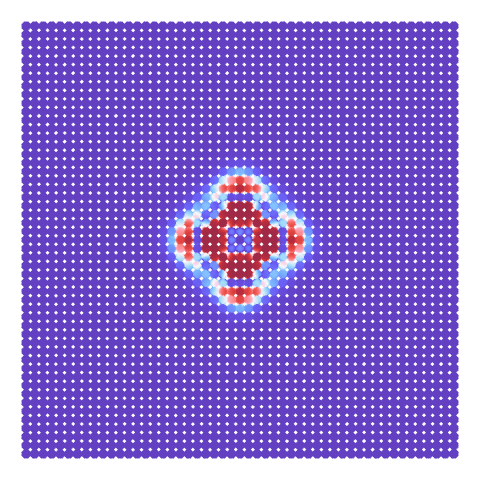}     &
        \includegraphics[width=\figsizeF]{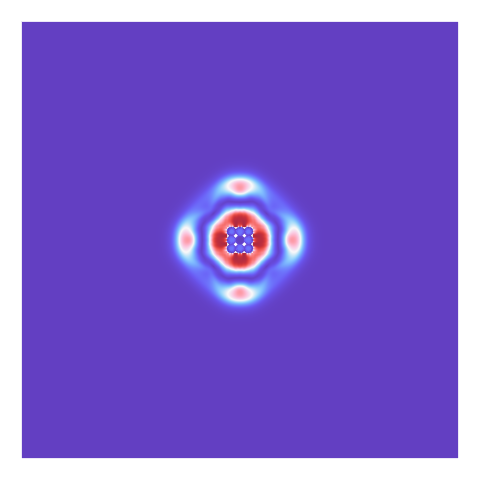}     &
        \multicolumn{1}{c|}{}
        \\ \cline{1-5}
        \multicolumn{1}{|c|}{\rotatebox{90}{\makebox[\figsizeF][c]{$t= 0.010$ s}}}                                &
        \includegraphics[width=\figsizeF]{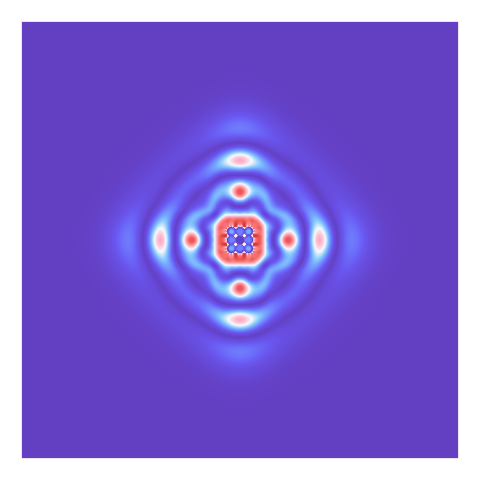}     &
        \includegraphics[width=\figsizeF]{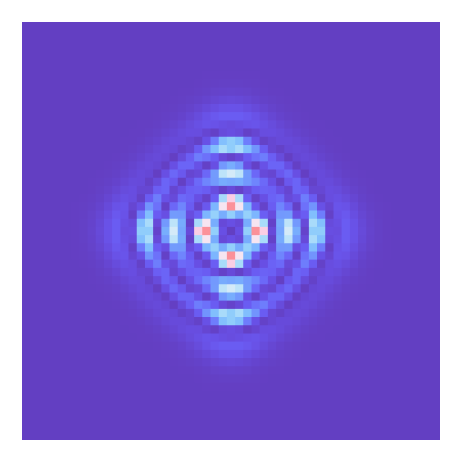} &
        \includegraphics[width=\figsizeF]{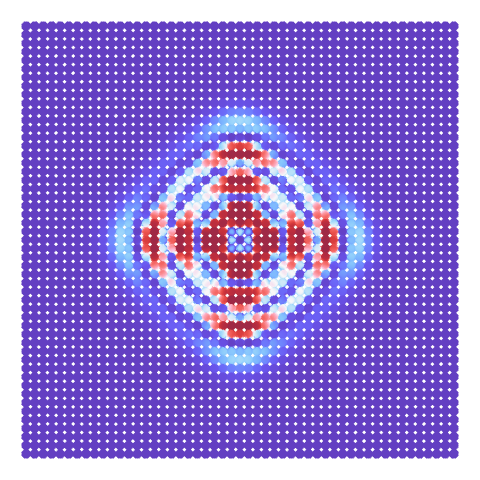}     &
        \includegraphics[width=\figsizeF]{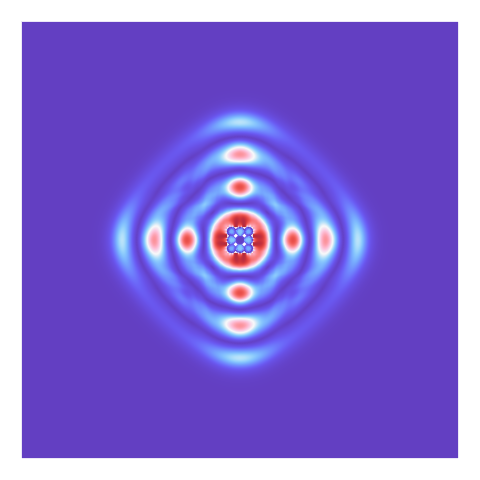}     &
        \multicolumn{1}{c|}{\rotatebox{90}{\makebox[\figsizeF][c]{Hydrostatic load - \fbox{3\texttimes 3} - $\omega = 200$ Hz}}}
        \\ \cline{1-5}
        \multicolumn{1}{|c|}{\rotatebox{90}{\makebox[\figsizeF][c]{$t= 0.015$ s}}}                                &
        \includegraphics[width=\figsizeF]{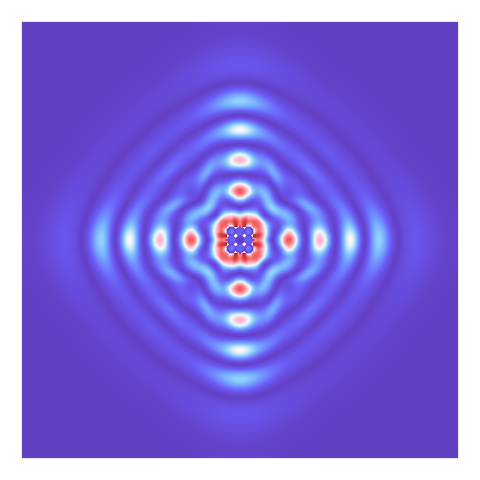}     &
        \includegraphics[width=\figsizeF]{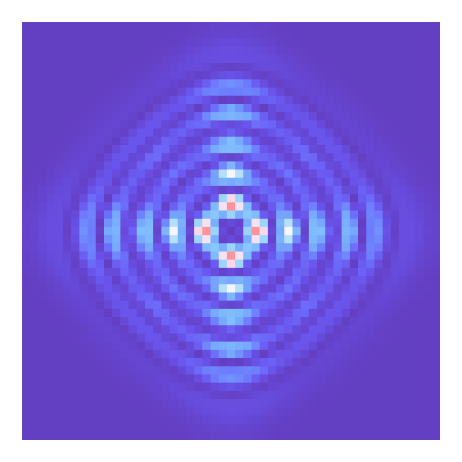} &
        \includegraphics[width=\figsizeF]{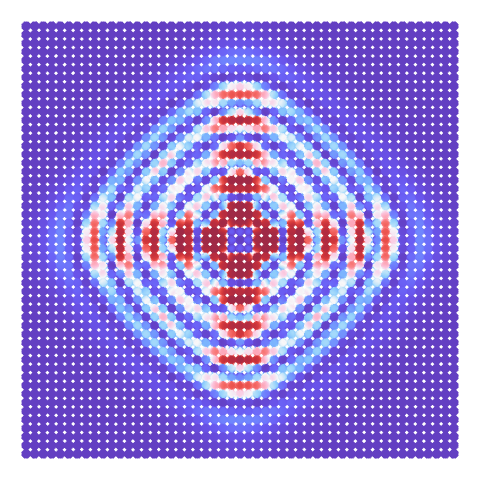}     &
        \includegraphics[width=\figsizeF]{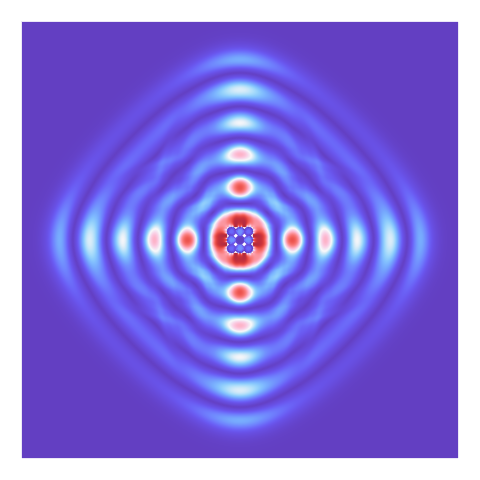}     &
        \multicolumn{1}{c|}{}
        \\ \hline
    \end{tabular}
    \caption{
        Dimensionless displacement norm ($\lVert \widetilde{u} \rVert$):
        (\textit{first column}) equivalent RRMM,
        (\textit{second column}) average metamaterial displacement ($\lVert\overline{u} \rVert$),
        (\textit{third column}) metamaterial $\mathcal{L}$,
        (\textit{fourth column}) macro Cauchy,
        for a hydrostatic load (Sec.~\ref{sec:load_boundary}) for $\omega = 200$ Hz with a 3\texttimes 3 central cluster.
    }
    \label{tab:maze_expa_figu_200_3x3}
\end{table}

\begin{table}[H]
    \centering
    \begin{tabular}{c|c|c|c|c}
        \cline{2-4}
                                                                                                               & RRMM
                                                                                                               & Microstructure
                                                                                                               & Macro Cauchy
                                                                                                               & \multicolumn{1}{c}{}
        \\ \hline
        \multicolumn{1}{|c|}{\rotatebox{90}{\makebox[\figsizeF][c]{$t= 0.005$ s}}}                             &
        \includegraphics[width=\figsizeF]{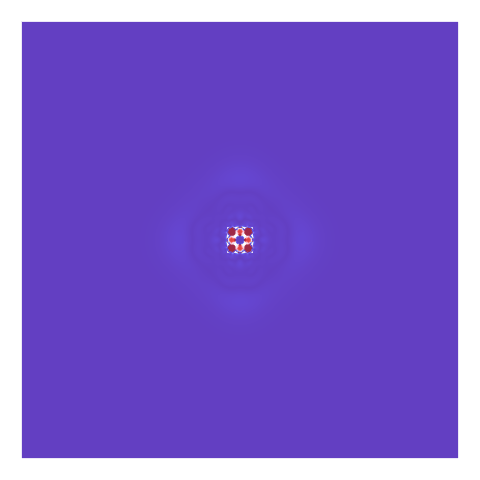} &
        \includegraphics[width=\figsizeF]{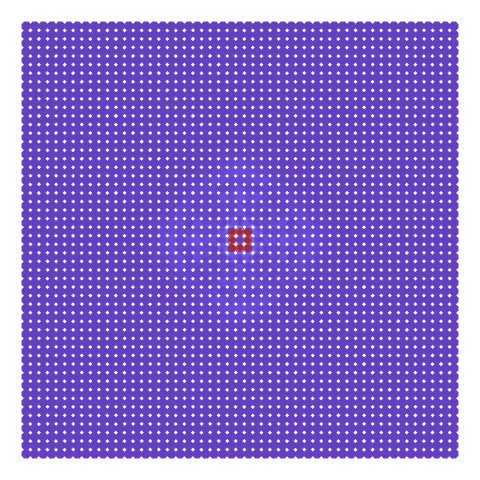} &
        \includegraphics[width=\figsizeF]{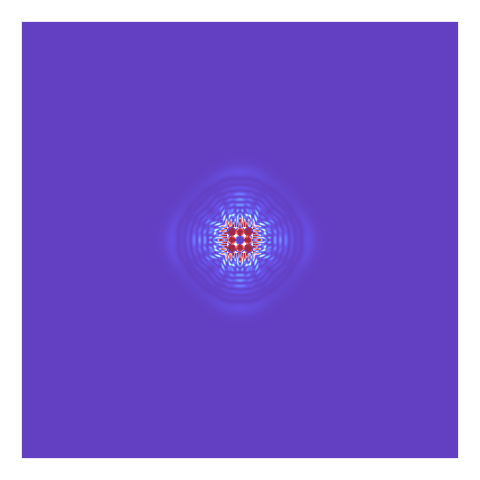} &
        \multicolumn{1}{c|}{}
        \\ \cline{1-4}
        \multicolumn{1}{|c|}{\rotatebox{90}{\makebox[\figsizeF][c]{$t= 0.010$ s}}}                             &
        \includegraphics[width=\figsizeF]{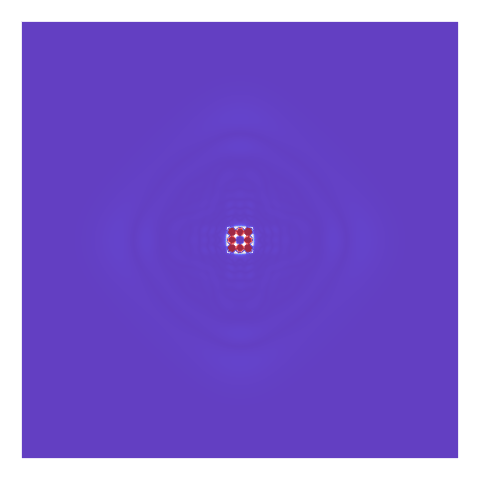} &
        \includegraphics[width=\figsizeF]{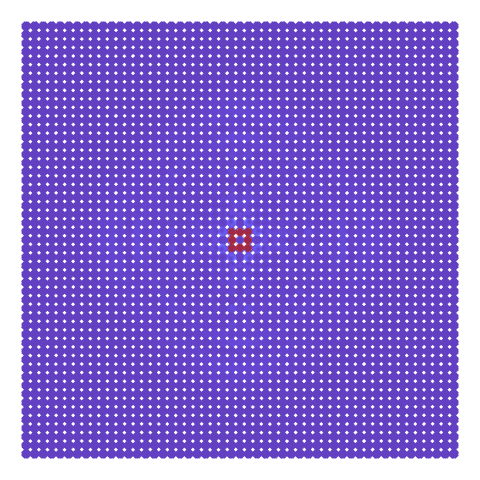} &
        \includegraphics[width=\figsizeF]{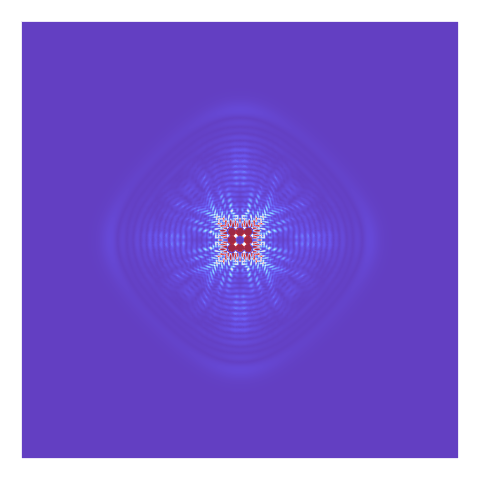} &
        \multicolumn{1}{c|}{\rotatebox{90}{\makebox[\figsizeF][c]{Hydrostatic load - \fbox{3\texttimes 3} - $\omega = 1000$ Hz}}}
        \\ \cline{1-4}
        \multicolumn{1}{|c|}{\rotatebox{90}{\makebox[\figsizeF][c]{$t= 0.015$ s}}}                             &
        \includegraphics[width=\figsizeF]{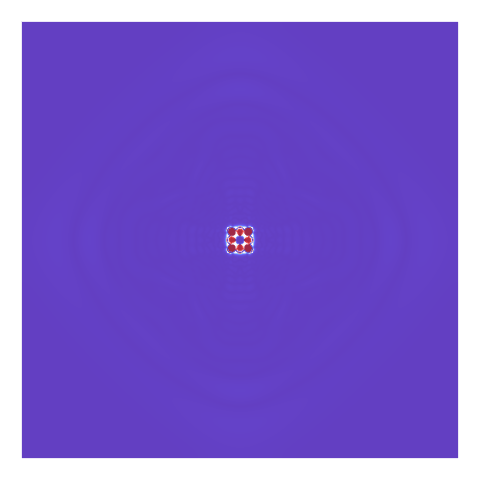} &
        \includegraphics[width=\figsizeF]{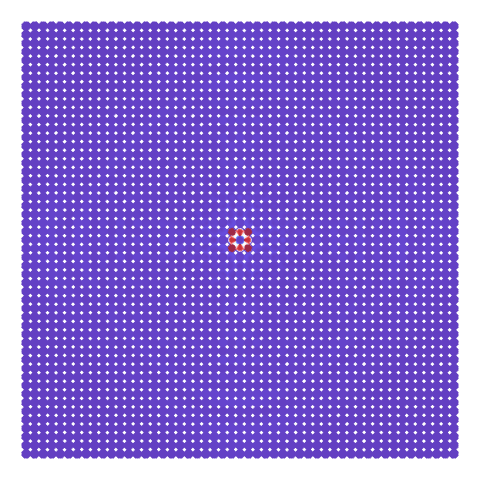} &
        \includegraphics[width=\figsizeF]{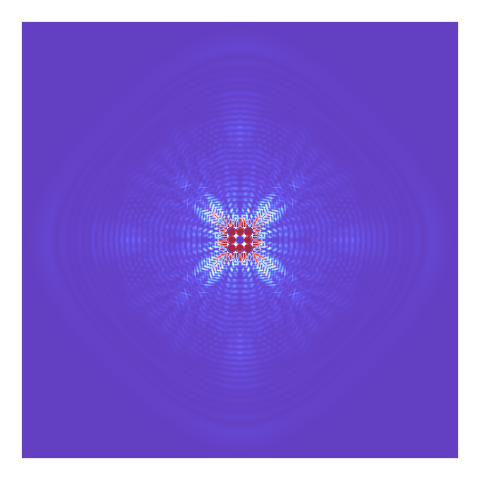} &
        \multicolumn{1}{c|}{}
        \\ \hline
    \end{tabular}
    \caption{
        Dimensionless displacement norm ($\lVert \widetilde{u} \rVert$):
        (\textit{first column}) equivalent RRMM,
        (\textit{second column}) metamaterial $\mathcal{L}$,
        (\textit{third column}) macro Cauchy,
        for a hydrostatic load (Sec.~\ref{sec:load_boundary}) for $\omega = 1000$ Hz with a 3\texttimes 3 central cluster.
    }
    \label{tab:maze_expa_figu_1000_3x3}
\end{table}

In Figs.~\ref{tab:maze_expa_figu_100_3x3}--\ref{tab:maze_expa_figu_1000_3x3}, we observe that the quantitative differences at 100 Hz are now almost eliminated, while they still persist at 200 Hz, although with lower magnitude.

\begin{table}[H]
    \centering
    \begin{tabular}{c|c|c|c|c|c}
        \cline{2-5}
                                                                                                              & RRMM
                                                                                                              & Avg. Microstr.
                                                                                                              & Microstructure
                                                                                                              & Macro Cauchy
                                                                                                              & \multicolumn{1}{c}{}
        \\ \hline
        \multicolumn{1}{|c|}{\rotatebox{90}{\makebox[\figsizeF][c]{$t= 0.005$ s}}}                            &
        \includegraphics[width=\figsizeF]{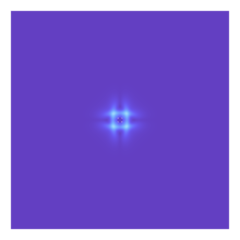}     &
        \includegraphics[width=\figsizeF]{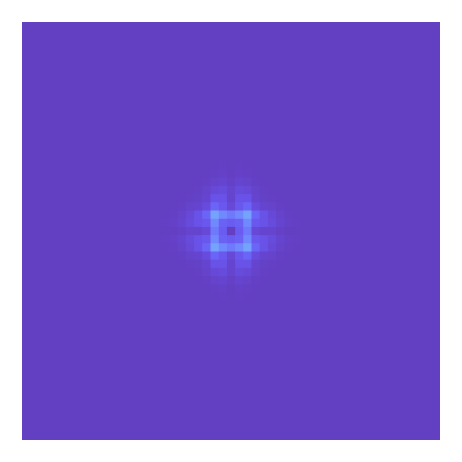} &
        \includegraphics[width=\figsizeF]{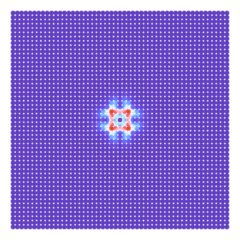}     &
        \includegraphics[width=\figsizeF]{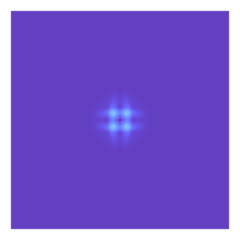}     &
        \multicolumn{1}{c|}{}
        \\ \cline{1-5}
        \multicolumn{1}{|c|}{\rotatebox{90}{\makebox[\figsizeF][c]{$t= 0.010$ s}}}                            &
        \includegraphics[width=\figsizeF]{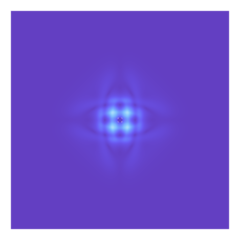}     &
        \includegraphics[width=\figsizeF]{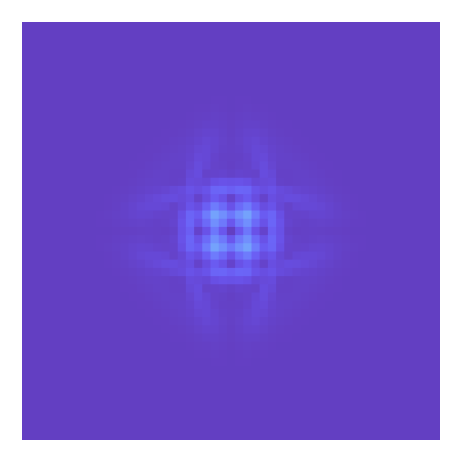} &
        \includegraphics[width=\figsizeF]{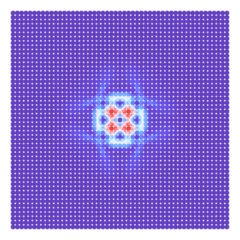}     &
        \includegraphics[width=\figsizeF]{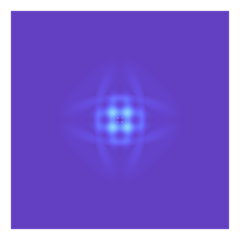}     &
        \multicolumn{1}{c|}{\rotatebox{90}{\makebox[\figsizeF][c]{Shear load - \fbox{1\texttimes 1} - $\omega = 100$ Hz}}}
        \\ \cline{1-5}
        \multicolumn{1}{|c|}{\rotatebox{90}{\makebox[\figsizeF][c]{$t= 0.015$ s}}}                            &
        \includegraphics[width=\figsizeF]{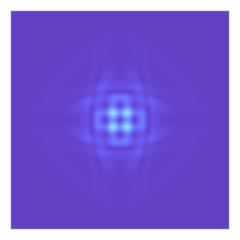}     &
        \includegraphics[width=\figsizeF]{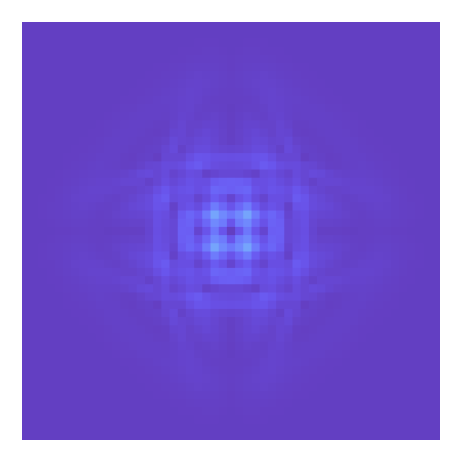} &
        \includegraphics[width=\figsizeF]{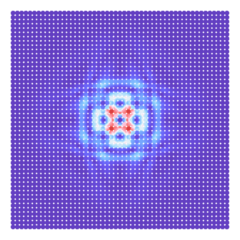}     &
        \includegraphics[width=\figsizeF]{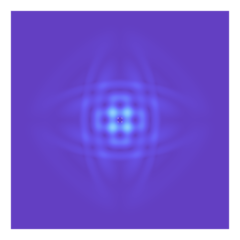}     &
        \multicolumn{1}{c|}{}
        \\ \hline
    \end{tabular}
    \caption{
        Dimensionless displacement norm ($\lVert \widetilde{u} \rVert$):
        (\textit{first column}) equivalent RRMM,
        (\textit{second column}) average metamaterial displacement ($\lVert\overline{u} \rVert$),
        (\textit{third column}) metamaterial $\mathcal{L}$,
        (\textit{fourth column}) macro Cauchy,
        for a shear load (Sec.~\ref{sec:load_boundary}) for $\omega = 100$ Hz with a single central unit cell.
    }
    \label{tab:maze_shea_figu_100}
\end{table}

\begin{table}[H]
    \centering
    \begin{tabular}{c|c|c|c|c|c}
        \cline{2-5}
                                                                                                              & RRMM
                                                                                                              & Avg. Microstr.
                                                                                                              & Microstructure
                                                                                                              & Macro Cauchy
                                                                                                              & \multicolumn{1}{c}{}
        \\ \hline
        \multicolumn{1}{|c|}{\rotatebox{90}{\makebox[\figsizeF][c]{$t= 0.005$ s}}}                            &
        \includegraphics[width=\figsizeF]{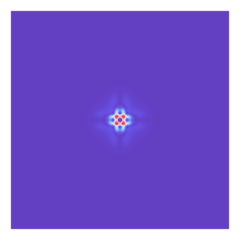}     &
        \includegraphics[width=\figsizeF]{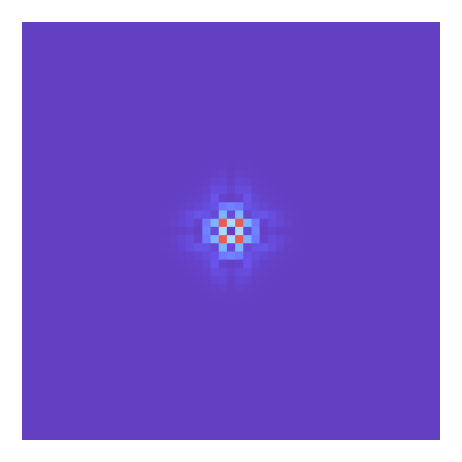} &
        \includegraphics[width=\figsizeF]{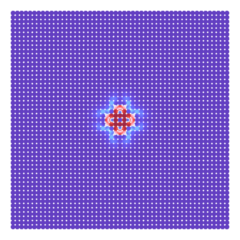}     &
        \includegraphics[width=\figsizeF]{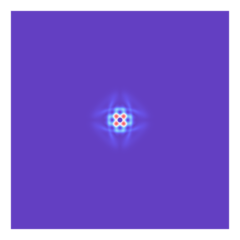}     &
        \multicolumn{1}{c|}{}
        \\ \cline{1-5}
        \multicolumn{1}{|c|}{\rotatebox{90}{\makebox[\figsizeF][c]{$t= 0.010$ s}}}                            &
        \includegraphics[width=\figsizeF]{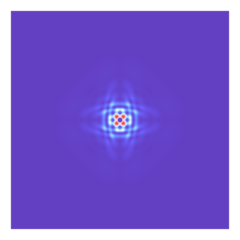}     &
        \includegraphics[width=\figsizeF]{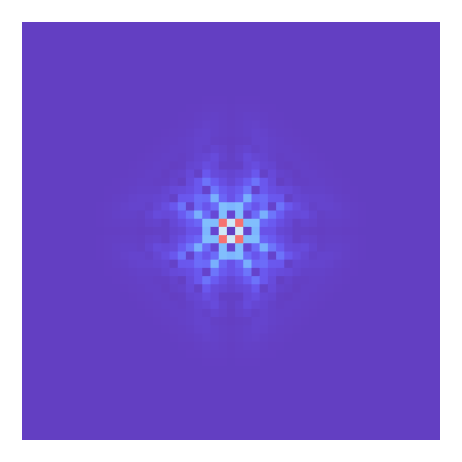} &
        \includegraphics[width=\figsizeF]{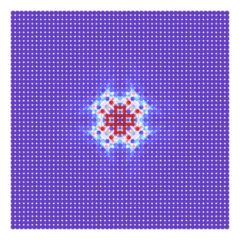}     &
        \includegraphics[width=\figsizeF]{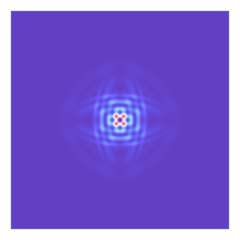}     &
        \multicolumn{1}{c|}{\rotatebox{90}{\makebox[\figsizeF][c]{Shear load - \fbox{1\texttimes 1} - $\omega = 200$ Hz}}}
        \\ \cline{1-5}
        \multicolumn{1}{|c|}{\rotatebox{90}{\makebox[\figsizeF][c]{$t= 0.015$ s}}}                            &
        \includegraphics[width=\figsizeF]{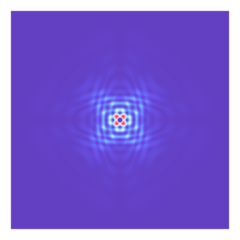}     &
        \includegraphics[width=\figsizeF]{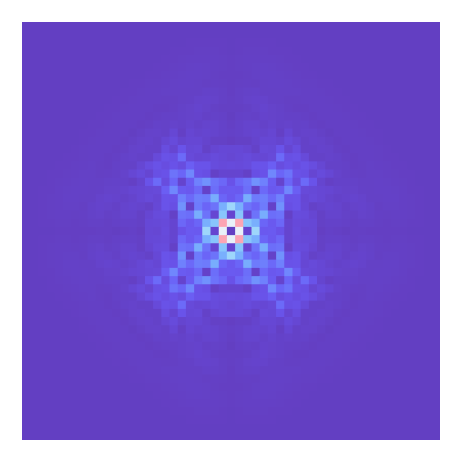} &
        \includegraphics[width=\figsizeF]{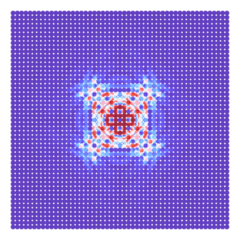}     &
        \includegraphics[width=\figsizeF]{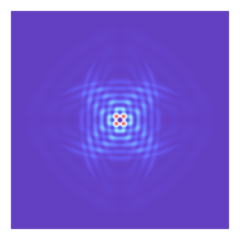}     &
        \multicolumn{1}{c|}{}
        \\ \hline
    \end{tabular}
    \caption{
        Dimensionless displacement norm ($\lVert \widetilde{u} \rVert$):
        (\textit{first column}) equivalent RRMM,
        (\textit{second column}) average metamaterial displacement ($\lVert\overline{u} \rVert$),
        (\textit{third column}) metamaterial $\mathcal{L}$,
        (\textit{fourth column}) macro Cauchy,
        for a shear load (Sec.~\ref{sec:load_boundary}) for $\omega = 200$ Hz with a single central unit cell.
    }
    \label{tab:maze_shea_figu_200}
\end{table}

\begin{table}[H]
    \centering
    \begin{tabular}{c|c|c|c|c}
        \cline{2-4}
                                                                                                           & RRMM
                                                                                                           & Microstructure
                                                                                                           & Macro Cauchy
                                                                                                           & \multicolumn{1}{c}{}
        \\ \hline
        \multicolumn{1}{|c|}{\rotatebox{90}{\makebox[\figsizeF][c]{$t= 0.005$ s}}}                         &
        \includegraphics[width=\figsizeF]{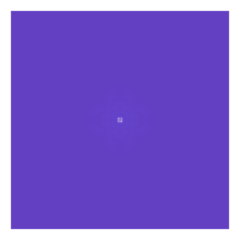} &
        \includegraphics[width=\figsizeF]{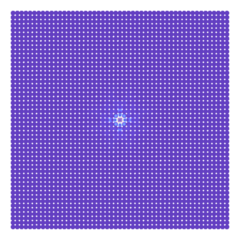} &
        \includegraphics[width=\figsizeF]{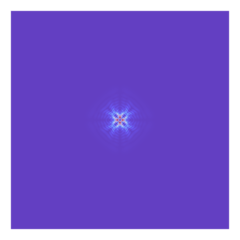} &
        \multicolumn{1}{c|}{}
        \\ \cline{1-4}
        \multicolumn{1}{|c|}{\rotatebox{90}{\makebox[\figsizeF][c]{$t= 0.010$ s}}}                         &
        \includegraphics[width=\figsizeF]{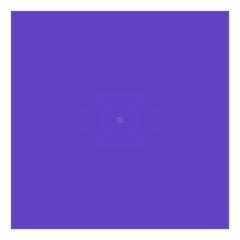} &
        \includegraphics[width=\figsizeF]{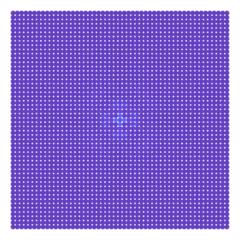} &
        \includegraphics[width=\figsizeF]{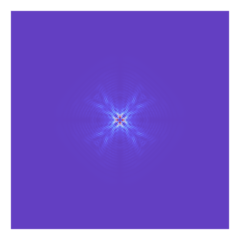} &
        \multicolumn{1}{c|}{\rotatebox{90}{\makebox[\figsizeF][c]{Shear load - \fbox{1\texttimes 1} - $\omega = 1000$ Hz}}}
        \\ \cline{1-4}
        \multicolumn{1}{|c|}{\rotatebox{90}{\makebox[\figsizeF][c]{$t= 0.015$ s}}}                         &
        \includegraphics[width=\figsizeF]{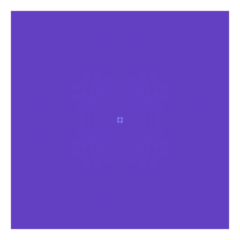} &
        \includegraphics[width=\figsizeF]{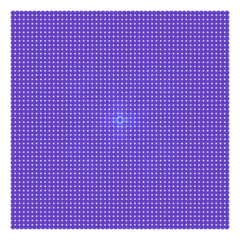} &
        \includegraphics[width=\figsizeF]{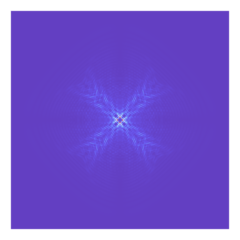} &
        \multicolumn{1}{c|}{}
        \\ \hline
    \end{tabular}
    \caption{
        Dimensionless displacement norm ($\lVert \widetilde{u} \rVert$):
        (\textit{first column}) equivalent RRMM,
        (\textit{second column}) metamaterial $\mathcal{L}$,
        (\textit{third column}) macro Cauchy,
        for a shear load (Sec.~\ref{sec:load_boundary}) for $\omega = 1000$ Hz with a single central unit cell.
    }
    \label{tab:maze_shea_figu_1000}
\end{table}

In Figs.~\ref{tab:maze_shea_figu_100}--\ref{tab:maze_shea_figu_1000}, as in the hydrostatic case, we observe that the RRMM shows good qualitative agreement (highlighted by the comparison with the average response of the microstructure) while exhibiting quantitative differences at all frequencies except 1000 Hz (the bandgap frequency), where the response is accurately captured.

\begin{table}[H]
    \centering
    \begin{tabular}{c|c|c|c|c}
        \cline{2-4}
                                                                                                              & RRMM
                                                                                                              & Microstructure
                                                                                                              & Macro Cauchy
                                                                                                              & \multicolumn{1}{c}{}
        \\ \hline
        \multicolumn{1}{|c|}{\rotatebox{90}{\makebox[\figsizeF][c]{$t= 0.005$ s}}}                            &
        \includegraphics[width=\figsizeF]{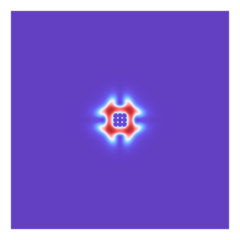} &
        \includegraphics[width=\figsizeF]{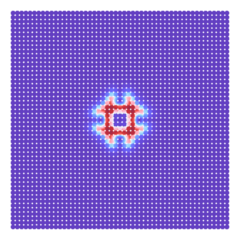} &
        \includegraphics[width=\figsizeF]{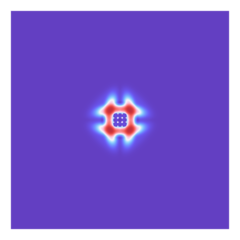} &
        \multicolumn{1}{c|}{}
        \\ \cline{1-4}
        \multicolumn{1}{|c|}{\rotatebox{90}{\makebox[\figsizeF][c]{$t= 0.010$ s}}}                            &
        \includegraphics[width=\figsizeF]{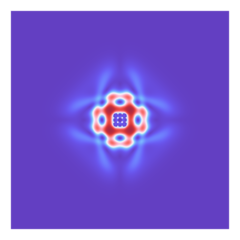} &
        \includegraphics[width=\figsizeF]{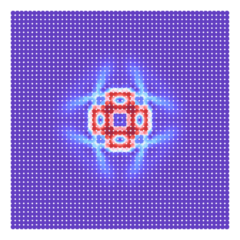} &
        \includegraphics[width=\figsizeF]{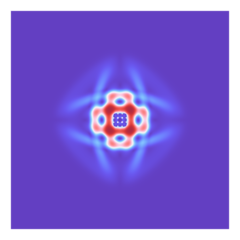} &
        \multicolumn{1}{c|}{\rotatebox{90}{\makebox[\figsizeF][c]{Shear load - \fbox{3\texttimes 3} - $\omega = 100$ Hz}}}
        \\ \cline{1-4}
        \multicolumn{1}{|c|}{\rotatebox{90}{\makebox[\figsizeF][c]{$t= 0.015$ s}}}                            &
        \includegraphics[width=\figsizeF]{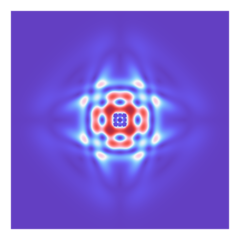} &
        \includegraphics[width=\figsizeF]{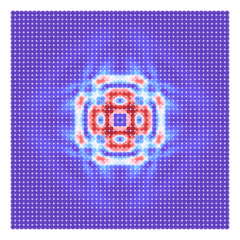} &
        \includegraphics[width=\figsizeF]{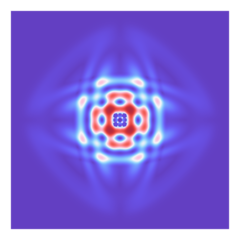} &
        \multicolumn{1}{c|}{}
        \\ \hline
    \end{tabular}
    \caption{
        Dimensionless displacement norm ($\lVert \widetilde{u} \rVert$):
        (\textit{first column}) equivalent RRMM,
        (\textit{second column}) metamaterial $\mathcal{L}$,
        (\textit{third column}) macro Cauchy,
        for a shear load (Sec.~\ref{sec:load_boundary}) for $\omega = 100$ Hz with a 3\texttimes 3 central cluster.
    }
    \label{tab:maze_shea_figu_100_3x3}
\end{table}

\begin{table}[H]
    \centering
    \begin{tabular}{c|c|c|c|c}
        \cline{2-4}
                                                                                                              & RRMM
                                                                                                              & Microstructure
                                                                                                              & Macro Cauchy
                                                                                                              & \multicolumn{1}{c}{}
        \\ \hline
        \multicolumn{1}{|c|}{\rotatebox{90}{\makebox[\figsizeF][c]{$t= 0.005$ s}}}                            &
        \includegraphics[width=\figsizeF]{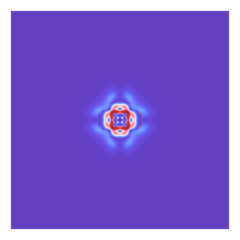} &
        \includegraphics[width=\figsizeF]{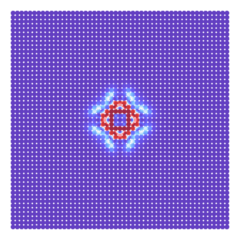} &
        \includegraphics[width=\figsizeF]{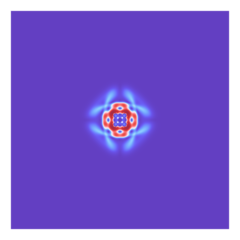} &
        \multicolumn{1}{c|}{}
        \\ \cline{1-4}
        \multicolumn{1}{|c|}{\rotatebox{90}{\makebox[\figsizeF][c]{$t= 0.010$ s}}}                            &
        \includegraphics[width=\figsizeF]{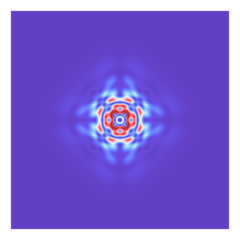} &
        \includegraphics[width=\figsizeF]{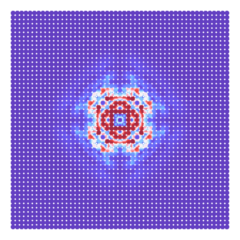} &
        \includegraphics[width=\figsizeF]{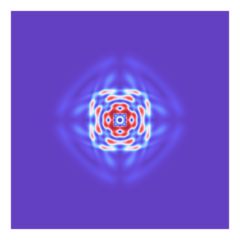} &
        \multicolumn{1}{c|}{\rotatebox{90}{\makebox[\figsizeF][c]{Shear load - \fbox{3\texttimes 3} - $\omega = 200$ Hz}}}
        \\ \cline{1-4}
        \multicolumn{1}{|c|}{\rotatebox{90}{\makebox[\figsizeF][c]{$t= 0.015$ s}}}                            &
        \includegraphics[width=\figsizeF]{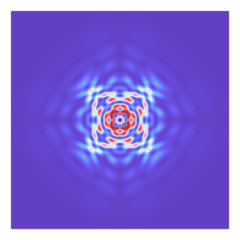} &
        \includegraphics[width=\figsizeF]{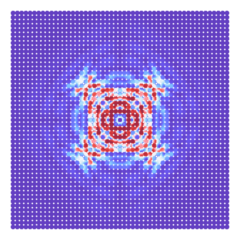} &
        \includegraphics[width=\figsizeF]{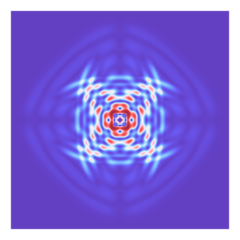} &
        \multicolumn{1}{c|}{}
        \\ \hline
    \end{tabular}
    \caption{
        Dimensionless displacement norm ($\lVert \widetilde{u} \rVert$):
        (\textit{first column}) equivalent RRMM,
        (\textit{second column}) metamaterial $\mathcal{L}$,
        (\textit{third column}) macro Cauchy,
        for a shear load (Sec.~\ref{sec:load_boundary}) for $\omega = 200$ Hz with a 3\texttimes 3 central cluster.
    }
    \label{tab:maze_shea_figu_200_3x3}
\end{table}

\begin{table}[H]
    \centering
    \begin{tabular}{c|c|c|c|c}
        \cline{2-4}
                                                                                                               & RRMM
                                                                                                               & Microstructure
                                                                                                               & Macro Cauchy
                                                                                                               & \multicolumn{1}{c}{}
        \\ \hline
        \multicolumn{1}{|c|}{\rotatebox{90}{\makebox[\figsizeF][c]{$t= 0.005$ s}}}                             &
        \includegraphics[width=\figsizeF]{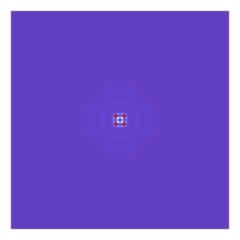} &
        \includegraphics[width=\figsizeF]{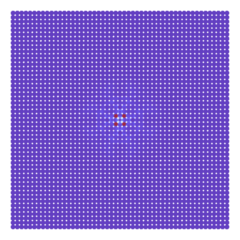} &
        \includegraphics[width=\figsizeF]{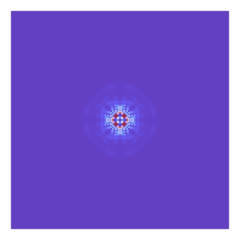} &
        \multicolumn{1}{c|}{}
        \\ \cline{1-4}
        \multicolumn{1}{|c|}{\rotatebox{90}{\makebox[\figsizeF][c]{$t= 0.010$ s}}}                             &
        \includegraphics[width=\figsizeF]{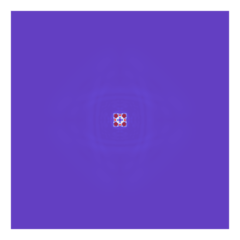} &
        \includegraphics[width=\figsizeF]{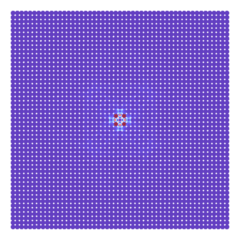} &
        \includegraphics[width=\figsizeF]{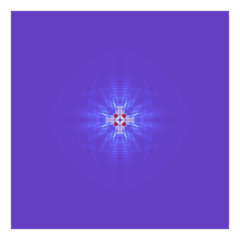} &
        \multicolumn{1}{c|}{\rotatebox{90}{\makebox[\figsizeF][c]{Shear load - \fbox{3\texttimes 3} - $\omega = 1000$ Hz}}}
        \\ \cline{1-4}
        \multicolumn{1}{|c|}{\rotatebox{90}{\makebox[\figsizeF][c]{$t= 0.015$ s}}}                             &
        \includegraphics[width=\figsizeF]{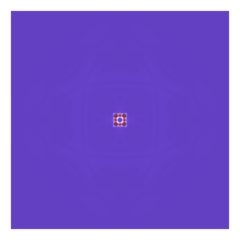} &
        \includegraphics[width=\figsizeF]{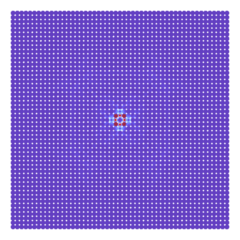} &
        \includegraphics[width=\figsizeF]{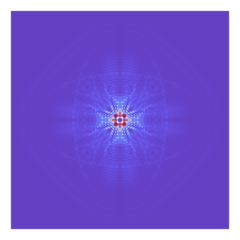} &
        \multicolumn{1}{c|}{}
        \\ \hline
    \end{tabular}
    \caption{
        Dimensionless displacement norm ($\lVert \widetilde{u} \rVert$):
        (\textit{first column}) equivalent RRMM,
        (\textit{second column}) metamaterial $\mathcal{L}$,
        (\textit{third column}) macro Cauchy,
        for a shear load (Sec.~\ref{sec:load_boundary}) for $\omega = 1000$ Hz with a 3\texttimes 3 central cluster.
    }
    \label{tab:maze_shea_figu_1000_3x3}
\end{table}

In Figs.~\ref{tab:maze_shea_figu_100_3x3}--\ref{tab:maze_shea_figu_1000_3x3}, we observe that the quantitative differences at 100 Hz are now severly reduced, while they still persist at 200 Hz, although with lower magnitude.

\subsection{\textit{Labyrinthine} unit cell's results discussion}

For the rotation test, no averaging operator is needed, regardless the frequency or the size of the central cluster: the RRMM is in this case capable of directly capturing well the displacement field (from Fig.~\ref{tab:maze_rota_figu_100}--\ref{tab:maze_shea_figu_1000_3x3}).
For the hydrostatic test, the average operator is required for the 1\texttimes 1 central unit cell case, at both 100 Hz and 200 Hz (Fig.~\ref{tab:maze_expa_figu_100}--\ref{tab:maze_expa_figu_200}).
Switching to a 3\texttimes 3 central cluster automatically resolves the need for averaging at 100 Hz (Fig.~\ref{tab:maze_expa_figu_100_3x3}), while it does not eliminate this necessity at 200 Hz (Fig.~\ref{tab:maze_expa_figu_200_3x3}).
As for the shear test, the averaging operator is only needed for the 1\texttimes 1 central unit cell, at both 100 Hz and 200 Hz, but it is already not necessary as soon as the cluster grows to a 3\texttimes 3 dimension.
It is generally true that increasing the size of the central cluster allows the load to be more uniformly distributed, which already helps reduce the appearance of local effects.
In summary, the RRMM proves to be effective in describing the response of the labyrinthine metamaterial and performs better than the macro-Cauchy model, especially at higher frequencies (near and within the bandgap).
Small quantitative differences between the RRMM and the microstructured solutions are observed in the 1 \texttimes 1 loading case.
These differences are mitigated in the 3\texttimes 3 loading case, suggesting that a boundary effect associated with the loaded interface may play a role.

\section{Time-dependent response for a \textit{four-resonator} unit cell}
\label{sec:reso}

\begin{table}[H]
    \centering
    \begin{tabular}{c|c|c|c|c|c}
        \cline{2-5}
                                                                                                               & RRMM
                                                                                                               & Avg. Microstr.
                                                                                                               & Microstructure
                                                                                                               & Macro Cauchy
                                                                                                               & \multicolumn{1}{c}{}
        \\ \hline
        \multicolumn{1}{|c|}{\rotatebox{90}{\makebox[\figsizeF][c]{$t= 0.0016$ s}}}                            &
        \includegraphics[width=\figsizeF]{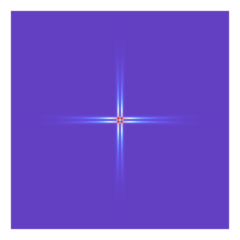}     &
        \includegraphics[width=\figsizeF]{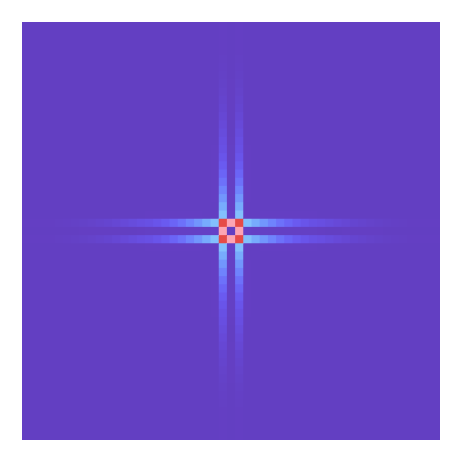} &
        \includegraphics[width=\figsizeF]{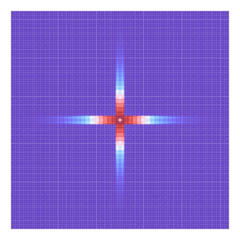}     &
        \includegraphics[width=\figsizeF]{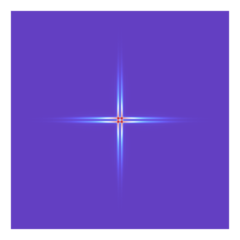}     &
        \multicolumn{1}{c|}{}
        \\ \cline{1-5}
        \multicolumn{1}{|c|}{\rotatebox{90}{\makebox[\figsizeF][c]{$t= 0.0022$ s}}}                            &
        \includegraphics[width=\figsizeF]{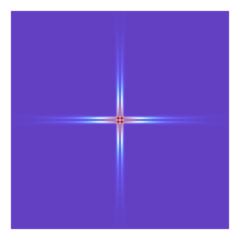}     &
        \includegraphics[width=\figsizeF]{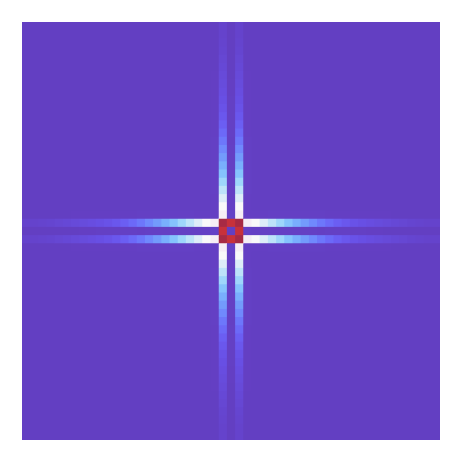} &
        \includegraphics[width=\figsizeF]{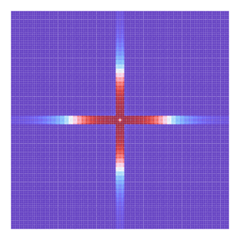}     &
        \includegraphics[width=\figsizeF]{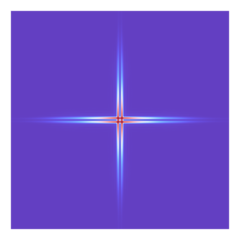}     &
        \multicolumn{1}{c|}{\rotatebox{90}{\makebox[\figsizeF][c]{Rotation load - \fbox{1\texttimes 1} - $\omega = 100$ Hz}}}
        \\ \cline{1-5}
        \multicolumn{1}{|c|}{\rotatebox{90}{\makebox[\figsizeF][c]{$t= 0.0028$ s}}}                            &
        \includegraphics[width=\figsizeF]{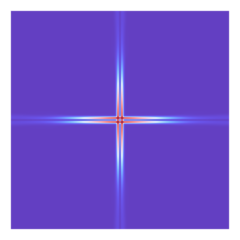}     &
        \includegraphics[width=\figsizeF]{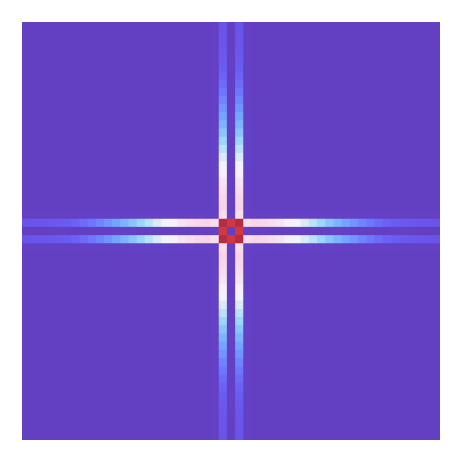} &
        \includegraphics[width=\figsizeF]{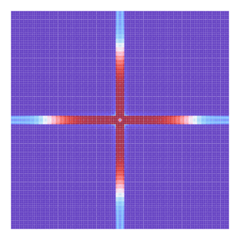}     &
        \includegraphics[width=\figsizeF]{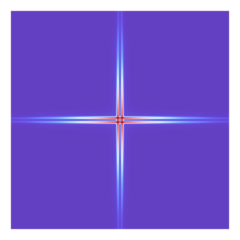}     &
        \multicolumn{1}{c|}{}
        \\ \hline
    \end{tabular}
    \caption{
        Dimensionless displacement norm ($\lVert \widetilde{u} \rVert$):
        (\textit{first column}) equivalent RRMM,
        (\textit{second column}) average metamaterial displacement ($\lVert\overline{u} \rVert$),
        (\textit{third column}) metamaterial $\mathcal{R}$,
        (\textit{fourth column}) macro Cauchy,
        for a rotation load (Sec.~\ref{sec:load_boundary}) for $\omega = 100$ Hz with a single central unit cell.
    }
    \label{tab:reso_rota_figu_100}
\end{table}

\begin{table}[H]
    \centering
    \begin{tabular}{c|c|c|c|c|c}
        \cline{2-5}
                                                                                                               & RRMM
                                                                                                               & Avg. Microstr.
                                                                                                               & Microstructure
                                                                                                               & Macro Cauchy
                                                                                                               & \multicolumn{1}{c}{}
        \\ \hline
        \multicolumn{1}{|c|}{\rotatebox{90}{\makebox[\figsizeF][c]{$t= 0.0016$ s}}}                            &
        \includegraphics[width=\figsizeF]{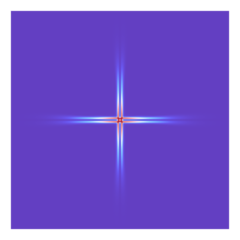}     &
        \includegraphics[width=\figsizeF]{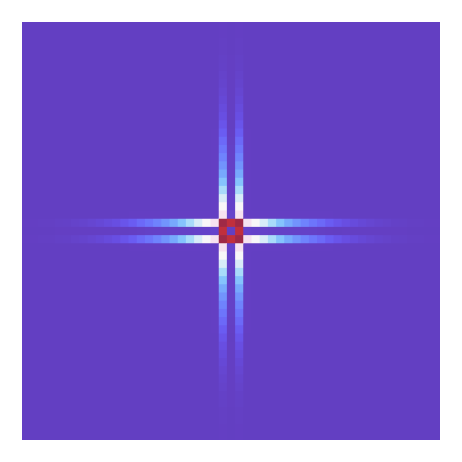} &
        \includegraphics[width=\figsizeF]{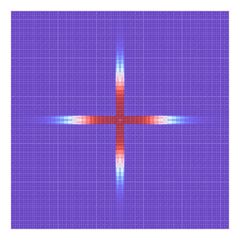}     &
        \includegraphics[width=\figsizeF]{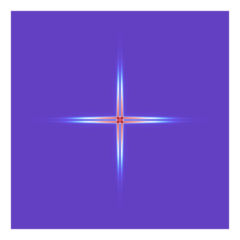}     &
        \multicolumn{1}{c|}{}
        \\ \cline{1-5}
        \multicolumn{1}{|c|}{\rotatebox{90}{\makebox[\figsizeF][c]{$t= 0.0022$ s}}}                            &
        \includegraphics[width=\figsizeF]{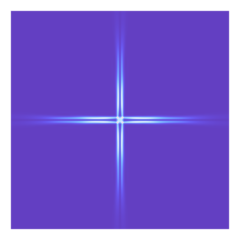}     &
        \includegraphics[width=\figsizeF]{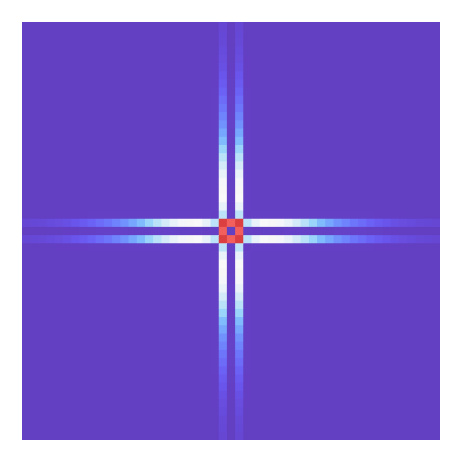} &
        \includegraphics[width=\figsizeF]{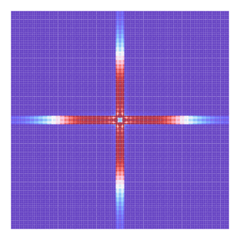}     &
        \includegraphics[width=\figsizeF]{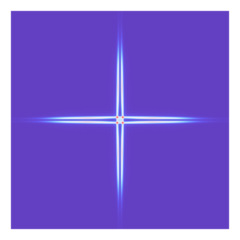}     &
        \multicolumn{1}{c|}{\rotatebox{90}{\makebox[\figsizeF][c]{Rotation load - \fbox{1\texttimes 1} - $\omega = 200$ Hz}}}
        \\ \cline{1-5}
        \multicolumn{1}{|c|}{\rotatebox{90}{\makebox[\figsizeF][c]{$t= 0.0028$ s}}}                            &
        \includegraphics[width=\figsizeF]{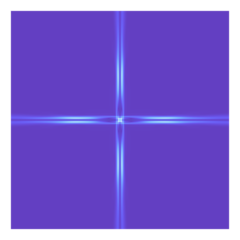}     &
        \includegraphics[width=\figsizeF]{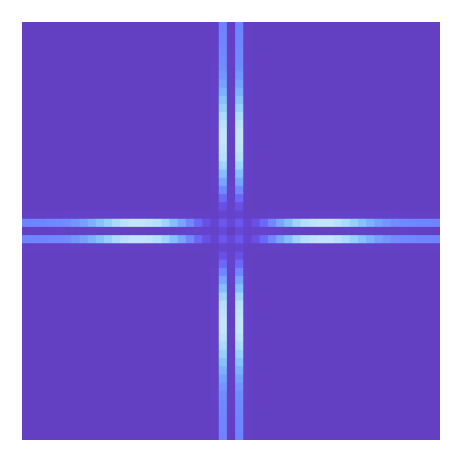} &
        \includegraphics[width=\figsizeF]{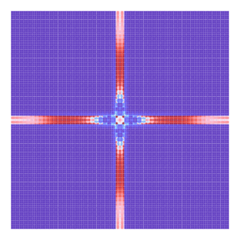}     &
        \includegraphics[width=\figsizeF]{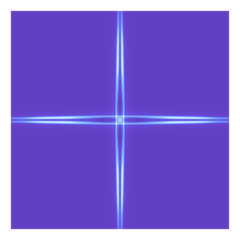}     &
        \multicolumn{1}{c|}{}
        \\ \hline
    \end{tabular}
    \caption{
        Dimensionless displacement norm ($\lVert \widetilde{u} \rVert$):
        (\textit{first column}) equivalent RRMM,
        (\textit{second column}) average metamaterial displacement ($\lVert\overline{u} \rVert$),
        (\textit{third column}) metamaterial $\mathcal{R}$,
        (\textit{fourth column}) macro Cauchy,
        for a rotation load (Sec.~\ref{sec:load_boundary}) for $\omega = 200$ Hz with a single central unit cell.
    }
    \label{tab:reso_rota_figu_200}
\end{table}

\begin{table}[H]
    \centering
    \begin{tabular}{c|c|c|c|c|c}
        \cline{2-5}
                                                                                                               & RRMM
                                                                                                               & Avg. Microstr.
                                                                                                               & Microstructure
                                                                                                               & Macro Cauchy
                                                                                                               & \multicolumn{1}{c}{}
        \\ \hline
        \multicolumn{1}{|c|}{\rotatebox{90}{\makebox[\figsizeF][c]{$t= 0.0016$ s}}}                            &
        \includegraphics[width=\figsizeF]{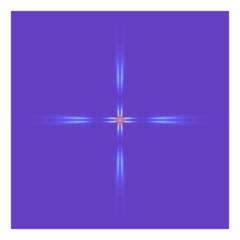}     &
        \includegraphics[width=\figsizeF]{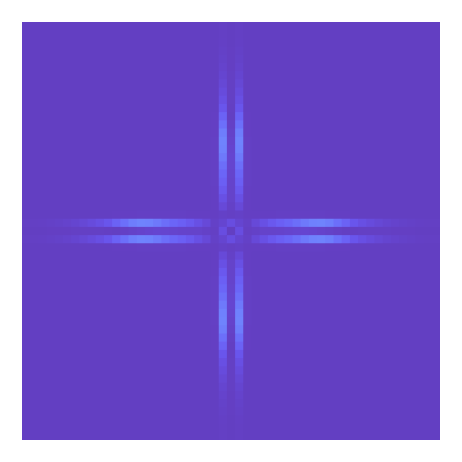} &
        \includegraphics[width=\figsizeF]{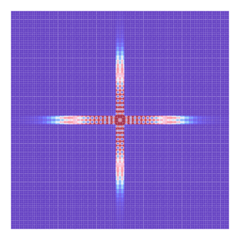}     &
        \includegraphics[width=\figsizeF]{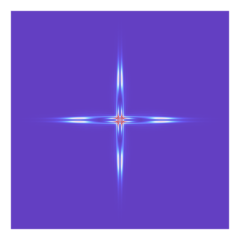}     &
        \multicolumn{1}{c|}{}
        \\ \cline{1-5}
        \multicolumn{1}{|c|}{\rotatebox{90}{\makebox[\figsizeF][c]{$t= 0.0022$ s}}}                            &
        \includegraphics[width=\figsizeF]{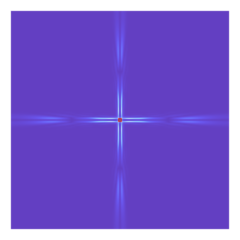}     &
        \includegraphics[width=\figsizeF]{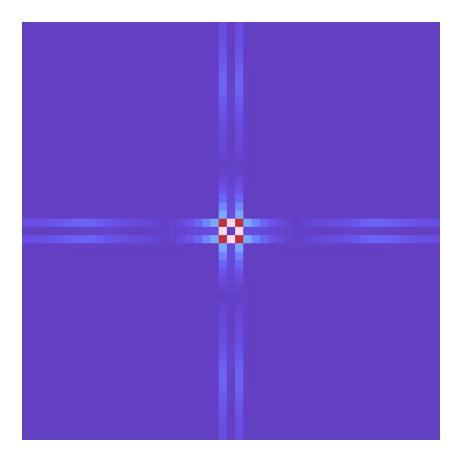} &
        \includegraphics[width=\figsizeF]{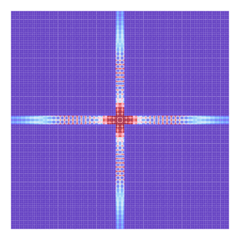}     &
        \includegraphics[width=\figsizeF]{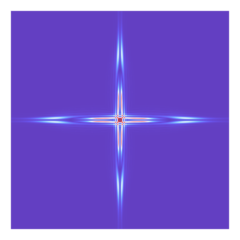}     &
        \multicolumn{1}{c|}{\rotatebox{90}{\makebox[\figsizeF][c]{Rotation load - \fbox{1\texttimes 1} - $\omega = 425$ Hz}}}
        \\ \cline{1-5}
        \multicolumn{1}{|c|}{\rotatebox{90}{\makebox[\figsizeF][c]{$t= 0.0028$ s}}}                            &
        \includegraphics[width=\figsizeF]{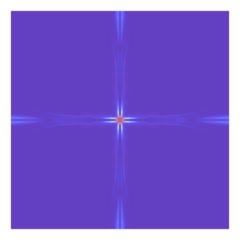}     &
        \includegraphics[width=\figsizeF]{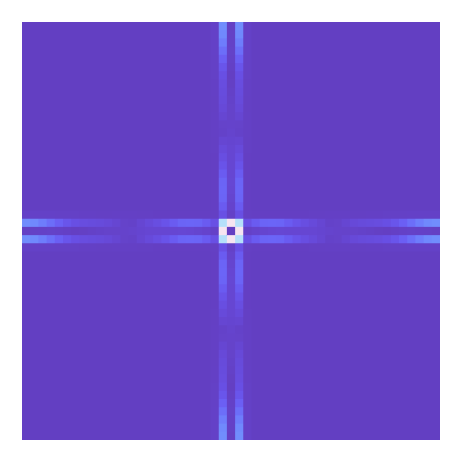} &
        \includegraphics[width=\figsizeF]{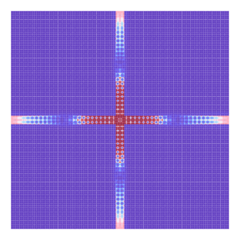}     &
        \includegraphics[width=\figsizeF]{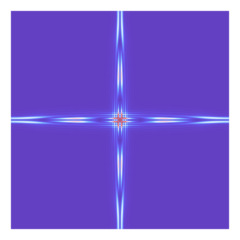}     &
        \multicolumn{1}{c|}{}
        \\ \hline
    \end{tabular}
    \caption{
        Dimensionless displacement norm ($\lVert \widetilde{u} \rVert$):
        (\textit{first column}) equivalent RRMM,
        (\textit{second column}) average metamaterial displacement ($\lVert\overline{u} \rVert$),
        (\textit{third column}) metamaterial $\mathcal{R}$,
        (\textit{fourth column}) macro Cauchy,
        for a rotation load (Sec.~\ref{sec:load_boundary}) for $\omega = 425$ Hz with a single central unit cell.
    }
    \label{tab:reso_rota_figu_425}
\end{table}

In Figs.~\ref{tab:reso_rota_figu_100}--\ref{tab:reso_rota_figu_425}, we observe that the RRMM shows good qualitative agreement across all frequencies, while significant quantitative differences still arise.
These differences can be mitigated by comparing the RRMM with the average response of the microstructured system.

\begin{table}[H]
    \centering
    \begin{tabular}{c|c|c|c|c|c}
        \cline{2-5}
                                                                                                                   & RRMM
                                                                                                                   & Avg. Microstr.
                                                                                                                   & Microstructure
                                                                                                                   & Macro Cauchy
                                                                                                                   & \multicolumn{1}{c}{}
        \\ \hline
        \multicolumn{1}{|c|}{\rotatebox{90}{\makebox[\figsizeF][c]{$t= 0.0016$ s}}}                                &
        \includegraphics[width=\figsizeF]{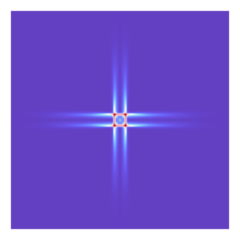}     &
        \includegraphics[width=\figsizeF]{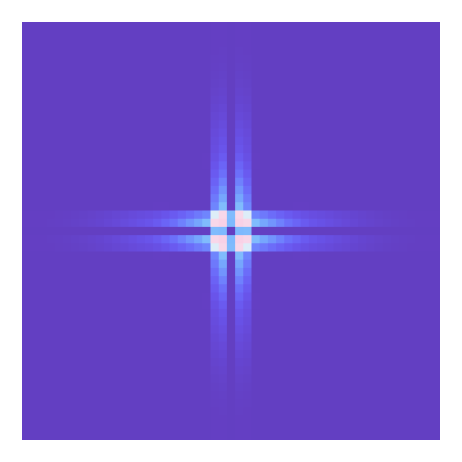} &
        \includegraphics[width=\figsizeF]{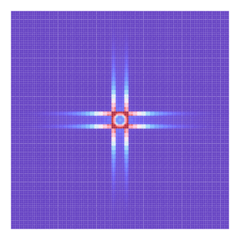}     &
        \includegraphics[width=\figsizeF]{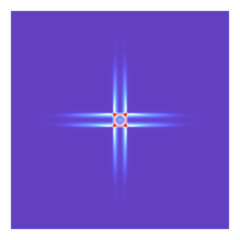}     &
        \multicolumn{1}{c|}{}
        \\ \cline{1-5}
        \multicolumn{1}{|c|}{\rotatebox{90}{\makebox[\figsizeF][c]{$t= 0.0022$ s}}}                                &
        \includegraphics[width=\figsizeF]{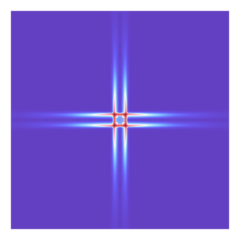}     &
        \includegraphics[width=\figsizeF]{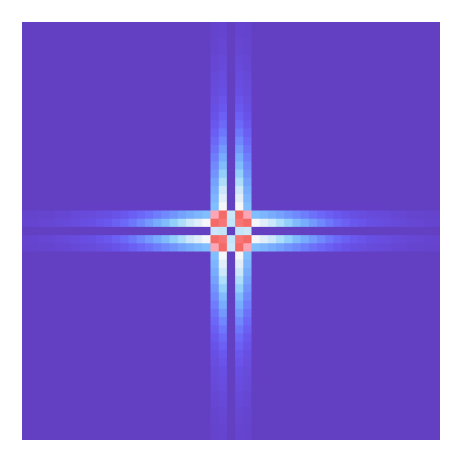} &
        \includegraphics[width=\figsizeF]{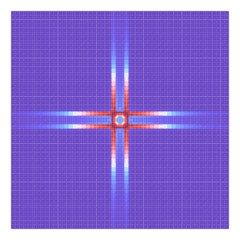}     &
        \includegraphics[width=\figsizeF]{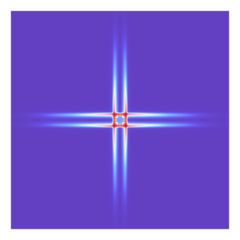}     &
        \multicolumn{1}{c|}{\rotatebox{90}{\makebox[\figsizeF][c]{Rotation load - \fbox{3\texttimes 3} - $\omega = 100$ Hz}}}
        \\ \cline{1-5}
        \multicolumn{1}{|c|}{\rotatebox{90}{\makebox[\figsizeF][c]{$t= 0.0028$ s}}}                                &
        \includegraphics[width=\figsizeF]{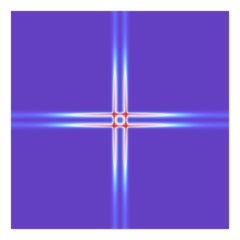}     &
        \includegraphics[width=\figsizeF]{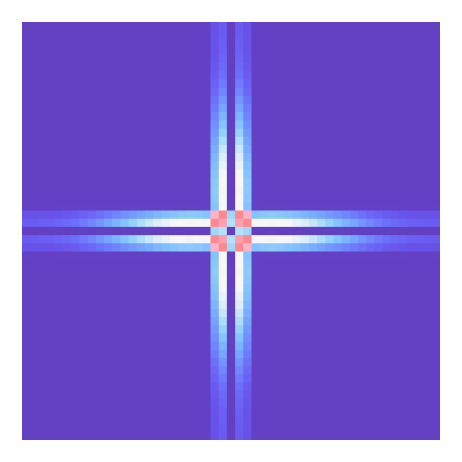} &
        \includegraphics[width=\figsizeF]{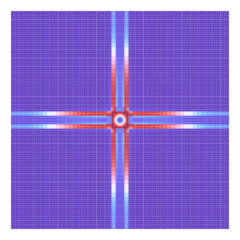}     &
        \includegraphics[width=\figsizeF]{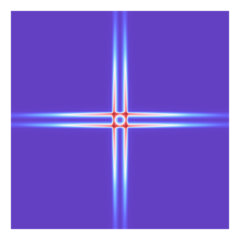}     &
        \multicolumn{1}{c|}{}
        \\ \hline
    \end{tabular}
    \caption{
        Dimensionless displacement norm ($\lVert \widetilde{u} \rVert$):
        (\textit{first column}) equivalent RRMM,
        (\textit{second column}) average metamaterial displacement ($\lVert\overline{u} \rVert$),
        (\textit{third column}) metamaterial $\mathcal{R}$,
        (\textit{fourth column}) macro Cauchy,
        for a rotation load (Sec.~\ref{sec:load_boundary}) for $\omega = 100$ Hz with a 3\texttimes 3 central cluster.
    }
    \label{tab:reso_rota_figu_100_3x3}
\end{table}

\begin{table}[H]
    \centering
    \begin{tabular}{c|c|c|c|c|c}
        \cline{2-5}
                                                                                                                   & RRMM
                                                                                                                   & Avg. Microstr.
                                                                                                                   & Microstructure
                                                                                                                   & Macro Cauchy
                                                                                                                   & \multicolumn{1}{c}{}
        \\ \hline
        \multicolumn{1}{|c|}{\rotatebox{90}{\makebox[\figsizeF][c]{$t= 0.0016$ s}}}                                &
        \includegraphics[width=\figsizeF]{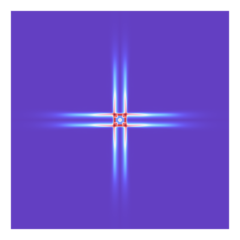}     &
        \includegraphics[width=\figsizeF]{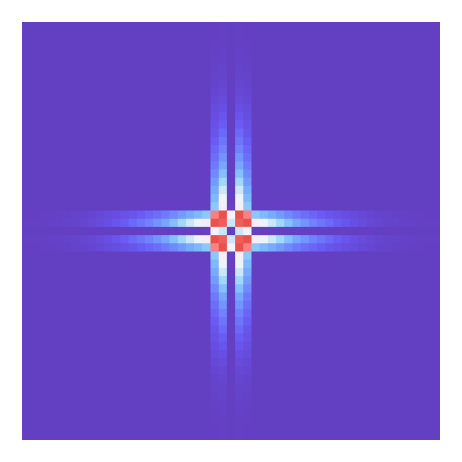} &
        \includegraphics[width=\figsizeF]{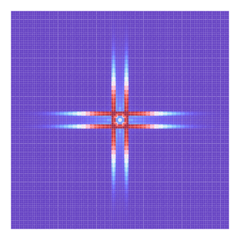}     &
        \includegraphics[width=\figsizeF]{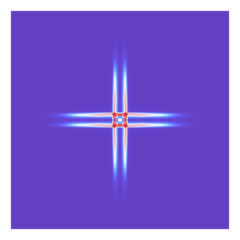}     &
        \multicolumn{1}{c|}{}
        \\ \cline{1-5}
        \multicolumn{1}{|c|}{\rotatebox{90}{\makebox[\figsizeF][c]{$t= 0.0022$ s}}}                                &
        \includegraphics[width=\figsizeF]{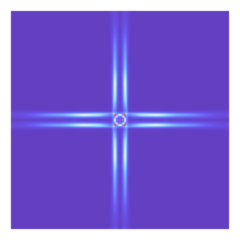}     &
        \includegraphics[width=\figsizeF]{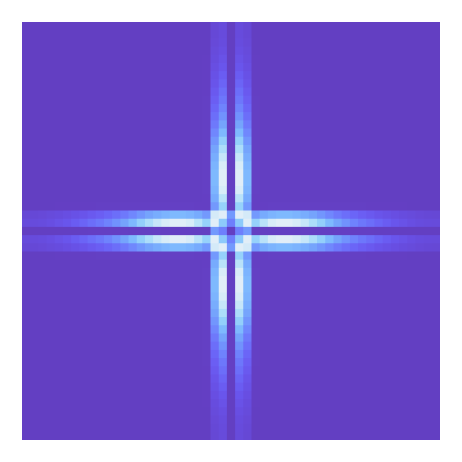} &
        \includegraphics[width=\figsizeF]{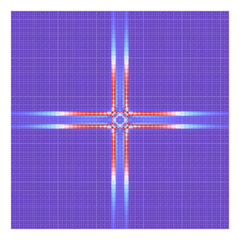}     &
        \includegraphics[width=\figsizeF]{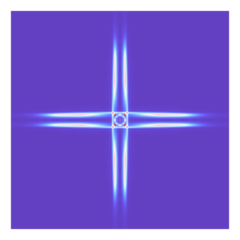}     &
        \multicolumn{1}{c|}{\rotatebox{90}{\makebox[\figsizeF][c]{Rotation load - \fbox{3\texttimes 3} - $\omega = 200$ Hz}}}
        \\ \cline{1-5}
        \multicolumn{1}{|c|}{\rotatebox{90}{\makebox[\figsizeF][c]{$t= 0.0028$ s}}}                                &
        \includegraphics[width=\figsizeF]{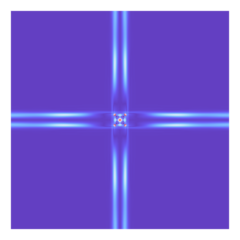}     &
        \includegraphics[width=\figsizeF]{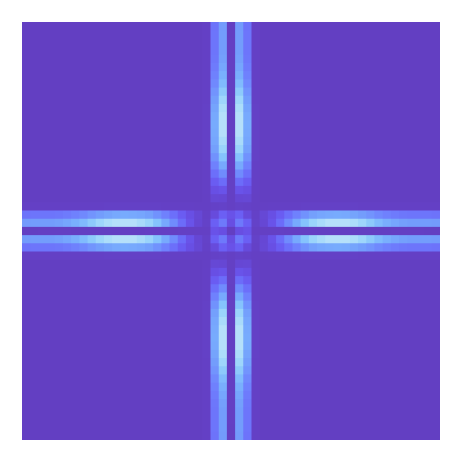} &
        \includegraphics[width=\figsizeF]{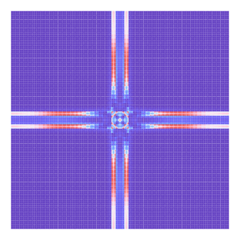}     &
        \includegraphics[width=\figsizeF]{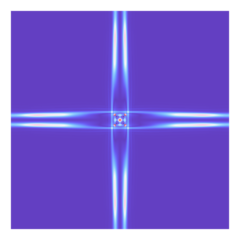}     &
        \multicolumn{1}{c|}{}
        \\ \hline
    \end{tabular}
    \caption{
        Dimensionless displacement norm ($\lVert \widetilde{u} \rVert$):
        (\textit{first column}) equivalent RRMM,
        (\textit{second column}) average metamaterial displacement ($\lVert\overline{u} \rVert$),
        (\textit{third column}) metamaterial $\mathcal{R}$,
        (\textit{fourth column}) macro Cauchy,
        for a rotation load (Sec.~\ref{sec:load_boundary}) for $\omega = 200$ Hz with a 3\texttimes 3 central cluster.
    }
    \label{tab:reso_rota_figu_200_3x3}
\end{table}

\begin{table}[H]
    \centering
    \begin{tabular}{c|c|c|c|c|c}
        \cline{2-5}
                                                                                                                   & RRMM
                                                                                                                   & Avg. Microstr.
                                                                                                                   & Microstructure
                                                                                                                   & Macro Cauchy
                                                                                                                   & \multicolumn{1}{c}{}
        \\ \hline
        \multicolumn{1}{|c|}{\rotatebox{90}{\makebox[\figsizeF][c]{$t= 0.0016$ s}}}                                &
        \includegraphics[width=\figsizeF]{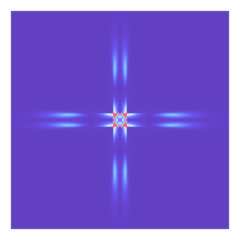}     &
        \includegraphics[width=\figsizeF]{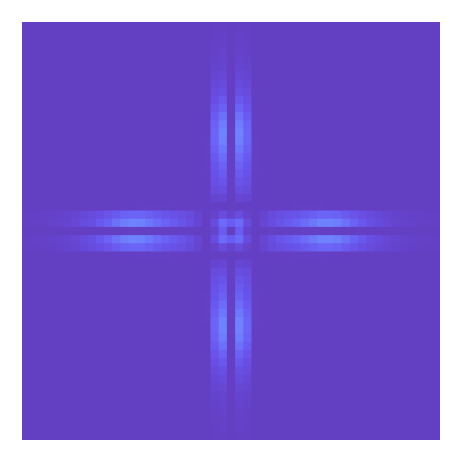} &
        \includegraphics[width=\figsizeF]{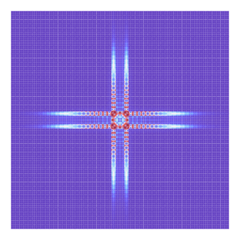}     &
        \includegraphics[width=\figsizeF]{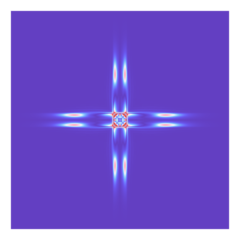}     &
        \multicolumn{1}{c|}{}
        \\ \cline{1-5}
        \multicolumn{1}{|c|}{\rotatebox{90}{\makebox[\figsizeF][c]{$t= 0.0022$ s}}}                                &
        \includegraphics[width=\figsizeF]{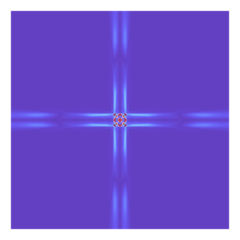}     &
        \includegraphics[width=\figsizeF]{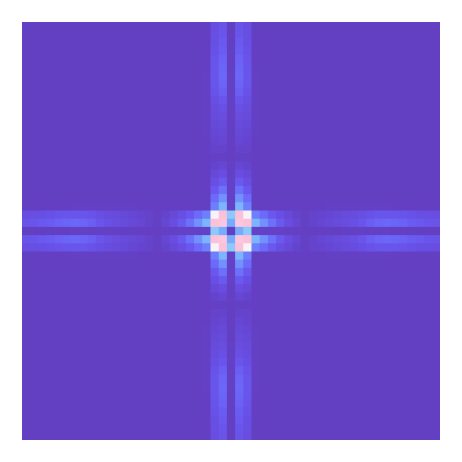} &
        \includegraphics[width=\figsizeF]{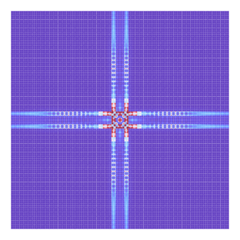}     &
        \includegraphics[width=\figsizeF]{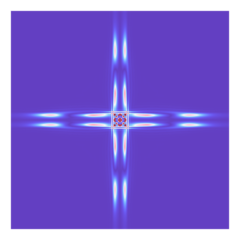}     &
        \multicolumn{1}{c|}{\rotatebox{90}{\makebox[\figsizeF][c]{Rotation load - \fbox{3\texttimes 3} - $\omega = 425$ Hz}}}
        \\ \cline{1-5}
        \multicolumn{1}{|c|}{\rotatebox{90}{\makebox[\figsizeF][c]{$t= 0.0028$ s}}}                                &
        \includegraphics[width=\figsizeF]{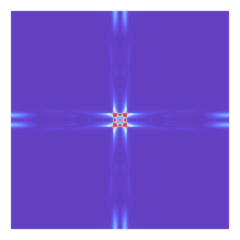}     &
        \includegraphics[width=\figsizeF]{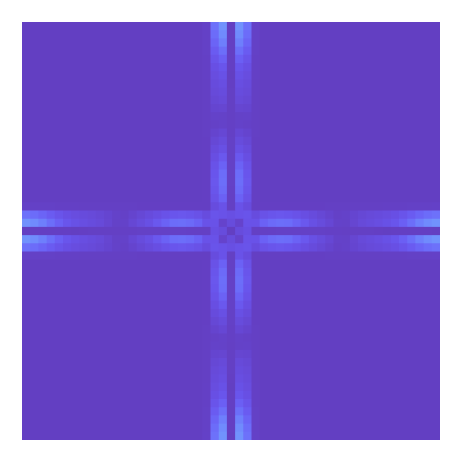} &
        \includegraphics[width=\figsizeF]{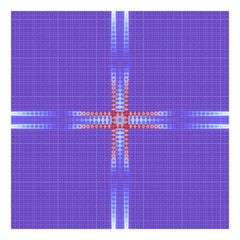}     &
        \includegraphics[width=\figsizeF]{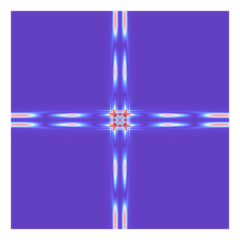}     &
        \multicolumn{1}{c|}{}
        \\ \hline
    \end{tabular}
    \caption{
        Dimensionless displacement norm ($\lVert \widetilde{u} \rVert$):
        (\textit{first column}) equivalent RRMM,
        (\textit{second column}) average metamaterial displacement ($\lVert\overline{u} \rVert$),
        (\textit{third column}) metamaterial $\mathcal{R}$,
        (\textit{fourth column}) macro Cauchy,
        for a rotation load (Sec.~\ref{sec:load_boundary}) for $\omega = 425$ Hz with a 3\texttimes 3 central cluster.
    }
    \label{tab:reso_rota_figu_425_3x3}
\end{table}

As in the 1\texttimes 1 cases, in Figs.~\ref{tab:reso_rota_figu_100_3x3}--\ref{tab:reso_rota_figu_425_3x3}, we observe that the RRMM shows good qualitative agreement across all frequencies, while moderate quantitative differences still arise. These differences can be mitigated by comparing the RRMM with the average response of the microstructured system.

\begin{table}[H]
    \centering
    \begin{tabular}{c|c|c|c|c|c|c}
        \cline{2-5}
                                                                                                               & RRMM
                                                                                                               & Avg. Microstr.
                                                                                                               & Microstructure
                                                                                                               & Macro Cauchy
                                                                                                               & \multicolumn{1}{c}{}
        \\ \hline
        \multicolumn{1}{|c|}{\rotatebox{90}{\makebox[\figsizeF][c]{$t= 0.0016$ s}}}                            &
        \includegraphics[width=\figsizeF]{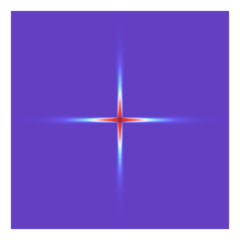}     &
        \includegraphics[width=\figsizeF]{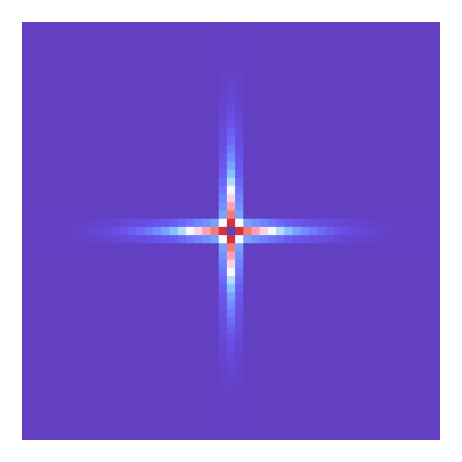} &
        \includegraphics[width=\figsizeF]{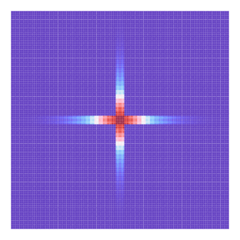}     &
        \includegraphics[width=\figsizeF]{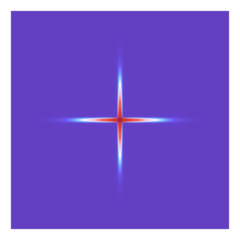}     &
        \multicolumn{1}{c|}{}
        \\ \cline{1-5}
        \multicolumn{1}{|c|}{\rotatebox{90}{\makebox[\figsizeF][c]{$t= 0.0022$ s}}}                            &
        \includegraphics[width=\figsizeF]{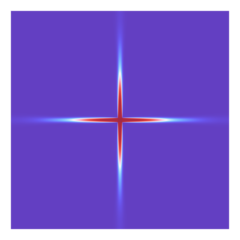}     &
        \includegraphics[width=\figsizeF]{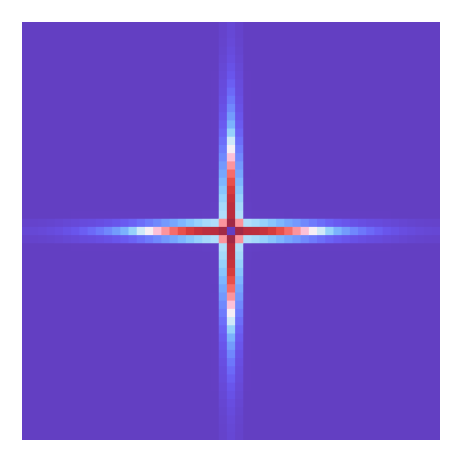} &
        \includegraphics[width=\figsizeF]{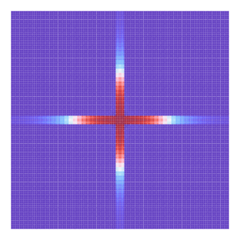}     &
        \includegraphics[width=\figsizeF]{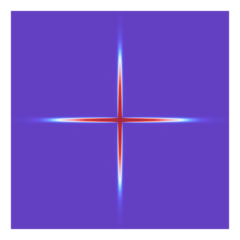}     &
        \multicolumn{1}{c|}{\rotatebox{90}{\makebox[\figsizeF][c]{Hydrostatic load - \fbox{1\texttimes 1} - $\omega = 100$ Hz}}}
        \\ \cline{1-5}
        \multicolumn{1}{|c|}{\rotatebox{90}{\makebox[\figsizeF][c]{$t= 0.0028$ s}}}                            &
        \includegraphics[width=\figsizeF]{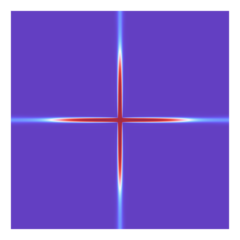}     &
        \includegraphics[width=\figsizeF]{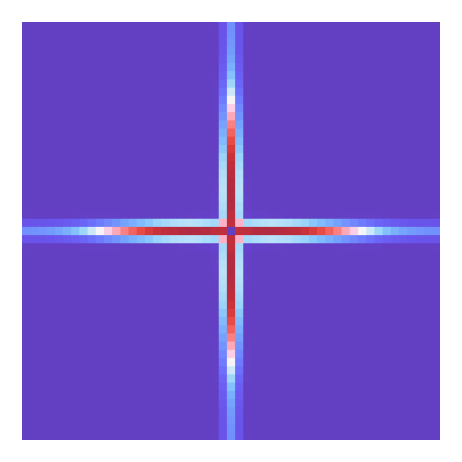} &
        \includegraphics[width=\figsizeF]{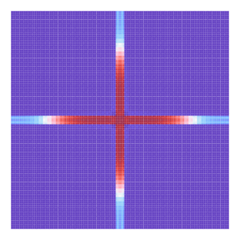}     &
        \includegraphics[width=\figsizeF]{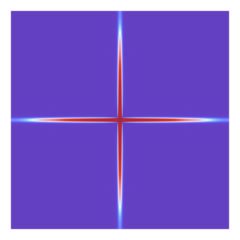}     &
        \multicolumn{1}{c|}{}
        \\ \hline
    \end{tabular}
    \caption{
        Dimensionless displacement norm ($\lVert \widetilde{u} \rVert$):
        (\textit{first column}) equivalent RRMM,
        (\textit{second column}) average metamaterial displacement ($\lVert\overline{u} \rVert$),
        (\textit{third column}) metamaterial $\mathcal{R}$,
        (\textit{fourth column}) macro Cauchy,
        for a hydrostatic load (Sec.~\ref{sec:load_boundary}) for $\omega = 100$ Hz with a single central unit cell.
    }
    \label{tab:reso_expa_figu_100}
\end{table}

\begin{table}[H]
    \centering
    \begin{tabular}{c|c|c|c|c|c}
        \cline{2-5}
                                                                                                               & RRMM
                                                                                                               & Avg. Microstr.
                                                                                                               & Microstructure
                                                                                                               & Macro Cauchy
                                                                                                               & \multicolumn{1}{c}{}
        \\ \hline
        \multicolumn{1}{|c|}{\rotatebox{90}{\makebox[\figsizeF][c]{$t= 0.0016$ s}}}                            &
        \includegraphics[width=\figsizeF]{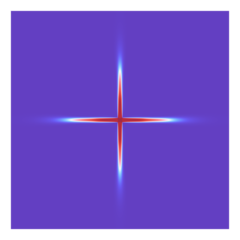}     &
        \includegraphics[width=\figsizeF]{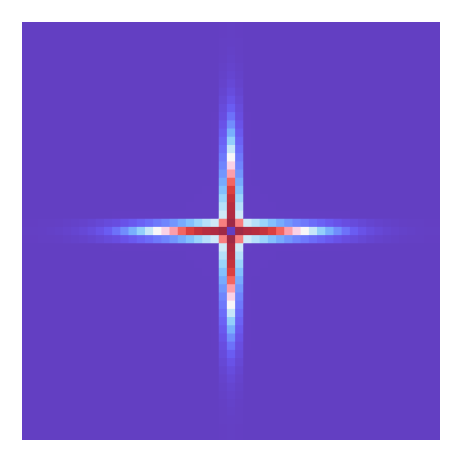} &
        \includegraphics[width=\figsizeF]{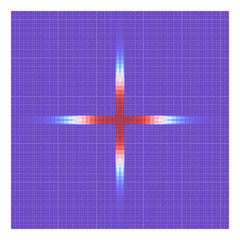}     &
        \includegraphics[width=\figsizeF]{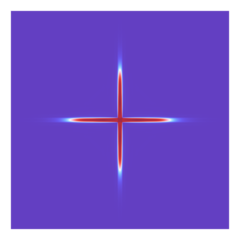}     &
        \multicolumn{1}{c|}{}
        \\ \cline{1-5}
        \multicolumn{1}{|c|}{\rotatebox{90}{\makebox[\figsizeF][c]{$t= 0.0022$ s}}}                            &
        \includegraphics[width=\figsizeF]{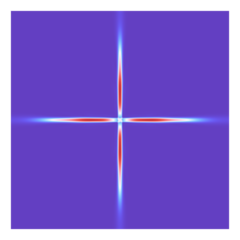}     &
        \includegraphics[width=\figsizeF]{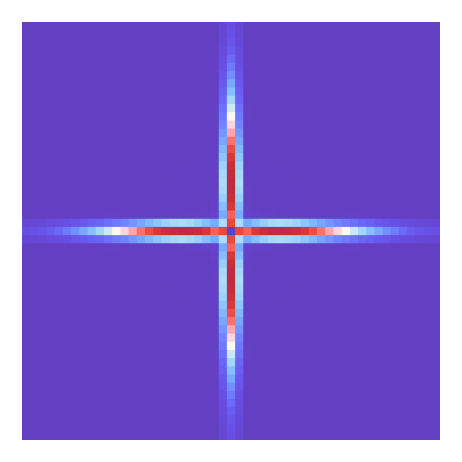} &
        \includegraphics[width=\figsizeF]{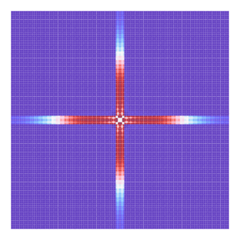}     &
        \includegraphics[width=\figsizeF]{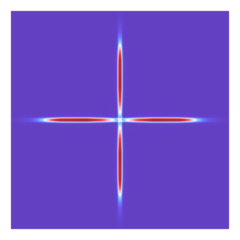}     &
        \multicolumn{1}{c|}{\rotatebox{90}{\makebox[\figsizeF][c]{Hydrostatic load - \fbox{1\texttimes 1} - $\omega = 200$ Hz}}}
        \\ \cline{1-5}
        \multicolumn{1}{|c|}{\rotatebox{90}{\makebox[\figsizeF][c]{$t= 0.0028$ s}}}                            &
        \includegraphics[width=\figsizeF]{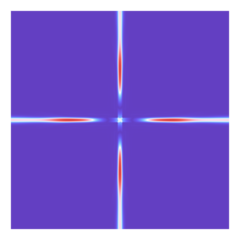}     &
        \includegraphics[width=\figsizeF]{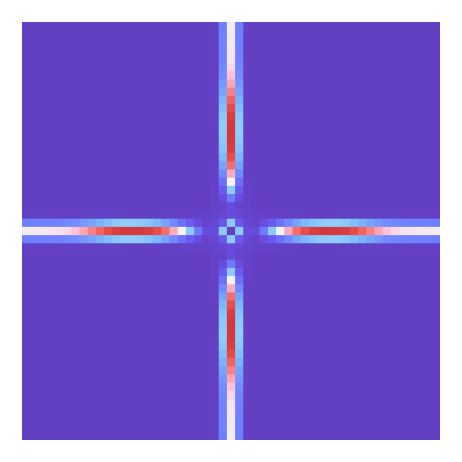} &
        \includegraphics[width=\figsizeF]{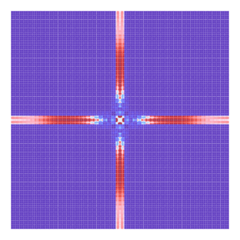}     &
        \includegraphics[width=\figsizeF]{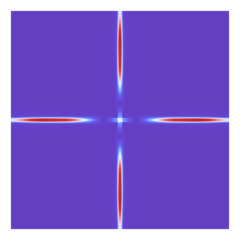}     &
        \multicolumn{1}{c|}{}
        \\ \hline
    \end{tabular}
    \caption{
        Dimensionless displacement norm ($\lVert \widetilde{u} \rVert$):
        (\textit{first column}) equivalent RRMM,
        (\textit{second column}) average metamaterial displacement ($\lVert\overline{u} \rVert$),
        (\textit{third column}) metamaterial $\mathcal{R}$,
        (\textit{fourth column}) macro Cauchy,
        for a hydrostatic load (Sec.~\ref{sec:load_boundary}) for $\omega = 200$ Hz with a single central unit cell.
    }
    \label{tab:reso_expa_figu_200}
\end{table}

\begin{table}[H]
    \centering
    \begin{tabular}{c|c|c|c|c|c}
        \cline{2-5}
                                                                                                               & RRMM
                                                                                                               & Avg. Microstr.
                                                                                                               & Microstructure
                                                                                                               & Macro Cauchy
                                                                                                               & \multicolumn{1}{c}{}
        \\ \hline
        \multicolumn{1}{|c|}{\rotatebox{90}{\makebox[\figsizeF][c]{$t= 0.0016$ s}}}                            &
        \includegraphics[width=\figsizeF]{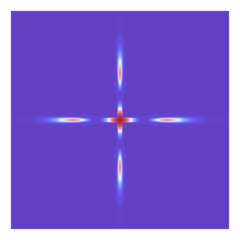}     &
        \includegraphics[width=\figsizeF]{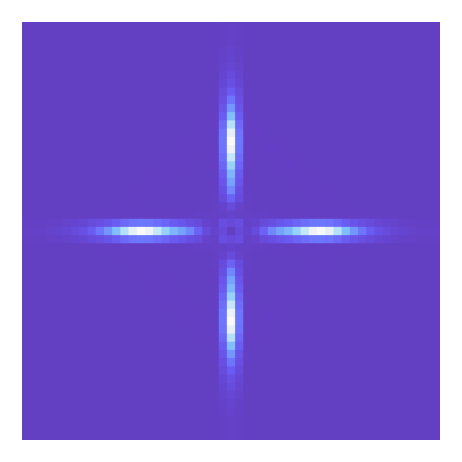} &
        \includegraphics[width=\figsizeF]{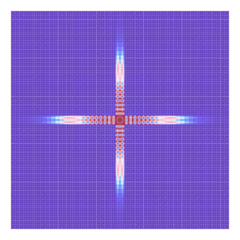}     &
        \includegraphics[width=\figsizeF]{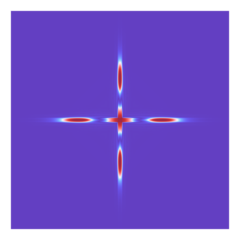}     &
        \multicolumn{1}{c|}{}
        \\ \cline{1-5}
        \multicolumn{1}{|c|}{\rotatebox{90}{\makebox[\figsizeF][c]{$t= 0.0022$ s}}}                            &
        \includegraphics[width=\figsizeF]{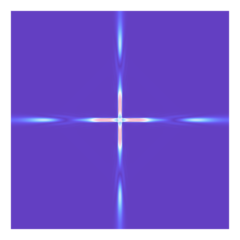}     &
        \includegraphics[width=\figsizeF]{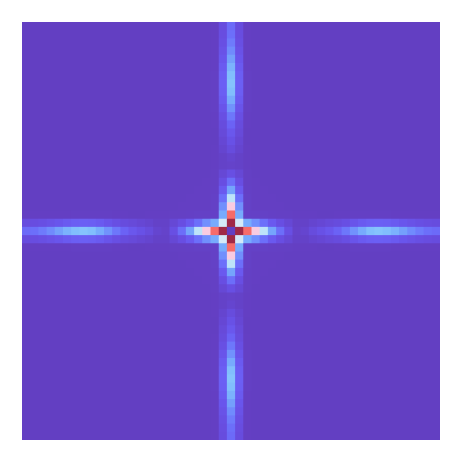} &
        \includegraphics[width=\figsizeF]{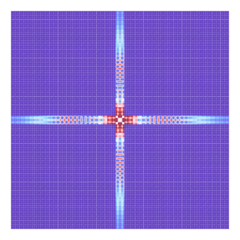}     &
        \includegraphics[width=\figsizeF]{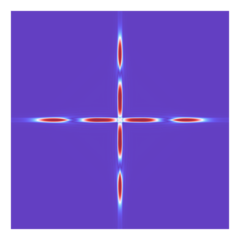}     &
        \multicolumn{1}{c|}{\rotatebox{90}{\makebox[\figsizeF][c]{Hydrostatic load - \fbox{1\texttimes 1} - $\omega = 425$ Hz}}}
        \\ \cline{1-5}
        \multicolumn{1}{|c|}{\rotatebox{90}{\makebox[\figsizeF][c]{$t= 0.0028$ s}}}                            &
        \includegraphics[width=\figsizeF]{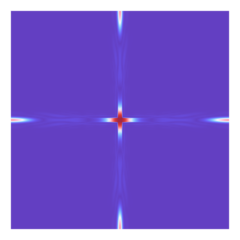}     &
        \includegraphics[width=\figsizeF]{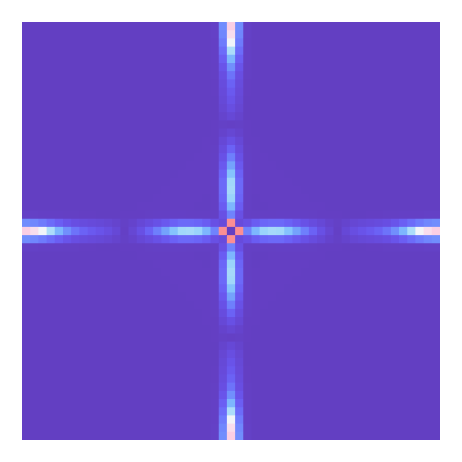} &
        \includegraphics[width=\figsizeF]{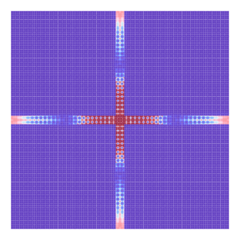}     &
        \includegraphics[width=\figsizeF]{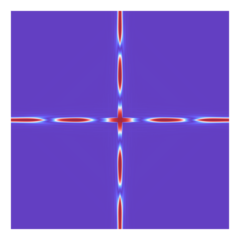}     &
        \multicolumn{1}{c|}{}
        \\ \hline
    \end{tabular}
    \caption{
        Dimensionless displacement norm ($\lVert \widetilde{u} \rVert$):
        (\textit{first column}) equivalent RRMM,
        (\textit{second column}) average metamaterial displacement ($\lVert\overline{u} \rVert$),
        (\textit{third column}) metamaterial $\mathcal{R}$,
        (\textit{fourth column}) macro Cauchy,
        for a hydrostatic load (Sec.~\ref{sec:load_boundary}) for $\omega = 425$ Hz with a single central unit cell.
    }
    \label{tab:reso_expa_figu_425}
\end{table}

As before, in Figs.~\ref{tab:reso_expa_figu_100}--\ref{tab:reso_expa_figu_425}, we observe that the RRMM shows good qualitative agreement across all frequencies, while quantitative differences still arise. These differences can be mitigated by comparing the RRMM with the average response of the microstructured system.

\begin{table}[H]
    \centering
    \begin{tabular}{c|c|c|c|c|c}
        \cline{2-5}
                                                                                                                   & RRMM
                                                                                                                   & Avg. Microstr.
                                                                                                                   & Microstructure
                                                                                                                   & Macro Cauchy
                                                                                                                   & \multicolumn{1}{c}{}
        \\ \hline
        \multicolumn{1}{|c|}{\rotatebox{90}{\makebox[\figsizeF][c]{$t= 0.0016$ s}}}                                &
        \includegraphics[width=\figsizeF]{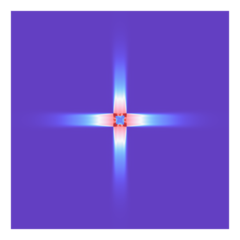}     &
        \includegraphics[width=\figsizeF]{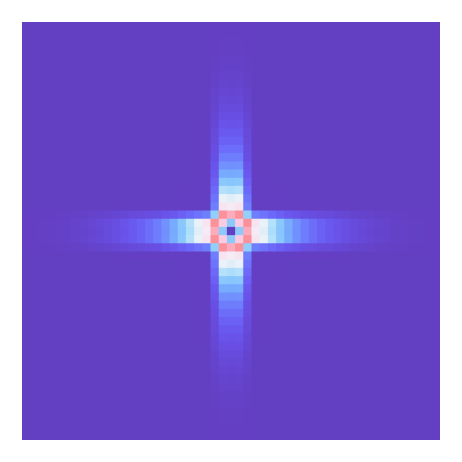} &
        \includegraphics[width=\figsizeF]{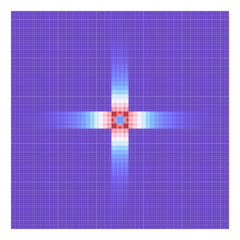}     &
        \includegraphics[width=\figsizeF]{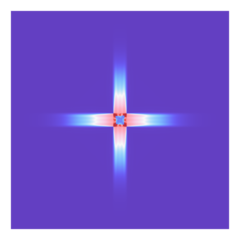}     &
        \multicolumn{1}{c|}{}
        \\ \cline{1-5}
        \multicolumn{1}{|c|}{\rotatebox{90}{\makebox[\figsizeF][c]{$t= 0.0022$ s}}}                                &
        \includegraphics[width=\figsizeF]{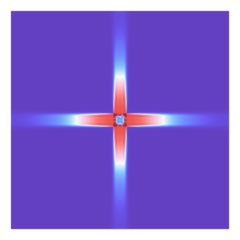}     &
        \includegraphics[width=\figsizeF]{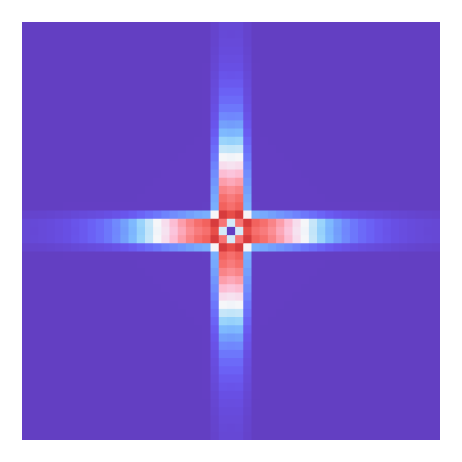} &
        \includegraphics[width=\figsizeF]{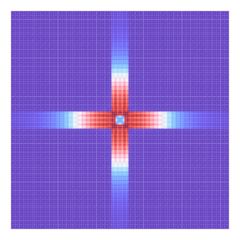}     &
        \includegraphics[width=\figsizeF]{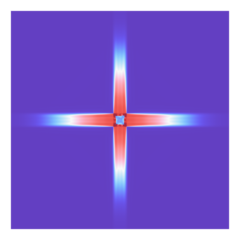}     &
        \multicolumn{1}{c|}{\rotatebox{90}{\makebox[\figsizeF][c]{Hydrostatic load - \fbox{3\texttimes 3} - $\omega = 100$ Hz}}}
        \\ \cline{1-5}
        \multicolumn{1}{|c|}{\rotatebox{90}{\makebox[\figsizeF][c]{$t= 0.0028$ s}}}                                &
        \includegraphics[width=\figsizeF]{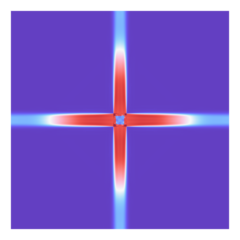}     &
        \includegraphics[width=\figsizeF]{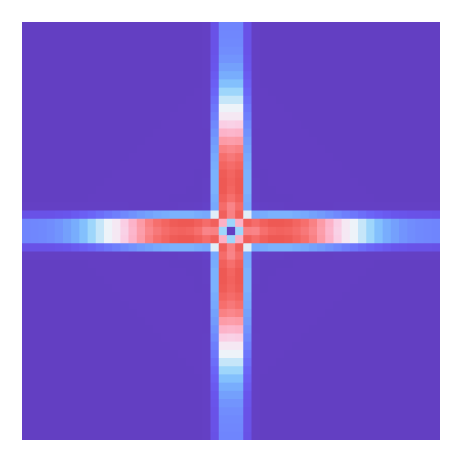} &
        \includegraphics[width=\figsizeF]{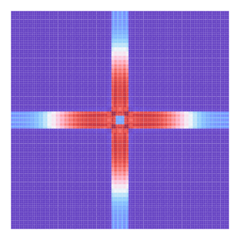}     &
        \includegraphics[width=\figsizeF]{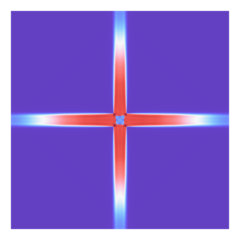}     &
        \multicolumn{1}{c|}{}
        \\ \hline
    \end{tabular}
    \caption{
        Dimensionless displacement norm ($\lVert \widetilde{u} \rVert$):
        (\textit{first column}) equivalent RRMM,
        (\textit{second column}) average metamaterial displacement ($\lVert\overline{u} \rVert$),
        (\textit{third column}) metamaterial $\mathcal{R}$,
        (\textit{fourth column}) macro Cauchy,
        for a hydrostatic load (Sec.~\ref{sec:load_boundary}) for $\omega = 100$ Hz with a 3\texttimes 3 central cluster.
    }
    \label{tab:reso_expa_figu_100_3x3}
\end{table}

\begin{table}[H]
    \centering
    \begin{tabular}{c|c|c|c|c|c}
        \cline{2-5}
                                                                                                                   & RRMM
                                                                                                                   & Avg. Microstr.
                                                                                                                   & Microstructure
                                                                                                                   & Macro Cauchy
                                                                                                                   & \multicolumn{1}{c}{}
        \\ \hline
        \multicolumn{1}{|c|}{\rotatebox{90}{\makebox[\figsizeF][c]{$t= 0.0016$ s}}}                                &
        \includegraphics[width=\figsizeF]{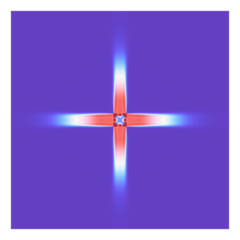}     &
        \includegraphics[width=\figsizeF]{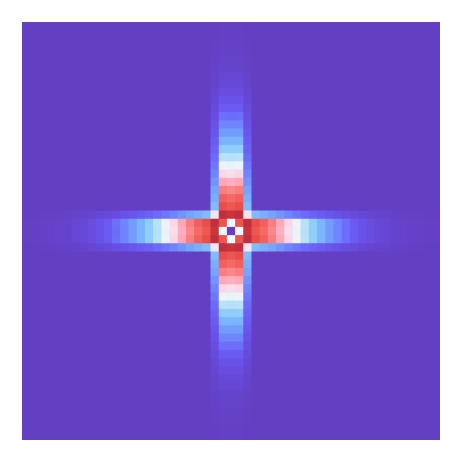} &
        \includegraphics[width=\figsizeF]{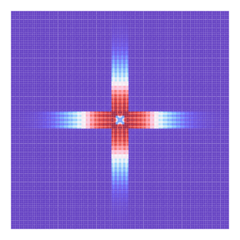}     &
        \includegraphics[width=\figsizeF]{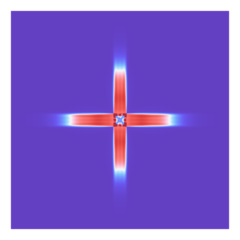}     &
        \multicolumn{1}{c|}{}
        \\ \cline{1-5}
        \multicolumn{1}{|c|}{\rotatebox{90}{\makebox[\figsizeF][c]{$t= 0.0022$ s}}}                                &
        \includegraphics[width=\figsizeF]{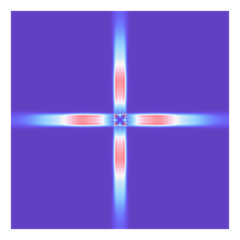}     &
        \includegraphics[width=\figsizeF]{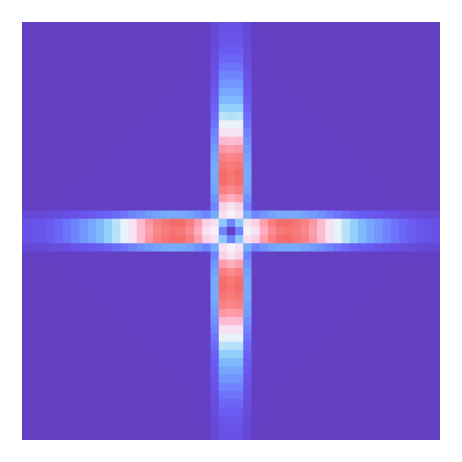} &
        \includegraphics[width=\figsizeF]{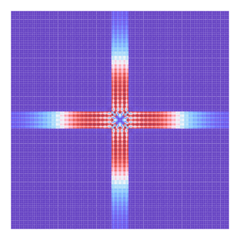}     &
        \includegraphics[width=\figsizeF]{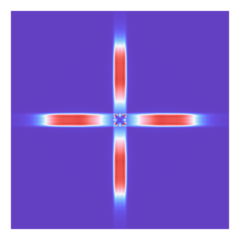}     &
        \multicolumn{1}{c|}{\rotatebox{90}{\makebox[\figsizeF][c]{Hydrostatic load - \fbox{3\texttimes 3} - $\omega = 200$ Hz}}}
        \\ \cline{1-5}
        \multicolumn{1}{|c|}{\rotatebox{90}{\makebox[\figsizeF][c]{$t= 0.0028$ s}}}                                &
        \includegraphics[width=\figsizeF]{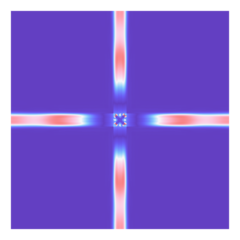}     &
        \includegraphics[width=\figsizeF]{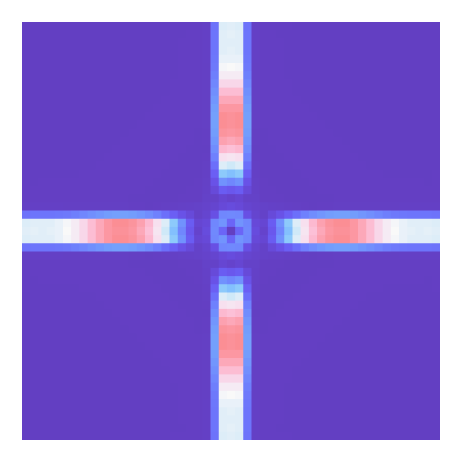} &
        \includegraphics[width=\figsizeF]{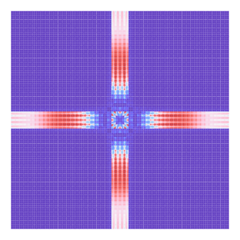}     &
        \includegraphics[width=\figsizeF]{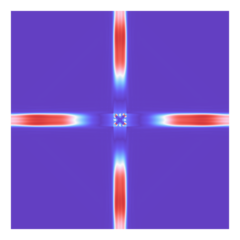}     &
        \multicolumn{1}{c|}{}
        \\ \hline
    \end{tabular}
    \caption{
        Dimensionless displacement norm ($\lVert \widetilde{u} \rVert$):
        (\textit{first column}) equivalent RRMM,
        (\textit{second column}) average metamaterial displacement ($\lVert\overline{u} \rVert$),
        (\textit{third column}) metamaterial $\mathcal{R}$,
        (\textit{fourth column}) macro Cauchy,
        for a hydrostatic load (Sec.~\ref{sec:load_boundary}) for $\omega = 200$ Hz with a 3\texttimes 3 central cluster.
    }
    \label{tab:reso_expa_figu_200_3x3}
\end{table}

\begin{table}[H]
    \centering
    \begin{tabular}{c|c|c|c|c|c}
        \cline{2-5}
                                                                                                                   & RRMM
                                                                                                                   & Avg. Microstr.
                                                                                                                   & Microstructure
                                                                                                                   & Macro Cauchy
                                                                                                                   & \multicolumn{1}{c}{}
        \\ \hline
        \multicolumn{1}{|c|}{\rotatebox{90}{\makebox[\figsizeF][c]{$t= 0.0016$ s}}}                                &
        \includegraphics[width=\figsizeF]{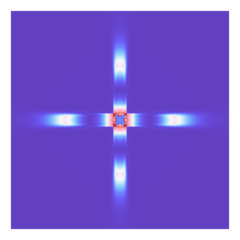}     &
        \includegraphics[width=\figsizeF]{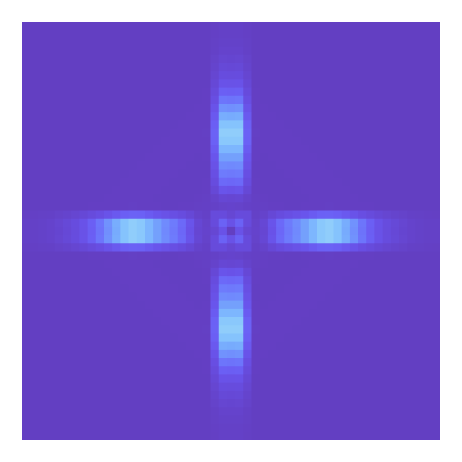} &
        \includegraphics[width=\figsizeF]{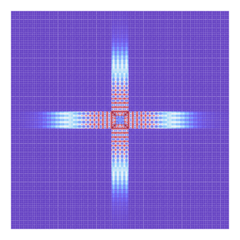}     &
        \includegraphics[width=\figsizeF]{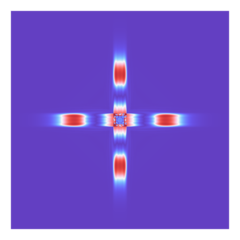}     &
        \multicolumn{1}{c|}{}
        \\ \cline{1-5}
        \multicolumn{1}{|c|}{\rotatebox{90}{\makebox[\figsizeF][c]{$t= 0.0022$ s}}}                                &
        \includegraphics[width=\figsizeF]{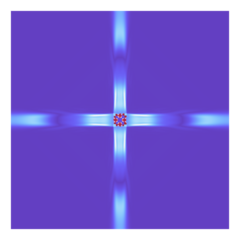}     &
        \includegraphics[width=\figsizeF]{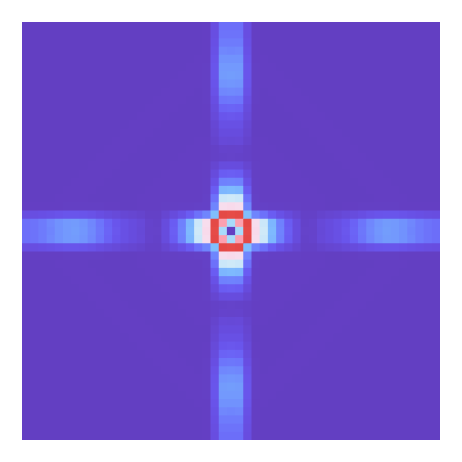} &
        \includegraphics[width=\figsizeF]{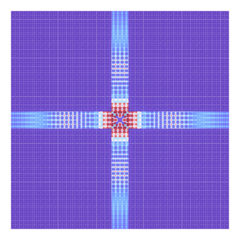}     &
        \includegraphics[width=\figsizeF]{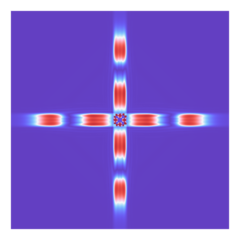}     &
        \multicolumn{1}{c|}{\rotatebox{90}{\makebox[\figsizeF][c]{Hydrostatic load - \fbox{3\texttimes 3} - $\omega = 425$ Hz}}}
        \\ \cline{1-5}
        \multicolumn{1}{|c|}{\rotatebox{90}{\makebox[\figsizeF][c]{$t= 0.0028$ s}}}                                &
        \includegraphics[width=\figsizeF]{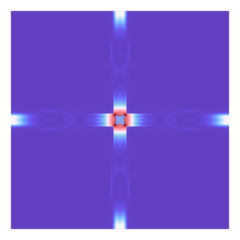}     &
        \includegraphics[width=\figsizeF]{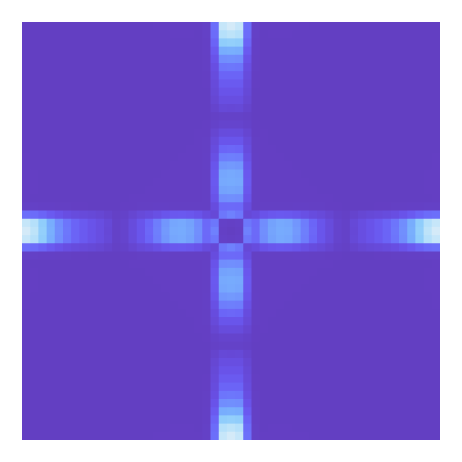} &
        \includegraphics[width=\figsizeF]{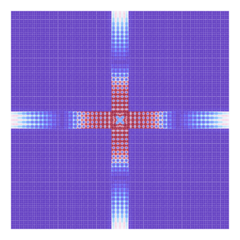}     &
        \includegraphics[width=\figsizeF]{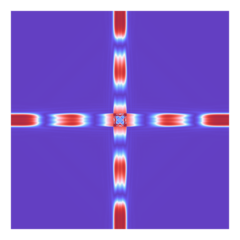}     &
        \multicolumn{1}{c|}{}
        \\ \hline
    \end{tabular}
    \caption{
        Dimensionless displacement norm ($\lVert \widetilde{u} \rVert$):
        (\textit{first column}) equivalent RRMM,
        (\textit{second column}) average metamaterial displacement ($\lVert\overline{u} \rVert$),
        (\textit{third column}) metamaterial $\mathcal{R}$,
        (\textit{fourth column}) macro Cauchy,
        for a hydrostatic load (Sec.~\ref{sec:load_boundary}) for $\omega = 425$ Hz with a 3\texttimes 3 central cluster.
    }
    \label{tab:reso_expa_figu_425_3x3}
\end{table}

Differently from the 1\texttimes 1 cases, in Figs.~\ref{tab:reso_expa_figu_100_3x3}--\ref{tab:reso_expa_figu_425_3x3}, we observe that the RRMM shows good qualitative agreement across all frequencies, while moderate quantitative differences still arise. These differences can be mitigated by comparing the RRMM with the average response of the microstructured system, especially for 425 Hz (bandgap frequency).

\begin{table}[H]
    \centering
    \begin{tabular}{c|c|c|c|c|c}
        \cline{2-5}
                                                                                                               & RRMM
                                                                                                               & Avg. Microstr.
                                                                                                               & Microstructure
                                                                                                               & Macro Cauchy
                                                                                                               & \multicolumn{1}{c}{}
        \\ \hline
        \multicolumn{1}{|c|}{\rotatebox{90}{\makebox[\figsizeF][c]{$t= 0.0016$ s}}}                            &
        \includegraphics[width=\figsizeF]{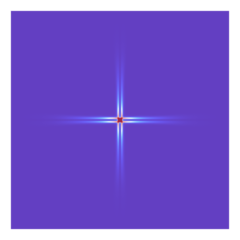}     &
        \includegraphics[width=\figsizeF]{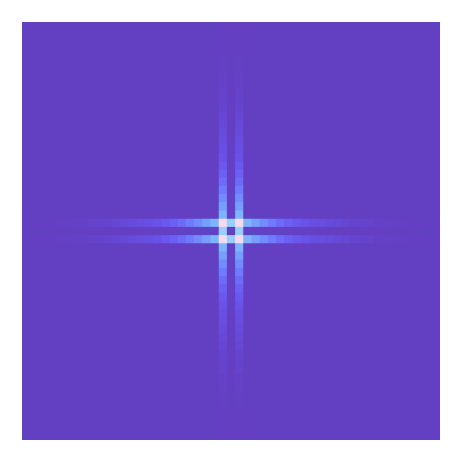} &
        \includegraphics[width=\figsizeF]{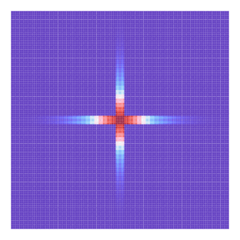}     &
        \includegraphics[width=\figsizeF]{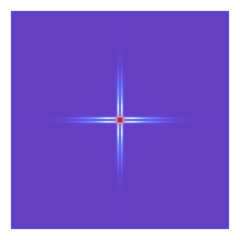}     &
        \multicolumn{1}{c|}{}
        \\ \cline{1-5}
        \multicolumn{1}{|c|}{\rotatebox{90}{\makebox[\figsizeF][c]{$t= 0.0022$ s}}}                            &
        \includegraphics[width=\figsizeF]{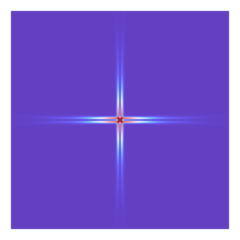}     &
        \includegraphics[width=\figsizeF]{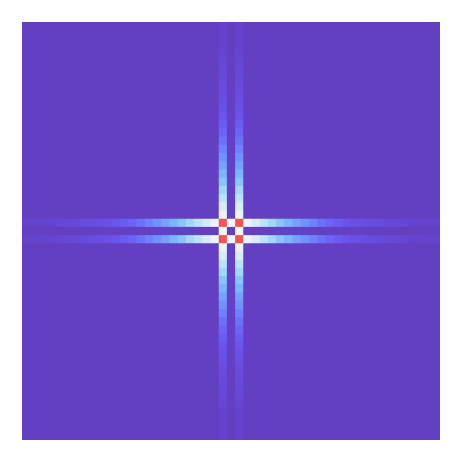} &
        \includegraphics[width=\figsizeF]{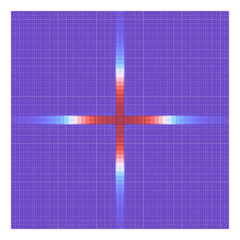}     &
        \includegraphics[width=\figsizeF]{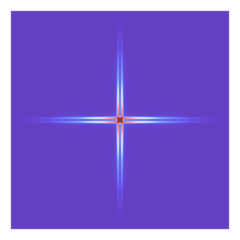}     &
        \multicolumn{1}{c|}{\rotatebox{90}{\makebox[\figsizeF][c]{Shear load - \fbox{1\texttimes 1} -  $\omega = 100$ Hz}}}
        \\ \cline{1-5}
        \multicolumn{1}{|c|}{\rotatebox{90}{\makebox[\figsizeF][c]{$t= 0.0028$ s}}}                            &
        \includegraphics[width=\figsizeF]{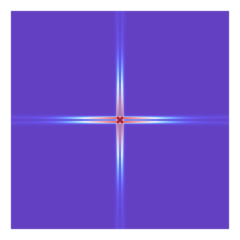}     &
        \includegraphics[width=\figsizeF]{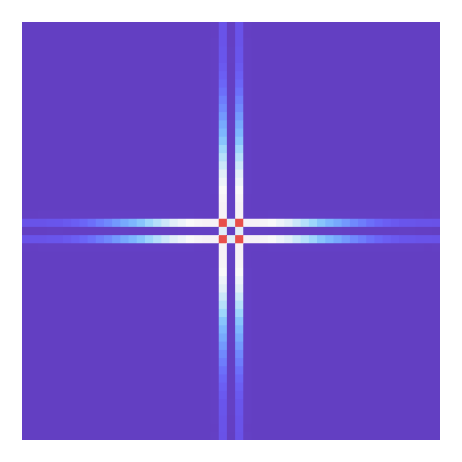} &
        \includegraphics[width=\figsizeF]{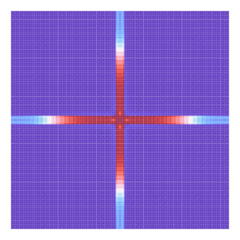}     &
        \includegraphics[width=\figsizeF]{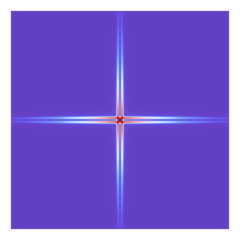}     &
        \multicolumn{1}{c|}{}
        \\ \hline
    \end{tabular}
    \caption{
        Dimensionless displacement norm ($\lVert \widetilde{u} \rVert$):
        (\textit{first column}) equivalent RRMM,
        (\textit{second column}) average metamaterial displacement ($\lVert\overline{u} \rVert$),
        (\textit{third column}) metamaterial $\mathcal{R}$,
        (\textit{fourth column}) macro Cauchy,
        for a shear load (Sec.~\ref{sec:load_boundary}) for $\omega = 100$ Hz with a single central unit cell.
    }
    \label{tab:reso_shea_figu_100}
\end{table}

\begin{table}[H]
    \centering
    \begin{tabular}{c|c|c|c|c|c}
        \cline{2-5}
                                                                                                               & RRMM
                                                                                                               & Avg. Microstr.
                                                                                                               & Microstructure
                                                                                                               & Macro Cauchy
                                                                                                               & \multicolumn{1}{c}{}
        \\ \hline
        \multicolumn{1}{|c|}{\rotatebox{90}{\makebox[\figsizeF][c]{$t= 0.0016$ s}}}                            &
        \includegraphics[width=\figsizeF]{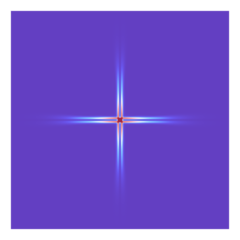}     &
        \includegraphics[width=\figsizeF]{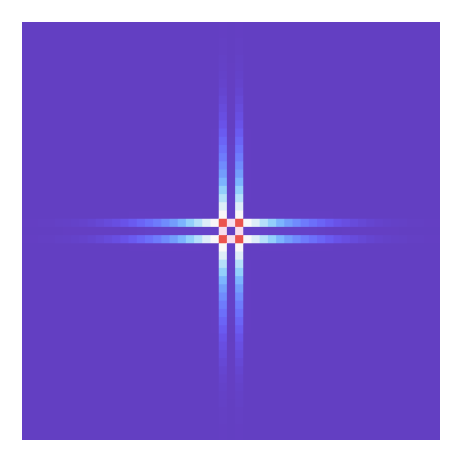} &
        \includegraphics[width=\figsizeF]{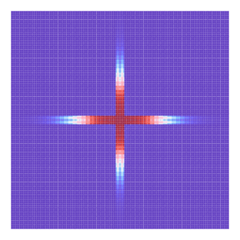}     &
        \includegraphics[width=\figsizeF]{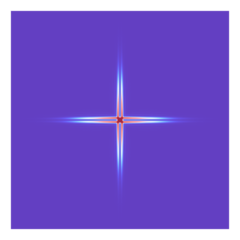}     &
        \multicolumn{1}{c|}{}
        \\ \cline{1-5}
        \multicolumn{1}{|c|}{\rotatebox{90}{\makebox[\figsizeF][c]{$t= 0.0022$ s}}}                            &
        \includegraphics[width=\figsizeF]{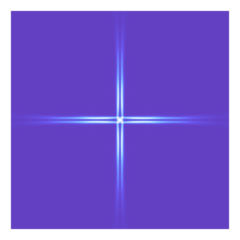}     &
        \includegraphics[width=\figsizeF]{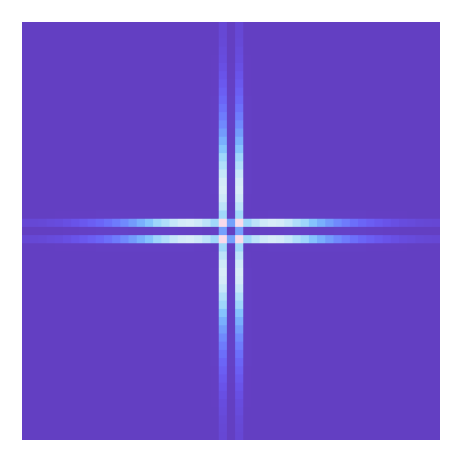} &
        \includegraphics[width=\figsizeF]{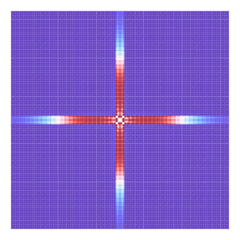}     &
        \includegraphics[width=\figsizeF]{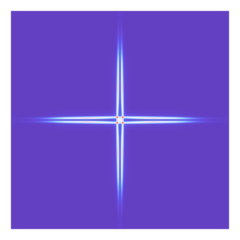}     &
        \multicolumn{1}{c|}{\rotatebox{90}{\makebox[\figsizeF][c]{Shear load - \fbox{1\texttimes 1} - $\omega = 200$ Hz}}}
        \\ \cline{1-5}
        \multicolumn{1}{|c|}{\rotatebox{90}{\makebox[\figsizeF][c]{$t= 0.0028$ s}}}                            &
        \includegraphics[width=\figsizeF]{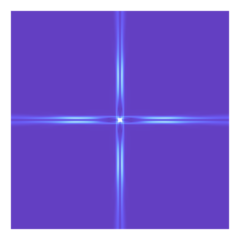}     &
        \includegraphics[width=\figsizeF]{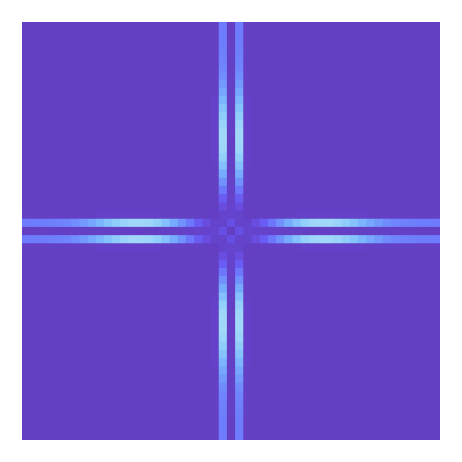} &
        \includegraphics[width=\figsizeF]{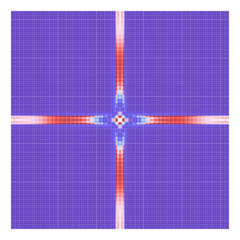}     &
        \includegraphics[width=\figsizeF]{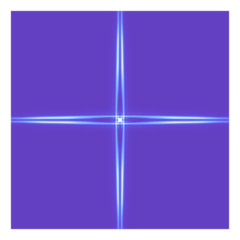}     &
        \multicolumn{1}{c|}{}
        \\ \hline
    \end{tabular}
    \caption{
        Dimensionless displacement norm ($\lVert \widetilde{u} \rVert$):
        (\textit{first column}) equivalent RRMM,
        (\textit{second column}) average metamaterial displacement ($\lVert\overline{u} \rVert$),
        (\textit{third column}) metamaterial $\mathcal{R}$,
        (\textit{fourth column}) macro Cauchy,
        for a shear load (Sec.~\ref{sec:load_boundary}) for $\omega = 200$ Hz with a single central unit cell.
    }
    \label{tab:reso_shea_figu_200}
\end{table}

\begin{table}[H]
    \centering
    \begin{tabular}{c|c|c|c|c|c}
        \cline{2-5}
                                                                                                               & RRMM
                                                                                                               & Avg. Microstr.
                                                                                                               & Microstructure
                                                                                                               & Macro Cauchy
                                                                                                               & \multicolumn{1}{c}{}
        \\ \hline
        \multicolumn{1}{|c|}{\rotatebox{90}{\makebox[\figsizeF][c]{$t= 0.0016$ s}}}                            &
        \includegraphics[width=\figsizeF]{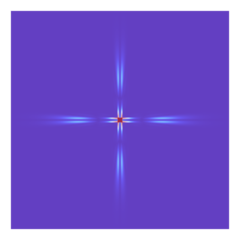}     &
        \includegraphics[width=\figsizeF]{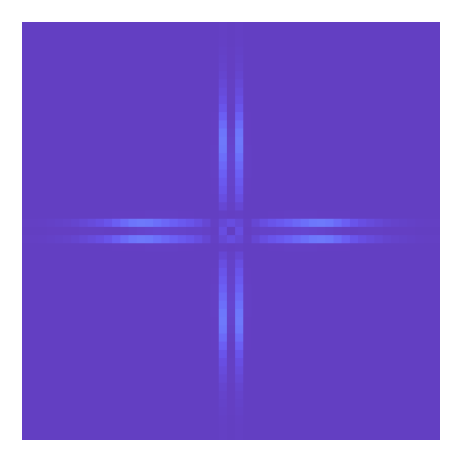} &
        \includegraphics[width=\figsizeF]{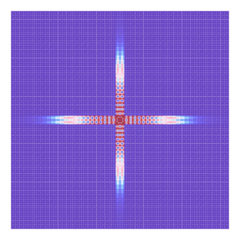}     &
        \includegraphics[width=\figsizeF]{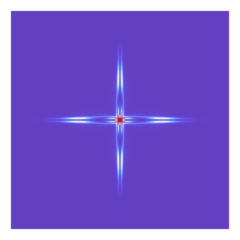}     &
        \multicolumn{1}{c|}{}
        \\ \cline{1-5}
        \multicolumn{1}{|c|}{\rotatebox{90}{\makebox[\figsizeF][c]{$t= 0.0022$ s}}}                            &
        \includegraphics[width=\figsizeF]{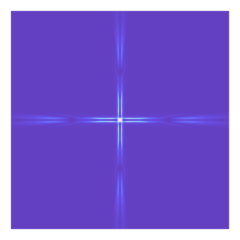}     &
        \includegraphics[width=\figsizeF]{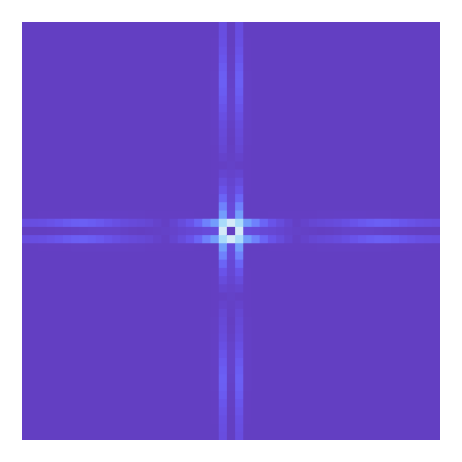} &
        \includegraphics[width=\figsizeF]{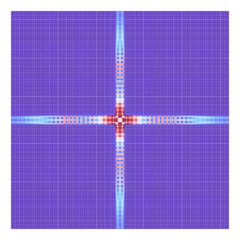}     &
        \includegraphics[width=\figsizeF]{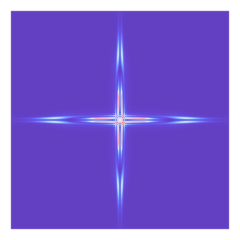}     &
        \multicolumn{1}{c|}{\rotatebox{90}{\makebox[\figsizeF][c]{Shear load - \fbox{1\texttimes 1} - $\omega = 425$ Hz}}}
        \\ \cline{1-5}
        \multicolumn{1}{|c|}{\rotatebox{90}{\makebox[\figsizeF][c]{$t= 0.0028$ s}}}                            &
        \includegraphics[width=\figsizeF]{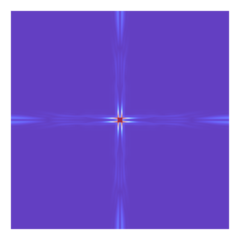}     &
        \includegraphics[width=\figsizeF]{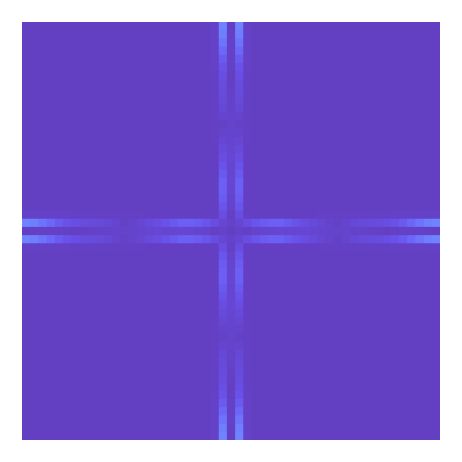} &
        \includegraphics[width=\figsizeF]{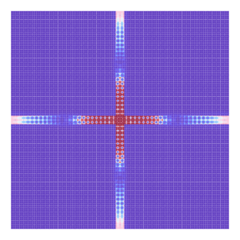}     &
        \includegraphics[width=\figsizeF]{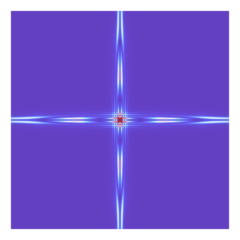}     &
        \multicolumn{1}{c|}{}
        \\ \hline
    \end{tabular}
    \caption{
        Dimensionless displacement norm ($\lVert \widetilde{u} \rVert$):
        (\textit{first column}) equivalent RRMM,
        (\textit{second column}) average metamaterial displacement ($\lVert\overline{u} \rVert$),
        (\textit{third column}) metamaterial $\mathcal{R}$,
        (\textit{fourth column}) macro Cauchy,
        for a shear load (Sec.~\ref{sec:load_boundary}) for $\omega = 425$ Hz with a single central unit cell.
    }
    \label{tab:reso_shea_figu_425}
\end{table}

Similarly, in Figs.~\ref{tab:reso_shea_figu_100}--\ref{tab:reso_shea_figu_425}, we observe that the RRMM shows good qualitative agreement across all frequencies, while quantitative differences still arise. These differences can be mitigated by comparing the RRMM with the average response of the microstructured system.

\begin{table}[H]
    \centering
    \begin{tabular}{c|c|c|c|c|c}
        \cline{2-5}
                                                                                                                   & RRMM
                                                                                                                   & Avg. Microstr.
                                                                                                                   & Microstructure
                                                                                                                   & Macro Cauchy
                                                                                                                   & \multicolumn{1}{c}{}
        \\ \hline
        \multicolumn{1}{|c|}{\rotatebox{90}{\makebox[\figsizeF][c]{$t= 0.0016$ s}}}                                &
        \includegraphics[width=\figsizeF]{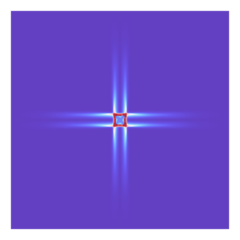}     &
        \includegraphics[width=\figsizeF]{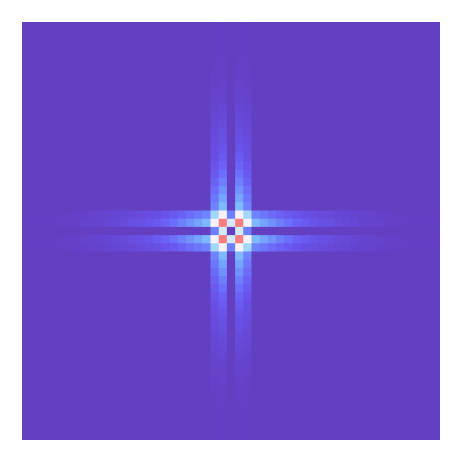} &
        \includegraphics[width=\figsizeF]{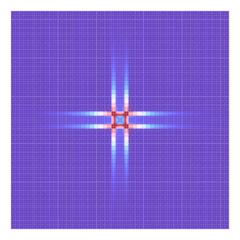}     &
        \includegraphics[width=\figsizeF]{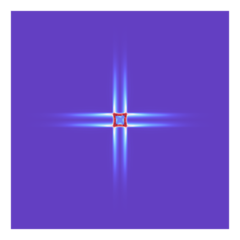}     &
        \multicolumn{1}{c|}{}
        \\ \cline{1-5}
        \multicolumn{1}{|c|}{\rotatebox{90}{\makebox[\figsizeF][c]{$t= 0.0022$ s}}}                                &
        \includegraphics[width=\figsizeF]{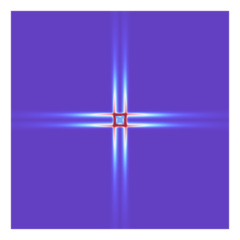}     &
        \includegraphics[width=\figsizeF]{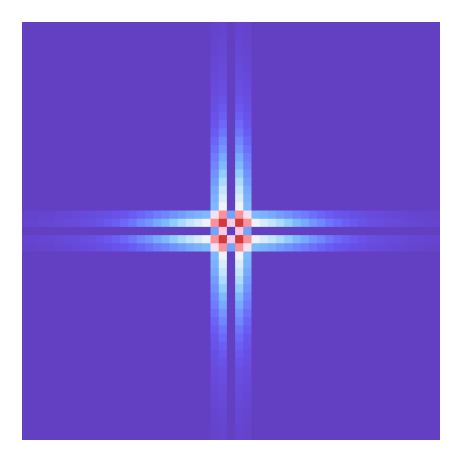} &
        \includegraphics[width=\figsizeF]{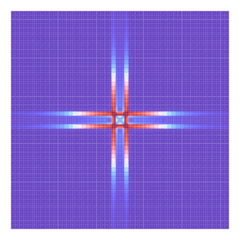}     &
        \includegraphics[width=\figsizeF]{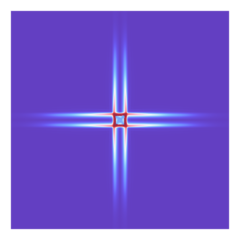}     &
        \multicolumn{1}{c|}{\rotatebox{90}{\makebox[\figsizeF][c]{Shear load - \fbox{3\texttimes 3} - $\omega = 100$ Hz}}}
        \\ \cline{1-5}
        \multicolumn{1}{|c|}{\rotatebox{90}{\makebox[\figsizeF][c]{$t= 0.0028$ s}}}                                &
        \includegraphics[width=\figsizeF]{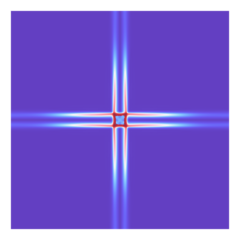}     &
        \includegraphics[width=\figsizeF]{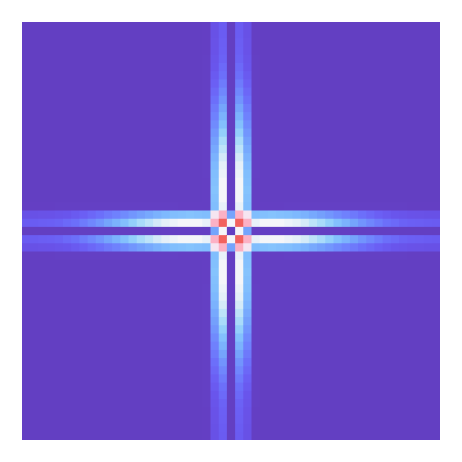} &
        \includegraphics[width=\figsizeF]{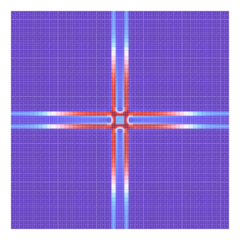}     &
        \includegraphics[width=\figsizeF]{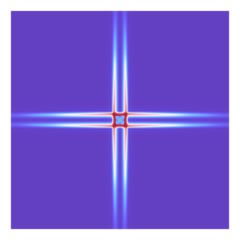}     &
        \multicolumn{1}{c|}{}
        \\ \hline
    \end{tabular}
    \caption{
        Dimensionless displacement norm ($\lVert \widetilde{u} \rVert$):
        (\textit{first column}) equivalent RRMM,
        (\textit{second column}) average metamaterial displacement ($\lVert\overline{u} \rVert$),
        (\textit{third column}) metamaterial $\mathcal{R}$,
        (\textit{fourth column}) macro Cauchy,
        for a shear load (Sec.~\ref{sec:load_boundary}) for $\omega = 100$ Hz with a 3\texttimes 3 central cluster.
    }
    \label{tab:reso_shea_figu_100_3x3}
\end{table}

\begin{table}[H]
    \centering
    \begin{tabular}{c|c|c|c|c|c}
        \cline{2-5}
                                                                                                                   & RRMM
                                                                                                                   & Avg. Microstr.
                                                                                                                   & Microstructure
                                                                                                                   & Macro Cauchy
                                                                                                                   & \multicolumn{1}{c}{}
        \\ \hline
        \multicolumn{1}{|c|}{\rotatebox{90}{\makebox[\figsizeF][c]{$t= 0.0016$ s}}}                                &
        \includegraphics[width=\figsizeF]{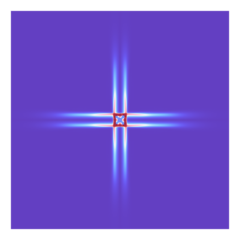}     &
        \includegraphics[width=\figsizeF]{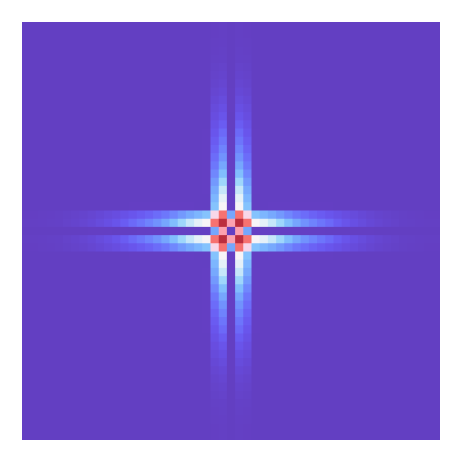} &
        \includegraphics[width=\figsizeF]{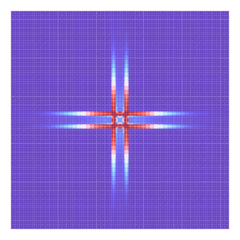}     &
        \includegraphics[width=\figsizeF]{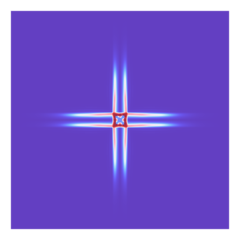}     &
        \multicolumn{1}{c|}{}
        \\ \cline{1-5}
        \multicolumn{1}{|c|}{\rotatebox{90}{\makebox[\figsizeF][c]{$t= 0.0022$ s}}}                                &
        \includegraphics[width=\figsizeF]{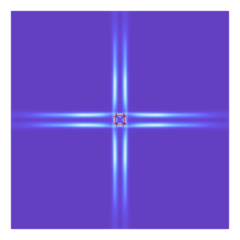}     &
        \includegraphics[width=\figsizeF]{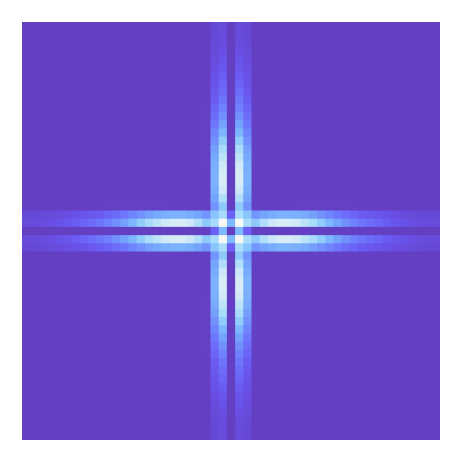} &
        \includegraphics[width=\figsizeF]{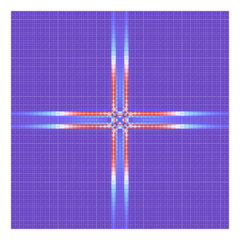}     &
        \includegraphics[width=\figsizeF]{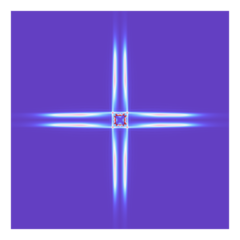}     &
        \multicolumn{1}{c|}{\rotatebox{90}{\makebox[\figsizeF][c]{Shear load - \fbox{3\texttimes 3} - $\omega = 200$ Hz}}}
        \\ \cline{1-5}
        \multicolumn{1}{|c|}{\rotatebox{90}{\makebox[\figsizeF][c]{$t= 0.0028$ s}}}                                &
        \includegraphics[width=\figsizeF]{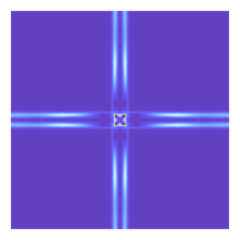}     &
        \includegraphics[width=\figsizeF]{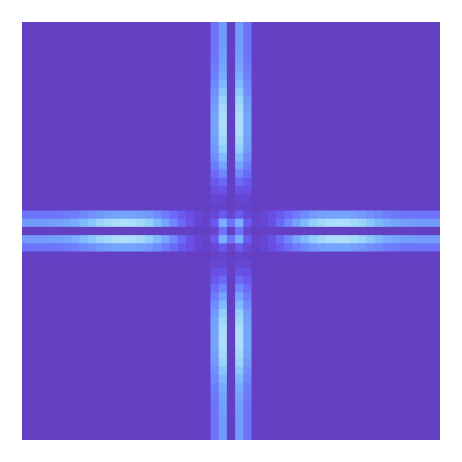} &
        \includegraphics[width=\figsizeF]{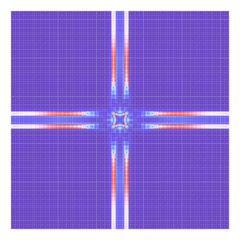}     &
        \includegraphics[width=\figsizeF]{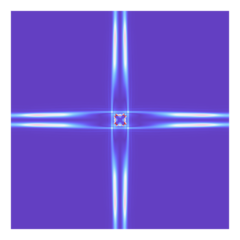}     &
        \multicolumn{1}{c|}{}
        \\ \hline
    \end{tabular}
    \caption{
        Dimensionless displacement norm ($\lVert \widetilde{u} \rVert$):
        (\textit{first column}) equivalent RRMM,
        (\textit{second column}) average metamaterial displacement ($\lVert\overline{u} \rVert$),
        (\textit{third column}) metamaterial $\mathcal{R}$,
        (\textit{fourth column}) macro Cauchy,
        for a shear load (Sec.~\ref{sec:load_boundary}) for $\omega = 200$ Hz with a 3\texttimes 3 central cluster.
    }
    \label{tab:reso_shea_figu_200_3x3}
\end{table}

\begin{table}[H]
    \centering
    \begin{tabular}{c|c|c|c|c|c}
        \cline{2-5}
                                                                                                                   & RRMM
                                                                                                                   & Avg. Microstr.
                                                                                                                   & Microstructure
                                                                                                                   & Macro Cauchy
                                                                                                                   & \multicolumn{1}{c}{}
        \\ \hline
        \multicolumn{1}{|c|}{\rotatebox{90}{\makebox[\figsizeF][c]{$t= 0.0016$ s}}}                                &
        \includegraphics[width=\figsizeF]{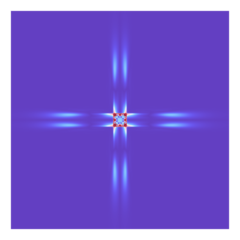}     &
        \includegraphics[width=\figsizeF]{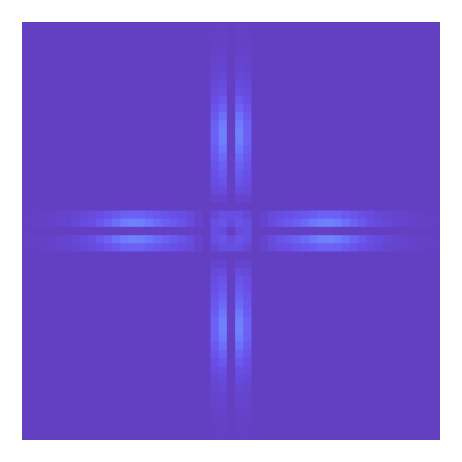} &
        \includegraphics[width=\figsizeF]{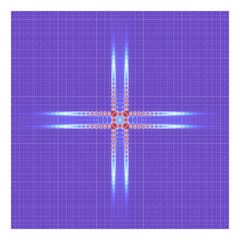}     &
        \includegraphics[width=\figsizeF]{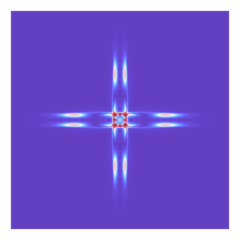}     &
        \multicolumn{1}{c|}{}
        \\ \cline{1-5}
        \multicolumn{1}{|c|}{\rotatebox{90}{\makebox[\figsizeF][c]{$t= 0.0022$ s}}}                                &
        \includegraphics[width=\figsizeF]{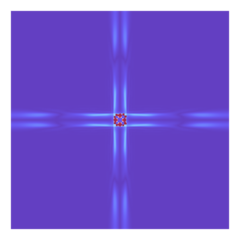}     &
        \includegraphics[width=\figsizeF]{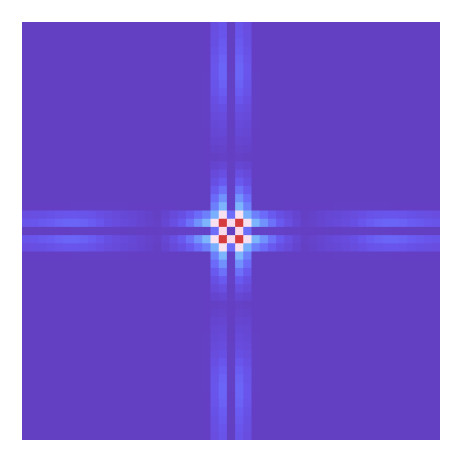} &
        \includegraphics[width=\figsizeF]{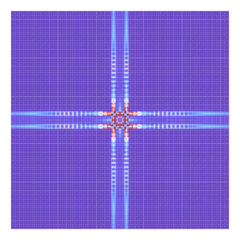}     &
        \includegraphics[width=\figsizeF]{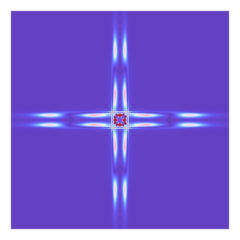}     &
        \multicolumn{1}{c|}{\rotatebox{90}{\makebox[\figsizeF][c]{Shear load - \fbox{3\texttimes 3} - $\omega = 425$ Hz}}}
        \\ \cline{1-5}
        \multicolumn{1}{|c|}{\rotatebox{90}{\makebox[\figsizeF][c]{$t= 0.0028$ s}}}                                &
        \includegraphics[width=\figsizeF]{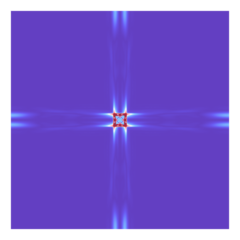}     &
        \includegraphics[width=\figsizeF]{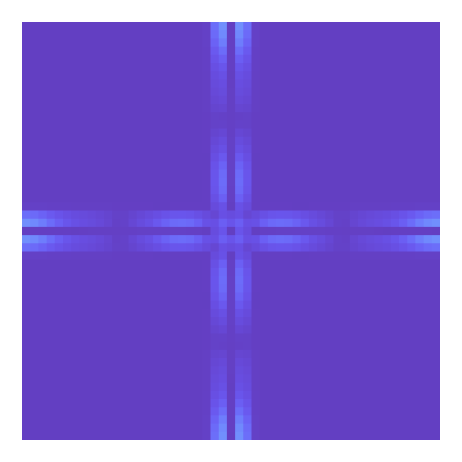} &
        \includegraphics[width=\figsizeF]{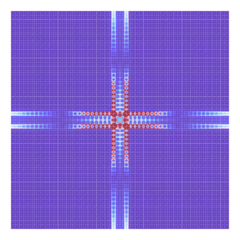}     &
        \includegraphics[width=\figsizeF]{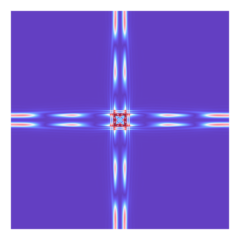}     &
        \multicolumn{1}{c|}{}
        \\ \hline
    \end{tabular}
    \caption{
        Dimensionless displacement norm ($\lVert \widetilde{u} \rVert$):
        (\textit{first column}) equivalent RRMM,
        (\textit{second column}) average metamaterial displacement ($\lVert\overline{u} \rVert$),
        (\textit{third column}) metamaterial $\mathcal{R}$,
        (\textit{fourth column}) macro Cauchy,
        for a shear load (Sec.~\ref{sec:load_boundary}) for $\omega = 425$ Hz with a 3\texttimes 3 central cluster.
    }
    \label{tab:reso_shea_figu_425_3x3}
\end{table}

Differently from the 1\texttimes 1 cases, in Figs.~\ref{tab:reso_shea_figu_100_3x3}--\ref{tab:reso_shea_figu_425_3x3}, we observe that the RRMM shows good qualitative agreement across all frequencies, while moderate and localised quantitative differences still arise. These differences can be significantly mitigated by comparing the RRMM with the average response of the microstructured system.

\subsection{\textit{Four-resonator} unit cell's results discussion}
For the rotation test, no averaging operator is needed, regardless the frequency or the size of the central cluster: the RRMM is in this case capable of directly capturing well the displacement field (from Fig.~\ref{tab:reso_rota_figu_100}--\ref{tab:reso_shea_figu_425_3x3}).
For the hydrostatic test, the average operator is required for the 1\texttimes 1 central unit cell case, at both 100 Hz and 200 Hz (Fig.~\ref{tab:reso_expa_figu_100}--\ref{tab:reso_expa_figu_200}).
Switching to a 3\texttimes 3 central cluster automatically resolves the need for averaging at 100 Hz (Fig.~\ref{tab:reso_expa_figu_100_3x3}), while it does not eliminate this necessity at 200 Hz (Fig.~\ref{tab:reso_expa_figu_200_3x3}).
As for the shear test, the averaging operator is only needed for the 1\texttimes 1 central unit cell, at both 100 Hz and 200 Hz, but it is already not necessary as soon as the cluster grows to a 3\texttimes 3 dimension.
It is generally true that increasing the size of the central cluster allows the load to be more uniformly distributed, which already helps reduce the appearance of local effects.
However, unlike the labyrinthine unit cell, some quantitative differences persist in certain cases, even when increasing the size of the central cluster.
This might indicate that, due to localised resonances, boundary effects at the loaded interface could still be active for the four-resonator unit cell.
In all cases, the overall trend is well captured by the RRMM.

\section{Conclusions}
In this study, we applied three different load cases to evaluate the capability of the Reduced Relaxed Micromorphic Model (RRMM) in capturing the dynamic behavior of two distinct metamaterials, each constructed through the periodic repetition of a single unit cell.
The model successfully reproduced the overall wave responses, although it showed limitations in resolving localised phenomena occurring at scales below the unit cell size.
It is also generally true that increasing the size of the central cluster allows the load to be more uniformly distributed, which already helps reduce the appearance of local effects.
To address this, we compared the averaged response of a single unit cell with the RRMM predictions, finding good agreement between the two.
The model consistently captured the bandgap range, regardless of the metamaterial type.
However, the manifestation of the bandgap differed due to the nature of the unit cells: in the case of the \textit{four-resonance} unit cell, the strong local resonance effects led to a mitigation of the bandgap within the studied domain size, whereas the \textit{labyrinthine} unit cell exhibited clear Bragg-scattering-induced bandgaps.
Finally, to isolate the intrinsic material behavior, outer boundary effects were minimized by limiting the simulation time such that wave reflections from the boundaries were either avoided or kept negligible.
However, small differences in the intensity between the RRMM and the microstructured solutions might still be related to slight effects occurring at the interface of the loaded unit cell (or cluster of unit cells).

    {
        \begin{spacing}{0.5}
            \footnotesize
            \noindent
            \textbf{\footnotesize Acknowledgements:}
            Angela Madeo and Gianluca Rizzi acknowledge support from the European Commission through the funding of the ERC Consolidator Grant META-LEGO, N$^\circ$ 101001759.
        \end{spacing}
    }


\bibliographystyle{abbrvnat}  
\bibliography{time_dependent_simulation.bib}

\end{document}